**emIAM v1.0: an emulator for Integrated Assessment Models using marginal abatement cost curves**

Weiwei Xiong[1,2,*], Katsumasa Tanaka[2,3,*], Philippe Ciais[2], Daniel J. A. Johansson[4], Mariliis Lehtveer[4]

[1] School of Economics and Management, China University of Geosciences, Wuhan, China

[2] Laboratoire des Sciences du Climat et de l'Environnement (LSCE), IPSL, CEA/CNRS/UVSQ, Université Paris-Saclay, Gif-sur-Yvette, France

[3] Earth System Division, National Institute for Environmental Studies (NIES), Tsukuba, Japan

[4] Division of Physical Resource Theory, Department of Space, Earth, and Environment, Chalmers University of Technology, Gothenburg, Sweden

*Corresponding authors: Weiwei Xiong (xww08012115@cug.edu.cn) and Katsumasa Tanaka (katsumasa.tanaka@lsce.ipsl.fr)

**Abstract**

We developed an emulator for Integrated Assessment Models (emIAM) based on a marginal abatement cost (MAC) curve approach. Using the output of IAMs in the ENGAGE Scenario Explorer and the GET model, we derived a large set of MAC curves: ten IAMs; global and eleven regions; three gases $CO_2$, $CH_4$, and $N_2O$; eight portfolios of available mitigation technologies; and two emission sources. We tested the performance of emIAM by coupling it with a simple climate model ACC2. We found that the optimizing climate-economy model emIAM-ACC2 adequately reproduced a majority of original IAM emission outcomes under similar conditions, allowing systematic explorations of IAMs with small computational resources. emIAM can expand the capability of simple climate models as a tool to calculate cost-effective pathways linked directly to a temperature target.

**Supplement:** The supplement related to this article is also available online.

**Acknowledgments**: K.T. dedicates this paper to the memory of Prof. Hiroshi Yanai (1937-2021) at Keio University, Tokyo, Japan, a pioneer of Operations Research and his bachelor thesis supervisor, by whom K.T. was taught the fundamentals of mathematical modeling and academic writing and the joy of intellectual pursuit. We are grateful for comments from Nico Bauer, Stéphane De Cara, Yann Gaucher, Xiangping Hu, Xuanming Su, Kiyoshi Takahashi, and Tokuta Yokohata, which were useful for this study.

**Author contributions**: Conceptualization of the IAM emulator, W.X. and K.T.; simulations using emIAM-ACC2, W.X. and K.T.; simulations using GET, K.T, D.J., and M.L.; analysis of simulation results, W.X., K.T., and D.J.; writing - original draft preparation, W.X. and K.T.; writing – revision and editing, W.X., K.T., P.C., D.J., and M.L. All authors have read and agreed to the submitted version of the manuscript.

**Funding**: W.X. acknowledges financial support from the China Scholarship Council. This research was conducted as part of the Achieving the Paris Agreement Temperature Targets after Overshoot (PRATO) Project under the Make Our Planet Great Again (MOPGA) Program and funded by the National Research Agency in France under the Programme d'Investissements d'Avenir, grant number ANR-19-MPGA-0008. We acknowledge the Environment Research and Technology Development Fund (JPMEERF20202002) of the Environmental Restoration and Conservation Agency (Japan).

**Data availability:** Data for the parameters in MAC curves and associated upper limits on abatement levels and their first and second derivatives are available on Zenodo with doi.org/XXX.

**Conflict of interest**: The authors declare no conflict of interest.

## 1. Introduction

Integrated Assessment Models (IAMs) combine economy, energy, and sometimes also land-use modeling approaches and are commonly used to evaluate climate policies under least-cost scenarios (Weyant, 2017). A variety of IAMs were integrated under common protocols in modeling intercomparison projects (MIPs) (O'Neill et al., 2016) and provided input to the series of the Intergovernmental Panel on Climate Change (IPCC) Assessment Reports. Simulating computationally expensive IAMs developed and maintained at different research institutes around the world, however, requires large coordination efforts. Here we propose a new methodological framework to i) emulate the emerging behavior of IAMs (i.e., emission abatement for a given carbon price) through marginal abatement cost (MAC) curves and then ii) reproduce the behavior of IAMs by using the MAC curves coupled with a simple climate model. We show that the MAC curves can be systematically applied to reproduce the behavior of IAMs as an emulator for IAMs (emIAM), paving a way to generate multi-IAM scenarios more easily than before, with small computational resources.

There is a burgeoning literature on MAC curves (Jiang et al., 2020) that can broadly fall into two categories (Kesicki and Ekins, 2012): i) data-based MAC curves (bottom-up) and ii) model-based MAC curves (top-down). First, a data-based MAC curve gives a relationship between the emission abatement potential from each of the mitigation measures considered and the associated marginal costs, in the order of low- to high-cost measures based on individual data. A prominent example of such data-based MAC curves is McKinsey & Company (2009). Second, a model-based MAC curve gives a relationship between the amount of emission abatement and the system-wide marginal costs based on simulation results of a model (e.g., an energy system model and a computational general equilibrium (CGE) model) perturbed under different carbon prices or carbon budgets. Our work takes the second approach, building on earlier studies (Nordhaus, 1991; Ellerman and Decaux, 1998; van Vuuren et al., 2004; Johansson et al., 2006; Klepper and Peterson, 2006; Johansson, 2011; Morris et al., 2012; Wagner et al., 2012; Tanaka et al., 2013; Su et al., 2017; Tanaka and O'Neill, 2018; Yue et al., 2020; Tanaka et al., 2021). While data-based MAC curves tend to be rich in the representation of technological details, they do not consider system-wide interactions that are captured by model-based MAC curves. Model-based MAC curves reflect such interactions, however, without much explicit technological detail. Advantages and disadvantages of MAC curves of different categories are discussed elsewhere (Vermont and De Cara, 2010; Kesicki and Strachan, 2011; Huang et al., 2016).

This study derives a large set of MAC curves from the simulation results of IAMs, couples them as an emulator (emIAM) with a simple climate model, and validates the simulation results with the original IAM results under similar conditions. Namely, we look up the ENGAGE Scenario Explorer hosted at IIASA, Austria (https://data.ene.iiasa.ac.at/engage), a publicly available database from the EU Horizon 2020 ENGAGE project (Riahi et al., 2021; Drouet et al., 2021), and extract total anthropogenic $CO_2$, $CH_4$, and $N_2O$ emission pathways until 2100 from nine IAMs under a range of carbon budget constraints. For each IAM, we derive a set of $CO_2$, $CH_4$, and $N_2O$ MAC curves as a function of the respective emission reduction in percentage relative to the baseline at global and regional levels (eleven regions). We then implement the sets of MAC curves (i.e., emIAM) into a simple climate model called the Aggregated Carbon Cycle, Atmospheric Chemistry, and Climate (ACC2) model (Tanaka et al., 2007; Tanaka and O'Neill, 2018; Xiong et al., 2022). emIAM-ACC2 works as a hard-linked optimizing climate-economy model where total cost of mitigation is optimized under a climate target or carbon budget. We validate to what extent the emission pathway derived from emIAM-ACC2 under a certain carbon budget or a temperature target can reproduce the corresponding pathway from the original IAM in the ENGAGE Scenario Explorer. We further apply the emIAM approach to the GET model (Lehtveer et al., 2019), an IAM that did not take part in the ENGAGE project: we obtain global energy-related $CO_2$ emission pathways under a range of carbon price projections but with several different portfolios of available mitigation technologies (e.g., differentiated Carbon Capture and Storage (CCS) capacity). We then derive a MAC curve for each technology portfolio. Although each MAC curve concerns only the total emission abatement without distinguishing individual mitigation measures, this approach allows us to explore the role of a specific mitigation measure by comparing the outcomes based on MAC curves with and without this mitigation measure. Note that all IAMs emulated in this study take a cost-effectiveness approach, in which least-cost emission pathways to achieve a climate-related target are calculated in terms of the cost of mitigation without considering climate damage.

To our knowledge, this study is one of the first attempts to apply the MAC curve approach extensively for developing an IAM emulator: we consider ten IAMs, global and eleven regions, three gases (i.e., $CO_2$, $CH_4$, and $N_2O$), eight technology portfolios, and two broad sources (i.e., total anthropogenic and energy-related emissions). We demonstrate the applicability of emIAM by implementing it to ACC2, but emIAM can be used also with other simple climate models (Joos et al., 2013; Nicholls et al., 2020). emIAM allows ACC2 and potentially other simple climate models to reproduce approximately global and regional cost-effective emission pathways from multiple IAMs under a range of carbon budgets and temperature targets. In recent years, efforts have been made to develop emulators of Earth System Models (ESMs) in CMIP6 and the use of ESM emulators was exploited in the latest IPCC Sixth Assessment Report (AR6) (Leach et al., 2021; Tsutsui, 2022); however, no emulator has yet been developed for IAMs.

The rest of the manuscript consists of four sections: Section 2 introduces the IAMs and the experiments used to derive MAC curves. Section 3 describes the methodology to derive MAC curves and presents the MAC curves that are derived (i.e., emIAM). Section 4 shows the validation results for emIAM-ACC2. The paper is concluded with general remarks on the use of emIAM in Section 5. Due to the large number of MAC curves spanning several dimensions, there are a vast amount of display items from our analysis. Results are only selectively shown in the main paper; they are more comprehensively and systematically presented in the Supplement and our Zenodo repository.

Following the common definitions of terminologies found in the literature (National Research Council, 2012; Mulugeta et al., 2018), we use "emulate" to indicate a process of identifying a reduced-complexity model (i.e., a MAC curve) that approximates the behavior of a complex model (i.e., an IAM), "reproduce" to refer to a process of generating an output (i.e., an emission pathway) from an emulator with the same input and constraints given to an IAM (i.e., a cumulative carbon budget or end-of-the-century temperature, for example), and "validate" to indicate a process of investigating the extent to which an emulator reproduces an intended outcome in comparison to the corresponding original outcome from an IAM. Regarding the units, we use the original units of each model (i.e., US$2010 and $tCO_2$-eq with 100-year Global Warming Potential (GWP100) for all IAMs emulated here) to keep the comparability with underlying data, unless noted otherwise.

## 2. IAMs to emulate

Our study uses the output from a total of ten IAMs: nine IAMs used in the ENGAGE project and another IAM GET. The subsections below describe these IAMs and their data used to derive MAC curves.

### 2.1 IAMs from the ENGAGE project

Nine IAMs are available in the database of the ENGAGE Scenario Explorer: AIM/CGE V2.2, COFFEE 1.1, GEM-E3 V2021, IMAGE 3.0, MESSAGEix-GLOBIOM 1.1, POLES-JRC ENGAGE, REMIND-MAgPIE 2.1-4.2, TIAM-ECN 1.1, and WITCH 5.0 (thereafter, shorter labels will be used as in Table 1). These IAMs are diverse in terms of solution concepts (general and partial equilibrium models) and solution methods (intertemporal optimization and recursive dynamic models) (Table 1), among many other perspectives (Guivarch et al., 2022). A series of scenarios following a carbon budget ranging from 200 to 3,000 $GtCO_2$ (for the period of 2019-2100), as well as baseline scenarios, are available from each IAM. All scenarios incorporate second marker baseline scenario from the Shared Socioeconomic Pathways (SSP2), which reflect middle-of-the-road socioeconomic conditions (Riahi et al., 2017).

There are two types of scenarios in the ENGAGE Scenario Explorer: i) net-negative emissions scenarios (implying a temperature overshoot; with "f" in the scenario name) and ii) net-zero $CO_2$ emissions scenarios (implying a limited or no temperature overshoot; without "f" in the scenario name) (Riahi et al., 2021). While the former type of scenarios is defined with a carbon budget till the end of this century including a possibility of temporarily overspending it before (i.e., a possibility of achieving net negative $CO_2$ emissions), the latter type of scenarios is defined with a carbon budget till the point of meeting net-zero $CO_2$ emissions without allowing a budget overspending. The distinction of the two sets of scenarios may have important near-term implications (Johansson, 2021) and are considered when MAC curves are derived. For each type of scenarios, there are another two types of scenarios: i) NPi2020 scenarios, which consider currently implemented national policies; ii) INDCi2030 scenarios, which further consider national emission pledges until 2030. The availability of scenarios depends on the types of scenarios and varies across IAMs. We used the NPi2100 scenario as the baseline scenario for all carbon budget scenarios in our analysis.

The ENGAGE Scenario Explorer contains emission data for many greenhouse gases (GHGs) and air pollutants from each IAM, including $CO_2$, $CH_4$, and $N_2O$ emissions analyzed in our study. Emission data are available at global and regional levels (for nine and five IAMs, respectively). There are two sets of regional emission data, with one for five regions and the other for ten regions, the latter of which was used in our study: that is, China (CHN), European Union and Western Europe (EUWE), Latin America (LATAME), Middle East (MIDEAST), North America (NORAM), Other Asian countries (OTASIAN), Pacific OECD (PACOECD), Reforming Economies (REFECO), South Asia (SOUASIA), Sub-Saharan Africa (SUBSAFR), and Rest of World (ROW). Although all ENGAGE IAMs are regionally disaggregated, only a subset of the IAMs provides data for ten regions in the ENGAGE Scenario Explore as shown in Table 1. Note that only the GEM model provides emissions for ROW in the ENGAGE Scenario Explorer. In other IAMs, we allocated emissions for ROW to account for the discrepancy between global emissions and the sum of regional emissions (e.g., 3% difference in $CO_2$ emissions in AIM/CGE). Regarding emission sources, total anthropogenic emissions and energy-related emissions (e.g., energy and industrial processes) were separately used to derive global MAC curves for three gases (only total anthropogenic emissions for regional MAC curves due to computational requirements for validating regional MAC curves). Non-energy-related emissions (e.g., agriculture, forestry, and land-use sector), the differences between the two, were not used for generating MAC curves because non-energy-related emissions did not appear to be strongly correlated with carbon prices in most IAMs and influenced by other factors (Diniz

Oliveira et al., 2021).

## 2.2 GET model

GET is a global energy system model designed to study climate mitigation and energy strategies to achieve long-term climate targets under exogenously given energy demand scenarios (Azar et al., 2003; Hedenus et al., 2010; Azar et al., 2013; Lehtveer and Hedenus, 2015; Lehtveer et al., 2019). It is an intertemporal optimization model that minimizes with perfect foresight the total energy system costs discounted over the simulation period till 2150 (5% discount rate by default). To do so, various technologies for converting and supplying energy are evaluated in the model. The model considers primary energy sources such as coal, natural gas, oil, biomass, solar, nuclear, wind, and hydropower. Energy carriers considered in the model are petroleum fuels (gasoline, diesel, and natural gas), synthetic fuels (e.g., methanol), hydrogen, and electricity. End-use sectors in the model are transport, feedstock, residential heat, industrial heat, and electricity. We employed GET version 10.0 (Lehtveer et al., 2019) with the representation of ten regions.

To develop global energy-related $CO_2$ MAC curves reflecting different sets of available mitigation measures, we set up the following eight technology portfolios: i) Base, ii) Optimistic, iii) Pessimistic, iv) No CCS+Carbon Capture and Utilization (CCU)+Direct Air Capture (DAC), v) Large bioenergy, vi) Large bioenergy + Small carbon storage, vii) Small bioenergy + Large carbon storage, and viii) No nuclear. The Base portfolio uses the default set of assumptions associated with mitigation options available in the model. The Optimistic portfolio combines the assumptions of Large bioenergy supply, Large carbon storage potential, CCS+CCU+DAC, and Nuclear power. The Pessimistic portfolio, on the contrary, combines those of Small bioenergy supply, Small carbon storage potential, No CCS+CCU+DAC, and No nuclear power. Large and Small bioenergy cases assume 100% larger and 50% smaller bioenergy, respectively, than default levels (134 EJ/year globally). Large and Small carbon storage cases assume 8,000 $GtCO_2$ and 1,000 $GtCO_2$, respectively (2,000 $GtCO_2$ by default). With each of these technology portfolios, we simulated the model under 22 different carbon price scenarios. In all carbon price scenarios, the carbon price grows 5% each year with a range of initial levels in 2010 (1, 2, 3, 5, 7, 10, …, 140 US$2010/tCO$_2$) (more details can be seen in Table S2), following the Hotelling rule (Hof et al., 2021). We assumed a discount rate of 5% for all portfolios and carbon price scenarios. Our analysis used a scenario with zero carbon prices as the baseline scenario. We derived only global energy-related $CO_2$ MAC curves from GET since the model did not explicitly describe processes related to non-energy related emissions.

## 3. MAC curves emulating IAMs

## 3.1 Deriving MAC curves

Our MAC curve approach aims to capture the relationship between the carbon price and the emission abatement in IAMs. For each IAM (i.e., ENGAGE IAMs and GET), we calculated the emission reduction level relative to the respective baseline level each year. Emission reductions can be expressed either in the absolute term (for example, in $GtCO_2$) or in the fractional term (in percentage relative to the baseline level) (Kesicki, 2013; Jiang et al., 2022), the latter of which is used in our analysis. When the emission is at the baseline level, the relative emission reduction is 0% by definition. When it is 100%, which can occur for $CO_2$, the emission is (net) zero. When it exceeds 100%, the emission turns (net) negative. If there are non-zero carbon prices in baseline scenarios (small carbon prices can be found in baseline scenarios from some IAMs), we subtracted them from the carbon prices in mitigation scenarios.

We then fitted a mathematical function $f(x)$ (equation (1); selected among several others as explained below) as a MAC curve to capture the emission abatement level for a given carbon price. We used a common time-invariant functional form of equation (1) for all cases (i.e., models, gases, regions, sources, and portfolios) for consistency, comparability, and simplicity of use.

$$f(x) = a * x^b + c * x^d \tag{1}$$

$a$, $b$, $c$, and $d$ are the parameters to optimize for each case. $x$ is a variable representing the emission abatement level in percentage relative to the assumed baseline level. The carbon price is in per ton of $CO_2$-equivalent emissions, in which GWP100 (28 and 265 for $CH_4$ and $N_2O$, respectively (IPCC., 2013)) is used to convert to the prices of $CH_4$ and $N_2O$, as assumed in the IAMs emulated here (Harmsen et al., 2016). GWP100 is practically a default emission metric used to convert non-$CO_2$ GHG emissions to the common scale of $CO_2$, which has been used for decades in multi-gas climate policies and assessments including the Paris Agreement (Lashof and Ahuja, 1990; Fuglestvedt et al., 2003; Tanaka et al., 2010; Tol et al., 2012; Levasseur et al., 2016; UNFCCC., 2018). MAC curves were obtained from the data for the period 2020-2100. There are three key assumptions in our approach: i) MAC curves are assumed time-independent, ii) abatement levels are assumed independent across gases, and iii) abatement levels are assumed independent across regions. While MAC curves are more commonly time-dependent or for a specific point in time, time-independent MAC curves have also been used for long-term pathway calculations (Johansson et al., 2006; Tanaka and O'Neill, 2018; Tanaka et al., 2021) and shorter-term assessments (De Cara and Jayet, 2011). The implications of the first assumption will be discussed later in this section. The second assumption indicates that co-reductions of GHG emissions (e.g., emission abatement of $CO_2$ and $CH_4$ from an early retirement of a coal-fired power plant (e.g. Tanaka et al., 2019)) are not explicitly considered in our MAC curve approach. The third assumption implies that the regional distribution of GHG abatements is determined primarily by the global cost-effectiveness (Su et al., 2022). The validities of these assumptions can be seen in Section 4, in which MAC curves are combined with a simple climate model to reproduce original IAM outcomes. There are further conditions applied to derive MAC curves from each model as summarized in Table 1. These conditions were identified based on visual inspection of data from each IAM.

| Model | Label | Solution concept | Solution method | Spatial resolution | Gas | Data range captured by MAC curves |
|---|---|---|---|---|---|---|
| AIM/CGE V2.2 | AIM | General equilibrium | Recursive dynamic | Global Regional | $CO_2$ $CH_4$ $N_2O$ | Carbon prices lower than $110/tCO_2 before 2040 and all data after 2040 |
| COFFEE 1.1 | COFFEE | Partial equilibrium | Intertemporal optimization | Global Regional | $CO_2$ $CH_4$ $N_2O$ | Carbon prices lower than $50/tCO_2 with abatement levels below 100% under scenarios without negative emissions |
| GEM-E3 V2021 | GEM | General equilibrium | Recursive dynamic | Global Regional | $CO_2$ $CH_4$ $N_2O$ | All data |
| IMAGE 3.0 | IMAGE | Partial equilibrium | Recursive dynamic | Global Regional | $CO_2$ $CH_4$ $N_2O$ | All data except: EN_INDCi2030_800f EN_NPi2020_600f EN_INDCi2030_1000f EN_Npi2020_800 |
| MESSAGEix-GLOBIOM 1.1 | MESSAGE | General equilibrium | Intertemporal optimization | Global Regional | $CO_2$ $CH_4$ $N_2O$ | All data except: EN_NPi2020_450 EN_NPi2020_500 |
| POLES-JRC ENGAGE | POLES | Partial equilibrium | Recursive dynamic | Global | $CO_2$ $CH_4$ $N_2O$ | Carbon prices lower than $5,000/tCO_2 |
| REMIND-MAgPIE 2.1-4.2 | REMIND | General equilibrium | Intertemporal optimization | Global | $CO_2$ $CH_4$ $N_2O$ | All data except: EN_INDCi2030_700 EN_INDCi2030_800 EN_NPi2020_400 EN_NPi2020_500 |
| TIAM-ECN 1.1 | TIAM | Partial equilibrium | Intertemporal optimization | Global | $CO_2$ $CH_4$ $N_2O$ | All data |
| WITCH 5.0 | WITCH | General equilibrium | Intertemporal optimization | Global | $CO_2$ $CH_4$ $N_2O$ | All data |
| GET 10.0 | GET | Partial equilibrium | Intertemporal optimization | Global | Energy $CO_2$ | Carbon prices lower than $5,000/tCO_2; excluded data for very high abatements with disproportionally low costs (found typically after 2100) |

**Table 1. Models and data considered for emIAM.** This table describes the features of models and the data (gases, regions (10 regions)) that were used to derive our MAC curves. "Solution concept" and "solution method" for ENGAGE IAMs (first nine IAMs in the table) are based on Riahi et al. (2021), Guivarch et al. (2022), and IAMC_wiki (2022). Total anthropogenic (and separately energy-related and non-energy-related) $CO_2$, $CH_4$, and $N_2O$ emissions were taken from ENGAGE IAMs; only energy-related $CO_2$ emissions were used from GET.

In selecting the functional form for fitting MAC curves (i.e., equation (1)), we needed to balance the competing requirements for i) capturing complex nonlinear relationships between the carbon price and the abatement level and ii) keeping the functional form at low complexity. Therefore, we tested the performance of several functional forms for fitting the data. The candidate functions, some of which are based on previous studies (Johansson, 2011; Su et al., 2017; Tanaka and O'Neill, 2018), are summarized in Table S1, with the ranges of parameters considered. To infer a good functional form, we further tried the symbolic regression approach by using the software HeuristicLab, but we were not able to obtain a functional form satisfactory for our purpose. Our results indicated that the polynomial function with two algebraic terms (equation (1)) gave the highest $R^2$ and adjusted $R^2$ among the equations tested for more than 50% of cases, performing consistently the best for all IAMs (see the Zenodo repository). Therefore, we applied equation (1) to generate MAC curves. A polynomial function with only one algebraic term was insufficient: two distinct algebraic terms were generally needed to capture the trend of our data (sometimes with a kink like a "reversed L" shape or with a plateau as shown later). It should be noted that we do not consider the parametric uncertainties in individual MAC curves, but a use of MAC curves from multiple IAMs can provide a sense of model uncertainties.

In addition to the derivation of MAC curves, we derived the maximum abatement level for each IAM from its simulation results under all carbon budgets or carbon prices, which reflected, for example, the limit of CCS capacity and hard-to-abate sectors. The minimum abatement level is, by definition, zero in all simulation periods. We also calculated for each gas and each IAM the maximum first and second derivatives of temporal changes in abatement levels, which corresponded roughly to the limit of the technological change rate and the socio-economic inertia, respectively. The limits on the first and second derivatives of abatement changes can prevent the use of deep mitigation levels in the MAC curve in early periods. These limits could be introduced also by more complex functional forms internally in the MAC curves (Ha-Duong et al., 1997; Schwoon and Tol, 2006; De Cara and Jayet, 2011; Hof et al., 2021), but we externally applied such limits on the MAC curves. "Learning by doing" and "learning with time," which reduces the mitigation cost with abatement and time, respectively (Hof et al., 2021), are not explicitly considered in our MAC curve approach but in part, albeit unintended, captured in our approach that describes percentage reduction rates relative to rising baseline scenarios (until 2080). For example, constant emission reductions in the absolute term can appear smaller with time in the percentage term and thus become less expensive in our approach.

For each gas and each IAM, we computed the rate of change in the abatement level at each time step from the previous time step (i.e., first derivatives) over the entire available period. We then approximated such data with a log-normal distribution and assumed the three-sigma level (upper side) as the maximum first derivative of abatement changes for each gas and each IAM. Likewise, we computed the rate of change of the rate of change in the abatement level (i.e., second derivatives), approximated the data with a normal distribution, and assumed the three-sigma level (upper side) as the maximum second derivative of abatement changes. We further assumed that the minimum first and second derivatives were at the opposite signs of the maximum first and second derivatives, respectively. These limits will be applied when MAC curves are coupled with ACC2 to generate cost-effective pathways (Section 4).

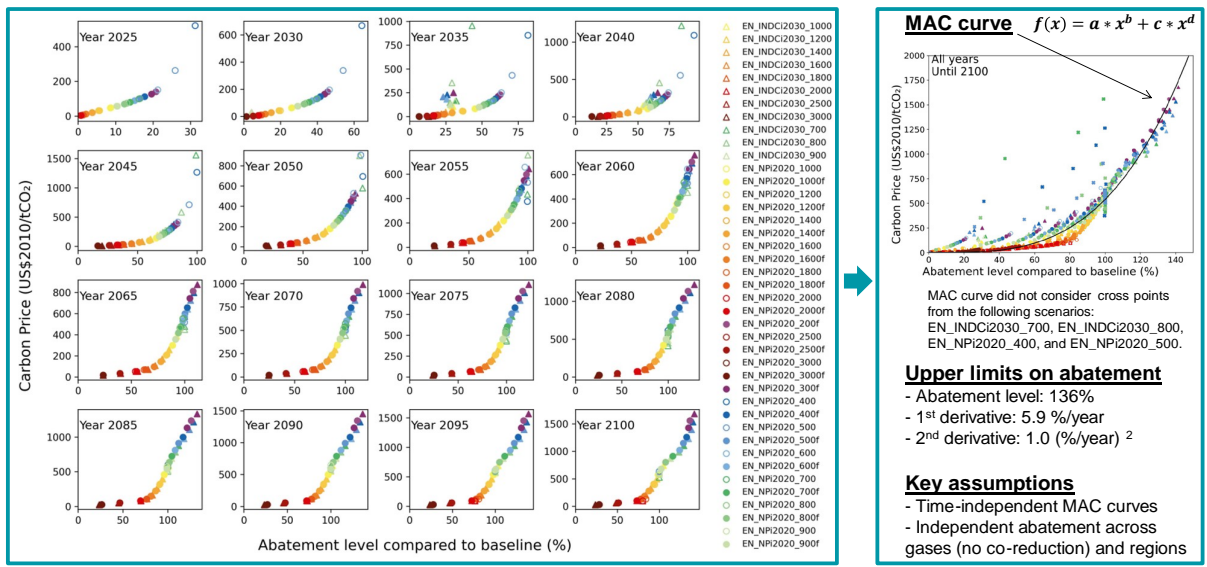

**Figure 1. Overview of the methods to derive MAC curves and limits on abatement (upper limits on abatement levels and their first and second derivatives).** The figure uses the data for global total anthropogenic $CO_2$ emissions from REMIND as an example. Scenario names indicate respective cumulative carbon budgets for the period 2019 – 2100 in $GtCO_2$. NPi2020 scenarios consider currently implemented national policies (circle); INDCi2030 scenarios further consider national emission pledges until 2030 (triangle). Among NPi2020 scenarios, those with "f" are net-negative emissions scenarios (filled circles); those without "f" are net-zero $CO_2$ emissions scenarios (open circles). Crosses indicate data points from scenarios that were not considered in the derivation of the MAC curve. In the equation of the MAC curve, $a$, $b$, $c$, and $d$ are the parameters to optimize; $x$ is the variable representing the abatement level in percentage relative to the assumed baseline level (i.e., NPi2100 (not shown)).

Figure 1 illustrates the approach described above by using the output from REMIND as an example

(corresponding figures for AIM and MESSAGE in Figures S1 and S2 of Supplement (note: figures and tables with "S" in numbering are in Supplement)). In sum, we combined a MAC curve with the upper limits on abatement levels and their first and second derivatives to emulate the behavior of an IAM. The series of panels in Figure 1 show the relationship between the carbon price and the abatement level at each point in time (every five years) as obtained from REMIND simulated with the range of carbon budgets. Data points can be seen only at low abatement levels in the near-term. With time, data points proceed to deeper abatement levels. Putting together across all years, the right panel of Figure 1 shows a secular relationship, which allows us to approximate with a time-independent MAC curve. Even though there are time-dependent processes due to technological changes and socio-economic inertia in this intertemporal optimization model, the same relationship can apply over time between the carbon price and the abatement level. There are outliers arising from very low carbon budget scenarios (crosses in the right panel of Figure 1). To capture the time-independent characteristics of the data, we identified outlier scenarios (if any) from each IAM and manually excluded them from the derivation of the MAC curve. However, it needs to be kept in mind that excluding such scenario(s) limits the range of applicability for the MAC curve.

But why does this time-independent approach work so well to capture IAM processes collectively that are time-dependent? The use of percentage reductions in our MAC curve approach goes some way in explaining this. Since most of the baseline scenarios are rising as pointed out earlier, the same amount of emission abatement in the absolute term can become smaller with time in the percentage term, which inadvertently but effectively captures "learning by doing" and "learning with time," at least partially. If the underlying data are shown in the absolute term, the data distribution does appear more dispersed (Figures S3-S5 for AIM, MESSAGE, and REMIND).

### 3.2 MAC curves from ENGAGE IAMs

### 3.2.1 Carbon price and abatement level

Figure 2 shows the relationships between the carbon price and the abatement level for global total anthropogenic $CO_2$ emissions obtained from nine ENGAGE IAMs. The results differ in terms of the range of carbon prices, the range of abatement levels, and the dispersion of data points. For example, the carbon prices of AIM and COFFEE stay below \$500/$tCO_2$, while the carbon prices of POLES and MESSAGE can exceed \$5,000/$tCO_2$. The maximum abatement levels of COFFEE and REMIND are above 150%, while others are in the range of 100%-120%. AIM offers a limited amount of data in the near term. IMAGE and POLES give more dispersed data distributions than other models, which may be related to the fact that these models are recursive dynamic models (Table 1). However, the other recursive dynamic models, AIM and GEM, produce less dispersed data distributions, which can be well captured by MAC curves. Nevertheless, on the whole, the relationships between the carbon price and the $CO_2$ abatement level are well captured by time-independent MAC curves for most IAMs here. Visual inspection of the data distributions reveals little differences between net-negative emissions scenarios and net-zero $CO_2$ emissions scenarios (except for WITCH), indicating that MAC curves are generally valid for both types of scenarios. If we consider in terms of the absolute amount of abatement, instead of percentage abatement, the data distributions become more dispersed (Figure S3-S5). Results for other gases and for energy-related emissions can be found in Figures S6-S37.

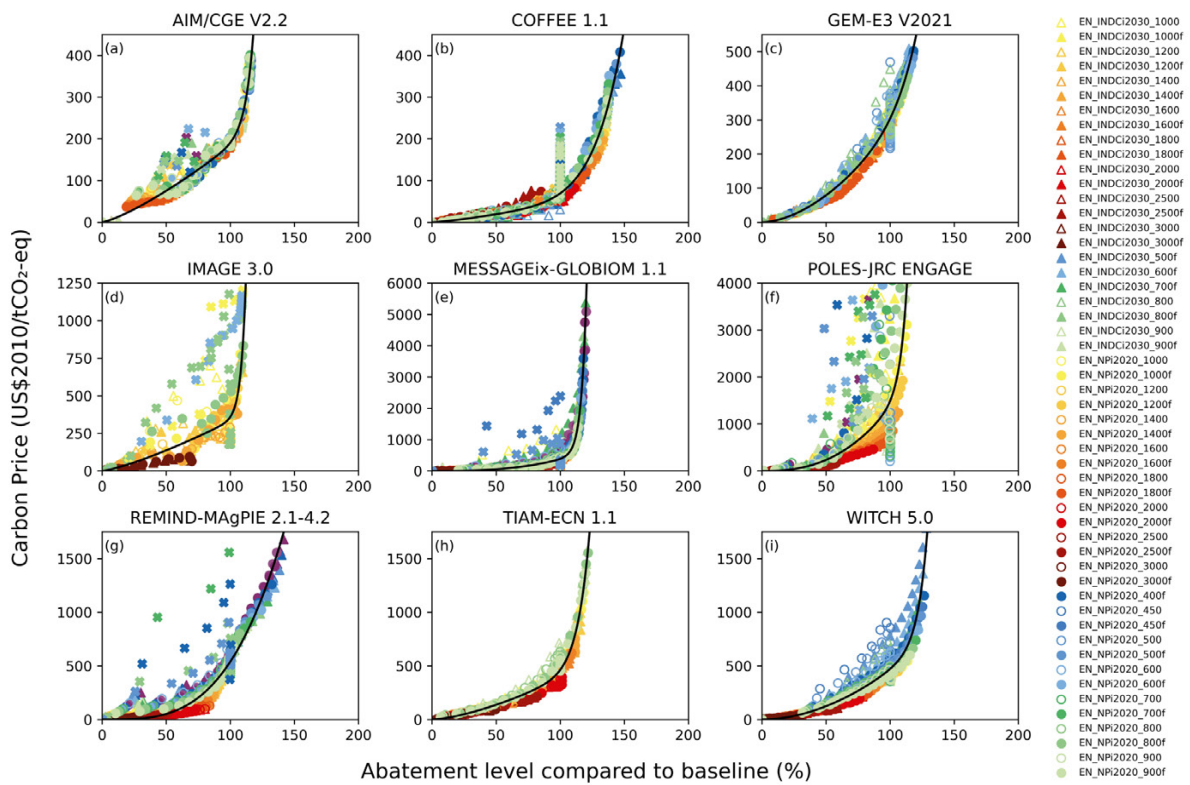

**Figure 2. Relationships between the carbon price and the global total anthropogenic $CO_2$ abatement level obtained from nine ENGAGE IAMs**. Each panel shows the results from each ENGAGE IAM. Data were obtained from the ENGAGE Scenario Explorer and are shown in colors and markers as designated in the legend. Black lines are the MAC curves. Crosses are the data points that were not considered in the derivation of MAC curves (Table 1).

### 3.2.2 First and second derivatives of abatement changes

The first and second derivatives of temporal changes in abatement levels for global total anthropogenic $CO_2$ emissions from each ENGAGE IAM are shown in Figure 3. Data for the first derivatives primarily distribute on the positive side and can be best captured by log-normal distributions, among other distributions tested. On the other hand, data for the second derivatives spread on both the positive and negative sides and can be approximated by normal distributions. On the basis of visual inspection, three-sigma ranges of distributions can largely capture data ranges. We thus use three-sigma ranges as the limits on the first and second derivatives of abatement changes. There are outliers (now shown) originating from net-zero $CO_2$ emissions scenarios, which we speculate are caused by sudden drops in carbon prices (Figure SI 1.1-6 of Riahi et al. (2021)). These outliers were effectively removed by considering three-sigma ranges (rather than the maxima and minima of original data points). For other gases and for energy-related emissions, see Figures S40-S87.

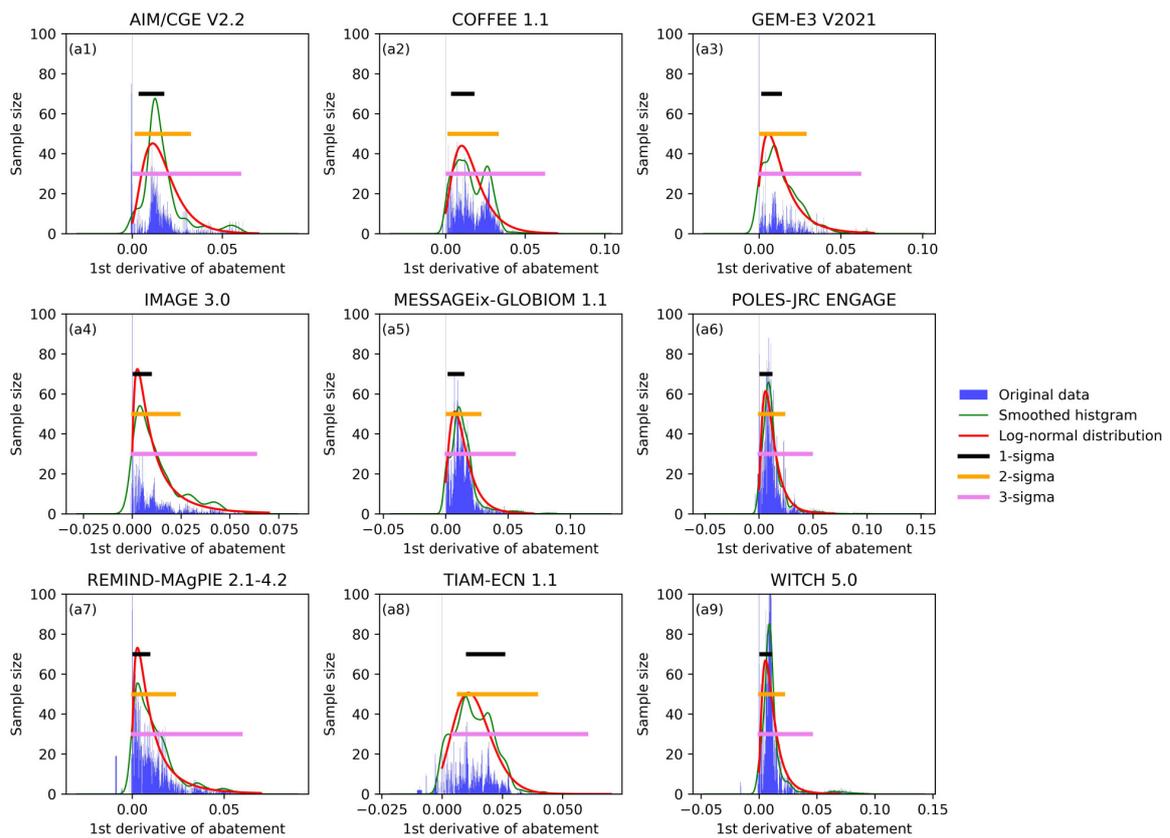

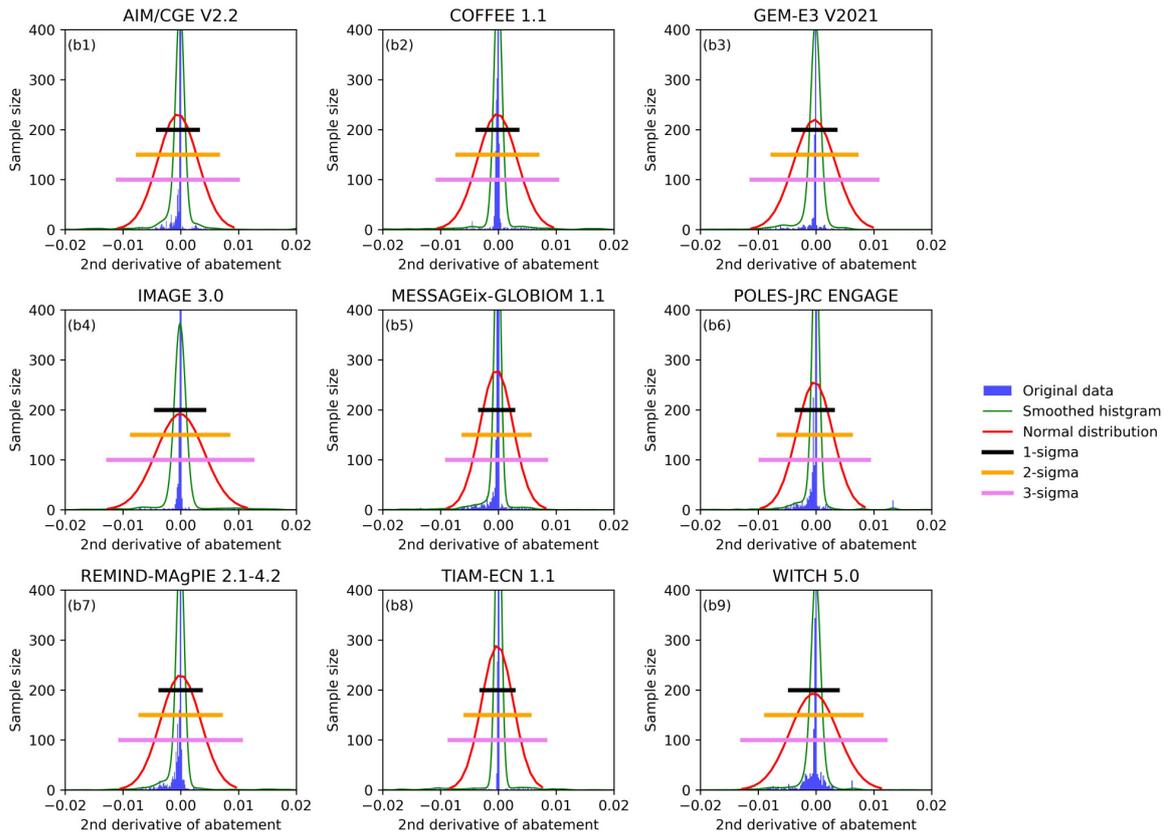

**Figure 3. The first and second derivatives of temporal changes in abatement levels for the global total anthropogenic CO₂ emissions from each ENGAGE IAM**. A log-normal distribution is applied to the data for the first derivatives of abatement changes obtained from each IAM. A normal distribution is applied to the data for the second derivatives of abatement changes obtained from each IAM.

The upper limits on the first and second derivatives of abatement changes estimated for ENGAGE IAMs are summarized in Table 2. Those for ACC2 were assumed to be 4.0 %/year and 0.4 (%/year)$^2$, respectively, for all three gases (CO₂, CH₄, and N₂O) (Tanaka and O'Neill, 2018; Tanaka et al., 2021). ENGAGE IAMs give higher upper limits on the first and second derivatives than ACC2 for CO₂. For other two gases, ENGAGE IAMs also give higher upper limits on the second derivatives but tend to indicate lower upper limits on the first derivatives.

The upper limits on the first and second derivatives of CO₂ abatement can determine the earliest possible year of achieving net zero CO₂ emissions (i.e., 100% abatement) for each IAM. In the case of ACC2, it is the year 2050 when net zero CO₂ emissions become first possible, if the abatement can start in 2020. Figure S88 compares earliest possible net zero years implied by the upper limits on the first and second derivatives with the years of net zero available in carbon budget scenarios from each ENGAGE IAM. The figure shows that the former precedes the latter in all IAMs, indicating that the upper limits based on three-sigma ranges are large enough to allow pathways to achieve net zero as shown by each IAM.

### 3.2.3 Global MAC curves

Figure 4 shows the global MAC curves for total anthropogenic and energy-related $CO_2$, $CH_4$, and $N_2O$ emissions from nine ENGAGE IAMs. The parameter values of these global MAC curves and associated limits on abatement are shown in Table 2 (for total anthropogenic emissions) and Table S3 (for energy-related emissions).

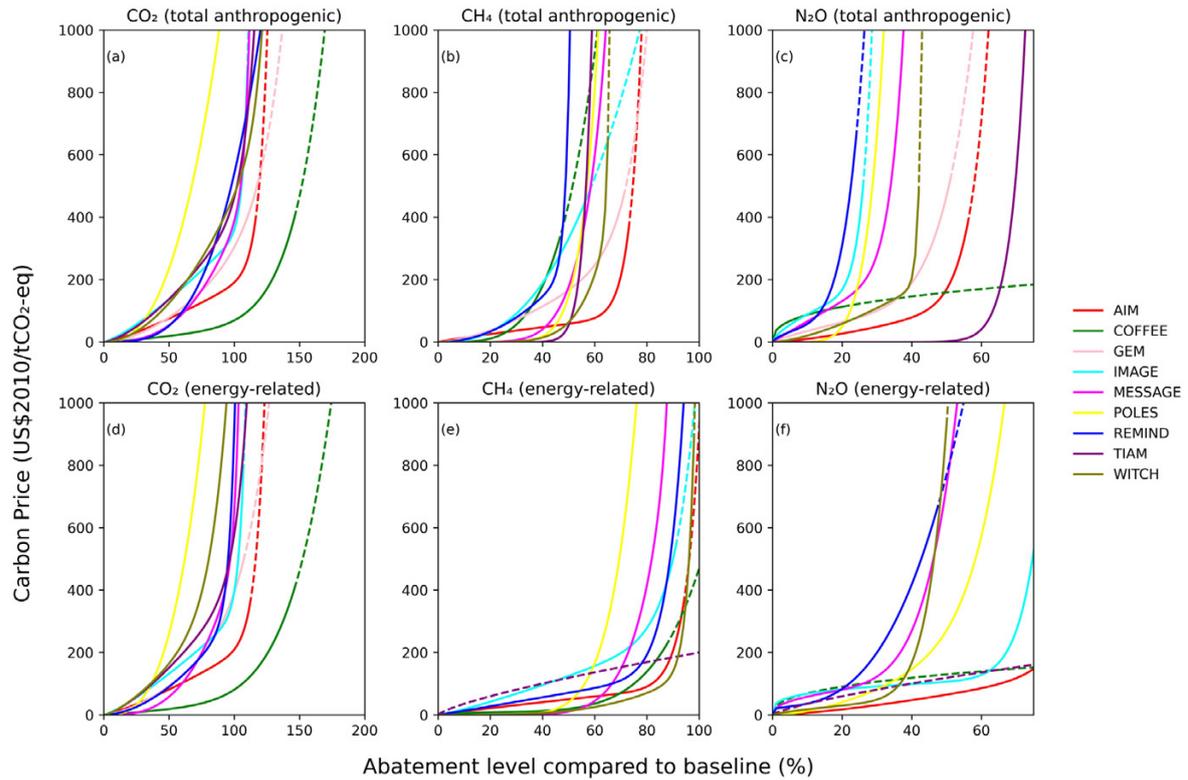

**Figure 4. Global MAC curves for total anthropogenic and energy-related $CO_2$, $CH_4$, and $N_2O$ emissions derived from nine ENGAGE IAMs.** In panels (a) to (f), the solid line indicates that the MAC curve is within the applicable range; the dashed line means that it is outside the applicable range (i.e., above the maximum abatement level indicated from underlying IAM simulation data or above the range of carbon prices considered for fitting the MAC curve; see Tables 1 and 2). Different colors indicate different IAMs.

MAC curves for total anthropogenic and energy-related $CO_2$ emissions resemble each other since total anthropogenic $CO_2$ emissions are predominantly energy-related $CO_2$ emissions. COFFEE gives the lowest carbon prices among all IAMs over a wide range of abatement levels; POLES shows the highest carbon prices. AIM has the second-lowest carbon prices at the abatement level of 63% and beyond. REMIND gives higher carbon prices than AIM beyond the abatement level of 60%.

The difference between MAC curves for total anthropogenic and energy-related emissions are more distinct for $CH_4$ and $N_2O$ than $CO_2$ because of disproportionally larger mitigation opportunities outside of the energy sector. $CH_4$ MAC curves generally rise sharply at lower abatement levels than $CO_2$ MAC curves. All MAC curves for energy-related $CH_4$ emissions are low up to about 50% abatement level, presumably reflecting low-cost abatement opportunities. AIM and WITCH give a low carbon price up to 80-90% abatement level for energy-related $CH_4$ emissions. Due to limited $N_2O$ abatement opportunities, $N_2O$ MAC curves rise steeply at low abatement levels, with the one from REMIND rising earliest.

| Model | Gas | a | b | c | d | MaxABL | Max1st | Max2nd |
|---|---|---|---|---|---|---|---|---|
| AIM | $CO_2$ | 182.14 | 1.27 | 8.68 | 19.71 | 116.2 | 5.9 | 1.0 |
| | $CH_4$ | 108.99 | 0.91 | 78686 | 17.91 | 73.6 | 6.1 | 1.3 |
| | $N_2O$ | 282.34 | 1.46 | 243642 | 11.84 | 56.1 | 4.5 | 1.0 |
| COFFEE | $CO_2$ | 46.66 | 1.29 | 22.59 | 7.01 | 146.7 | 6.1 | 1.0 |
| | $CH_4$ | 3658.91 | 4.05 | 3658.91 | 4.05 | 47.0 | 2.3 | 1.1 |
| | $N_2O$ | 102.75 | 0.37 | 102.75 | 0.37 | 20.2 | 3.6 | 1.3 |
| GEM | $CO_2$ | 267.14 | 1.76 | 36.85 | 8.53 | 118.2 | 6.1 | 1.1 |
| | $CH_4$ | 7133.48 | 10.70 | 486.16 | 1.59 | 71.9 | 4.3 | 0.9 |
| | $N_2O$ | 240.14 | 0.83 | 31072 | 6.54 | 51.1 | 3.8 | 0.7 |
| IMAGE | $CO_2$ | 28.57 | 29.83 | 330.58 | 1.27 | 110.1 | 6.3 | 1.2 |
| | $CH_4$ | 959.11 | 2.53 | 959.11 | 2.53 | 58.3 | 3.1 | 0.6 |
| | $N_2O$ | 1.54E+08 | 9.70 | 426.52 | 0.68 | 26.3 | 2.4 | 0.5 |
| MESSAGE | $CO_2$ | 18.30 | 30.24 | 368.79 | 2.78 | 120.9 | 5.4 | 0.8 |
| | $CH_4$ | 3.29E+07 | 29.08 | 16789 | 6.57 | 73.3 | 3.5 | 0.6 |
| | $N_2O$ | 610.67 | 0.97 | 7909596 | 9.47 | 45.2 | 1.9 | 0.3 |
| POLES | $CO_2$ | 1347.98 | 2.52 | 144.57 | 21.87 | 131.9 | 4.8 | 0.9 |
| | $CH_4$ | 48160 | 9.36 | 48160 | 9.36 | 75.7 | 4.2 | 0.9 |
| | $N_2O$ | 1513291 | 94.73 | 1512842 | 6.42 | 37.3 | 2.2 | 0.5 |
| REMIND | $CO_2$ | 269.52 | 3.38 | 269.52 | 3.38 | 136.2 | 5.9 | 1.0 |
| | $CH_4$ | 1.61E+11 | 28.11 | 1002.16 | 2.11 | 51.2 | 3.2 | 0.9 |
| | $N_2O$ | 633401 | 4.92 | 224.21 | 0.65 | 24.8 | 1.6 | 0.8 |
| TIAM | $CO_2$ | 78.52 | 13.31 | 384.32 | 1.48 | 121.7 | 6.0 | 0.8 |
| | $CH_4$ | 1.23E+07 | 17.81 | 157.83 | 100 | 59.5 | 3.9 | 1.0 |
| | $N_2O$ | 215121 | 16.79 | 99.08 | 100 | 73.3 | 4.3 | 2.3 |
| WITCH | $CO_2$ | 462.12 | 1.89 | 10.13 | 18.05 | 126.5 | 4.5 | 1.2 |
| | $CH_4$ | 6658.29 | 6.72 | 2.78E+15 | 69.59 | 65.6 | 3.7 | 2.0 |
| | $N_2O$ | 681.73 | 1.52 | 9.13E+18 | 43.78 | 42.7 | 3.0 | 1.0 |

**Table 2. Parameter values of global MAC curves for total anthropogenic $CO_2$, $CH_4$, and $N_2O$ emissions derived from nine ENGAGE IAMs and associated limits on abatement.** See equation (1) for parameters $a$, $b$, $c$, and $d$. MaxABL denotes the maximum abatement level (%) of each gas indicated from IAM simulation data. The units for $a$ and $c$ are UD\$2010/t$CO_2$. Max1st and Max2nd represent the maximum first and second derivatives ((%/year) and (%/year)$^2$), respectively, of abatement changes of each gas also indicated from IAM simulation data. For those of global MAC curves

for energy-related $CO_2$, $CH_4$, and $N_2O$ emissions, see Table S3. For those of regional MAC curves, see the Zenodo repository.

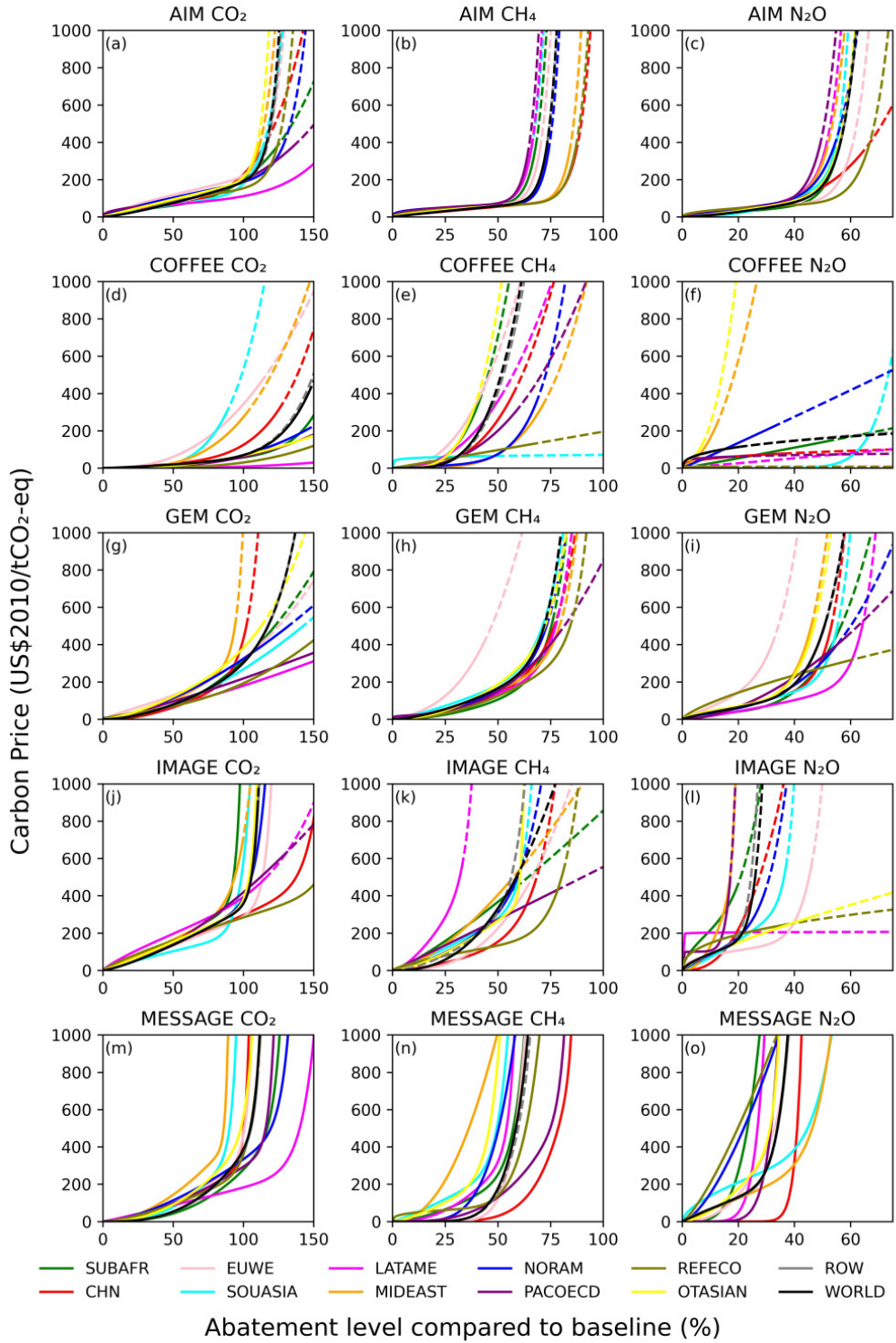

**Figure 5. Regional MAC curves for total anthropogenic $CO_2$, $CH_4$, and $N_2O$ emissions derived from five ENGAGE IAMs.** The solid line indicates that the MAC curve is within the applicable range; the dashed line means that it is outside

the applicable range (i.e., above the maximum abatement level indicated from underlying IAM simulation data or above the range of carbon prices considered for fitting the MAC curve; see Tables 1 and 2). Different colors indicate different regions: China (CHN), European Union and Western Europe (EUWE), Latin America (LATAME), Middle East (MIDEAST), North America (NORAM), Other Asian countries (OTASIAN), Pacific OECD (PACOECD), Reforming Economies (REFECO), South Asia (SOUASIA), Sub-Saharan Africa (SUBSAFR), and Rest of World (ROW).

### 3.2.4 Regional MAC curves

Figure 5 shows the regional MAC curves for total anthropogenic $CO_2$, $CH_4$, and $N_2O$ emissions from five ENGAGE IAMs. The parameter values of the regional MAC curves and associated limits on abatement can be found in our Zenodo repository. While various inter-model and inter-regional differences can be seen in Figure 5, the regional variations of AIM MAC curves look smallest for all three gases.

MIDEST generally shows a high $CO_2$ MAC curve relative to other regions. LATAM gives the lowest MAC curve at abatement levels above approximately 79% in all IAMs considered here, except for the IMAGE model with SOUASIA and REFECO being the lowest MAC curve at the abatement level of above and below 90%, respectively. LATAM also indicates very deep $CO_2$ abatement potentials exceeding 150% in some models. The $CH_4$ MAC curves from AIM indicate low-cost $CH_4$ abatement opportunities up to abatement levels of approximately 50% in all regions, while such opportunities appear less abundant in the $CH_4$ MAC curves from other models. REFECO exhibits a very low $CH_4$ MAC curve in all five models. MIDEST gives either a high or a low $CH_4$ MAC curve, depending on the IAM. The $N_2O$ MAC curves generally rise sharply earlier than the $CH_4$ MAC curves.

### 3.3 MAC curves from GET

Figure 6 shows the relationships between the carbon price and the abatement level of global energy-related $CO_2$ emissions and their dependency on underlying technology portfolios considered in GET. MAC curves from different technology portfolios are compared in Figure 7. They are further compared with the Global MAC curves for energy-related $CO_2$ emissions from ENGAGE IAMs. The parameter values of these global MAC curves and associated limits on abatement are in Table 3. Further details of the first and second derivatives of abatement changes from GET can be found in Figures S38 and S39.

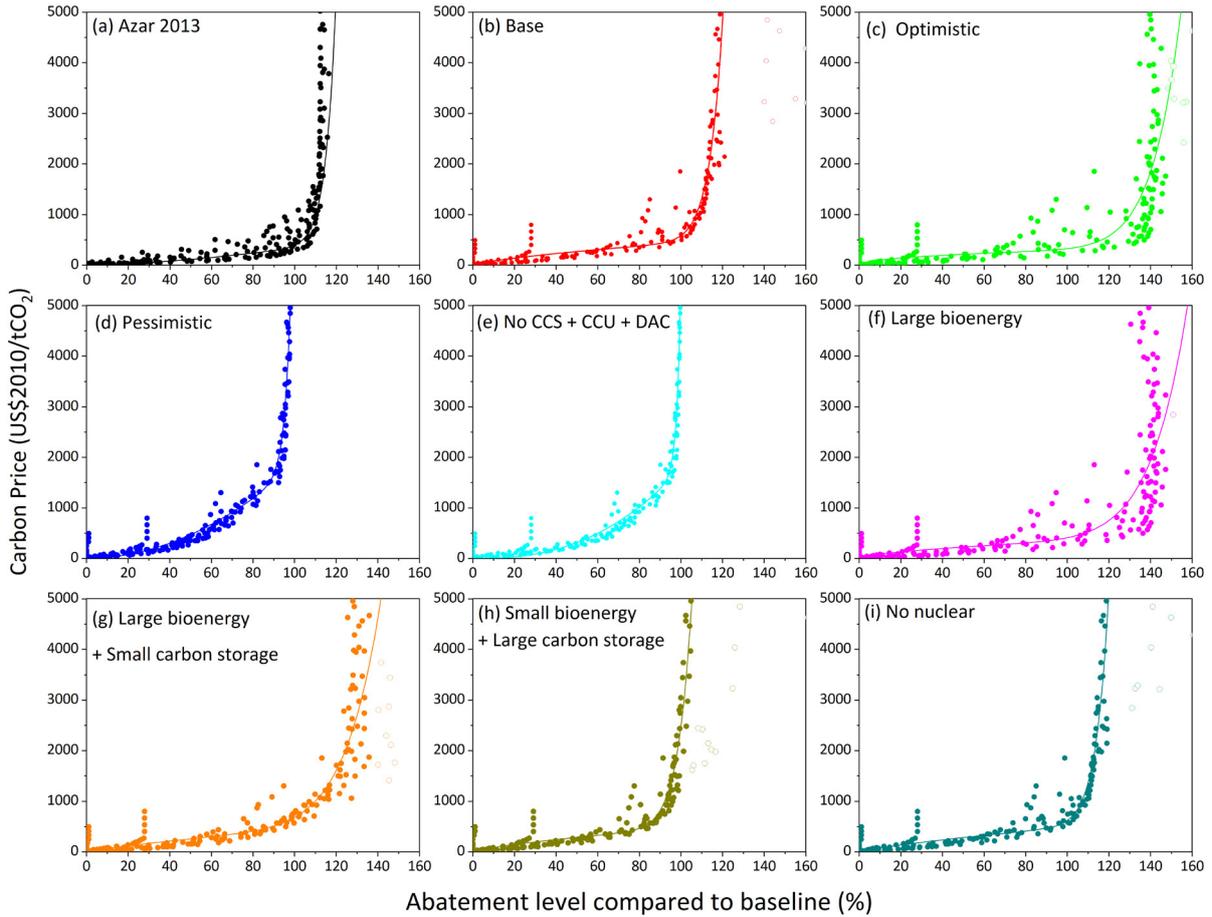

**Figure 6. Relationships between the carbon price and the global energy-related CO$_2$ abatement level obtained from GET with different portfolios of available mitigation technologies.** Panel (a) shows the results obtained from an older version of GET (Azar et al., 2013) for the sake of comparison. Panels (b) to (i) show the results from GET (Lehtveer et al., 2019) with different technology portfolios. See Section 2.2 for the definitions of technology portfolios. Points are the data obtained from GET; lines are the MAC curves calculated based on our approach. Open circles are the data that were not considered in the derivation of MAC curves (Table 1) and are typically found after 2100, in some cases above the abatement level of 160% (not shown). Note that we have converted the unit in Panel (a) from US\$2010/tC, which is used in the older version of GET, to UD\$2010/tCO$_2$, the commonly used unit here.

Global MAC curves for energy-related CO$_2$ emissions from different technology portfolios span a wide range. The range is nearly as wide as that from ENGAGE IAMs (i.e., inter-technology portfolio range ≈ inter-model range), if we disregard the MAC curve from COFFEE (Figure 2d). The MAC curve from the Base portfolio is generally higher than the MAC curve based on the previous version of the model (Azar et al., 2013; Tanaka and O'Neill, 2018), reflecting the biomass supply potential being smaller in GET used in our analysis (i.e., 134 EJ/year) than in the previous version (approximately 200 EJ/year), among other reasons. The maximum abatement level of the Base portfolio is about 120%, which is slightly higher than the estimate of 112% based on the previous model version. The Optimistic portfolio generally gives lower carbon prices and deeper mitigation potentials than the Base portfolio. Conversely, the Pessimistic portfolio shows higher carbon prices and more limited mitigation potentials than the Base portfolio. The Optimistic and Large bioenergy portfolios yield more than 150% $CO_2$ abatement levels at maximum. The Large bioenergy + Low carbon storage portfolio gives lower maximum abatement levels than the previous two portfolios due to the assumed lower carbon storage potential. The Low bioenergy + Large carbon storage portfolio limits the maximum $CO_2$ abatement levels at only slightly above 100%. With the Pessimistic portfolio, the maximum $CO_2$ abatement levels do not exceed 100% (i.e., no net negative $CO_2$ emissions) primarily because no carbon capture technologies such as CCS, CCU, and DAC are available. Likewise, the No CCS+CCU+DAC portfolio also gives a maximum abatement level below 100%. The No nuclear portfolio gives a similar relationship to the one from the Base portfolio, indicating a limited role of nuclear energy here. Finally, the results are somewhat but not strongly sensitive to the choice of discount rate (5% by default), as indicated from the results based on alternative discount rates of 3% and 7%, in which the growth rate of carbon price is fixed at the value of the respective discount rate based on the Hotelling rule (Figure S89).

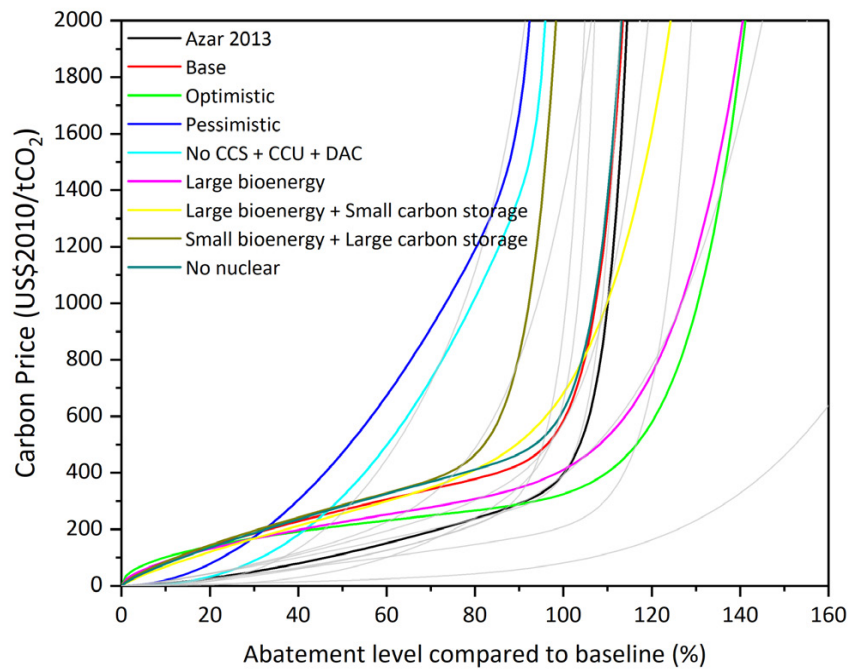

**Figure 7. Global MAC curves for energy-related $CO_2$ emissions derived from the GET model with different portfolios of available mitigation technologies.** Different colors indicate different technology portfolios (see Section 2.2 for details). Global MAC curves for energy-related $CO_2$ emissions from ENGAGE IAMs are shown as a comparison in gray lines.

| Technology portfolio | Gas | $a$ | $b$ | $c$ | $d$ | MaxABL | Max1st | Max2nd |
|---|---|---|---|---|---|---|---|---|
| Azar 2013 | $CO_2$ | 338.61 | 1.58 | 57.08 | 24.59 | 112 | 5.6 | 0.9 |
| Base | $CO_2$ | 441.86 | 0.72 | 142.54 | 18.73 | 121 | 7.4 | 1.3 |
| Optimistic | $CO_2$ | 292.67 | 0.46 | 32.43 | 11.41 | 148 | 11.5 | 2.1 |
| Pessimistic | $CO_2$ | 1839.19 | 1.97 | 6716.35 | 34.62 | 100 | 4.5 | 0.8 |
| No CCS + CCU + DAC | $CO_2$ | 3707.48 | 53.90 | 1775.74 | 2.49 | 100 | 5.4 | 0.9 |
| Large bioenergy | $CO_2$ | 340.99 | 0.59 | 69.68 | 9.17 | 148 | 11.3 | 2.0 |
| Large bioenergy + Small carbon storage | $CO_2$ | 229.12 | 8.52 | 452.10 | 0.82 | 140 | 7.6 | 1.5 |
| Small bioenergy + Large carbon storage | $CO_2$ | 480.65 | 0.75 | 1992.76 | 15.93 | 105 | 6.1 | 1.1 |
| No nuclear | $CO_2$ | 489.97 | 0.80 | 131.23 | 19.52 | 120 | 7.2 | 1.3 |

**Table 3. Parameter values of global MAC curves for energy-related $CO_2$ emissions derived from GET and associated limits on abatement.** See equation (1) for parameters $a$, $b$, $c$, and $d$. The units for $a$ and $c$ are UD\$2010/t$CO_2$. MaxABL denotes the maximum abatement level (%) of each gas indicated from GET simulation data. Max1st and Max2nd represent the maximum first and second derivatives ((%/year) and (%/year)$^2$), respectively, of abatement changes.

## 4. Validation tests for emIAM-ACC2

### 4.1 ACC2 model

To validate the performance of our MAC curves emulating IAM responses (i.e., emIAM), we couple emIAM with the ACC2 model. ACC2 dates back impulse response functions of the global carbon cycle and climate system (Hasselmann et al., 1997; Hooss et al., 2001; Bruckner et al., 2003). The model was later developed to a simple climate model with a full set of climate forcers (Tanaka et al., 2007) and then the current form (Tanaka et al., 2013; Tanaka and O'Neill, 2018; Tanaka et al., 2021): a simple climate-economy model that consists of i) carbon cycle, ii) atmospheric chemistry, iii) physical climate, and iv) mitigation modules.

The representations of natural Earth system processes in the first three modules of ACC2 are at the global-annual-mean level as in other simple climate models (Joos et al., 2013; Nicholls et al., 2020). The carbon cycle module falls into the category of box models (Mackenzie and Lerman, 2006) and the physical climate module is a heat diffusion model DOECLIM (Kriegler, 2005). ACC2 covers a comprehensive set of direct and indirect climate forcers: $CO_2$, $CH_4$, $N_2O$, $O_3$, $SF_6$, 29 species of halocarbons, OH, $NO_x$, CO, VOC, aerosols (both radiative and cloud interactions), and stratospheric $H_2O$. The model captures key nonlinearities, for example, those associated with $CO_2$ fertilization, tropospheric $O_3$ production from $CH_4$, and ocean heat diffusion. Uncertain parameters are optimized (Tanaka et al., 2009a, b; Tanaka and Raddatz, 2011) based on an inverse estimation theory (Tarantola, 2005). The equilibrium climate sensitivity is assumed at 3 °C, the best estimate of IPCC (2021).

The mitigation module contains a set of global MAC curves for $CO_2$, $CH_4$, and $N_2O$ (Johansson, 2011; Azar et al., 2013), which is a previous version of MAC curves to be replaced with the MAC curves discussed in this study. ACC2 can be used to derive $CO_2$, $CH_4$, and $N_2O$ emission pathways based on a cost-effectiveness approach. That is, the model can calculate least-cost emission pathways for the three gases since year 2020, while meeting a climate target (e.g., 2 °C warming target) with the lowest total cumulative mitigation costs in terms of the net present value. The model is written in GAMS and numerically solved using CONOPT3 and CONOPT4, solvers for nonlinear programming or nonlinear optimization problems.

In this study, we replace the existing set of MAC curves with the variety of global and regional MAC curves obtained from our study. We further replace the limits on abatement (i.e., upper limits on abatement levels and their first and second derivatives) with those obtained from this study. We assume a 5% discount rate in the validation tests, a rate commonly assumed in IAMs (Emmerling et al., 2019), which is also consistent with some of the IAMs analyzed here such as MESSAGE and GET. In fact, we were not able to find the estimates of discount rates used in ENGAGE IAMs, but we inferred the discount rate used in MESSAGE by comparing Figures SI 1.2-1 and 1.2-2 of Riahi et al. (2021) with data in ENGAGE Scenario Explorer. Note that a 4% discount rate was used as default in recent studies using ACC2 (Tanaka and O'Neill, 2018; Tanaka et al., 2021). We discuss the results until 2100 thus consider the mitigation costs until 2100 in scenario optimizations. With the updates described above, we generate cost-effective pathways through emIAM-ACC2.

emIAM-ACC2 (and ACC2 with the previous version of MAC curves) can be regarded broadly as an IAM, that is, a simple cost-effective IAM considering global mitigation costs relative to an assumed baseline. In terms of the level of simplicity, emIAM-ACC2 is akin to the DICE model (Nordhaus, 2017) and other simple cost-benefit IAMs informing the social cost of carbon (Errickson et al., 2021; Rennert et al., 2022). However, emIAM-ACC2 does not have an economic growth model and does not consider climate damage. In this study, emIAM-ACC2 is characterized as a climate-economy model, but not an IAM, to distinguish itself from the more complex IAMs emulated by the MAC curves. emIAM-ACC2 is also different from these complex IAMs, which are usually not directly coupled with a climate model, with a previous version of GET (Azar et al., 2013) being an exception. ACC2 itself is a hybrid of a simple climate model and a climate-economy model, depending on how the model is used (i.e., with or without the mitigation module). For the sake of discussion, ACC2 is characterized as a simple climate model in this paper when it is coupled with MAC curves obtained from this study.

## 4.2 Experimental setups for the validation tests

The emission pathways of ENGAGE IAMs were generated under a series of cumulative carbon budgets (Section 2.1). Those of GET were calculated with a series of carbon price pathways (Section 2.2). All these pathways are not directly related to a temperature target, which is typically used as a constraint for ACC2. Given this, we validated the performance of emIAM-ACC2 successively by applying a constraint first on the cumulative emission budget (Test 1) and then on the global-mean temperature (Tests 2 to 4). Four types of experiments were progressively performed as summarized in Table 4. Test 1 mimics a condition for how the original IAM simulations were carried out. Thus, it can be regarded as a test for MAC curves, strictly speaking. Tests 2 to 4 are more practical test to check how MAC curves can work with a simple climate model in an applied setting. However, the settings of these three tests deviate from how the original IAM simulations were performed. The outcomes of these three tests are influenced by how the temperature target is set.

|  | Test 1 | Test 2 | Test 3 | Test 4 |
|---|---|---|---|---|
| **Target** | Emission budget | 2100 temperature | 2100 temperature | 2100 temperature Peak temperature |
| **Variable** | Separately gas by gas | Separately gas by gas | Simultaneously all three gases | Simultaneously all three gases |

**Table 4. Experimental designs of the validation tests for emIAM-ACC2.** See text for details.

- Test 1: Constraint on the cumulative emission budget of each gas. We generate least-cost emission pathways with a cap on cumulative emissions of each gas separately (total anthropogenic $CO_2$, $CH_4$, and $N_2O$ emissions for ENGAGE IAMs; energy-related $CO_2$ emissions for GET). The cap on $CO_2$ for an ENGAGE IAM is equivalent to the cumulative carbon budget as specified in each ENGAGE IAM simulation. The cap on $CO_2$ for GET was calculated from the output of GET, which was simulated under carbon price pathways. The caps on $CH_4$ and $N_2O$ for ENGAGE IAMs were obtained by calculating respective cumulative emissions from 2019 to 2100. Note that the cumulative $CH_4$ budget, or an emission budget of short-lived gases in general, does not offer any useful physical interpretation, while the cumulative $CO_2$ budget, or an emission budget of long-lived gases, can be an indicator of the global-mean temperature change (Matthews et al., 2009; Allen et al., 2022). It should also be noted that this experiment does not directly make use of the carbon cycle, atmospheric chemistry, physical climate modules of ACC2 (i.e., simple climate model) since these modules do not influence the results. But this test is about the way how the cumulative emission budget can be distributed over time, which depends on the MAC curves and the limits on abatement (i.e., upper limits on abatements and their first and second derivatives), with the total abatement costs being minimized.

- Test 2: Constraint on the end-of-the-century warming for one gas at a time. We first use ACC2 to calculate the temperature pathway from each carbon budget scenario of each IAM. The end-of-the-century temperature is used as a constraint on emIAM-ACC2. To keep consistency with the emission budget, this test does not use temperature data found in the ENGAGE Scenario Explorer, which were calculated using different simple climate models (Xiong et al., 2022). We calculate least-cost emission pathways only for one gas at a time ($CO_2$, $CH_4$, or $N_2O$ for ENGAGE IAMs). For example, when we compute a least-cost emission pathway for $CO_2$, the $CH_4$ and $N_2O$ emissions follow the respective pathways from the corresponding carbon budget scenario available in the ENGAGE Scenario Explorer. This test validates the temporal distribution of emissions under an end-of-the-century warming target with global MAC curves and additionally the trade-off among different regions with regional MAC curves; however, it does not validate the trade-off among different gases.

- Test 3: Constraint on the end-of-the-century warming for three gases simultaneously. This test is the same as Test 2, except that least-cost emission pathways are calculated simultaneously for three gases ($CO_2$, $CH_4$, and $N_2O$ for ENGAGE IAMs). This test validates not only the aspects described for Test 2 but also the trade-offs among different gases. It should be noted that we do not use GWP100 in emIAM-ACC2 to generate least-cost emissions pathways for $CO_2$, $CH_4$, and $N_2O$. In other words, abatement levels among the three gases are determined directly through the MAC curves without being constrained by GWP100. It is well-known that a use of GWP100 in an IAM leads to a deviation from the cost-effective solution (O'Neill, 2003; Reisinger et al., 2013; van den Berg et al., 2015; Tanaka et al., 2021). Although the deviation is probably not very large in the scenarios simulated with the IAMs here, this can be a small source of discrepancy between the original and reproduced emission pathways.

- Test 4: Constraint on the end-of-the-century warming and the mid-century peak warming for three gases simultaneously. This test is the same as Test 3, except that the maximum temperature in mid-century is used as an additional constraint on emIAM-ACC2. The peak temperature was taken from the temperature calculation using ACC2 as done in Test 2 for each carbon budget scenario of each IAM. The constraint of the mid-century peak warming aims to influence near-term $CH_4$ emissions, which are known to have strong impacts on peak temperatures in mid-century but little impacts on end-of-the-century temperatures (Shoemaker et al., 2013; Sun et al., 2021; Xiong et al., 2022; McKeough, 2022).

There are further technical notes applied to all four tests above. When the scenario allows only a limited or no temperature overshoot (i.e., scenarios without 'f'; see Section 2.1), we impose a condition prohibiting net negative $CO_2$ emissions on emIAM-ACC2. In Test 1, when the scenario allows a temperature overshoot (i.e., scenarios with 'f'), we assume that a carbon budget can be interpreted simply as a net budget as commonly assumed in the IAM community, although such an assumption may not hold under large temperature overshoot scenarios (Tachiiri et al., 2019; Melnikova et al., 2021; Zickfeld et al., 2021). For INDCi2030 scenarios, which follow NDCs until 2030, we impose the original scenarios until 2030 and perform the optimization from 2030. For NPi2020 scenarios, on the other hand, we perform the optimization from 2020. For emissions scenarios of all GHGs and air pollutants other than the three gases, we prescribe corresponding scenarios from ENGAGE Scenario Explore or the most proximate SSP in the case of GET.

It is important to note that the outcome of the tests described above needs to be interpreted differently, depending on whether the IAM is an intertemporal optimization model or a recursive dynamic model (Table 1) (Babiker et al., 2009; Guivarch and Rogelj, 2017; Melnikov et al., 2021). While the temporal distribution of emission abatement is internally calculated in an intertemporal optimization model, it is usually an a priori assumption in a recursive dynamic model and determined either by a given emission pathway or by a given carbon price pathway. In a recursive dynamic model, the underlying economic and energy-related relationships that determine the temporal distribution of emission abatement may not be consistent with those used to allocate emission abatement across sectors and regions at each time step.

### 4.3 Results from the validation tests

Figure 8 provides an overview of the validation results, using REMIND as an example. Overall, emIAM-ACC2 has closely reproduced original $CO_2$ emission pathways from REMIND in the series of four tests. The outcomes for $CH_4$ and $N_2O$ were generally also satisfactory if not as successful as those for $CO_2$ in general. When the test was performed for each gas with the emission budget (Test 1), the results were good for all three gases. The results were similar with the 2100 temperature target for each gas (Test 2), except for a minor discrepancy arising from a small rise in emissions at the end of the century. A small rise in emission is known to occur in ACC2 before a temperature target is achieved after an overshoot due to the inertia of the system (Tanaka et al., 2021). However, when such a test was performed simultaneously for three gases (Test 3), the results indicated discrepancies in near-term $CH_4$ pathways from low carbon budget cases and late-century $CH_4$ and $N_2O$ pathways from high carbon budget cases. The discrepancy of near-term $CH_4$ emissions seemed to have been caused by the trade-off between $CH_4$ and $N_2O$: the $N_2O$ MAC curve underestimates the prices at high abatement levels (above 20% for $N_2O$) (Figure S20), which might have led to an overestimate of $N_2O$ abatements and, in turn, an underestimate of $CH_4$ abatements. The discrepancy for near-term $CH_4$ emissions was narrowed down with the additional constraint on peak temperatures in mid-century (Test 4). $CH_4$ abatements tend to be incentivized later in the century in the cost optimization of ACC2 with the high discount rate of 5% (Tanaka et al., 2021). This effect can be compensated by the additional constraint on peak temperature in mid-century because near-term $CH_4$ emissions can strongly influence mid-century temperatures (Shoemaker et al., 2013; Sun et al., 2021; Xiong et al., 2022; McKeough, 2022).

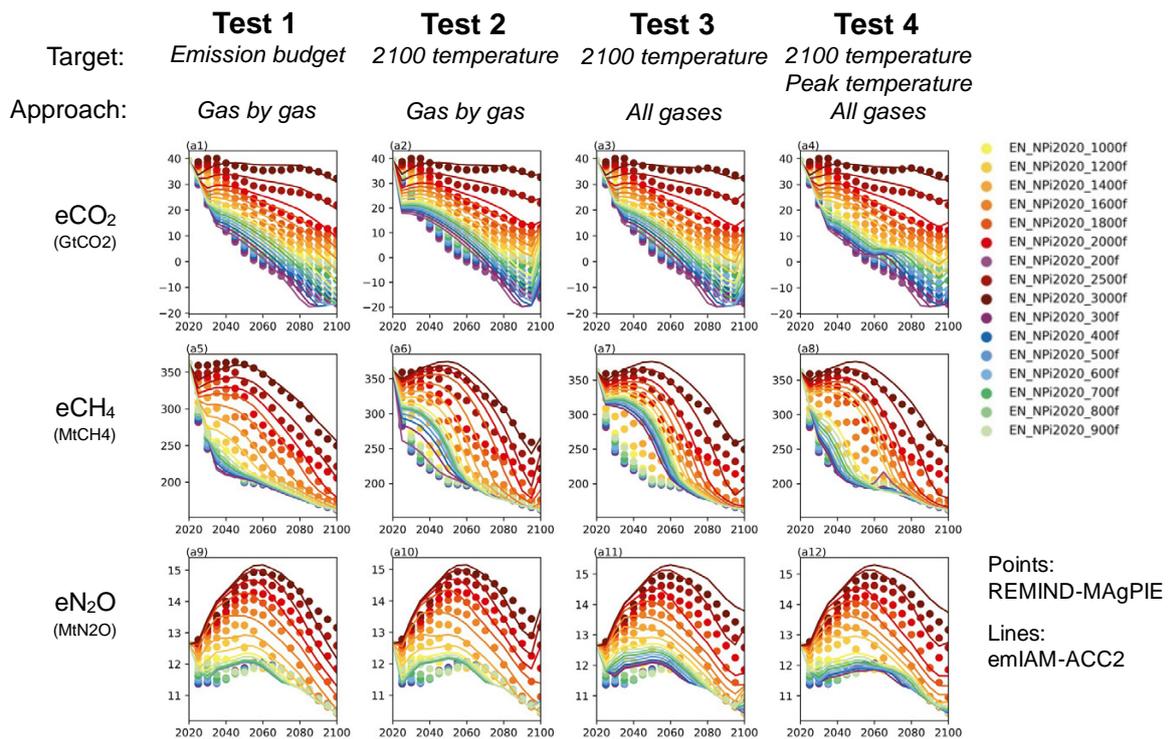

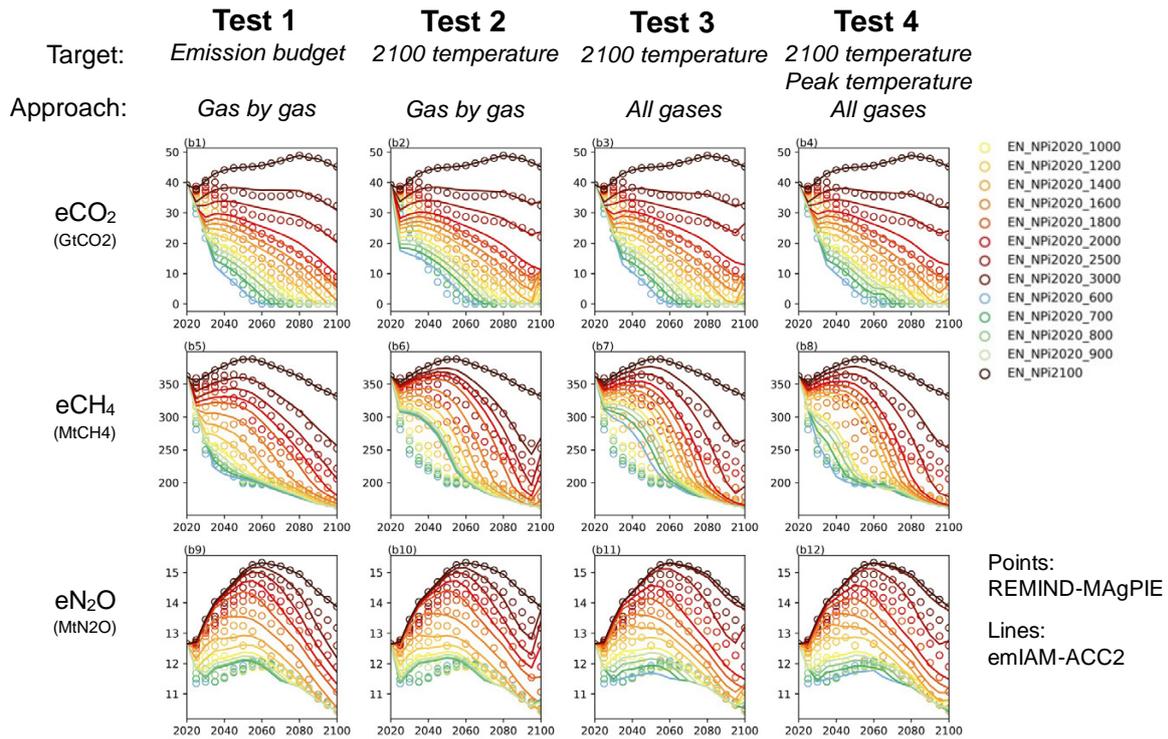

**Figure 8. Overview of the validation results for emIAM-ACC2 with REMIND as an example.** The outcomes for scenarios with "f" (i.e., net-negative emissions scenarios (filled circles)) are shown in the upper set of panels; those for scenarios without "f" (i.e., net-zero $CO_2$ emissions scenarios (open circles)) are in the lower set of panels. The points show the original emission pathways from REMIND obtained from the ENGAGE Scenario Explorer; the lines show the emission pathways reproduced from emIAM-ACC2. The same color is used for each pair of original and reproduced pathways. Scenario names indicate respective cumulative carbon budgets for the period 2019 – 2100 in $GtCO_2$. NPi2100 is the baseline scenario for our analysis (black open circles). For the sake of presentation, only the outcomes of NPi2020 scenarios, which consider currently implemented national policies, are presented; the outcomes of INDCi2030 scenarios, which further consider national emission pledges until 2030, are not shown here.

Figure 9 shows the validation results from Test 4 for all nine ENGAGE IAMs (global total anthropogenic $CO_2$, $CH_4$, and $N_2O$ emissions) and GET with different technology portfolios (global energy-related $CO_2$, emissions). The entire validation results from Tests 1 to 4 can be found in Figures S90-S108, Figures S128-S146, Figures S166-S183, and Figures S202-S216, respectively. $CO_2$ emission pathways were generally well reproduced through emIAM-ACC2 for all ENGAGE IAMs. The outcomes for $CH_4$ and $N_2O$ were not as good as those for $CO_2$: only a subset of ENGAGE IAMs such as REMIND and WITCH was reasonably well captured by emIAM-ACC2. Some of the mismatches can be explained by the poor fits of $N_2O$ MAC curves from COFFEE and TIAM (Figure S11). The general difficulty in capturing IMAGE through MAC curves (Figure S17) can be seen in the mismatches for IMAGE in Figure 9. It is also worth mentioning that, in spite of very good fits of MAC curves of GEM (Figure S16), $CH_4$ and $N_2O$ emission pathways of GEM were not well reproduced. The results for GET were also generally good, but the Large bioenergy + Small carbon storage portfolio gave a relatively poor result. This might have been caused by the relatively poor fit of the MAC curve for this technology portfolio, compared to those from other portfolios (Figure 6).

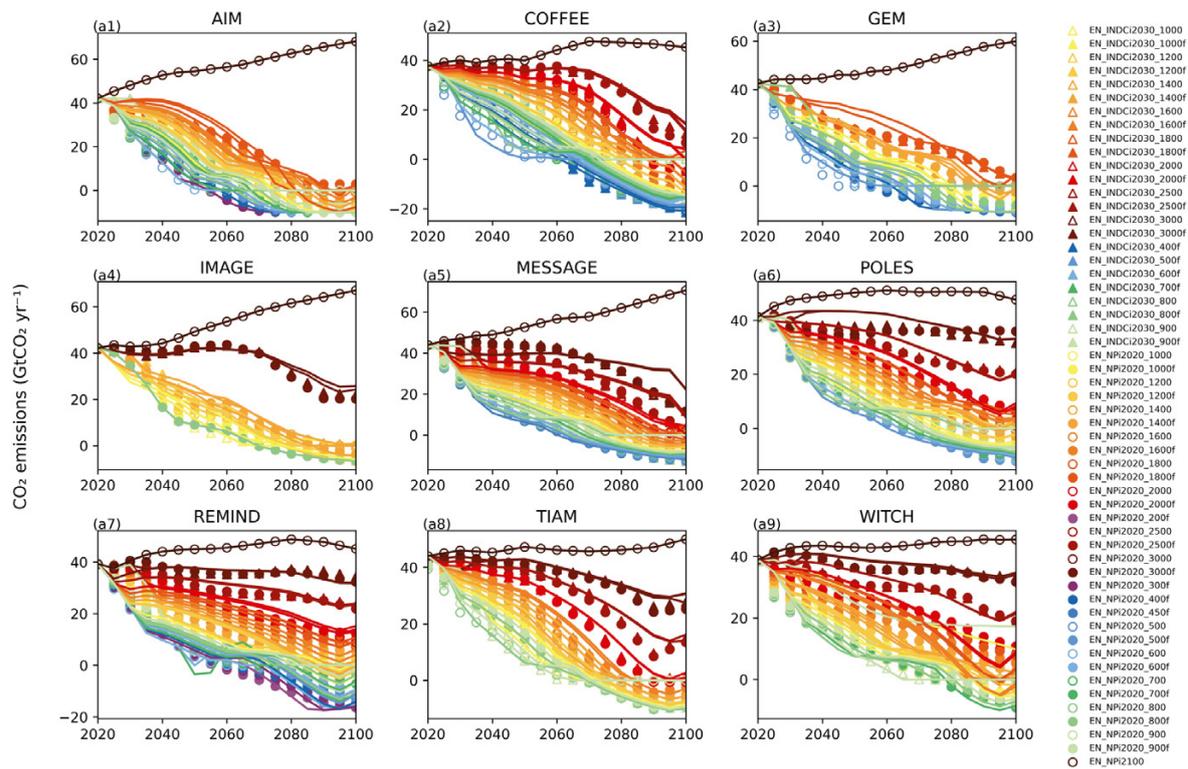

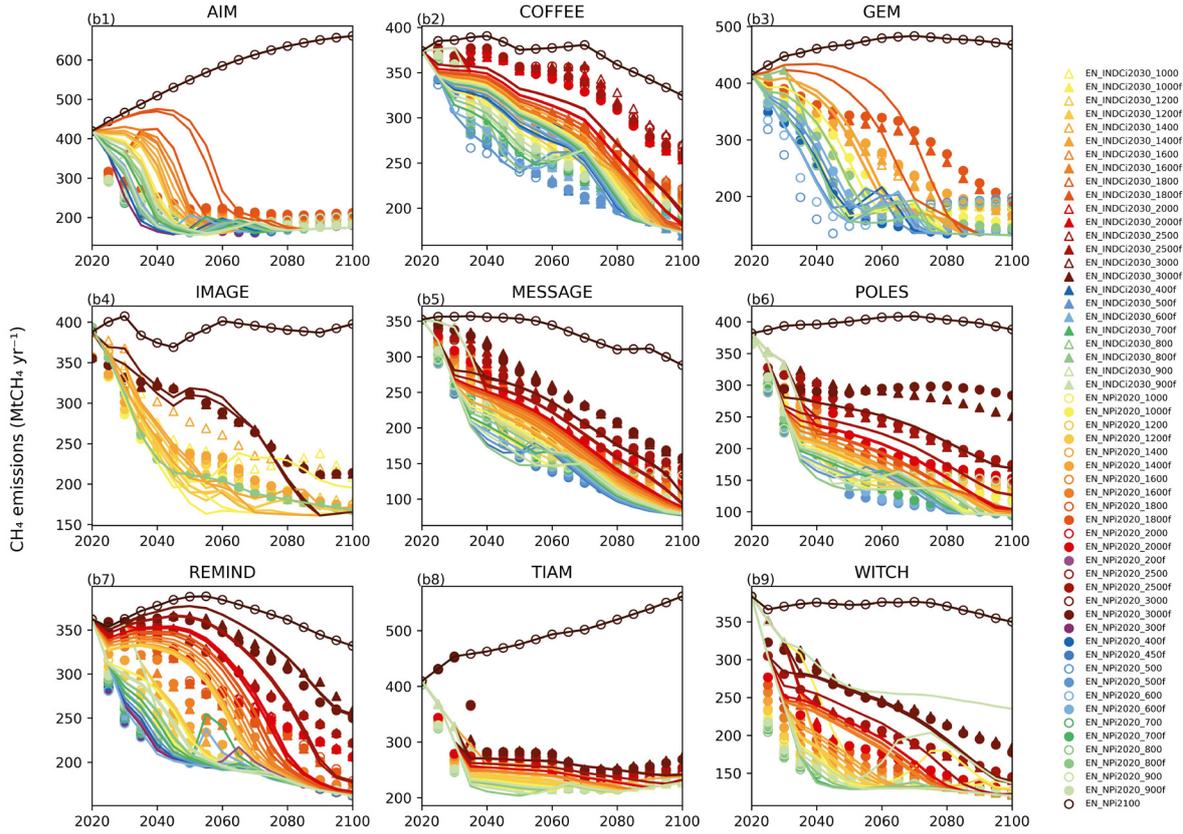

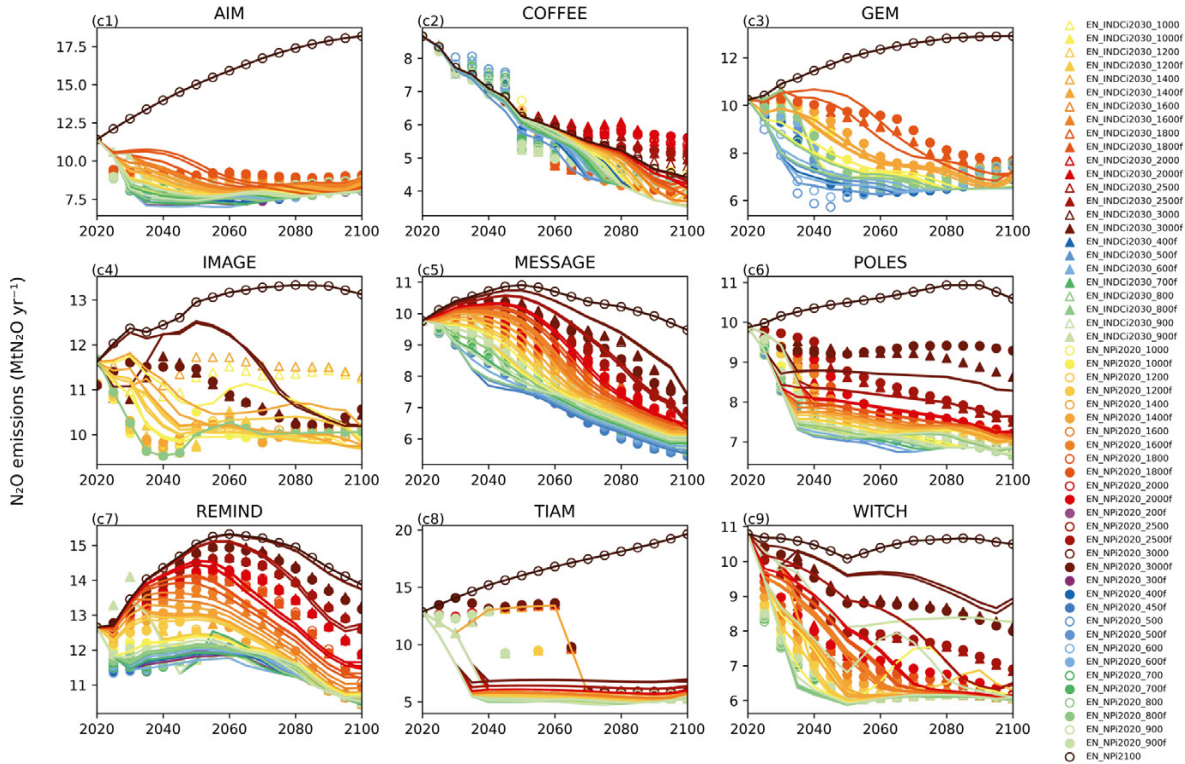

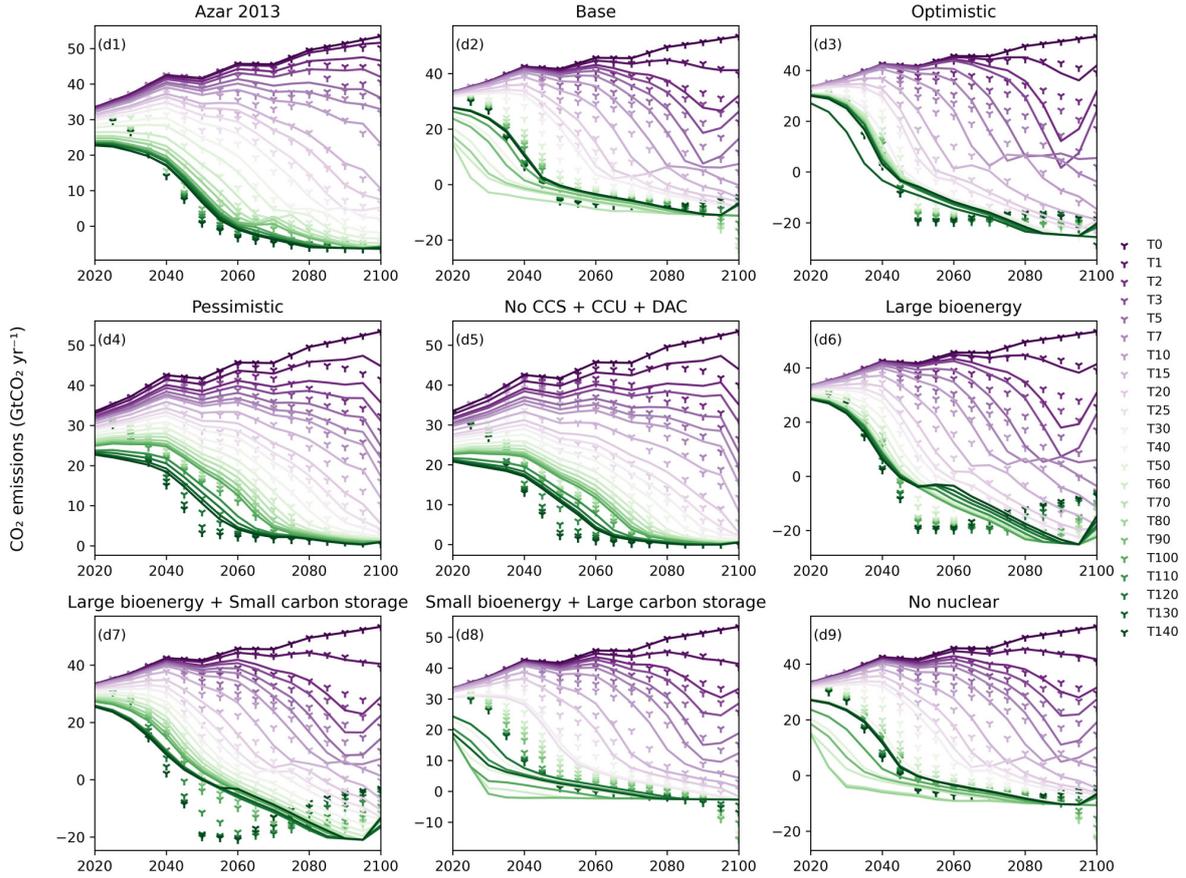

**Figure 9. Original and reproduced global emission pathways from Test 4 for nine ENGAGE IAMs (total anthropogenic $CO_2$, $CH_4$, and $N_2O$ emissions) and GET (energy-related $CO_2$ emissions) with different technology portfolios.** The first three sets of panels are from the nine ENGAGE IAMs. The last set of panels is from GET with different technology portfolios. The points show the original emission pathways from ENGAGE IAMs and GET; the lines show the emission pathways reproduced from emIAM-ACC2. The same color is used for each pair of original and reproduced pathways. For the legend of the panels for ENGAGE IAMs, see the caption of Figure 8. For the legend of the panel for GET, the number indicates the initial carbon price (US$2010/t$CO_2$), from which the carbon price grows 5% each year.

## 4.4 Statistics of the validation tests

To measure to what extent emission pathways obtained from emIAM-ACC2, denoted as $y$, agree with original pathways from ENGAGE IAMs and GET, denoted as $x$, we calculate the following two different indicators for samples: i) ordinary Pearson's correlation coefficient $r_P$ and ii) Lin's concordance coefficient $r_C$. Each of these indicators is discussed below.

First, because of the prevalent use of $r_P$ and its square form (i.e., coefficient of simple determination, so-called $r^2$) in numerous applications, we use $r_P$ as a reference for comparison, although $r_P$ is known to be inappropriate for testing agreement: it is suited to test the strength of linear relationship, but not the strength of agreement (Martin Bland and Altman, 1986; Cox, 2006). More specifically, $r_P$ (and $r^2$) shows the strength of linear regression line $\acute{y} = \alpha\acute{x} + \beta$, not necessarily $\acute{y} = \acute{x}$, a special case of agreement. Note that it is possible to calculate $r^2$ based on $\acute{y} = \acute{x}$ by using the sum of square of residuals and the total sum of squares (i.e., not equation (2)); however, if $\acute{y} = \acute{x}$ is a very poor regression line, $r^2$ can become negative (page 21 of Hayashi (2000)) and cannot be interpreted as a square of $r_P$. Other arguments that suggest a more restricted use of $r_P$ can be found elsewhere (Ricker, 1973; Laws, 1997; Tanaka and Mackenzie, 2005). For our application, $r_P$ is defined as below.

$$r_P = \frac{\sum_{i=1}^{l}\sum_{j=1}^{m}(x_{i,j} - \bar{x})(y_{i,j} - \bar{y})}{\sqrt{\sum_{i=1}^{l}\sum_{j=1}^{m}(x_{i,j} - \bar{x})^2}\sqrt{\sum_{i=1}^{l}\sum_{j=1}^{m}(y_{i,j} - \bar{y})^2}} \tag{2}$$

where $x_{i,j}$ and $y_{i,j}$ are the original and reproduced emission, respectively, for year $i$ (for $i = 1, \dots, l$) under scenario $j$ (for $j = 1, \dots, m$). $\bar{x}$ and $\bar{y}$ are the mean of $x_{i,j}$ and $y_{i,j}$, respectively, over $i$ and $j$. $r_P$ can change between -1 and 1. When it is 1, the samples have a perfect linear relationship, which is a necessary condition for a perfect agreement. When it is 0, there is no linear relationship in the samples.

Second, $r_C$ is a more appropriate indicator for measuring agreement than $r_P$ (Lin, 1989; Barnhart et al., 2007; Lin et al., 2012). $r_C$ is defined as follows.

$$r_C = \frac{2s_{xy}}{s_x^2 + s_y^2 + (\bar{x} - \bar{y})^2} \tag{3}$$

where $s_x^2$ and $s_y^2$ are the variance of $x_{i,j}$ and $y_{i,j}$, respectively. That is, $s_x^2 = \frac{1}{l \times m}\sum_{i=1}^{l}\sum_{j=1}^{m}(x_{i,j} - \bar{x})^2$ and $s_y^2 = \frac{1}{l \times m}\sum_{i=1}^{l}\sum_{j=1}^{m}(y_{i,j} - \bar{y})^2$, respectively. $s_{xy}$ is the covariance of $x_{i,j}$ and $y_{i,j}$. That is, $s_{xy} = \frac{1}{l \times m}\sum_{i=1}^{l}\sum_{j=1}^{m}(x_{i,j} - \bar{x})(y_{i,j} - \bar{y})$. $r_C$ also distributes between -1 and 1. When it is 1, 0, and -1, it indicates a perfect concordance, no concordance, and a perfect discordance (or reverse concordance), respectively. $r_C$ is commonly interpreted either similar to $r_P$ or in the following way: >0.99, almost perfect; 0.95 to 0.99, substantial; 0.90 to 0.95, moderate; <0.90, poor (Akoglu, 2018). An underlying assumption for this parametric statistic is that the population follows Gaussian distributions.

| | | | AIM | COFFEE | GEM | IMAGE | MESSAGE | POLES | REMIND | TIAM | WITCH |
|---|---|---|---|---|---|---|---|---|---|---|---|
| $r_P$ | Test 1 | $CO_2$ | 0.986 | 0.990 | 0.983 | 0.975 | 0.981 | 0.960 | 0.976 | 0.985 | 0.937 |
| | | $CH_4$ | 0.962 | 0.977 | 0.964 | 0.973 | 0.975 | 0.967 | 0.976 | 0.927 | 0.985 |
| | | $N_2O$ | 0.921 | 0.966 | 0.948 | 0.875 | 0.985 | 0.964 | 0.962 | 0.466 | 0.975 |
| | Test 2 | $CO_2$ | 0.984 | 0.996 | 0.982 | 0.979 | 0.991 | 0.985 | 0.962 | 0.997 | 0.980 |
| | | $CH_4$ | 0.667 | 0.943 | 0.805 | 0.890 | 0.965 | 0.954 | 0.941 | 0.951 | 0.954 |
| | | $N_2O$ | 0.908 | 0.962 | 0.777 | 0.875 | 0.985 | 0.969 | 0.962 | 0.531 | 0.978 |
| | Test 3 | $CO_2$ | 0.983 | 0.989 | 0.989 | 0.974 | 0.974 | 0.981 | 0.973 | 0.879 | 0.973 |
| | | $CH_4$ | 0.678 | 0.914 | 0.898 | 0.886 | 0.968 | 0.934 | 0.903 | 0.740 | 0.934 |
| | | $N_2O$ | 0.933 | 0.950 | 0.978 | 0.600 | 0.977 | 0.962 | 0.945 | 0.549 | 0.956 |
| | Test 4 | $CO_2$ | 0.989 | 0.990 | 0.985 | 0.990 | 0.981 | 0.990 | 0.981 | 0.995 | 0.953 |
| | | $CH_4$ | 0.879 | 0.916 | 0.951 | 0.954 | 0.978 | 0.951 | 0.955 | 0.953 | 0.940 |
| | | $N_2O$ | 0.954 | 0.955 | 0.964 | 0.713 | 0.977 | 0.957 | 0.960 | 0.607 | 0.935 |
| $r_C$ | Test 1 | $CO_2$ | 0.981 | 0.985 | 0.981 | 0.974 | 0.980 | 0.955 | 0.975 | 0.983 | 0.918 |
| | | $CH_4$ | 0.957 | 0.977 | 0.960 | 0.968 | 0.974 | 0.966 | 0.976 | 0.927 | 0.984 |
| | | $N_2O$ | 0.916 | 0.964 | 0.946 | 0.863 | 0.981 | 0.963 | 0.962 | 0.385 | 0.972 |
| | Test 2 | $CO_2$ | 0.979 | 0.995 | 0.980 | 0.978 | 0.991 | 0.984 | 0.962 | 0.997 | 0.977 |
| | | $CH_4$ | 0.549 | 0.932 | 0.730 | 0.833 | 0.960 | 0.942 | 0.919 | 0.949 | 0.941 |
| | | $N_2O$ | 0.901 | 0.958 | 0.764 | 0.863 | 0.982 | 0.969 | 0.961 | 0.454 | 0.975 |
| | Test 3 | $CO_2$ | 0.976 | 0.986 | 0.988 | 0.973 | 0.971 | 0.978 | 0.972 | 0.877 | 0.964 |
| | | $CH_4$ | 0.558 | 0.908 | 0.880 | 0.858 | 0.962 | 0.930 | 0.877 | 0.715 | 0.919 |
| | | $N_2O$ | 0.914 | 0.947 | 0.975 | 0.572 | 0.975 | 0.948 | 0.944 | 0.494 | 0.945 |
| | Test 4 | $CO_2$ | 0.987 | 0.987 | 0.982 | 0.990 | 0.978 | 0.987 | 0.981 | 0.993 | 0.949 |
| | | $CH_4$ | 0.852 | 0.911 | 0.934 | 0.931 | 0.964 | 0.933 | 0.944 | 0.946 | 0.935 |
| | | $N_2O$ | 0.935 | 0.954 | 0.956 | 0.699 | 0.972 | 0.940 | 0.960 | 0.482 | 0.920 |

| | GET technology portfolio | Azar2013 | Base | Optimistic | Pessimistic | No_cap | L_bio | L_bio/S_str | S_bio/L_str | No_nc |
|---|---|---|---|---|---|---|---|---|---|---|
| $r_P$ | Test 1 | 0.992 | 0.983 | 0.973 | 0.991 | 0.981 | 0.964 | 0.964 | 0.984 | 0.984 |
| | Test 2 | 0.985 | 0.943 | 0.973 | 0.982 | 0.968 | 0.967 | 0.965 | 0.904 | 0.938 |
| | Test 4 | 0.988 | 0.945 | 0.979 | 0.982 | 0.967 | 0.971 | 0.967 | 0.910 | 0.939 |
| $r_C$ | Test 1 | 0.992 | 0.983 | 0.972 | 0.991 | 0.979 | 0.963 | 0.964 | 0.983 | 0.984 |
| | Test 2 | 0.985 | 0.938 | 0.973 | 0.981 | 0.963 | 0.966 | 0.964 | 0.885 | 0.932 |
| | Test 4 | 0.988 | 0.940 | 0.979 | 0.981 | 0.963 | 0.971 | 0.967 | 0.894 | 0.933 |

**Figure 10. Statistical validation of global emission pathways reproduced from emIAM-ACC2 with original emission pathways from nine ENGAGE IAMs and GET.** The upper and lower panels are the results for ENGAGE IAMs (global total anthropogenic $CO_2$, $CH_4$, and $N_2O$ emissions) and GET (global energy-related $CO_2$ emissions), respectively. The figure shows two indicators: i) ordinary Pearson's correlation coefficient $r_P$ and ii) Lin's concordance coefficient $r_C$. The higher the value of the indicator is, the darker the color of the cell is. See text for the details of these statistical indicators.

| | | | Test 1 | | | | | Test 2 | | | | | Test 3 | | | | | Test 4 | | | | |
|---|---|---|---|---|---|---|---|---|---|---|---|---|---|---|---|---|---|---|---|---|---|---|
| | | | AIM | COFFEE | GEM | IMAGE | MESSAGE | AIM | COFFEE | GEM | IMAGE | MESSAGE | AIM | COFFEE | GEM | IMAGE | MESSAGE | AIM | COFFEE | GEM | IMAGE | MESSAGE |
| $r_P$ | $CO_2$ | SUBSAFR | 0.954 | 0.528 | 0.978 | 0.963 | 0.979 | 0.951 | 0.960 | 0.973 | 0.966 | 0.989 | 0.954 | 0.951 | 0.971 | 0.956 | 0.989 | 0.961 | 0.952 | 0.982 | 0.978 | 0.988 |
| | | CHN | 0.974 | 0.893 | 0.978 | 0.966 | 0.990 | 0.970 | 0.991 | 0.973 | 0.965 | 0.995 | 0.975 | 0.988 | 0.969 | 0.957 | 0.992 | 0.982 | 0.989 | 0.980 | 0.984 | 0.994 |
| | | EUWE | 0.979 | 0.706 | 0.974 | 0.973 | 0.994 | 0.977 | 0.930 | 0.969 | 0.979 | 0.996 | 0.984 | 0.924 | 0.975 | 0.974 | 0.993 | 0.982 | 0.923 | 0.973 | 0.988 | 0.995 |
| | | SQUASIA | 0.965 | 0.578 | 0.964 | 0.957 | 0.958 | 0.961 | 0.914 | 0.952 | 0.956 | 0.974 | 0.964 | 0.917 | 0.956 | 0.953 | 0.957 | 0.976 | 0.915 | 0.966 | 0.970 | 0.982 |
| | | LATAME | 0.964 | 0.500 | 0.962 | 0.969 | 0.964 | 0.962 | 0.965 | 0.948 | 0.906 | 0.975 | 0.968 | 0.964 | 0.952 | 0.904 | 0.968 | 0.966 | 0.964 | 0.960 | 0.964 | 0.983 |
| | | MIDEAST | 0.983 | 0.575 | 0.964 | 0.952 | 0.979 | 0.982 | 0.947 | 0.946 | 0.957 | 0.986 | 0.985 | 0.946 | 0.941 | 0.952 | 0.981 | 0.987 | 0.945 | 0.957 | 0.976 | 0.977 |
| | | NORAM | 0.981 | 0.838 | 0.959 | 0.963 | 0.990 | 0.980 | 0.962 | 0.957 | 0.965 | 0.993 | 0.986 | 0.958 | 0.958 | 0.992 | 0.982 | 0.981 | 0.957 | 0.961 | 0.977 | 0.989 |
| | | PACOECD | 0.969 | 0.329 | 0.966 | 0.944 | 0.993 | 0.967 | 0.979 | 0.960 | 0.944 | 0.995 | 0.973 | 0.976 | 0.964 | 0.937 | 0.993 | 0.976 | 0.976 | 0.968 | 0.954 | 0.994 |
| | | REFECO | 0.986 | 0.592 | 0.944 | 0.969 | 0.993 | 0.985 | 0.968 | 0.934 | 0.971 | 0.996 | 0.987 | 0.965 | 0.943 | 0.974 | 0.993 | 0.988 | 0.964 | 0.950 | 0.982 | 0.996 |
| | | OTASIAN | 0.979 | 0.385 | 0.970 | 0.957 | 0.988 | 0.977 | 0.982 | 0.962 | 0.957 | 0.995 | 0.983 | 0.978 | 0.959 | 0.947 | 0.987 | 0.983 | 0.978 | 0.972 | 0.972 | 0.988 |
| | | ROW | 0.772 | 0.040 | 0.922 | 0.873 | 0.659 | 0.767 | 0.823 | 0.890 | 0.883 | 0.616 | 0.790 | 0.836 | 0.882 | 0.870 | 0.629 | 0.802 | 0.835 | 0.935 | 0.888 | 0.624 |
| | $CH_4$ | SUBSAFR | 0.952 | 0.890 | 0.949 | 0.838 | 0.956 | 0.556 | 0.861 | 0.680 | 0.691 | 0.776 | 0.567 | 0.838 | 0.786 | 0.669 | 0.838 | 0.878 | 0.859 | 0.931 | 0.750 | 0.927 |
| | | CHN | 0.980 | 0.988 | 0.962 | 0.984 | 0.980 | 0.792 | 0.979 | 0.783 | 0.921 | 0.976 | 0.803 | 0.982 | 0.858 | 0.942 | 0.985 | 0.932 | 0.986 | 0.958 | 0.972 | 0.957 |
| | | EUWE | 0.935 | 0.989 | 0.920 | 0.923 | 0.940 | 0.695 | 0.976 | 0.758 | 0.770 | 0.953 | 0.693 | 0.982 | 0.794 | 0.828 | 0.955 | 0.884 | 0.985 | 0.879 | 0.927 | 0.957 |
| | | SQUASIA | 0.906 | 0.533 | 0.947 | 0.938 | 0.930 | 0.566 | 0.520 | 0.717 | 0.782 | 0.795 | 0.567 | 0.513 | 0.810 | 0.829 | 0.852 | 0.810 | 0.567 | 0.947 | 0.901 | 0.932 |
| | | LATAME | 0.969 | 0.975 | 0.962 | 0.900 | 0.981 | 0.752 | 0.956 | 0.739 | 0.811 | 0.936 | 0.760 | 0.976 | 0.838 | 0.813 | 0.964 | 0.913 | 0.983 | 0.962 | 0.846 | 0.985 |
| | | MIDEAST | 0.966 | 0.953 | 0.958 | 0.801 | 0.877 | 0.689 | 0.953 | 0.785 | 0.718 | 0.844 | 0.697 | 0.949 | 0.857 | 0.721 | 0.852 | 0.897 | 0.958 | 0.962 | 0.832 | 0.913 |
| | | NORAM | 0.924 | 0.942 | 0.952 | 0.933 | 0.979 | 0.634 | 0.906 | 0.698 | 0.750 | 0.982 | 0.630 | 0.949 | 0.808 | 0.797 | 0.979 | 0.864 | 0.955 | 0.952 | 0.906 | 0.983 |
| | | PACOECD | 0.905 | 0.909 | 0.849 | 0.882 | 0.958 | 0.547 | 0.864 | 0.471 | 0.766 | 0.920 | 0.544 | 0.884 | 0.574 | 0.784 | 0.953 | 0.818 | 0.897 | 0.825 | 0.857 | 0.972 |
| | | REFECO | 0.957 | 0.801 | 0.968 | 0.939 | 0.968 | 0.737 | 0.701 | 0.823 | 0.749 | 0.834 | 0.736 | 0.730 | 0.883 | 0.787 | 0.889 | 0.906 | 0.770 | 0.967 | 0.861 | 0.956 |
| | | OTASIAN | 0.964 | 0.970 | 0.953 | 0.973 | 0.994 | 0.695 | 0.960 | 0.807 | 0.905 | 0.956 | 0.701 | 0.964 | 0.861 | 0.931 | 0.973 | 0.895 | 0.972 | 0.953 | 0.972 | 0.990 |
| | | ROW | 0.895 | 0.000 | 0.945 | 0.000 | 0.000 | 0.665 | 0.000 | 0.776 | 0.000 | 0.999 | 0.672 | 0.000 | 0.836 | 0.000 | 0.999 | 0.782 | 0.000 | 0.947 | 0.000 | 0.999 |
| | $N_2O$ | SUBSAFR | 0.969 | 0.953 | 0.867 | 0.859 | 0.902 | 0.965 | 0.924 | 0.885 | 0.857 | 0.905 | 0.979 | 0.974 | 0.905 | 0.776 | 0.896 | 0.982 | 0.973 | 0.897 | 0.803 | 0.906 |
| | | CHN | 0.982 | 0.933 | 0.967 | 0.971 | 0.952 | 0.977 | 0.954 | 0.967 | 0.970 | 0.952 | 0.984 | 0.965 | 0.980 | 0.977 | 0.967 | 0.982 | 0.964 | 0.983 | 0.983 | 0.966 |
| | | EUWE | 0.924 | 0.967 | 0.828 | 0.851 | 0.971 | 0.908 | 0.961 | 0.805 | 0.839 | 0.972 | 0.951 | 0.959 | 0.810 | 0.703 | 0.980 | 0.938 | 0.959 | 0.959 | 0.795 | 0.762 |
| | | SQUASIA | 0.913 | 0.603 | 0.947 | 0.863 | 0.930 | 0.900 | 0.880 | 0.946 | 0.860 | 0.932 | 0.948 | 0.826 | 0.966 | 0.841 | 0.945 | 0.939 | 0.824 | 0.977 | 0.843 | 0.959 |
| | | LATAME | 0.951 | 0.807 | 0.953 | 0.068 | 0.930 | 0.940 | 0.873 | 0.954 | 0.090 | 0.934 | 0.968 | 0.867 | 0.969 | -0.153 | 0.943 | 0.957 | 0.867 | 0.973 | -0.122 | 0.941 |
| | | MIDEAST | 0.874 | 0.735 | 0.963 | 0.986 | 0.935 | 0.859 | 0.793 | 0.963 | 0.983 | 0.940 | 0.929 | 0.762 | 0.971 | 0.990 | 0.965 | 0.922 | 0.770 | 0.968 | 0.992 | 0.974 |
| | | NORAM | 0.915 | 0.899 | 0.874 | 0.864 | 0.930 | 0.898 | 0.906 | 0.903 | 0.863 | 0.934 | 0.948 | 0.978 | 0.925 | 0.807 | 0.962 | 0.948 | 0.978 | 0.974 | 0.839 | 0.974 |
| | | PACOECD | 0.848 | 0.811 | 0.873 | 0.038 | 0.971 | 0.811 | 0.876 | 0.906 | 0.013 | 0.971 | 0.916 | 0.848 | 0.913 | -0.106 | 0.974 | 0.925 | 0.845 | 0.917 | -0.023 | 0.972 |
| | | REFECO | 0.926 | 0.631 | 0.389 | 0.430 | 0.774 | 0.910 | 0.656 | 0.452 | 0.431 | 0.787 | 0.956 | 0.558 | 0.518 | 0.207 | 0.885 | 0.956 | 0.558 | 0.531 | 0.263 | 0.873 |
| | | OTASIAN | 0.925 | 0.725 | 0.933 | 0.739 | 0.953 | 0.908 | 0.830 | 0.935 | 0.717 | 0.957 | 0.955 | 0.797 | 0.949 | 0.532 | 0.973 | 0.952 | 0.800 | 0.939 | 0.571 | 0.979 |
| | | ROW | 0.000 | 0.000 | 0.862 | 0.000 | 0.998 | 0.000 | 0.000 | 0.879 | 0.000 | 0.998 | 0.000 | 0.000 | 0.895 | 0.000 | 0.999 | 0.000 | 0.000 | 0.920 | 0.000 | 0.999 |
| $r_C$ | $CO_2$ | SUBSAFR | 0.945 | 0.430 | 0.976 | 0.959 | 0.978 | 0.941 | 0.960 | 0.970 | 0.980 | 0.989 | 0.949 | 0.950 | 0.962 | 0.953 | 0.988 | 0.958 | 0.952 | 0.972 | 0.978 | 0.985 |
| | | CHN | 0.963 | 0.878 | 0.976 | 0.964 | 0.989 | 0.957 | 0.989 | 0.970 | 0.978 | 0.994 | 0.964 | 0.985 | 0.958 | 0.954 | 0.990 | 0.976 | 0.985 | 0.972 | 0.980 | 0.993 |
| | | EUWE | 0.973 | 0.704 | 0.941 | 0.973 | 0.994 | 0.971 | 0.922 | 0.936 | 0.991 | 0.996 | 0.981 | 0.912 | 0.955 | 0.972 | 0.992 | 0.982 | 0.911 | 0.951 | 0.986 | 0.994 |
| | | SQUASIA | 0.958 | 0.578 | 0.957 | 0.957 | 0.952 | 0.953 | 0.885 | 0.946 | 0.984 | 0.973 | 0.959 | 0.892 | 0.941 | 0.956 | 0.974 | 0.972 | 0.889 | 0.951 | 0.971 | 0.981 |
| | | MIDEAST | 0.958 | 0.531 | 0.961 | 0.951 | 0.974 | 0.957 | 0.939 | 0.943 | 0.997 | 0.984 | 0.983 | 0.940 | 0.924 | 0.951 | 0.980 | 0.986 | 0.938 | 0.939 | 0.974 | 0.962 |
| | | NORAM | 0.976 | 0.767 | 0.957 | 0.962 | 0.985 | 0.974 | 0.955 | 0.956 | 0.977 | 0.991 | 0.983 | 0.949 | 0.952 | 0.956 | 0.991 | 0.981 | 0.948 | 0.953 | 0.976 | 0.986 |
| | | PACOECD | 0.962 | 0.222 | 0.958 | 0.943 | 0.993 | 0.959 | 0.975 | 0.955 | 0.941 | 0.995 | 0.968 | 0.971 | 0.957 | 0.935 | 0.992 | 0.975 | 0.970 | 0.964 | 0.952 | 0.993 |
| | | REFECO | 0.982 | 0.522 | 0.940 | 0.961 | 0.993 | 0.980 | 0.960 | 0.929 | 0.957 | 0.996 | 0.985 | 0.957 | 0.932 | 0.965 | 0.993 | 0.987 | 0.956 | 0.941 | 0.978 | 0.995 |
| | | OTASIAN | 0.975 | 0.273 | 0.968 | 0.963 | 0.987 | 0.972 | 0.978 | 0.961 | 0.971 | 0.991 | 0.982 | 0.974 | 0.954 | 0.941 | 0.986 | 0.983 | 0.973 | 0.968 | 0.968 | 0.986 |
| | | ROW | 0.705 | 0.002 | 0.891 | 0.561 | 0.300 | 0.699 | 0.658 | 0.877 | 0.953 | 0.417 | 0.746 | 0.687 | 0.866 | 0.580 | 0.432 | 0.751 | 0.684 | 0.931 | 0.591 | 0.431 |
| | $CH_4$ | SUBSAFR | 0.944 | 0.885 | 0.942 | 0.793 | 0.953 | 0.512 | 0.853 | 0.537 | 0.621 | 0.713 | 0.514 | 0.826 | 0.727 | 0.628 | 0.814 | 0.876 | 0.856 | 0.921 | 0.625 | 0.911 |
| | | CHN | 0.973 | 0.986 | 0.955 | 0.978 | 0.980 | 0.681 | 0.977 | 0.666 | 0.846 | 0.969 | 0.684 | 0.981 | 0.810 | 0.898 | 0.984 | 0.912 | 0.985 | 0.943 | 0.957 | 0.989 |
| | | EUWE | 0.913 | 0.989 | 0.824 | 0.914 | 0.932 | 0.552 | 0.974 | 0.757 | 0.586 | 0.932 | 0.536 | 0.979 | 0.755 | 0.783 | 0.812 | 0.772 | 0.938 | 0.929 | 0.848 | 0.913 |
| | | SQUASIA | 0.961 | 0.972 | 0.955 | 0.877 | 0.981 | 0.630 | 0.953 | 0.614 | 0.724 | 0.917 | 0.629 | 0.972 | 0.783 | 0.757 | 0.958 | 0.889 | 0.980 | 0.948 | 0.787 | 0.984 |
| | | LATAME | 0.960 | 0.950 | 0.953 | 0.766 | 0.857 | 0.566 | 0.950 | 0.693 | 0.638 | 0.802 | 0.570 | 0.948 | 0.821 | 0.616 | 0.872 | 0.872 | 0.957 | 0.952 | 0.774 | 0.891 |
| | | NORAM | 0.903 | 0.934 | 0.949 | 0.959 | 0.969 | 0.492 | 0.881 | 0.588 | 0.678 | 0.982 | 0.474 | 0.938 | 0.766 | 0.759 | 0.974 | 0.821 | 0.948 | 0.941 | 0.850 | 0.973 |
| | | PACOECD | 0.888 | 0.907 | 0.841 | 0.841 | 0.957 | 0.409 | 0.846 | 0.413 | 0.697 | 0.902 | 0.394 | 0.886 | 0.568 | 0.724 | 0.950 | 0.776 | 0.892 | 0.799 | 0.749 | 0.965 |
| | | REFECO | 0.947 | 0.721 | 0.965 | 0.919 | 0.962 | 0.615 | 0.623 | 0.733 | 0.621 | 0.783 | 0.607 | 0.666 | 0.851 | 0.700 | 0.859 | 0.878 | 0.714 | 0.958 | 0.829 | 0.947 |
| | | OTASIAN | 0.955 | 0.966 | 0.948 | 0.969 | 0.993 | 0.564 | 0.958 | 0.728 | 0.844 | 0.937 | 0.562 | 0.963 | 0.830 | 0.899 | 0.963 | 0.866 | 0.971 | 0.935 | 0.956 | 0.986 |
| | | ROW | 0.767 | 0.000 | 0.923 | 0.000 | 0.000 | 0.452 | 0.000 | 0.707 | 0.000 | 0.999 | 0.461 | 0.000 | 0.776 | 0.000 | 0.999 | 0.675 | 0.000 | 0.948 | 0.000 | 0.999 |
| | $N_2O$ | SUBSAFR | 0.950 | 0.768 | 0.857 | 0.832 | 0.854 | 0.952 | 0.905 | 0.874 | 0.839 | 0.856 | 0.971 | 0.973 | 0.903 | 0.768 | 0.882 | 0.973 | 0.972 | 0.896 | 0.796 | 0.889 |
| | | CHN | 0.969 | 0.925 | 0.961 | 0.958 | 0.926 | 0.963 | 0.954 | 0.960 | 0.958 | 0.927 | 0.970 | 0.961 | 0.978 | 0.961 | 0.959 | 0.971 | 0.960 | 0.983 | 0.972 | 0.958 |
| | | EUWE | 0.916 | 0.959 | 0.693 | 0.835 | 0.970 | 0.871 | 0.939 | 0.685 | 0.824 | 0.971 | 0.939 | 0.929 | 0.660 | 0.684 | 0.978 | 0.946 | 0.920 | 0.775 | 0.975 | 0.820 |
| | | SQUASIA | 0.910 | 0.545 | 0.939 | 0.834 | 0.909 | 0.898 | 0.865 | 0.938 | 0.837 | 0.917 | 0.920 | 0.776 | 0.963 | 0.817 | 0.936 | 0.920 | 0.775 | 0.975 | 0.820 | 0.958 |
| | | LATAME | 0.939 | 0.618 | 0.948 | 0.067 | 0.903 | 0.927 | 0.804 | 0.948 | 0.089 | 0.907 | 0.949 | 0.796 | 0.967 | -0.152 | 0.929 | 0.941 | 0.796 | 0.967 | -0.121 | 0.926 |
| | | MIDEAST | 0.873 | 0.726 | 0.957 | 0.979 | 0.910 | 0.858 | 0.775 | 0.956 | 0.980 | 0.919 | 0.904 | 0.681 | 0.968 | 0.987 | 0.959 | 0.901 | 0.696 | 0.966 | 0.990 | 0.966 |
| | | NORAM | 0.891 | 0.873 | 0.871 | 0.854 | 0.913 | 0.870 | 0.954 | 0.896 | 0.853 | 0.916 | 0.916 | 0.977 | 0.925 | 0.795 | 0.959 | 0.919 | 0.978 | 0.922 | 0.836 | 0.966 |
| | | PACOECD | 0.836 | 0.800 | 0.830 | 0.024 | 0.968 | 0.791 | 0.875 | 0.879 | 0.038 | 0.957 | 0.879 | 0.769 | 0.879 | -0.069 | 0.966 | 0.910 | 0.816 | 0.881 | -0.014 | 0.964 |
| | | REFECO | 0.901 | 0.484 | 0.334 | 0.370 | 0.691 | 0.877 | 0.573 | 0.414 | 0.371 | 0.703 | 0.947 | 0.438 | 0.485 | 0.174 | 0.865 | 0.947 | 0.438 | 0.519 | 0.234 | 0.853 |
| | | OTASIAN | 0.913 | 0.610 | 0.927 | 0.700 | 0.941 | 0.895 | 0.798 | 0.929 | 0.698 | 0.946 | 0.932 | 0.746 | 0.943 | 0.497 | 0.967 | 0.932 | 0.748 | 0.931 | 0.545 | 0.974 |
| | | ROW | 0.000 | 0.000 | 0.806 | 0.000 | 0.998 | 0.000 | 0.000 | 0.840 | 0.000 | 0.998 | 0.000 | 0.000 | 0.833 | 0.000 | 0.999 | 0.000 | 0.000 | 0.866 | 0.000 | 0.999 |

**Figure 11. Statistical validation of regional emission pathways reproduced from emIAM-ACC2 with original emission pathways from five ENGAGE IAMs.** Ordinary Pearson's correlation coefficient $r_P$ and Lin's concordance coefficient $r_C$ are shown in the figure. The higher the value of the indicator is, the darker the color of the cell is.

The statistics of the validation tests for global MAC curves are shown in Figure 10. Those for regional MAC curves are in Figure 11. The values of $r_C$ are generally lower than the corresponding values of $r_P$, as expected. The reproducibility is generally higher for $CO_2$ than for $CH_4$ and $N_2O$. Certain models tend to have higher values for such indicators than other models. In the global case, AIM tends to show relatively low values for $CH_4$. IMAGE and TIAM tend to show low values for $N_2O$. In the regional results, COFFEE gives lowest values for $CO_2$ for Test 1, but in other Tests, it gives similar values with other models. The outcome for $CH_4$ and $N_2O$ are diverse and difficult to be generalized. Finally, ROW is marked with low values in many models and from most of the Tests.

## 5. Conclusions

We have developed emIAM, a novel modeling approach to emulating IAMs by using a large set of MAC curves: ten IAMs (nine ENGAGE IAMs and GET); global and eleven regions; three gases ($CO_2$, $CH_4$, and $N_2O$); eight portfolios of available mitigation technologies; and two emission sources (total anthropogenic and energy-related). A series of four validation tests were performed using emIAM-ACC2, the hard-linked optimizing climate-economy model, to reproduce original IAM outcomes. The results showed that the original emission pathways were reproduced reasonably well in a majority of cases. However, if one is interested in using emIAM, the goodness of fit of the MAC curves to the original IAM data and the results of validation tests should be carefully examined. We do not provide specific recommendations on the appropriateness of the use of each MAC curve and leave the users to decide which MAC curves to apply. Materials that are required for making such decisions are systematically presented in Supplement and our Zenodo repository. Some IAMs were more easily emulated than other IAMs. The goodness of fit of the MAC curves depends on gases and regions.

This study demonstrated 1) a methodological framework to generate MAC curves from multiple IAMs simulated under a range of carbon budgets and carbon price scenarios and 2) another methodological framework to assess the performance of MAC curves with a simple climate model to reproduce original IAM outcomes. Our methods are generic and transparent, providing an avenue for extending simple climate models to hard-linked climate-economy models. Future studies may emulate specific IAMs with more tailored parameterization approaches. We also open up an avenue for performing a quasi-multiple IAM analysis with a small computational cost. In view of the diversity of IAMs available today, insights from multiple IAMs are indispensable for creating robust findings. Finally, simple models are complementary to complex models; modeling is an art that can shed light into the fundamental laws of complex systems (Yanai, 2009). In similar vein, emIAM can further pave an avenue for understanding the general behavior of IAMs.

Supplementary for

# emIAM v1.0: an emulator for Integrated Assessment Models using marginal abatement cost curves

Weiwei Xiong[1,2,*], Katsumasa Tanaka[2,3,*], Philippe Ciais[2], Daniel J. A. Johansson[4], Mariliis Lehtveer[4]

[1] School of Economics and Management, China University of Geosciences, Wuhan, China

[2] Laboratoire des Sciences du Climat et de l'Environnement (LSCE), IPSL, CEA/CNRS/UVSQ, Université Paris-Saclay, Gif-sur-Yvette, France

[3] Earth System Division, National Institute for Environmental Studies (NIES), Tsukuba, Japan

[4] Division of Physical Resource Theory, Department of Space, Earth, and Environment, Chalmers University of Technology, Gothenburg, Sweden

*Corresponding authors: Weiwei Xiong (xww08012115@cug.edu.cn) and Katsumasa Tanaka (katsumasa.tanaka@lsce.ipsl.fr)

# List

**Table S1. Equation forms and boundary of parameters for fitting MAC curves**

| | AIM | GEM | MESSAGE | IMAGE | COFFEE | TIAM | REMIND | WITCH | POLES | GET |
|---|---|---|---|---|---|---|---|---|---|---|
| **Equation 1** | $f(x)=a*x^b+c*x^d$ | | | | | | | | | |
| a | [0,+inf] | [0,+inf] | [0,+inf] | [0,+inf] | [0,+inf] | [0,+inf] | [0,+inf] | [0,+inf] | [0,+inf] | [0,+inf] |
| b | [0.01,100] | [0.01,100] | [0.01,100] | [0.01,100] | [0.01,100] | [0.01,100] | [0.01,100] | [0.01,100] | [0.01,100] | [0.01,100] |
| c | [0,+inf] | [0,+inf] | [0,+inf] | [0,+inf] | [0,+inf] | [0,+inf] | [0,+inf] | [0,+inf] | [0,+inf] | [0,+inf] |
| d | [0.01,100] | [0.01,100] | [0.01,100] | [0.01,100] | [0.01,100] | [0.01,100] | [0.01,100] | [0.01,100] | [0.01,100] | [0.01,100] |
| **Equation 2** | $f(x)=a*x+b*(exp(c*x)-1)$ | | | | | | | | | |
| a | - | - | - | - | - | - | - | - | - | - |
| b | - | - | - | - | - | - | - | - | - | - |
| c | (-inf,50] | (-inf,50] | (-inf,50] | (-inf,50] | (-inf,50] | (-inf,50] | (-inf,50] | (-inf,50] | (-inf,50] | (-inf,50] |
| **Equation 3** | $f(x)=a*x+b*x^2+c*x^3+d*x^4$ | | | | | | | | | |
| a | - | - | - | - | - | - | - | - | - | - |
| b | [0.01,100] | [0.01,100] | [0.01,100] | [0.01,100] | [0.01,100] | [0.01,100] | [0.01,100] | [0.01,100] | [0.01,100] | [0.01,100] |
| c | - | - | - | - | - | - | - | - | - | - |
| d | [0.01,100] | [0.01,100] | [0.01,100] | [0.01,100] | [0.01,100] | [0.01,100] | [0.01,100] | [0.01,100] | [0.01,100] | [0.01,100] |
| **Equation 4** | $f(x)=a*(b^{(c*x)}-1)$ | | | | | | | | | |
| a | [0,+inf] | [0,+inf] | [0,+inf] | [0,+inf] | [0,+inf] | [0,+inf] | [0,+inf] | [0,+inf] | [0,+inf] | [0,+inf] |
| b | [0.001,+inf] | [0.001,+inf] | [0.001,+inf] | [0.001,+inf] | [0.001,+inf] | [0.001,+inf] | [0.001,+inf] | [0.001,+inf] | [0.001,+inf] | [0.001,+inf] |
| c | (-inf,100] | (-inf,100] | (-inf,100] | (-inf,100] | (-inf,100] | (-inf,100] | (-inf,100] | (-inf,100] | (-inf,100] | (-inf,100] |

**Table S2. Carbon price pathways of different initial levels with a 5% of growth rate**

| Scenarios | 2010 | 2015 | 2020 | 2025 | 2030 | 2035 | 2040 | 2045 | 2050 | 2055 | 2060 | 2065 | 2070 | 2075 | 2080 | 2085 | 2090 | 2095 | 2100 |
|---|---|---|---|---|---|---|---|---|---|---|---|---|---|---|---|---|---|---|---|
| T0 | 0 | 0 | 0 | 0 | 0 | 0 | 0 | 0 | 0 | 0 | 0 | 0 | 0 | 0 | 0 | 0 | 0 | 0 | 0 |
| T1 | 1 | 1.3 | 1.6 | 2.1 | 2.7 | 3.4 | 4.3 | 5.5 | 7.0 | 9.0 | 11.5 | 14.6 | 18.7 | 23.8 | 30.4 | 38.8 | 49.6 | 63.3 | 80.7 |
| T2 | 2 | 2.6 | 3.3 | 4.2 | 5.3 | 6.8 | 8.6 | 11.0 | 14.1 | 18.0 | 22.9 | 29.3 | 37.4 | 47.7 | 60.9 | 77.7 | 99.1 | 126.5 | 161.5 |
| T3 | 3 | 3.8 | 4.9 | 6.2 | 8.0 | 10.2 | 13.0 | 16.5 | 21.1 | 27.0 | 34.4 | 43.9 | 56.0 | 71.5 | 91.3 | 116.5 | 148.7 | 189.8 | 242.2 |
| T5 | 5 | 6.4 | 8.1 | 10.4 | 13.3 | 16.9 | 21.6 | 27.6 | 35.2 | 44.9 | 57.3 | 73.2 | 93.4 | 119.2 | 152.1 | 194.2 | 247.8 | 316.3 | 403.7 |
| T7 | 7 | 8.9 | 11.4 | 14.6 | 18.6 | 23.7 | 30.3 | 38.6 | 49.3 | 62.9 | 80.3 | 102.4 | 130.8 | 166.9 | 213.0 | 271.8 | 346.9 | 442.8 | 565.1 |
| T10 | 10 | 12.8 | 16.3 | 20.8 | 26.5 | 33.9 | 43.2 | 55.2 | 70.4 | 89.9 | 114.7 | 146.4 | 186.8 | 238.4 | 304.3 | 388.3 | 495.6 | 632.5 | 807.3 |
| T15 | 15 | 19.1 | 24.4 | 31.2 | 39.8 | 50.8 | 64.8 | 82.7 | 105.6 | 134.8 | 172.0 | 219.5 | 280.2 | 357.6 | 456.4 | 582.5 | 743.4 | 948.8 | 1211.0 |
| T20 | 20 | 25.5 | 32.6 | 41.6 | 53.1 | 67.7 | 86.4 | 110.3 | 140.8 | 179.7 | 229.3 | 292.7 | 373.6 | 476.8 | 608.5 | 776.7 | 991.2 | 1265.1 | 1614.6 |
| T25 | 25 | 31.9 | 40.7 | 52.0 | 66.3 | 84.7 | 108.0 | 137.9 | 176.0 | 224.6 | 286.7 | 365.9 | 467.0 | 596.0 | 760.7 | 970.8 | 1239.0 | 1581.4 | 2018.3 |
| T30 | 30 | 38.3 | 48.9 | 62.4 | 79.6 | 101.6 | 129.7 | 165.5 | 211.2 | 269.6 | 344.0 | 439.1 | 560.4 | 715.2 | 912.8 | 1165.0 | 1486.8 | 1897.6 | 2421.9 |
| T40 | 40 | 51.1 | 65.2 | 83.2 | 106.1 | 135.5 | 172.9 | 220.6 | 281.6 | 359.4 | 458.7 | 585.4 | 747.2 | 953.6 | 1217.1 | 1553.3 | 1982.5 | 2530.2 | 3229.2 |
| T50 | 50 | 63.8 | 81.4 | 103.9 | 132.7 | 169.3 | 216.1 | 275.8 | 352.0 | 449.3 | 573.4 | 731.8 | 934.0 | 1192.0 | 1521.3 | 1941.6 | 2478.1 | 3162.7 | 4036.5 |
| T60 | 60 | 76.6 | 97.7 | 124.7 | 159.2 | 203.2 | 259.3 | 331.0 | 422.4 | 539.1 | 688.0 | 878.1 | 1120.8 | 1430.4 | 1825.6 | 2330.0 | 2973.7 | 3795.3 | 4843.8 |
| T70 | 70 | 89.3 | 114.0 | 145.5 | 185.7 | 237.0 | 302.5 | 386.1 | 492.8 | 629.0 | 802.7 | 1024.5 | 1307.5 | 1668.8 | 2129.8 | 2718.3 | 3469.3 | 4427.8 | 5651.1 |
| T80 | 80 | 102.1 | 130.3 | 166.3 | 212.3 | 270.9 | 345.8 | 441.3 | 563.2 | 718.8 | 917.4 | 1170.9 | 1494.3 | 1907.2 | 2434.1 | 3106.6 | 3964.9 | 5060.3 | 6458.4 |
| T90 | 90 | 114.9 | 146.6 | 187.1 | 238.8 | 304.8 | 389.0 | 496.4 | 633.6 | 808.7 | 1032.1 | 1317.2 | 1681.1 | 2145.6 | 2738.4 | 3494.9 | 4460.5 | 5692.9 | 7265.7 |
| T100 | 100 | 127.6 | 162.9 | 207.9 | 265.3 | 338.6 | 432.2 | 551.6 | 704.0 | 898.5 | 1146.7 | 1463.6 | 1867.9 | 2384.0 | 3042.6 | 3883.3 | 4956.1 | 6325.4 | 8073.0 |
| T110 | 110 | 140.4 | 179.2 | 228.7 | 291.9 | 372.5 | 475.4 | 606.8 | 774.4 | 988.4 | 1261.4 | 1609.9 | 2054.7 | 2622.4 | 3346.9 | 4271.6 | 5451.8 | 6958.0 | 8880.3 |
| T120 | 120 | 153.2 | 195.5 | 249.5 | 318.4 | 406.4 | 518.6 | 661.9 | 844.8 | 1078.2 | 1376.1 | 1756.3 | 2241.5 | 2860.8 | 3651.2 | 4659.9 | 5947.4 | 7590.5 | 9687.6 |
| T130 | 130 | 165.9 | 211.8 | 270.3 | 344.9 | 440.2 | 561.9 | 717.1 | 915.2 | 1168.1 | 1490.8 | 1902.6 | 2428.3 | 3099.2 | 3955.4 | 5048.2 | 6443.0 | 8223.1 | 10494.9 |
| T140 | 140 | 178.7 | 228.0 | 291.0 | 371.5 | 474.1 | 605.1 | 772.2 | 985.6 | 1257.9 | 1605.4 | 2049.0 | 2615.1 | 3337.6 | 4259.7 | 5436.6 | 6938.6 | 8855.6 | 11302.3 |

**Table S3. Parameter values in global MAC curves for energy-related CO₂, CH₄, and N₂O emissions derived from nine ENGAGE IAMs and GET.** No data are available for energy-related CH₄, and N₂O emissions from GEM and TIAM.

| Model | Variable | a | b | c | d | MaxABL | Max1st | Max2nd |
|-------|----------|-----|-----|-----|-----|--------|--------|--------|
| AIM | $CO_2$ | 192.98 | 1.25 | 16.51 | 18.28 | 112.7 | 6.4 | 1.0 |
| AIM | $CH_4$ | 94.24 | 0.91 | 822.81 | 18.27 | 94.3 | 6.2 | 1.7 |
| AIM | $N_2O$ | 171.87 | 1.41 | 1249.37 | 12.65 | 87.2 | 5.7 | 1.1 |
| COFFEE | $CO_2$ | 40.32 | 1.15 | 40.48 | 5.63 | 146.4 | 5.9 | 0.9 |
| COFFEE | $CH_4$ | 455.91 | 5.77 | 14.38 | 0.35 | 87.9 | 4.9 | 3.1 |
| COFFEE | $N_2O$ | 85.37 | 0.39 | 85.37 | 0.39 | 38.4 | 6.2 | 2.3 |
| GEM | $CO_2$ | 272.42 | 1.57 | 119.95 | 6.82 | 108.0 | 5.7 | 1.1 |
| GEM | $CH_4$ | | | | | | | |
| GEM | $N_2O$ | | | | | | | |
| IMAGE | $CO_2$ | 309.98 | 1.23 | 83.50 | 24.63 | 107.6 | 6.0 | 1.1 |
| IMAGE | $CH_4$ | 879.73 | 11.96 | 283.45 | 1.18 | 91.8 | 4.8 | 1.0 |
| IMAGE | $N_2O$ | 22900.43 | 14.02 | 126.54 | 0.27 | 78.1 | 5.3 | 1.3 |
| MESSAGE | $CO_2$ | 471.55 | 3.02 | 179.97 | 30.24 | 112.0 | 5.0 | 0.8 |
| MESSAGE | $CH_4$ | 103852.03 | 49.10 | 2332.36 | 7.75 | 93.7 | 5.5 | 1.4 |
| MESSAGE | $N_2O$ | 38175.32 | 5.95 | 155.77 | 0.43 | 62.0 | 3.8 | 0.9 |
| POLES | $CO_2$ | 1785.75 | 16.24 | 2092.84 | 3.01 | 110.2 | 4.4 | 0.8 |
| POLES | $CH_4$ | 4016.39 | 7.61 | 4016.39 | 7.61 | 97.2 | 5.8 | 1.1 |
| POLES | $N_2O$ | 630.22 | 1.71 | 14691.35 | 7.56 | 87.4 | 5.3 | 0.8 |
| REMIND | $CO_2$ | 316.94 | 1.82 | 591.39 | 21.77 | 103.5 | 6.1 | 0.7 |
| REMIND | $CH_4$ | 143.80 | 1.02 | 2139.05 | 14.81 | 97.4 | 5.3 | 1.3 |
| REMIND | $N_2O$ | 5558.24 | 2.93 | 44.00 | 0.17 | 47.8 | 2.9 | 0.9 |
| TIAM | $CO_2$ | 183.59 | 11.93 | 394.27 | 1.39 | 116.2 | 4.6 | 0.8 |
| TIAM | $CH_4$ | | | | | | | |
| TIAM | $N_2O$ | | | | | | | |
| WITCH | $CO_2$ | 421.02 | 1.40 | 971.12 | 7.56 | 100.0 | 3.8 | 1.1 |
| WITCH | $CH_4$ | 1528.23 | 36.27 | 153.56 | 3.52 | 98.2 | 5.8 | 3.4 |
| WITCH | $N_2O$ | 97.19 | 0.73 | 437940.23 | 8.98 | 50.8 | 3.3 | 1.3 |

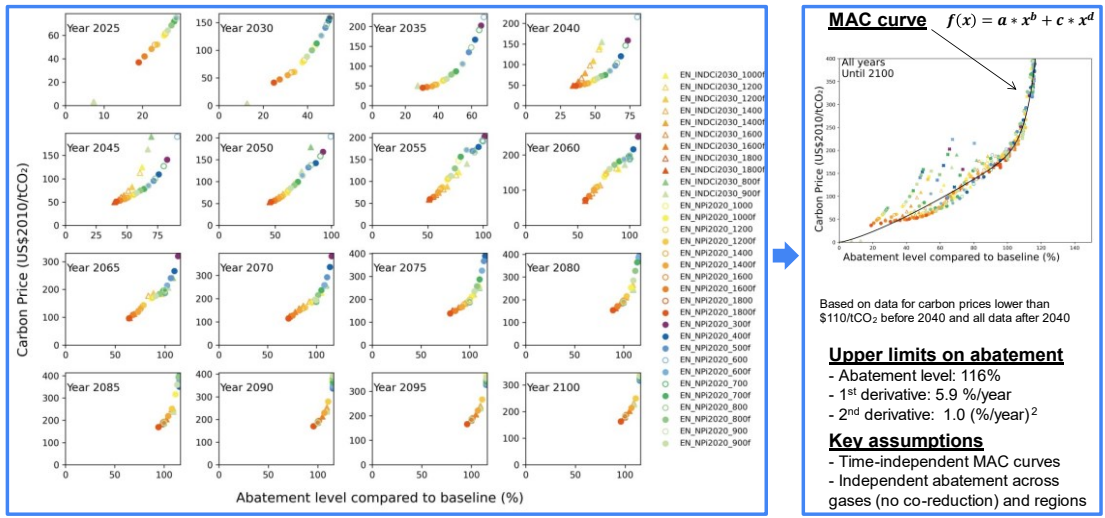

**Figure S1. Overview of the methods to derive AIM MAC curves and limits on abatement.** The description of the figure can be found in Figure 1 of the main text.

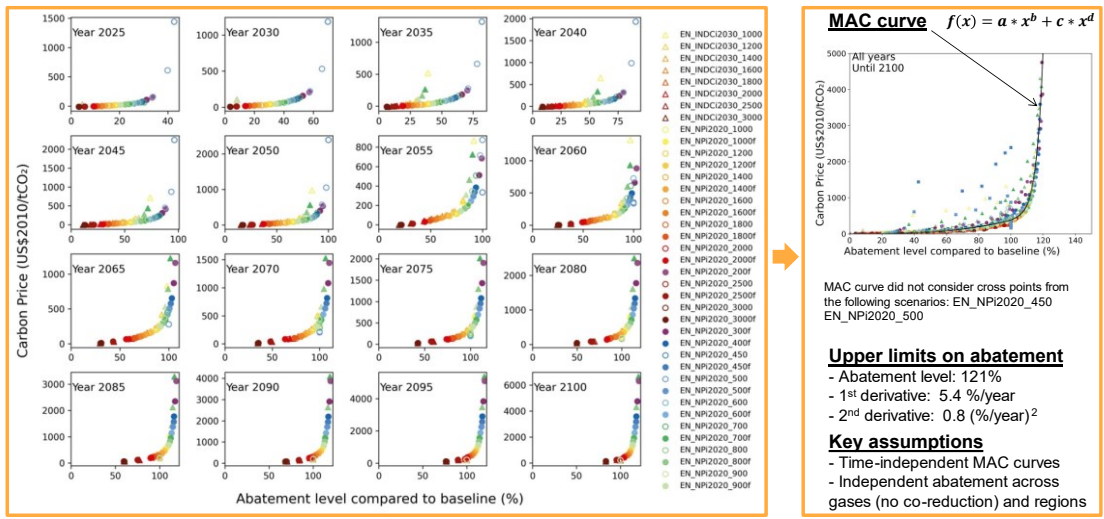

**Figure S2. Overview of the methods to derive MESSAGE MAC curves and limits on abatement.**

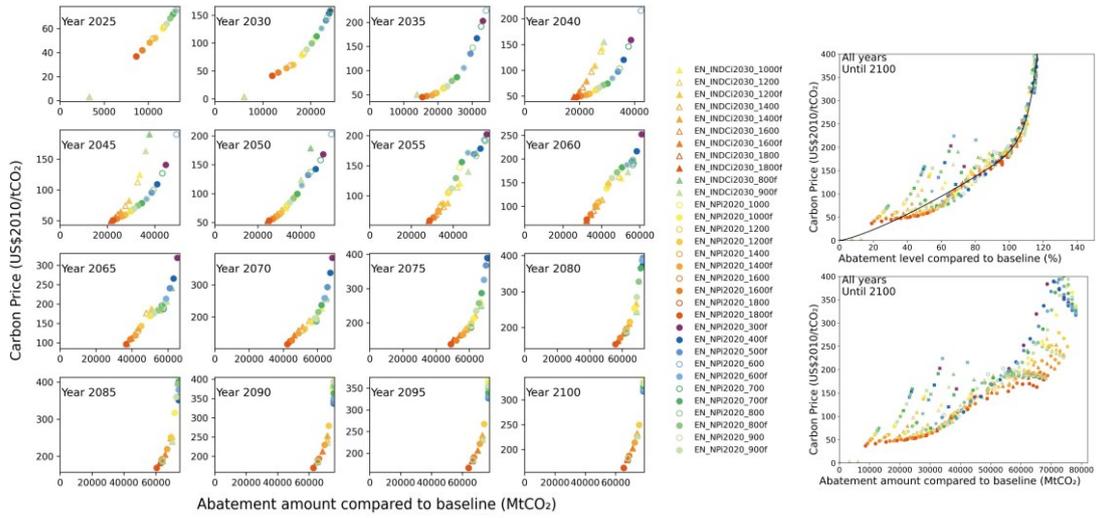

**Figure S3. Overview of the methods to derive AIM MAC curves and limits on abatement.** The results is in the absolute term.

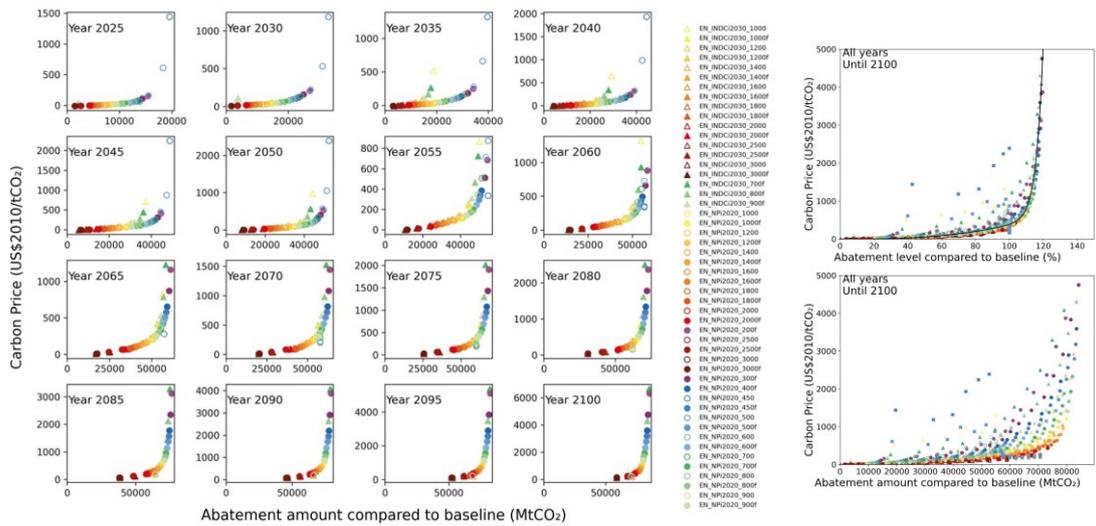

**Figure S4. Overview of the methods to derive MESSAGE MAC curves and limits on abatement.** The results is in the absolute term.

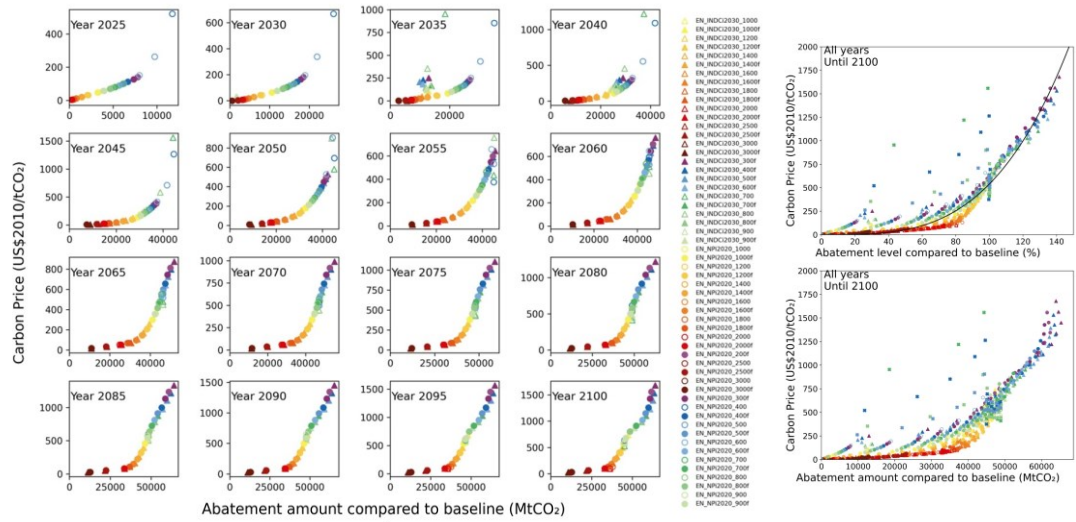

**Figure S5. Overview of the methods to derive AIM MAC curves and limits on abatement.** The results is in the absolute term.

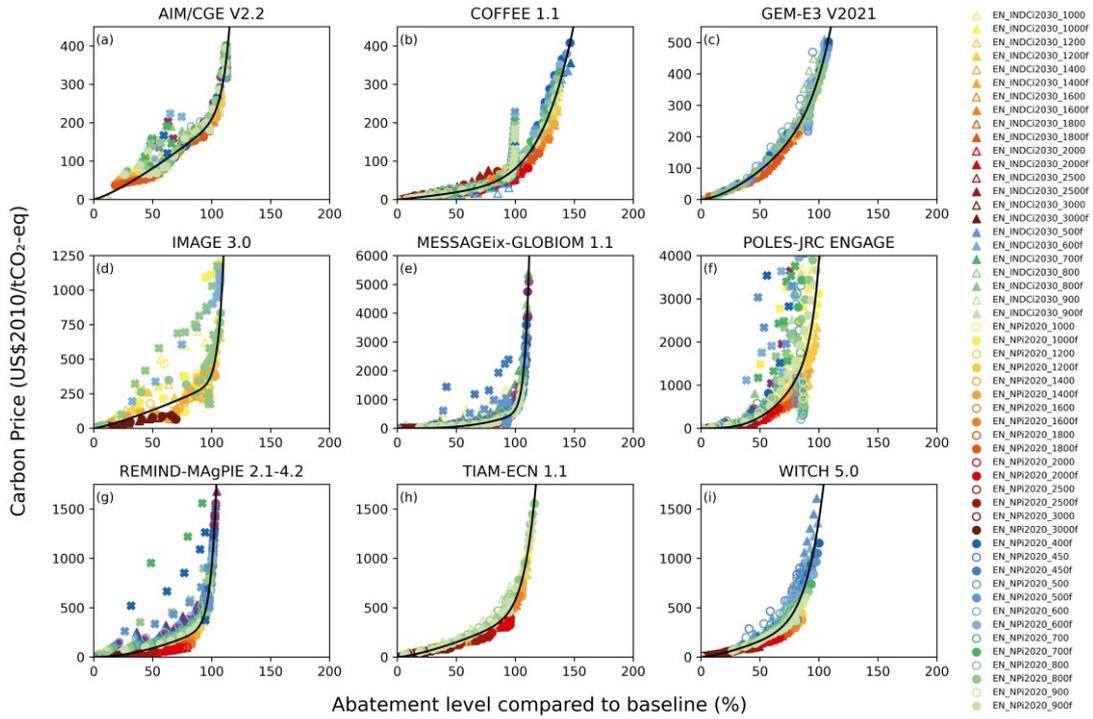

**Figure S6. Global energy-related CO₂ MAC curve from nine ENGAGE IAMs.** Each panel shows the results from each IAM from the ENGAGE Scenario Explorer. Points are the data obtained from the ENGAGE Scenario Explorer shown in colors and markers as designated in the legend. Black lines are the MAC curves. Open circles are the data that were not considered in the derivation of MAC curves.

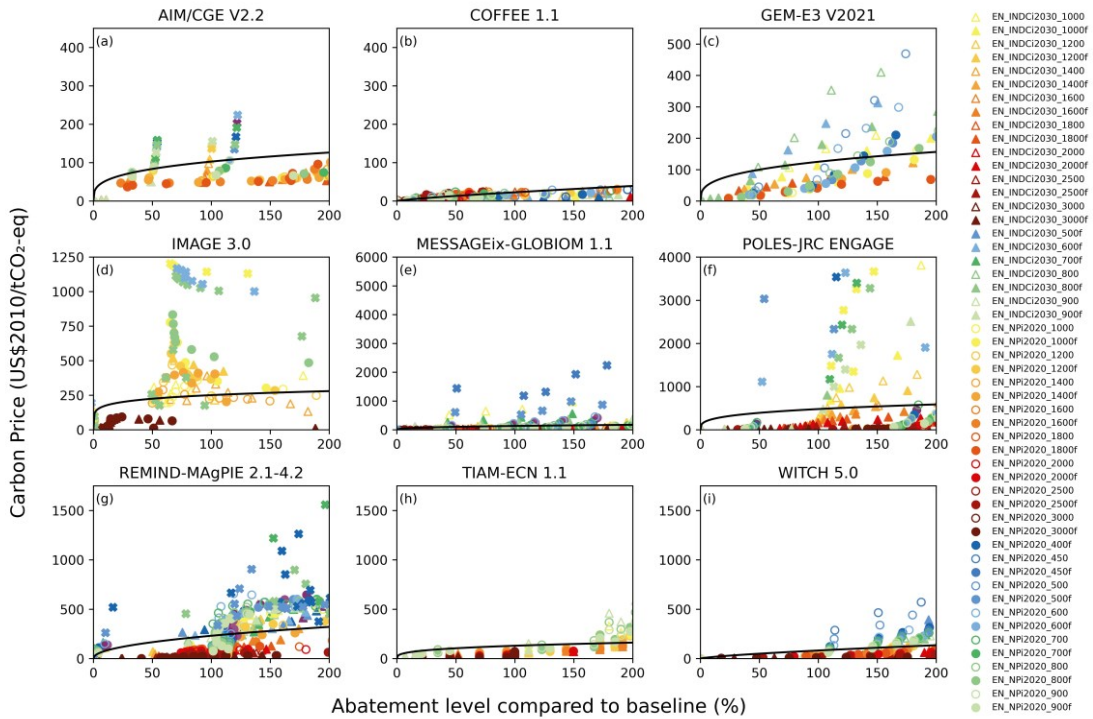

**Figure S7. Global non-energy-related CO₂ MAC curve from nine ENGAGE IAMs**

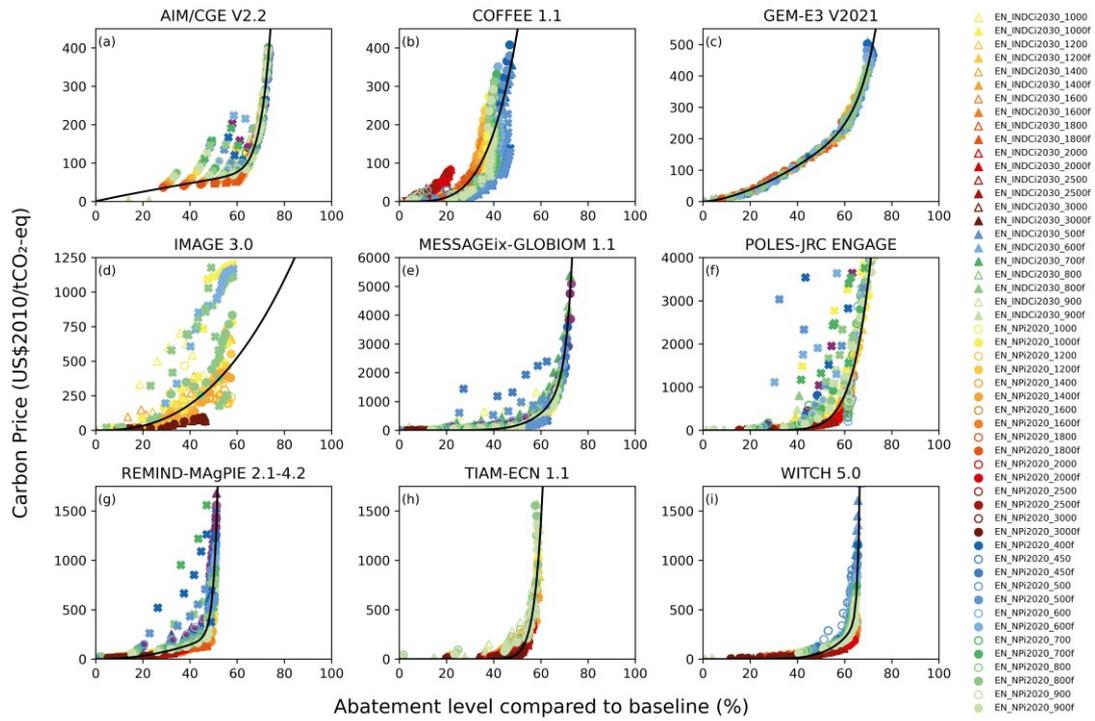

**Figure S8. Global total anthropogenic CH₄ MAC curve from nine ENGAGE IAMs**

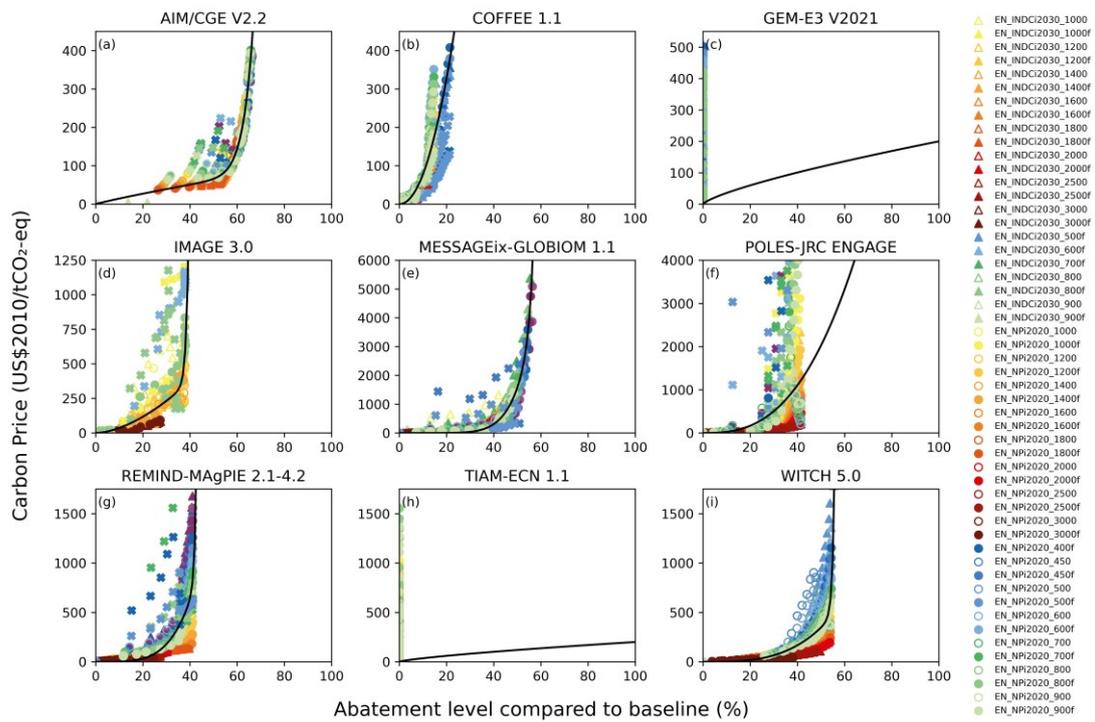

**Figure S9. Global energy-related CH₄ MAC curve from nine ENGAGE IAMs**

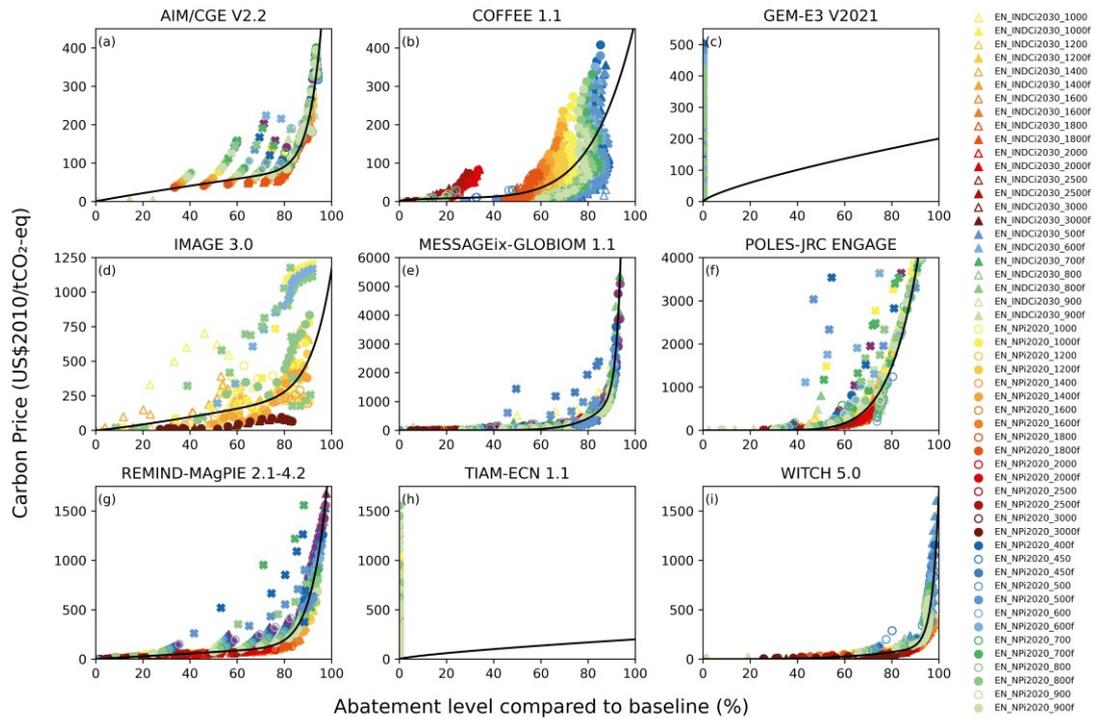

**Figure S10. Global non-energy-related CH$_4$ MAC curve from nine ENGAGE IAMs**

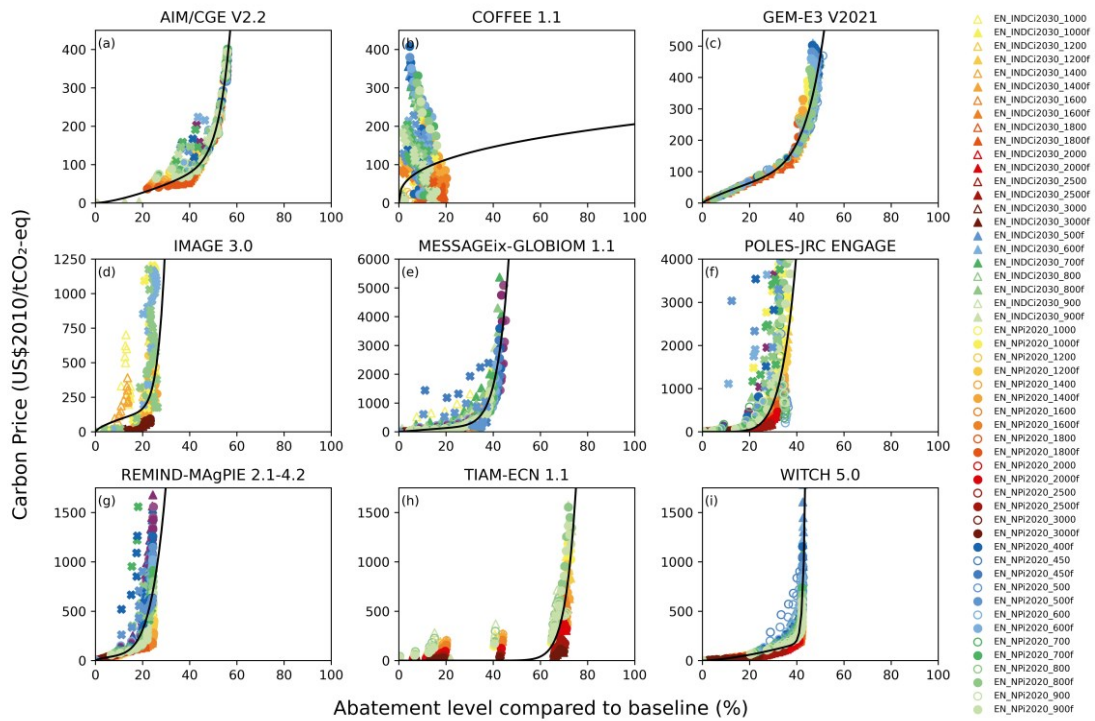

**Figure S11. Global total anthropogenic N$_2$O MAC curve from nine ENGAGE IAMs**

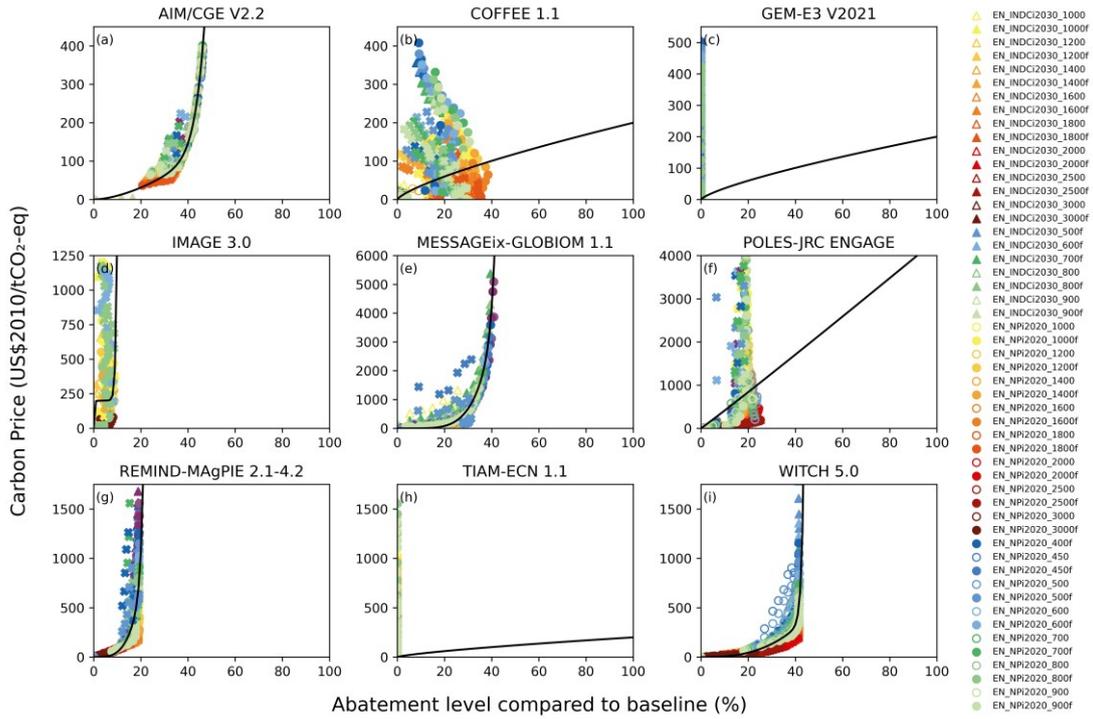

**Figure S12. Global energy-related N₂O MAC curve from nine ENGAGE IAMs**

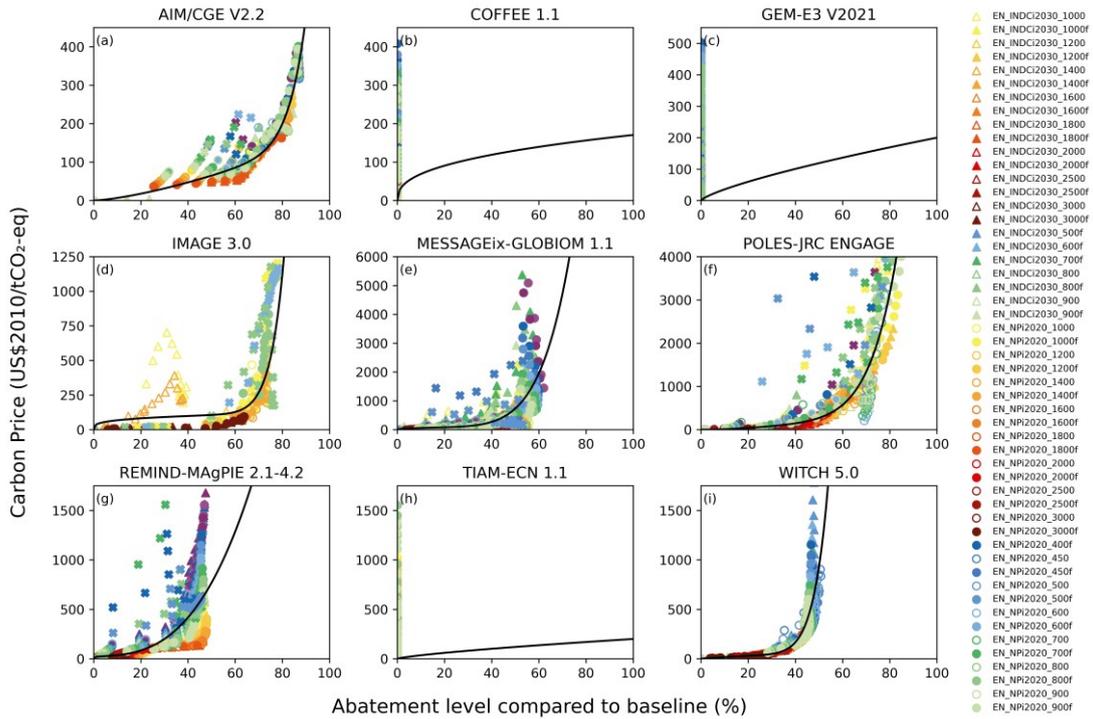

**Figure S13. Global non-energy-related N₂O MAC curve from nine ENGAGE IAMs**

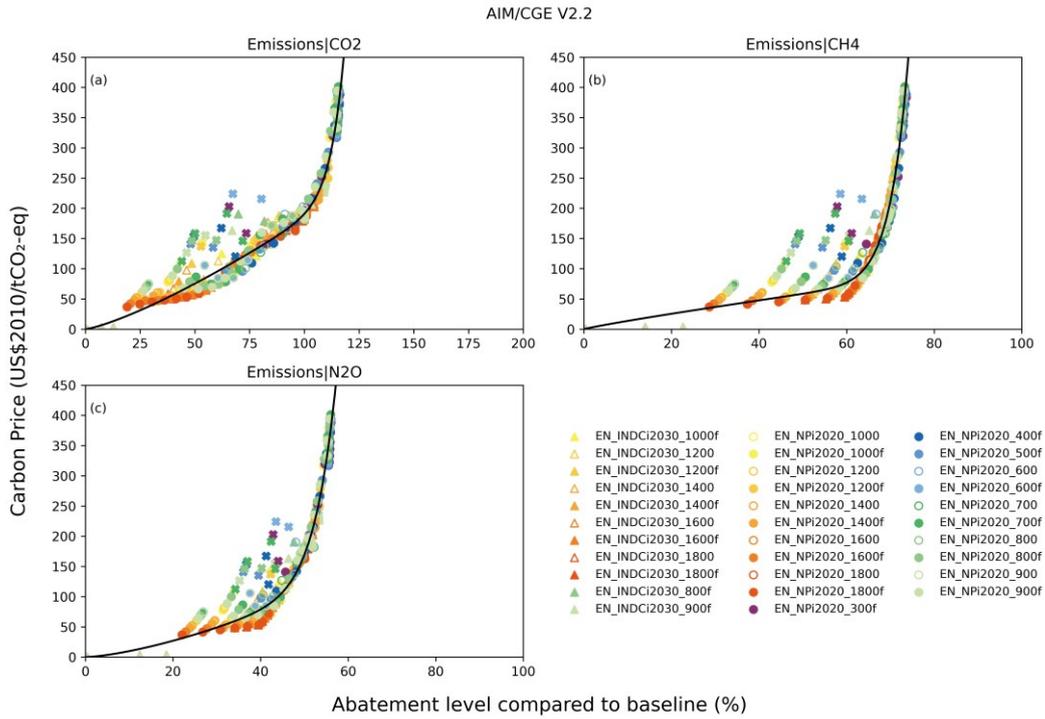

**Figure S14. Global AIM MAC curve**

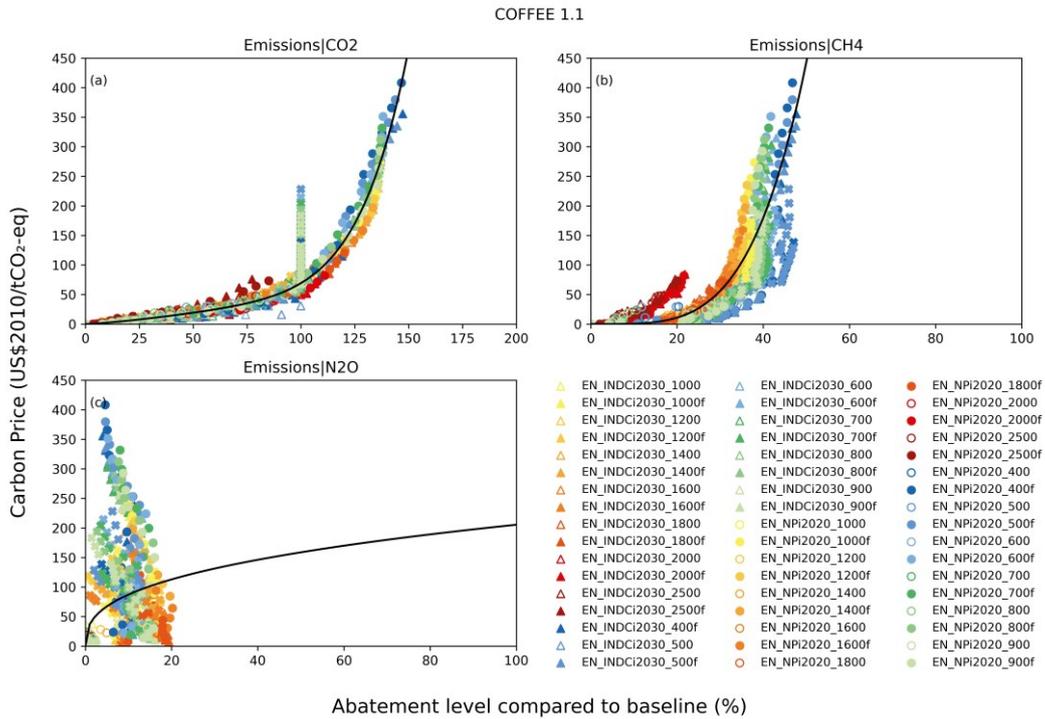

**Figure S15. Global COFFEE MAC curve**

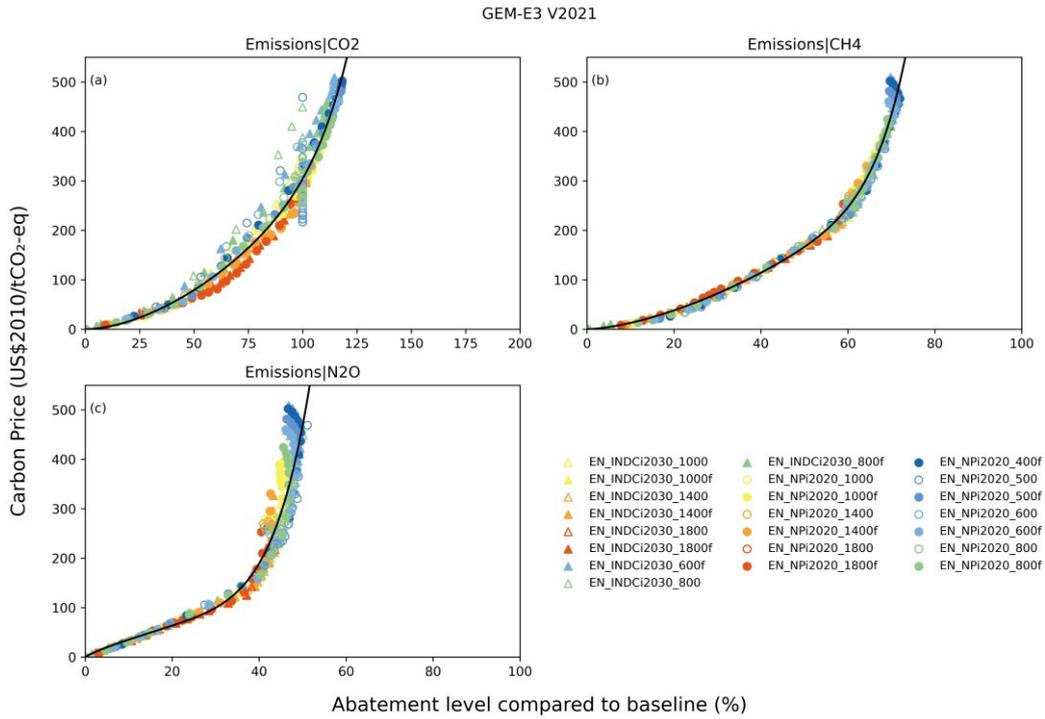

**Figure S16. Global GEM MAC curve**

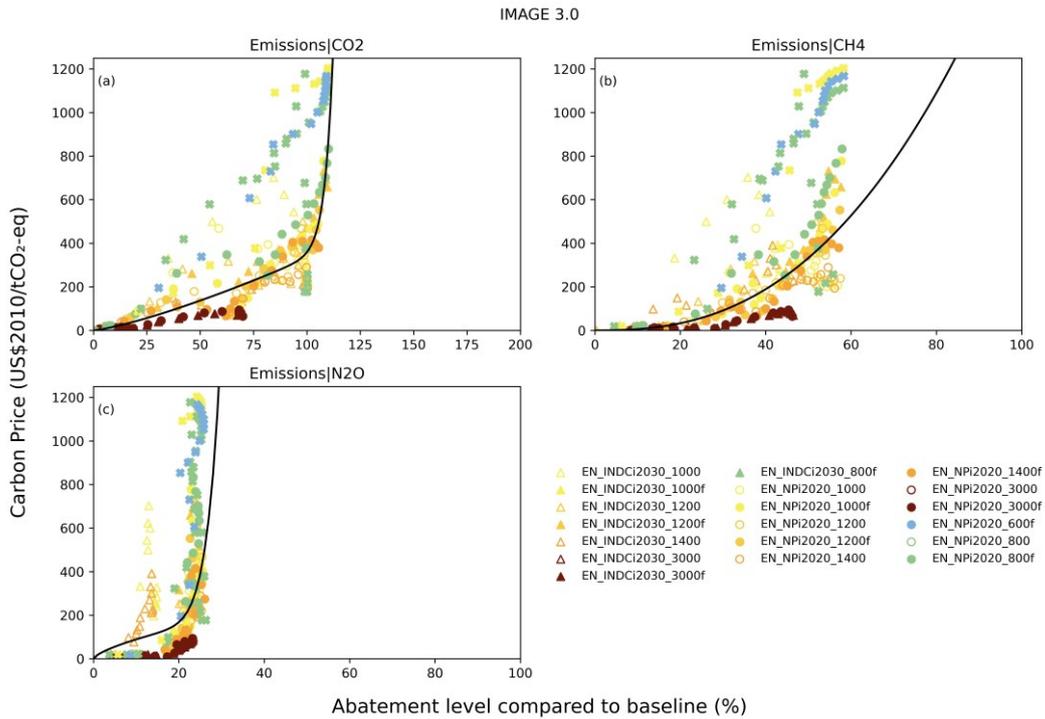

**Figure S17. Global IMAGE MAC curve**

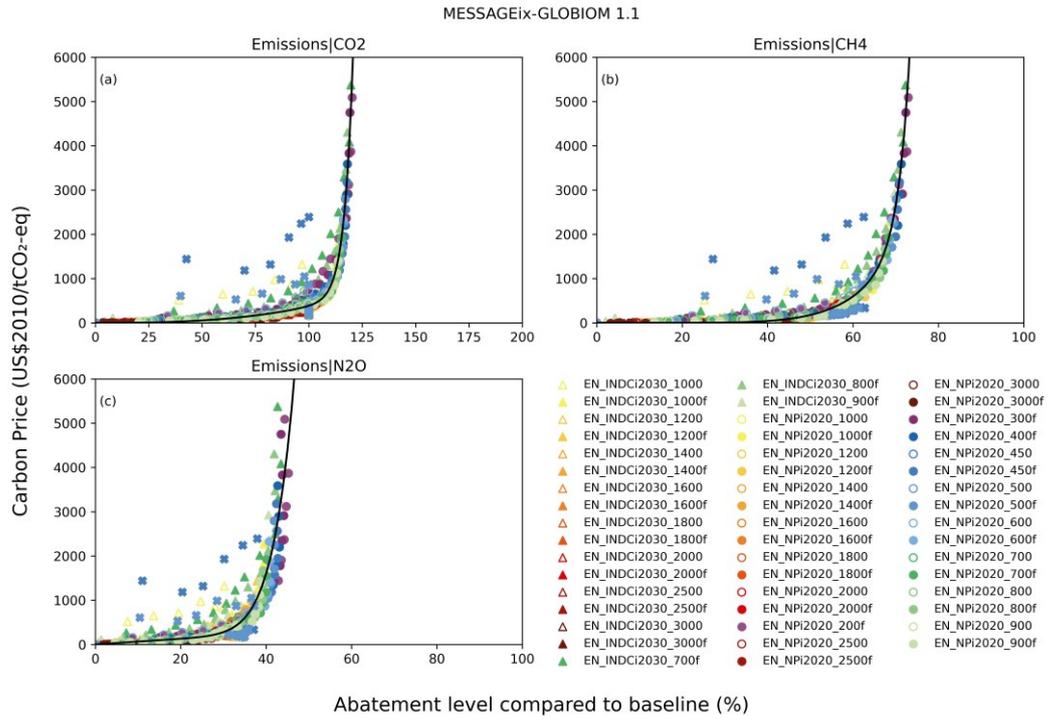

**Figure S18. Global MESSAGE MAC curve**

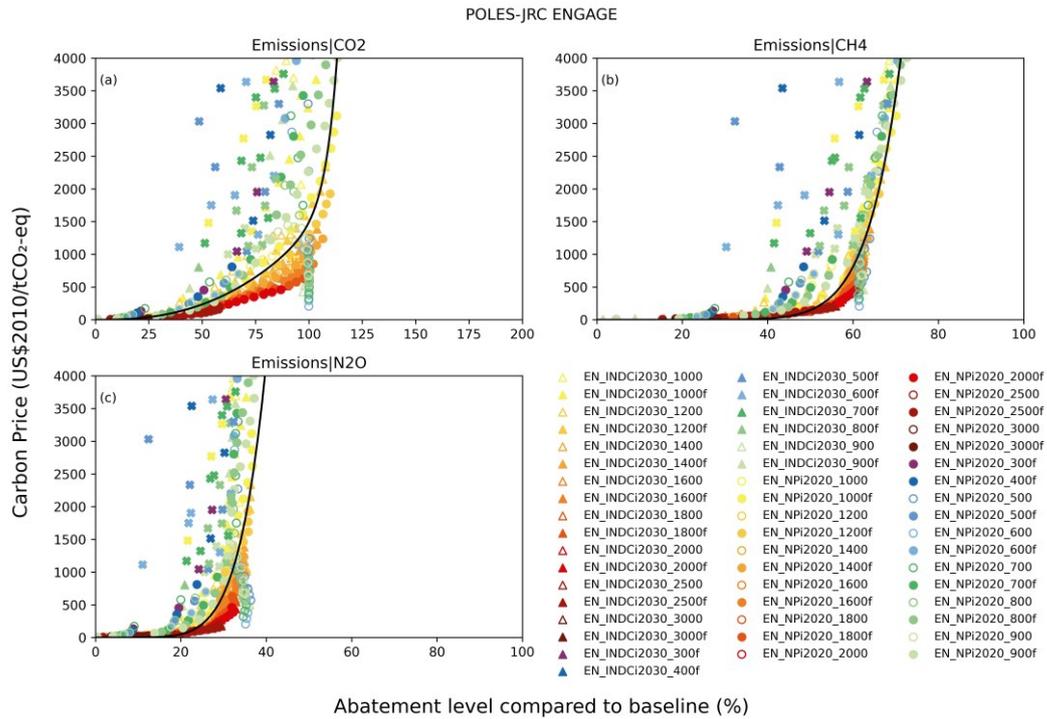

**Figure S19. Global POLES MAC curve**

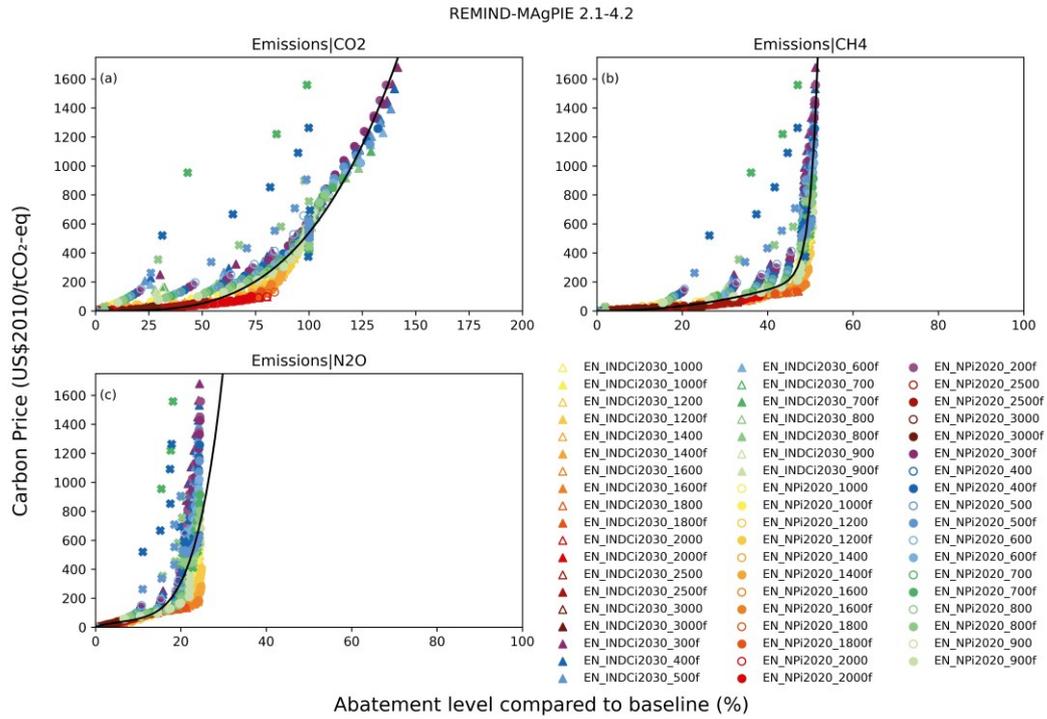

**Figure S20. Global REMIND MAC curve**

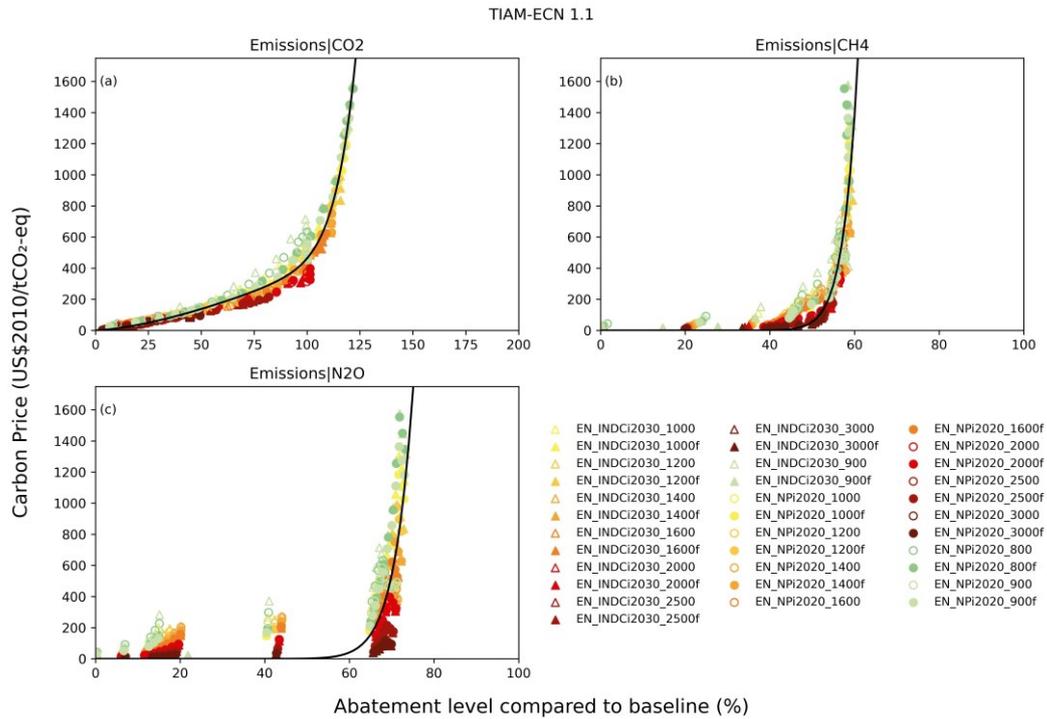

**Figure S21. Global TIAM MAC curve**

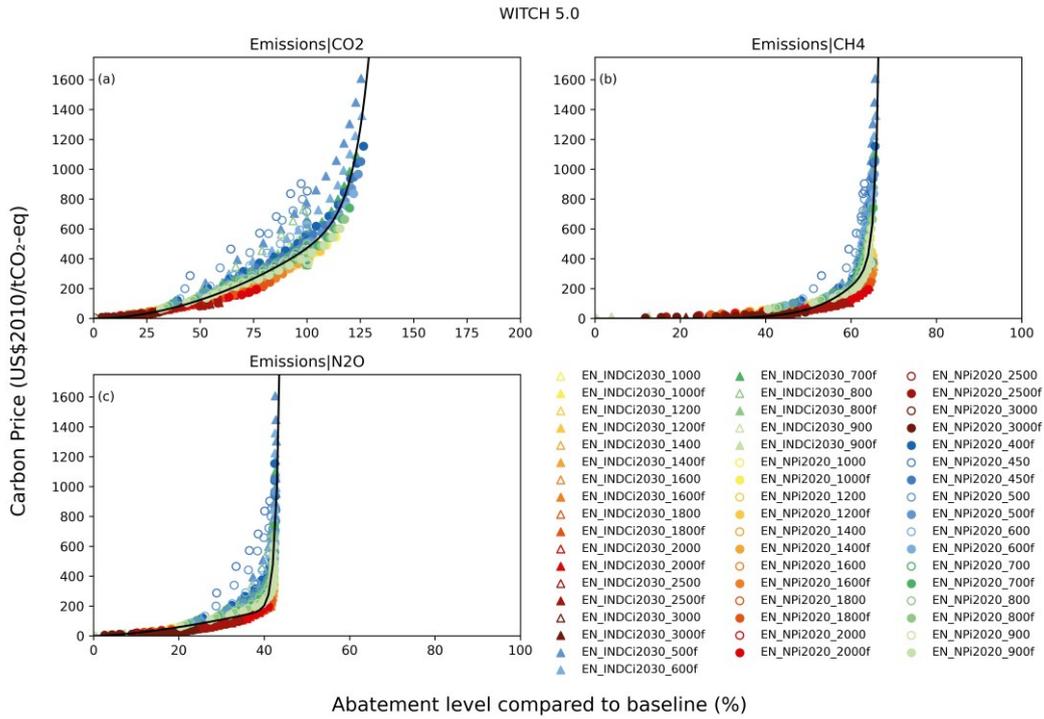

**Figure S22. Global WITCH MAC curve**

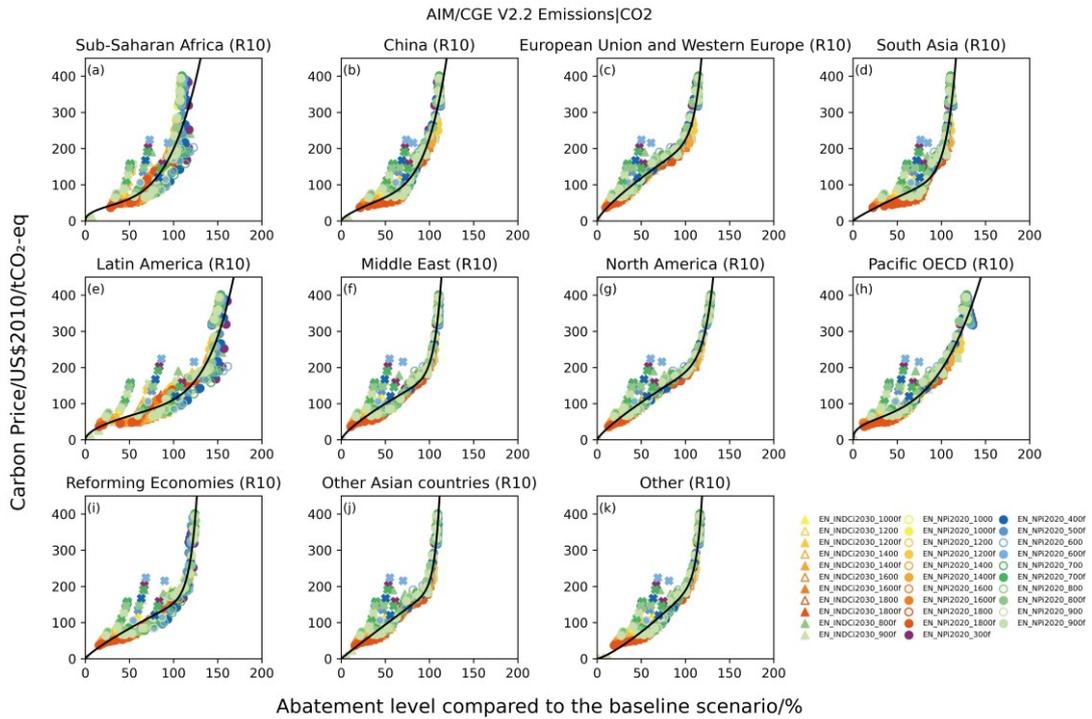

**Figure S23. Regional AIM total anthropogenic CO$_2$ MAC curve**

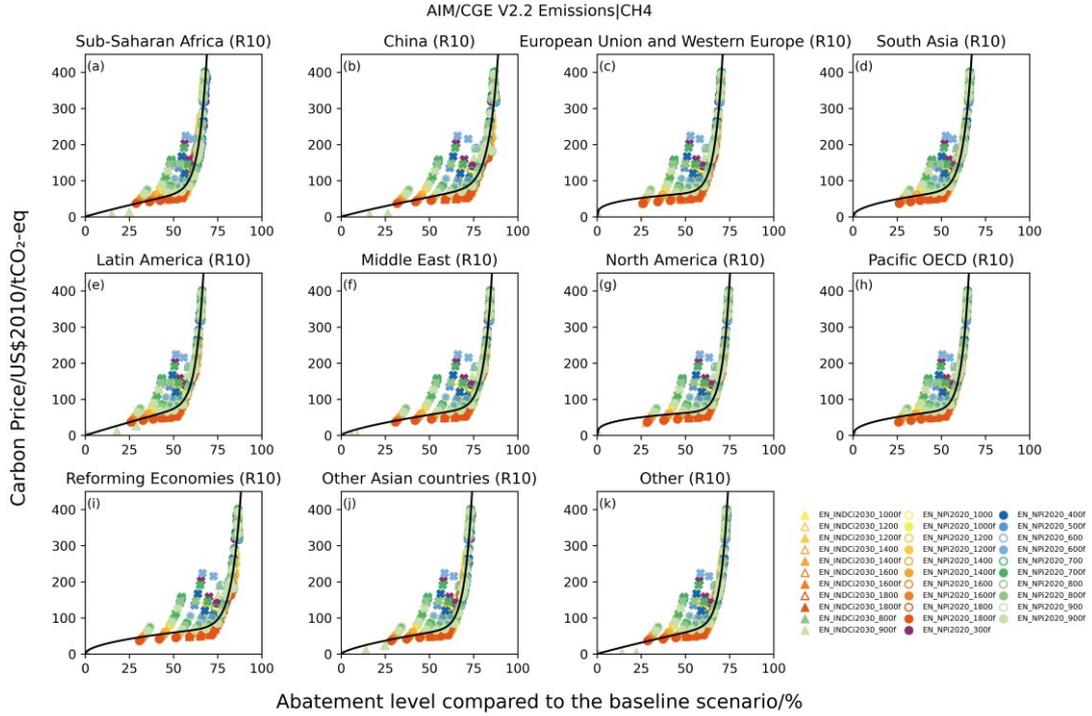

**Figure S24. Regional AIM total anthropogenic CH₄ MAC curve**

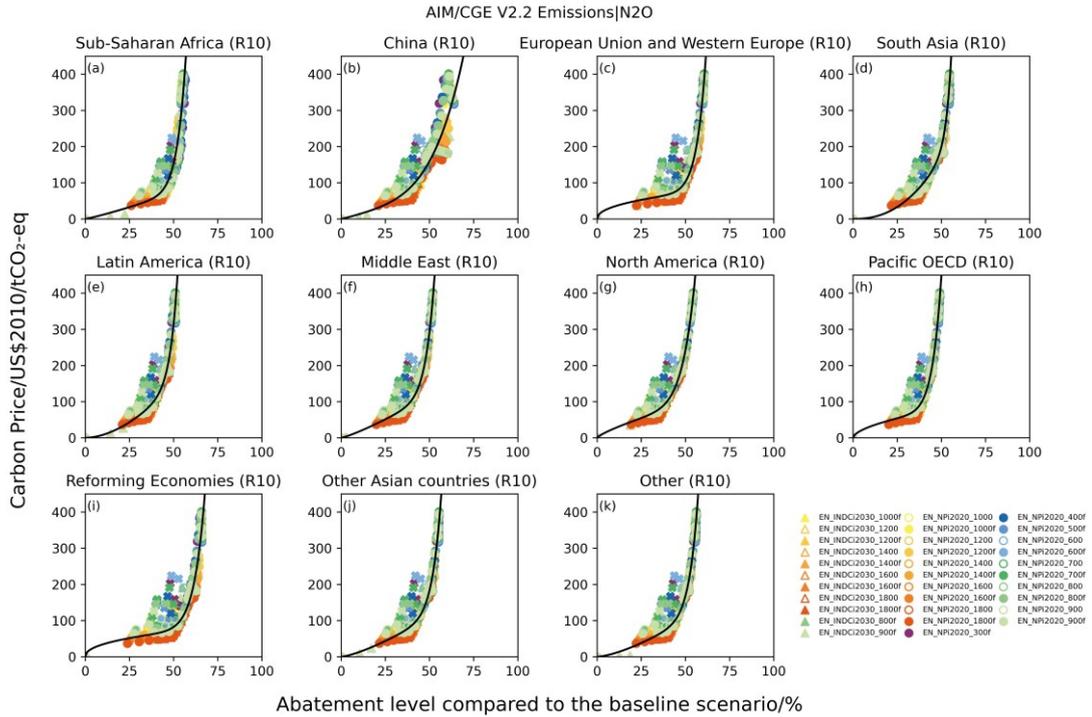

**Figure S25. Regional AIM total anthropogenic N₂O MAC curve**

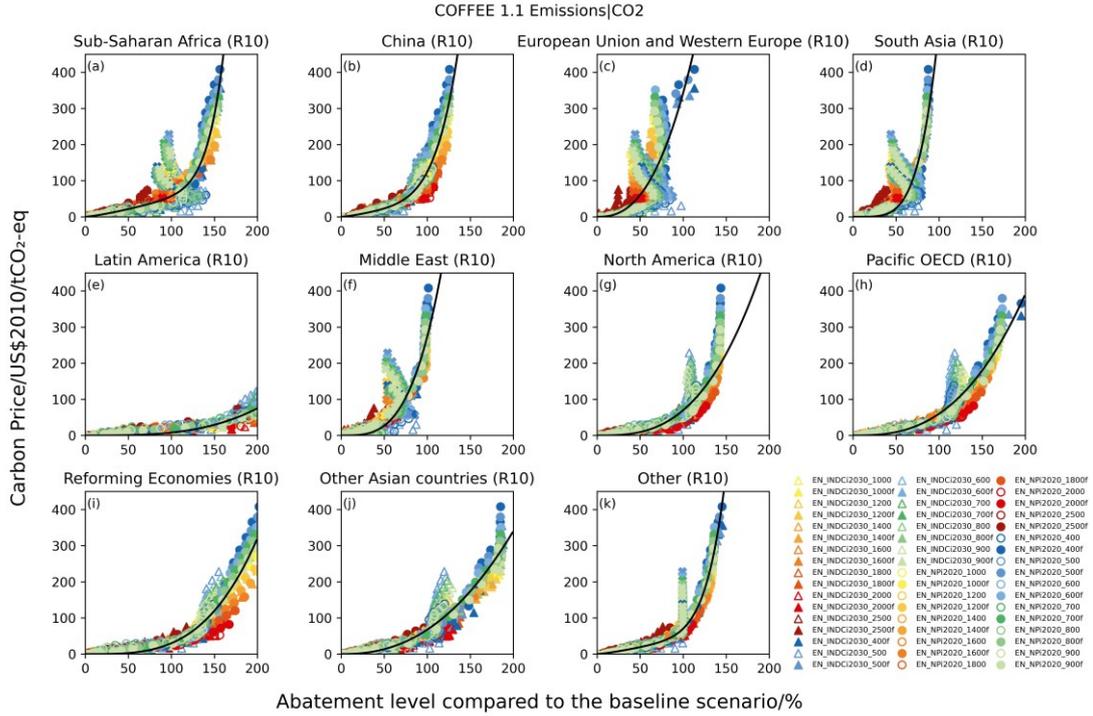

**Figure S26. Regional COFFEE total anthropogenic CO₂ MAC curve**

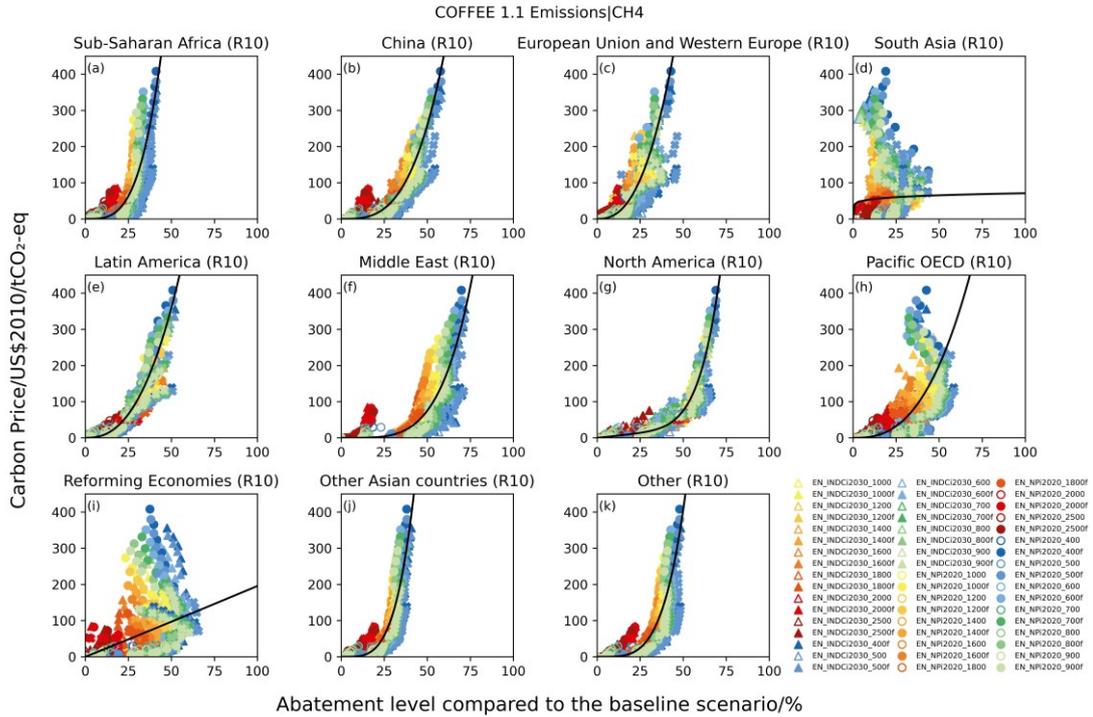

**Figure S27. Regional COFFEE total anthropogenic CH₄ MAC curve**

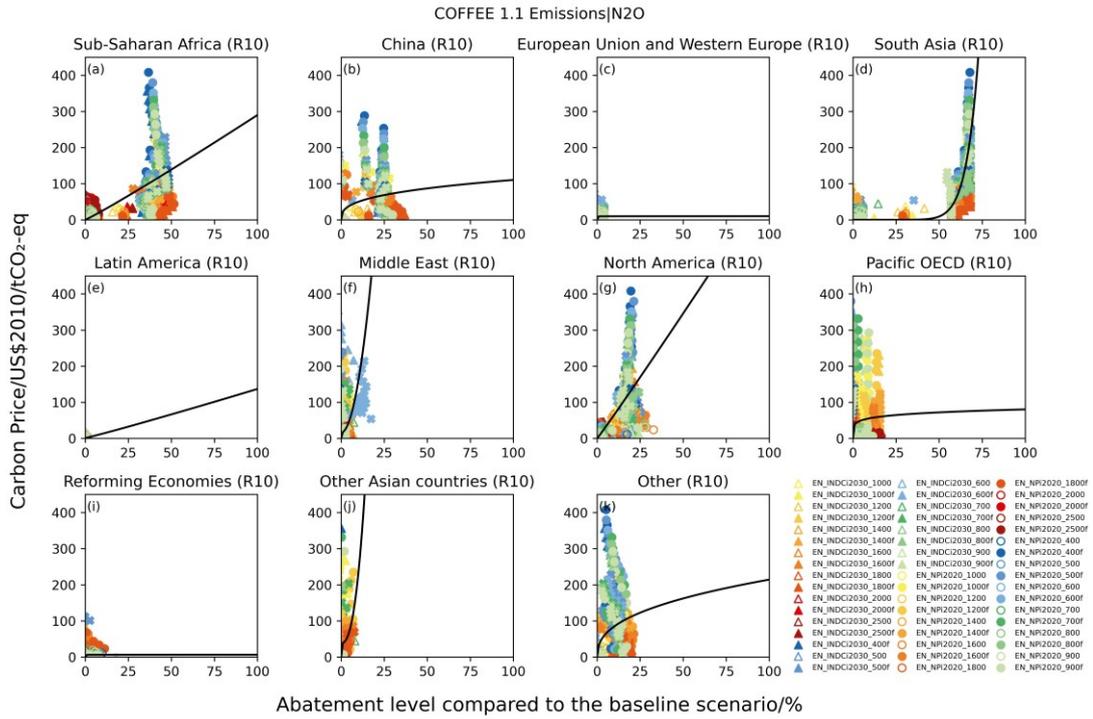

**Figure S28. Regional COFFEE total anthropogenic N₂O MAC curve**

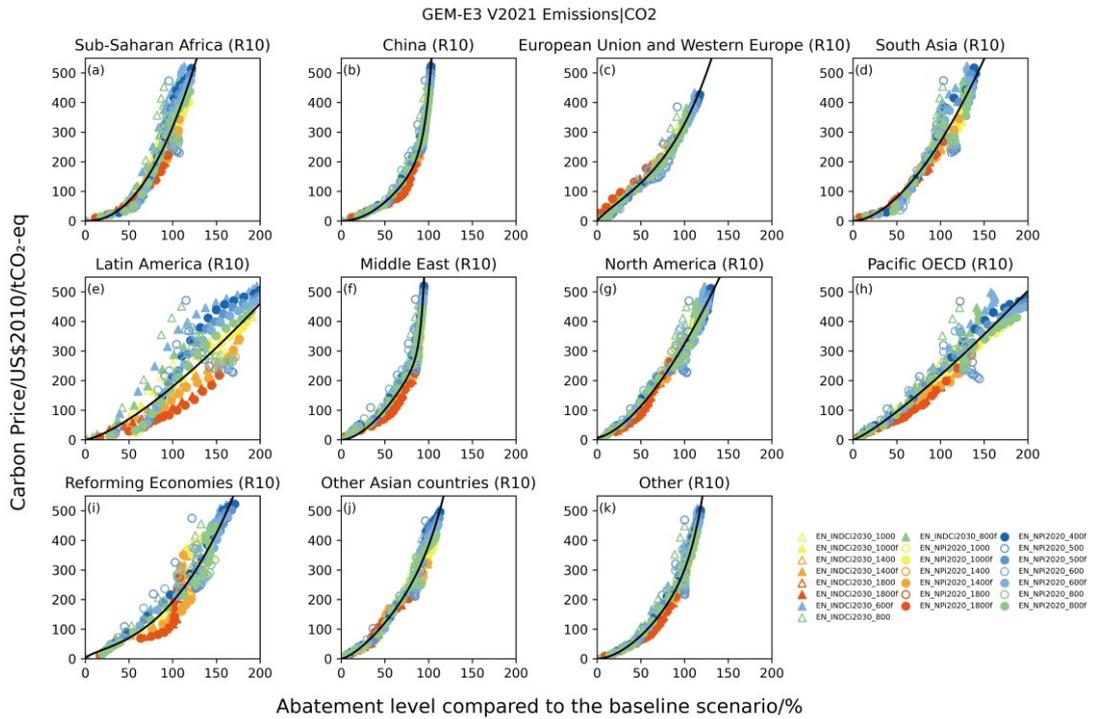

**Figure S29. Regional GEM total anthropogenic CO₂ MAC curve**

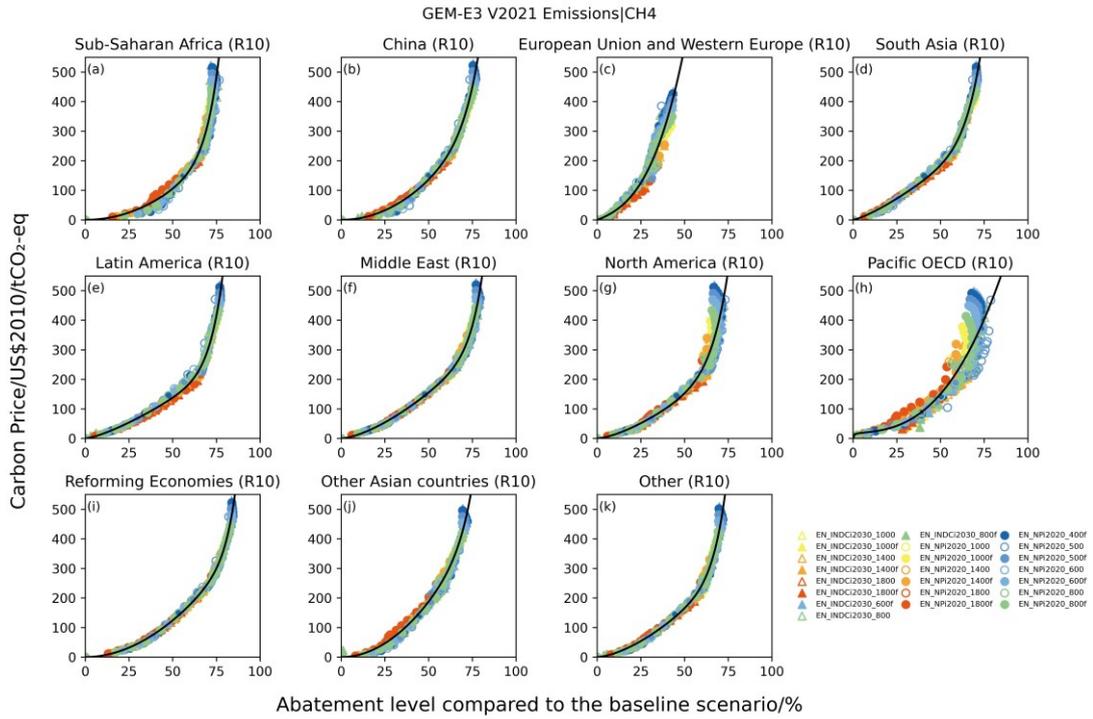

**Figure S30. Regional GEM CH4 MAC curve**

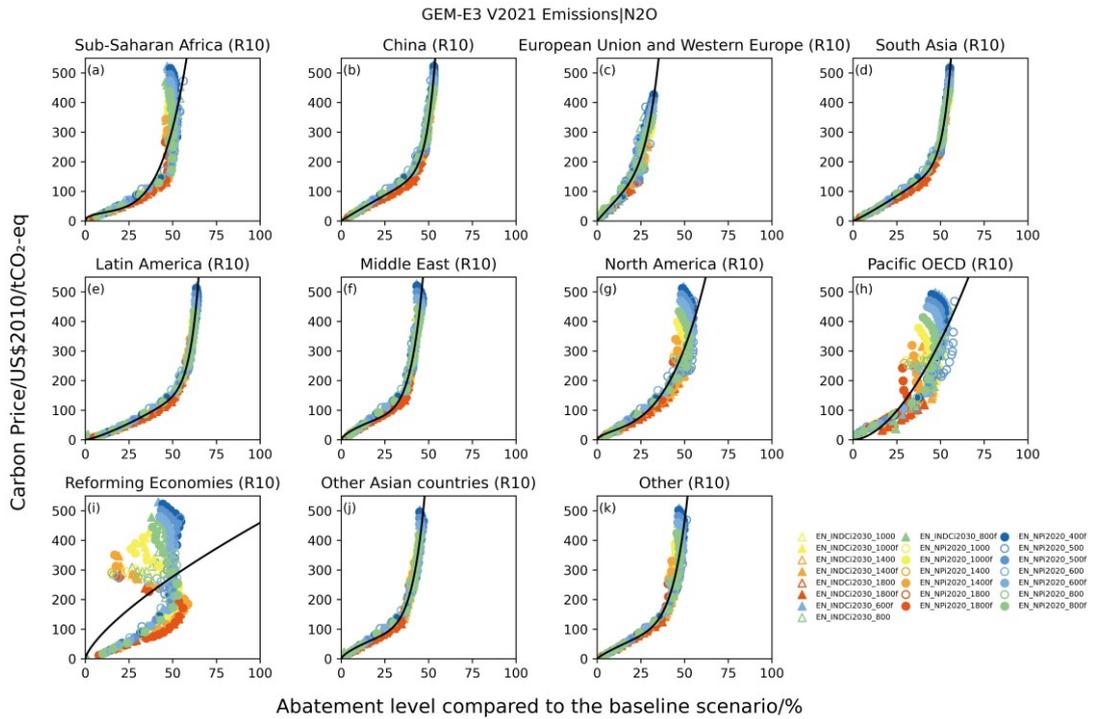

**Figure S31. Regional GEM total anthropogenic N2O MAC curve**

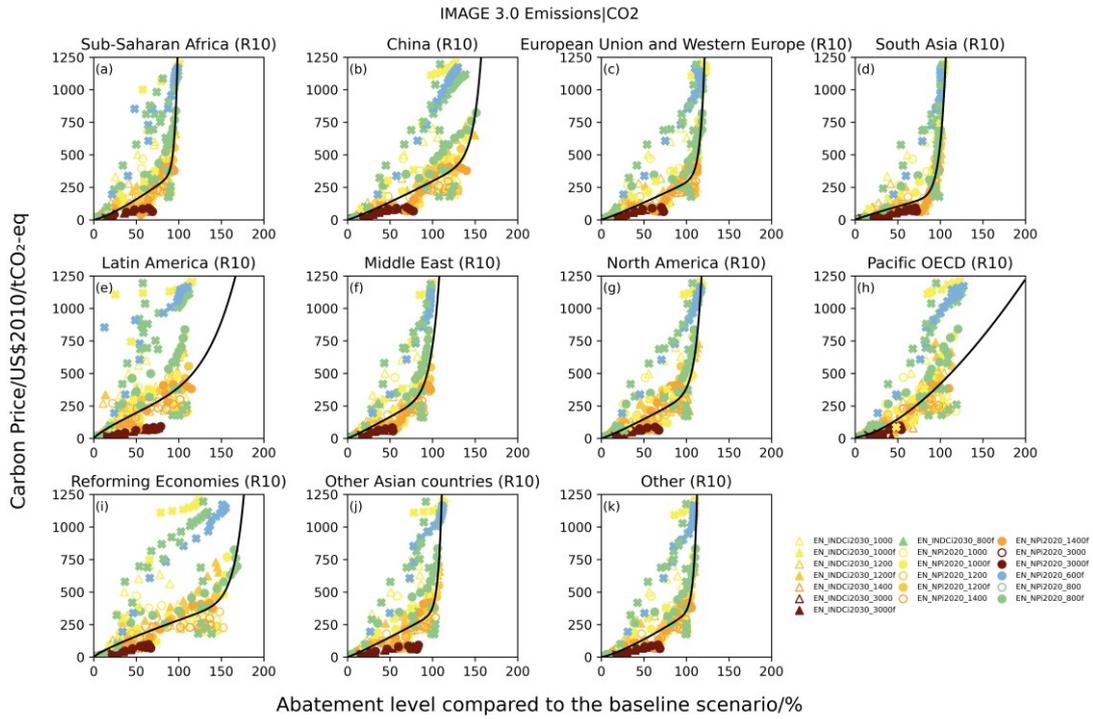

**Figure S32. Regional IMAGE total anthropogenic CO₂ MAC curve**

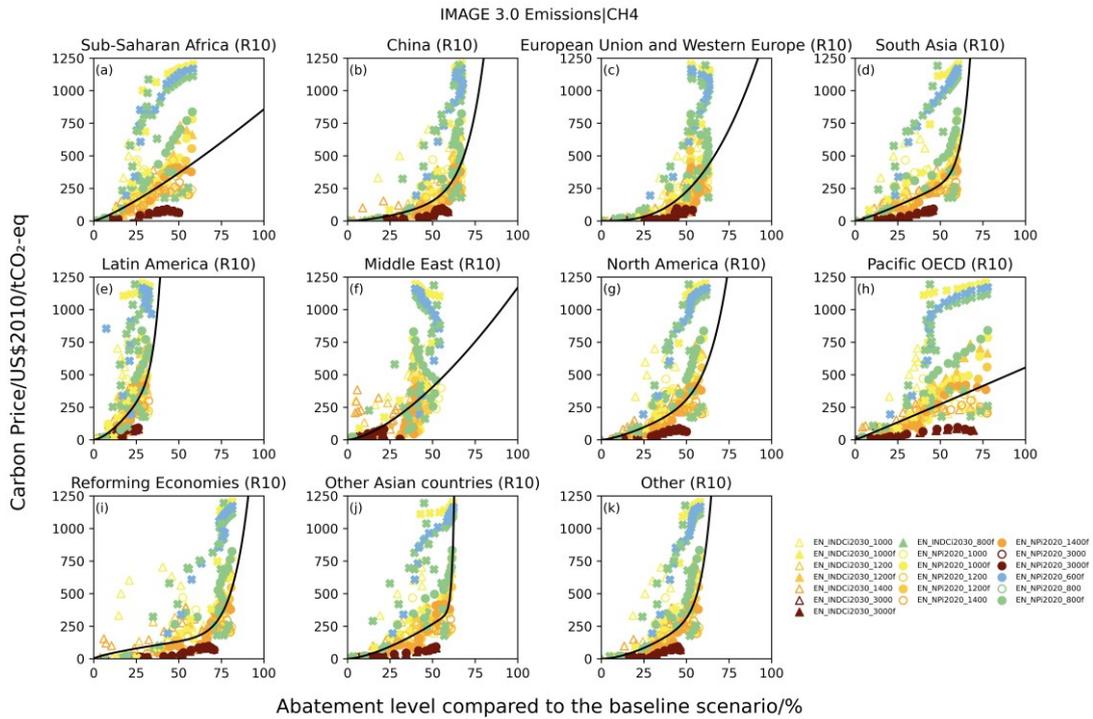

**Figure S33. Regional IMAGE total anthropogenic CH₄ MAC curve**

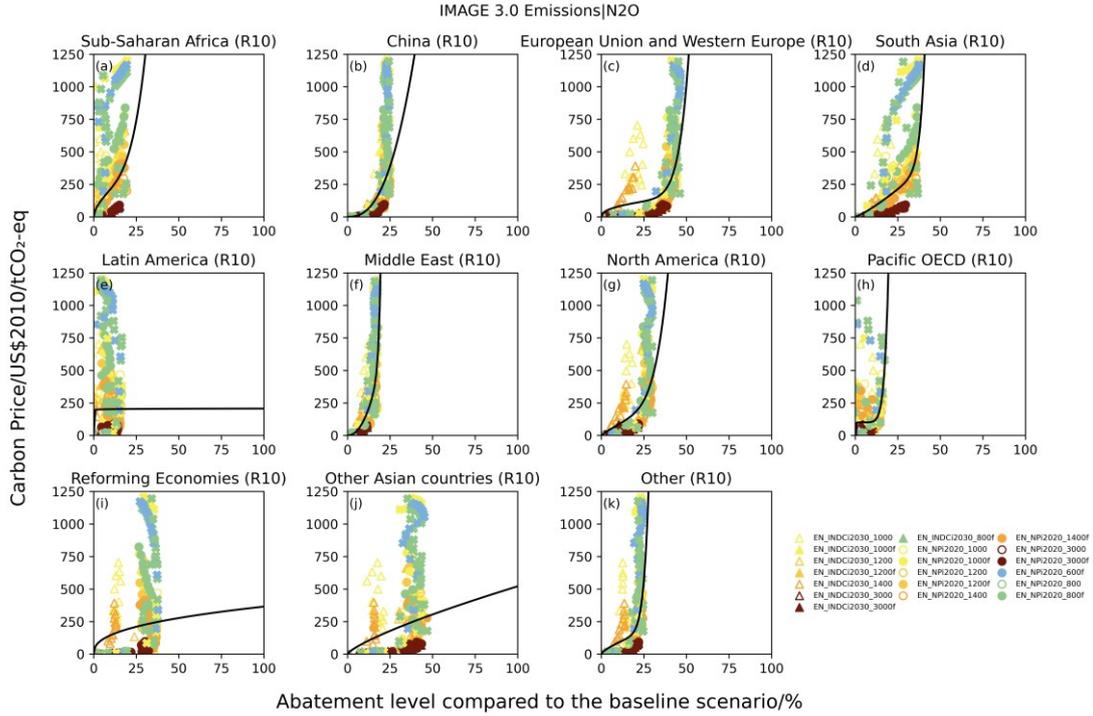

**Figure S34. Regional IMAGE total anthropogenic N₂O MAC curve**

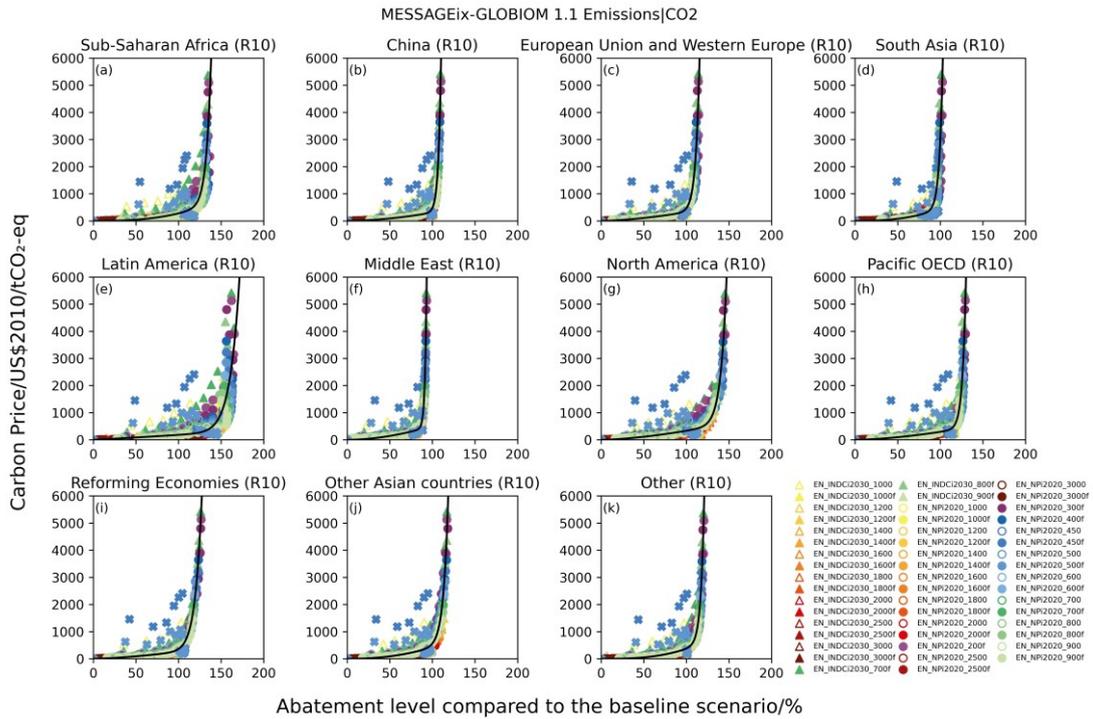

**Figure S35. Regional MESSAGE total anthropogenic CO₂ MAC curve**

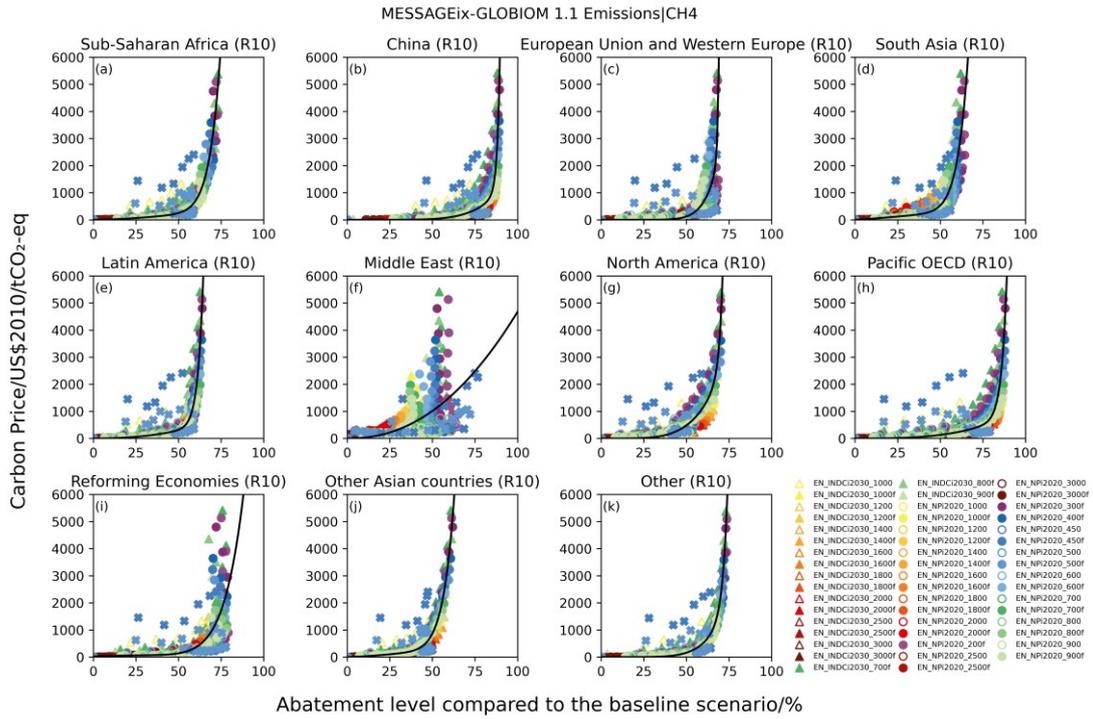

**Figure S36. Regional MESSAGE total anthropogenic CH₄ MAC curve**

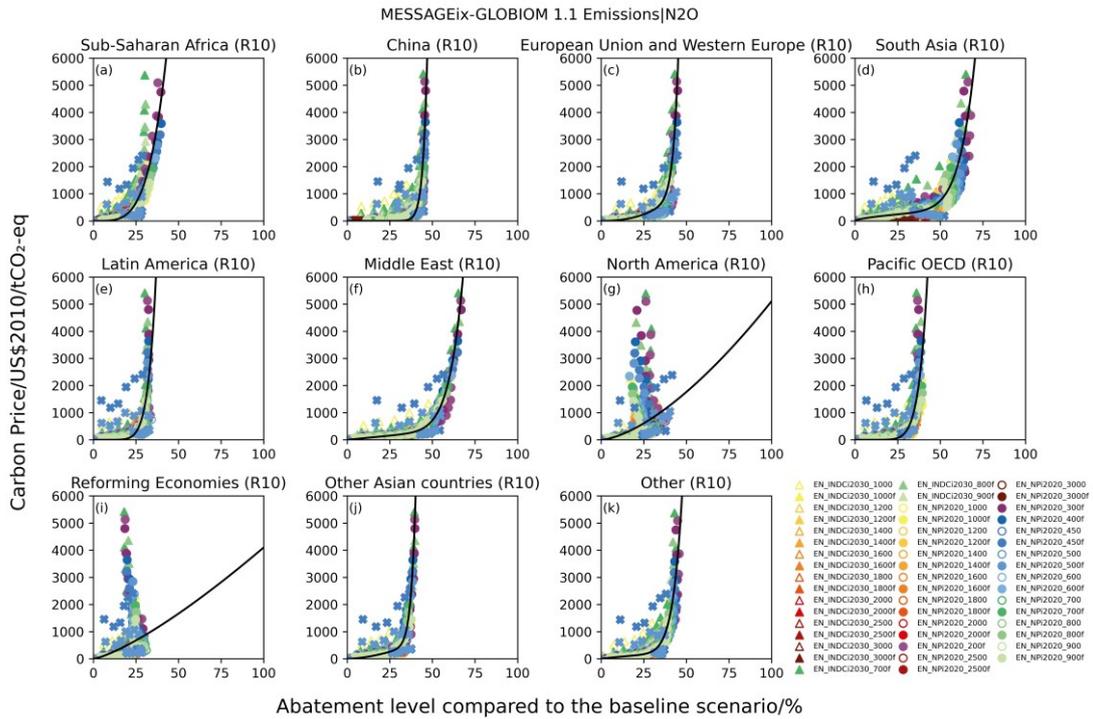

**Figure S37. Regional MESSAGE total anthropogenic N₂O MAC curve**

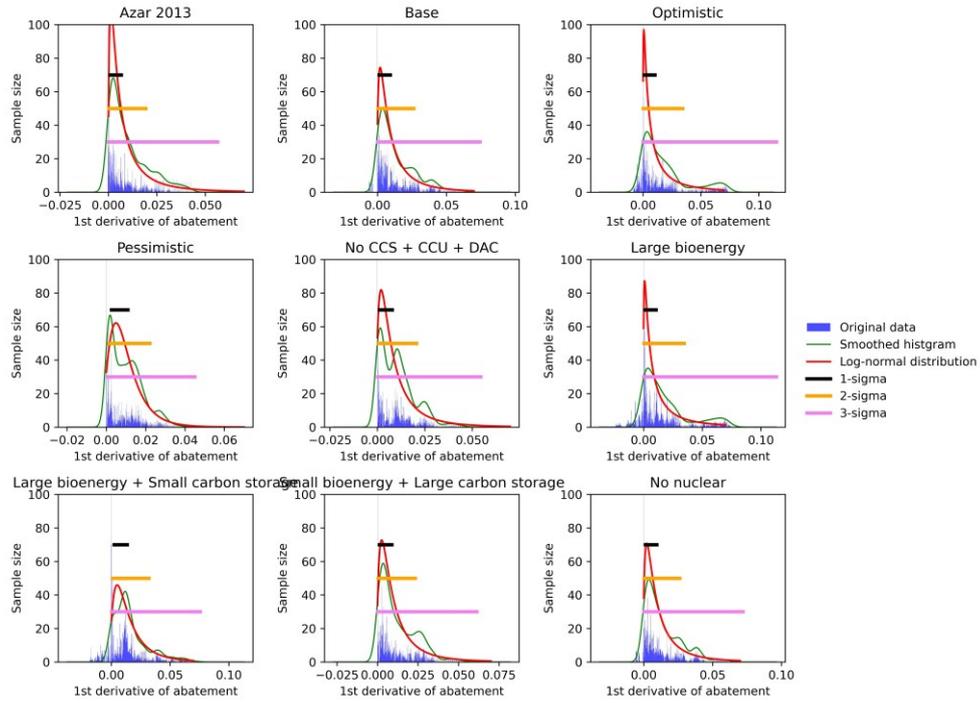

**Figure S38. Global GET - Distribution of first derivative of abatement levels**

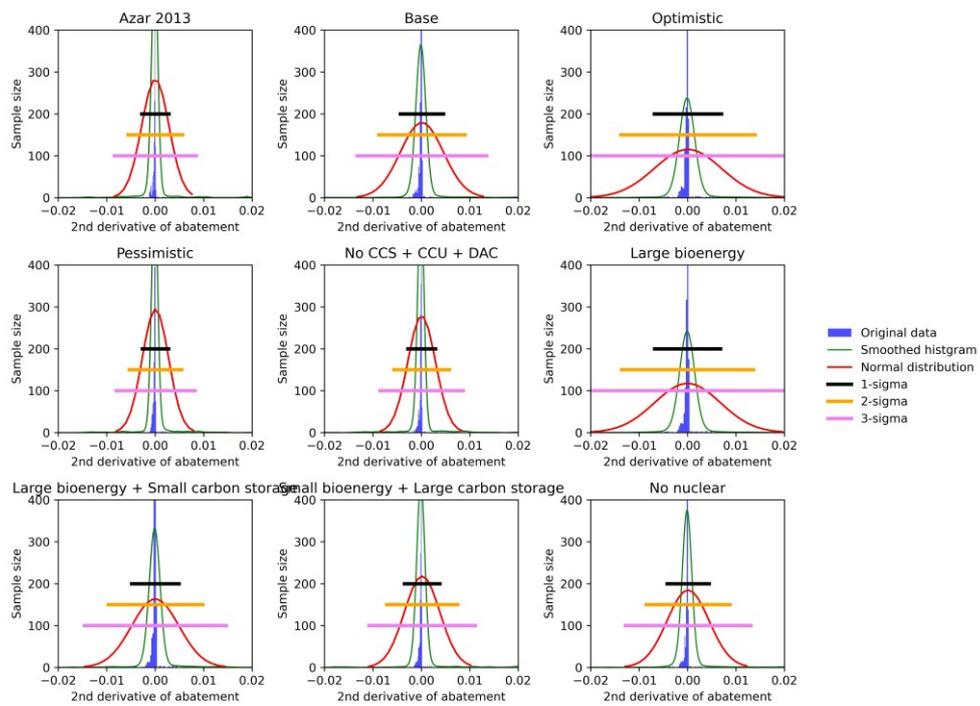

**Figure S39. Global GET - Distribution of second derivative of abatement levels**

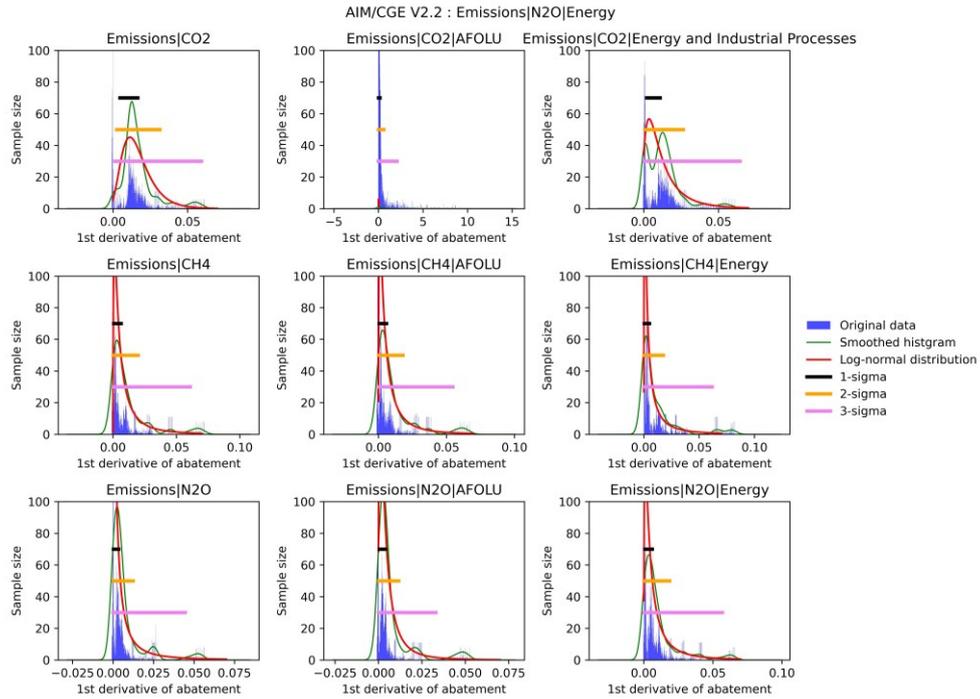

**Figure S40. Global AIM - Distribution of first derivative of abatement levels**

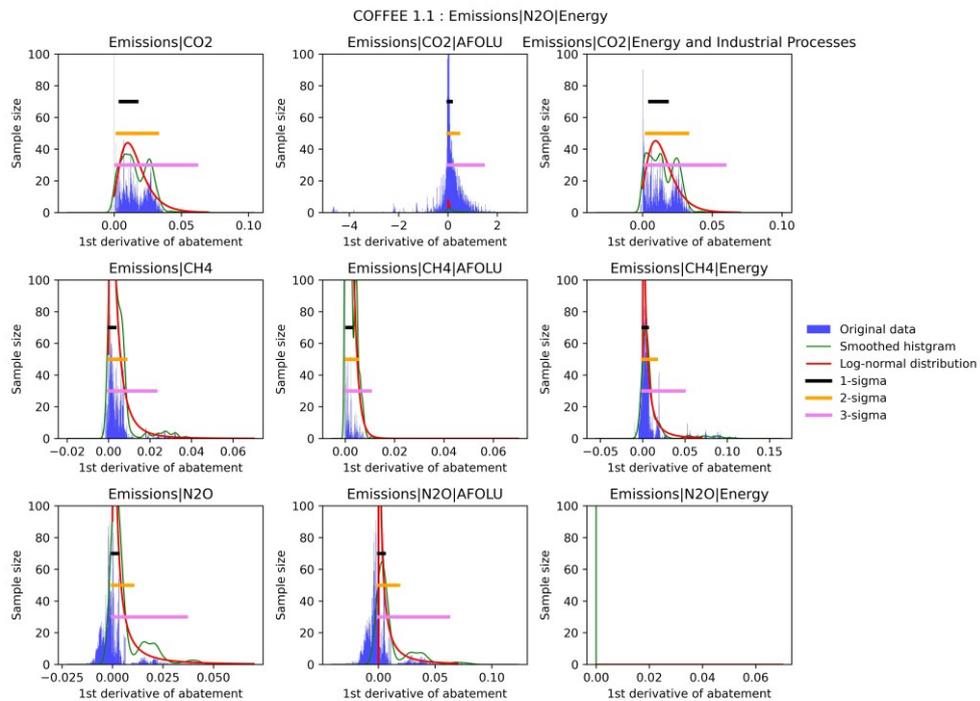

**Figure S41. Global COFFEE - Distribution of first derivative of abatement levels**

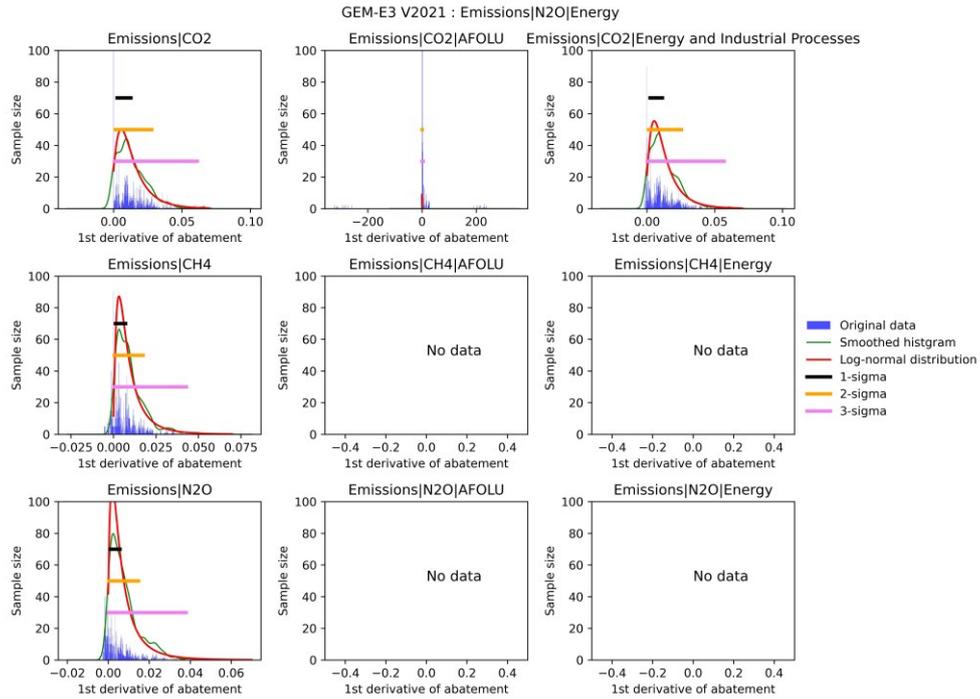

**Figure S42. Global GEM - Distribution of first derivative of abatement levels**

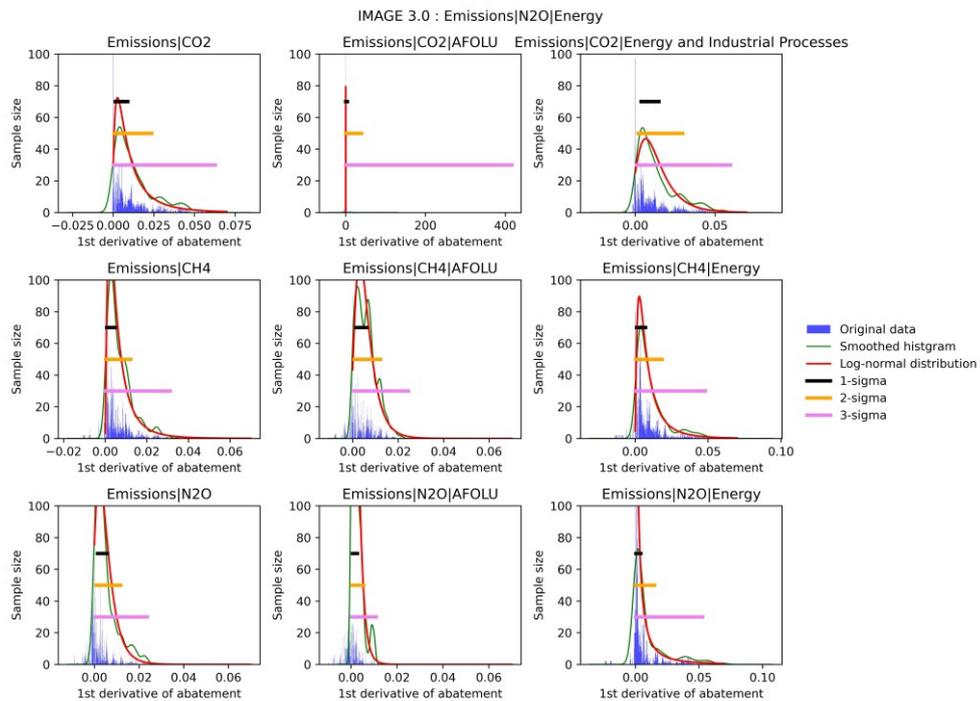

**Figure S43. Global IMAGE - Distribution of first derivative of abatement levels**

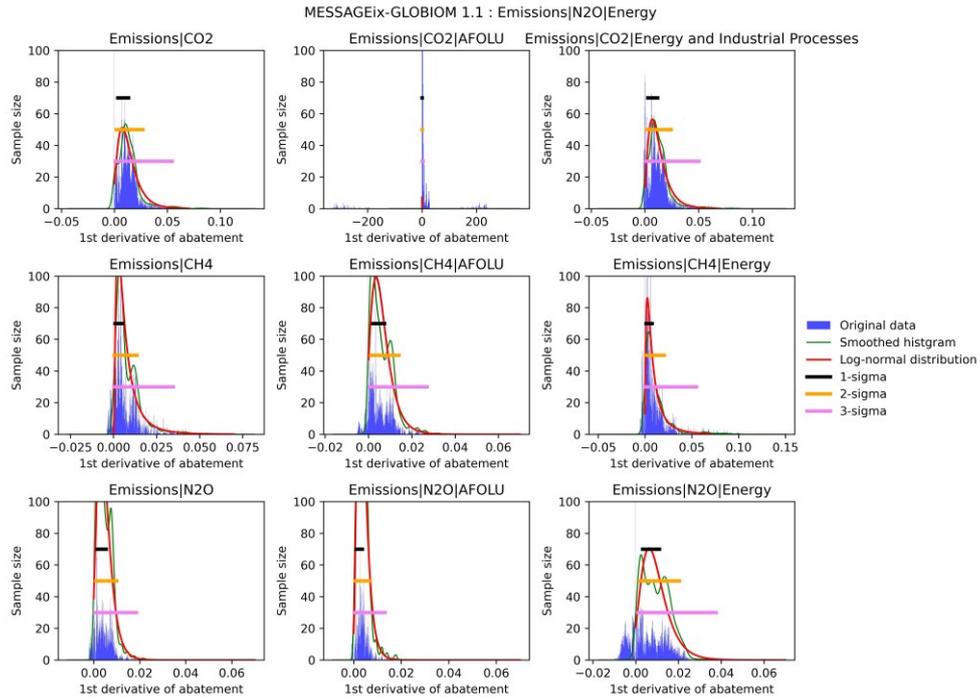

**Figure S44. Global MESSAGE - Distribution of first derivative of abatement levels**

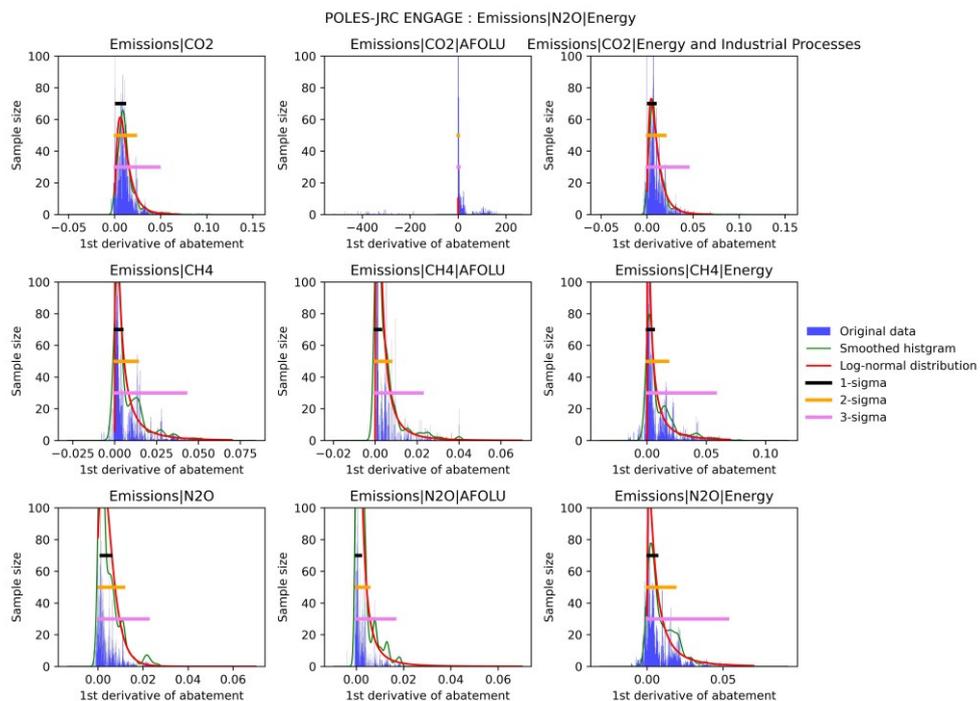

**Figure S45. Global POLES - Distribution of first derivative of abatement levels**

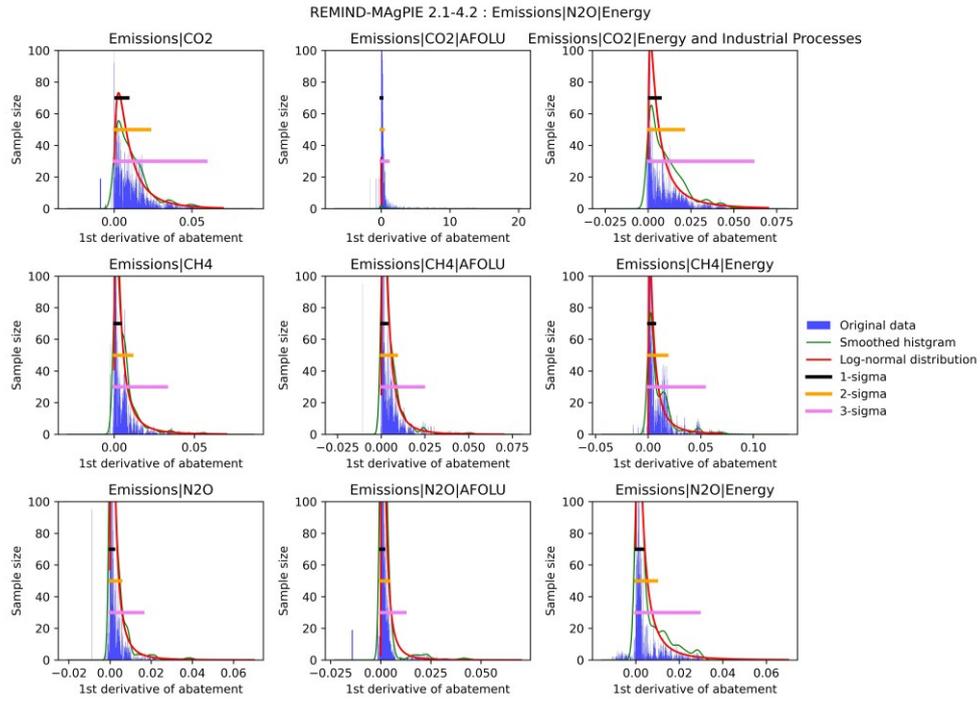

**Figure S46.    Global REMIND - Distribution of first derivative of abatement levels**

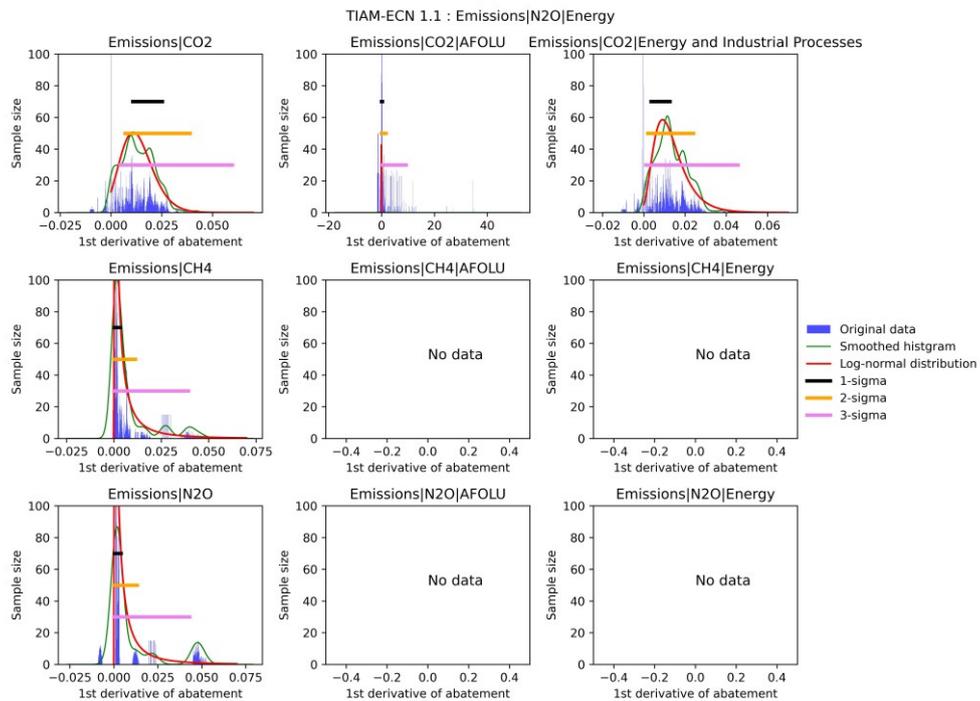

**Figure S47. Global TIAM - Distribution of first derivative of abatement levels**

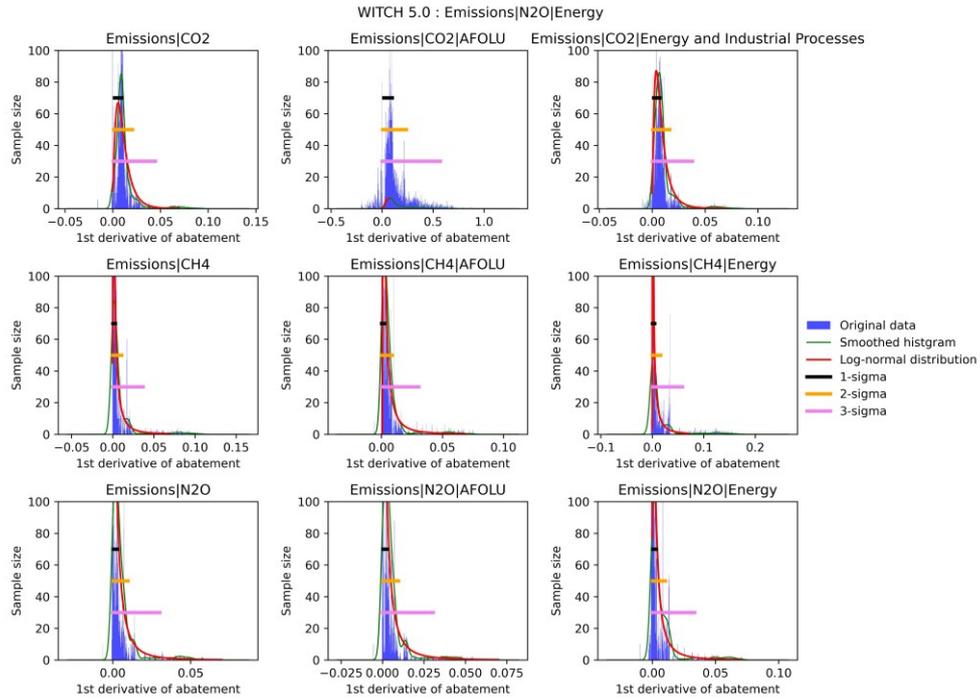

**Figure S48. Global WITCH - Distribution of first derivative of abatement levels**

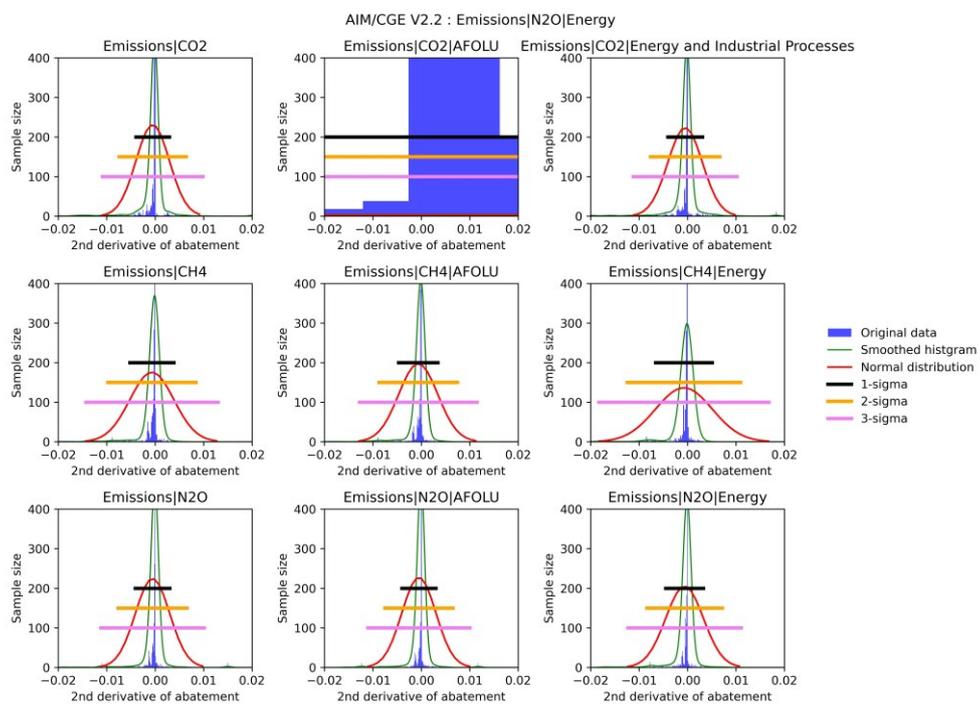

**Figure S49. Global AIM - Distribution of second derivative of abatement levels**

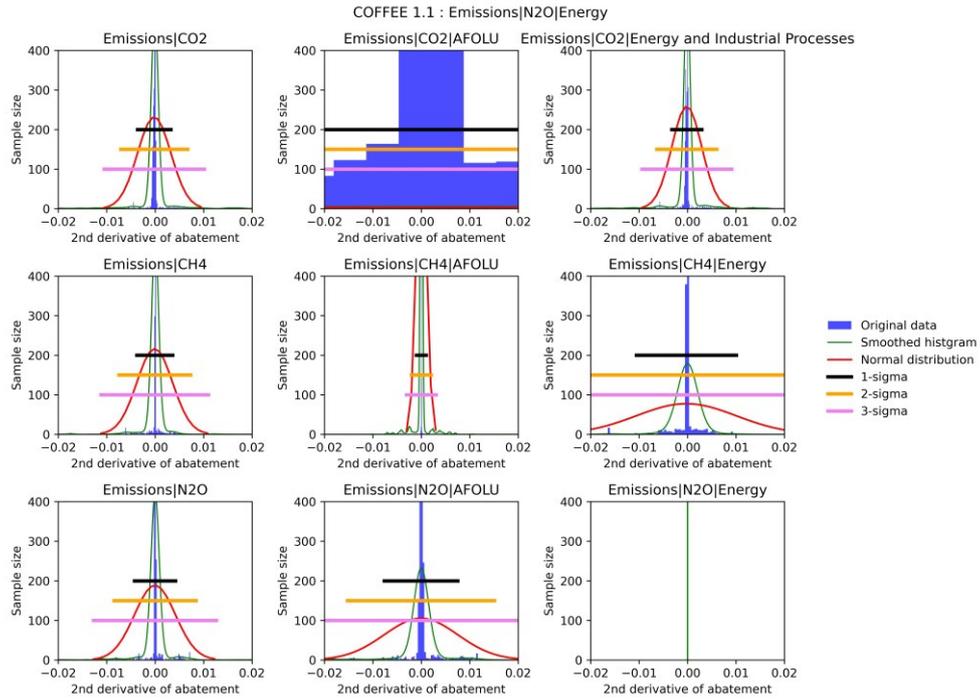

**Figure S50. Global COFFEE - Distribution of second derivative of abatement levels**

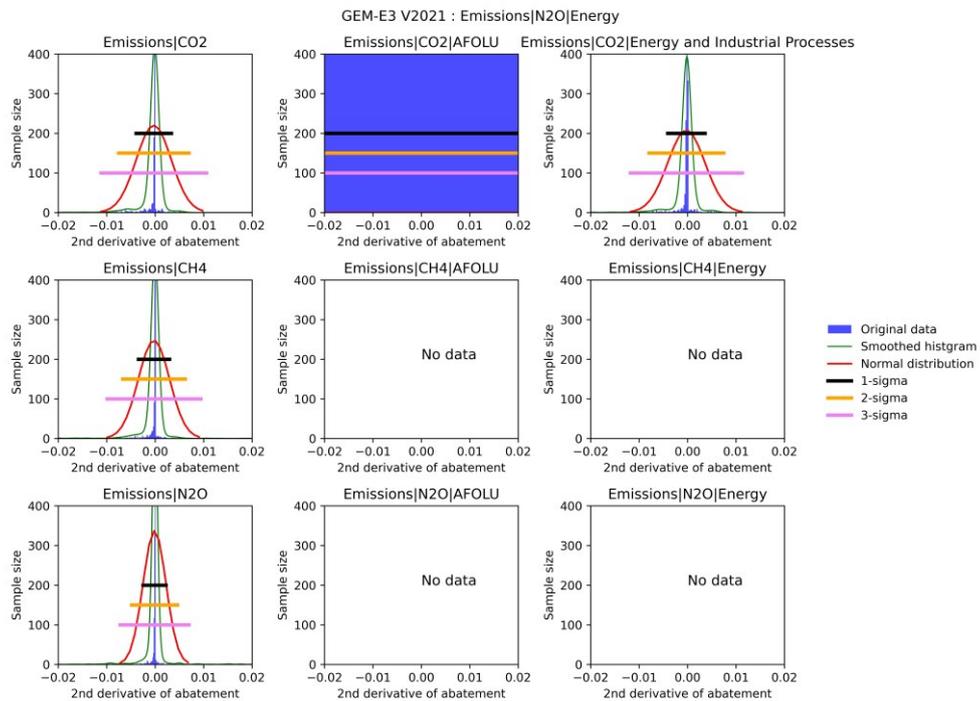

**Figure S51. Global GEM - Distribution of second derivative of abatement levels**

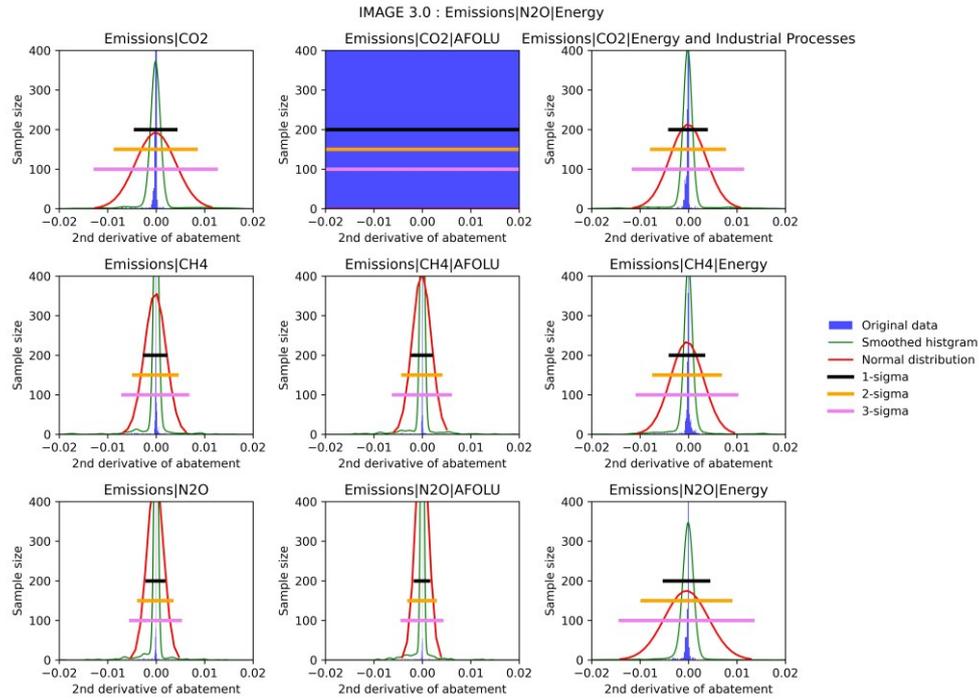

**Figure S52. Global IMAGE - Distribution of second derivative of abatement levels**

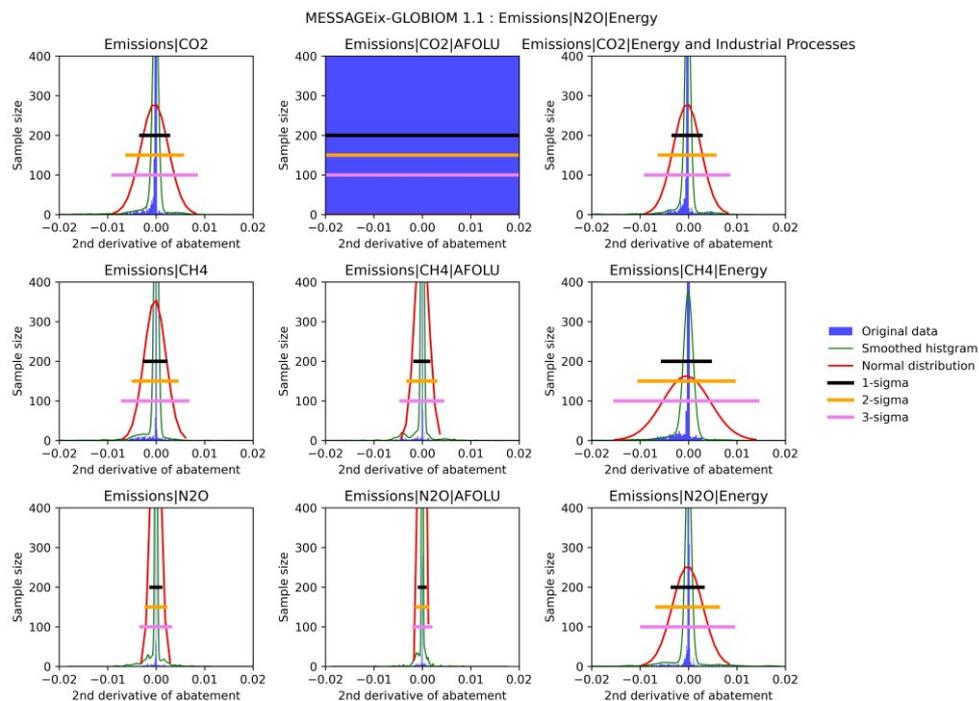

**Figure S53. Global MESSAGE - Distribution of second derivative of abatement levels**

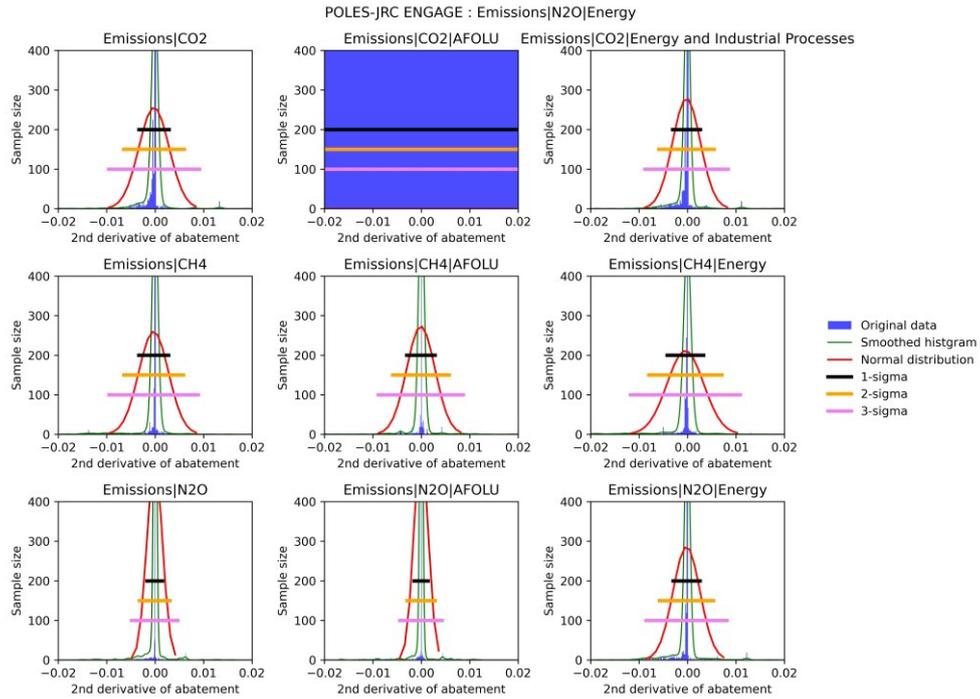

**Figure S54. Global POLES - Distribution of second derivative of abatement levels**

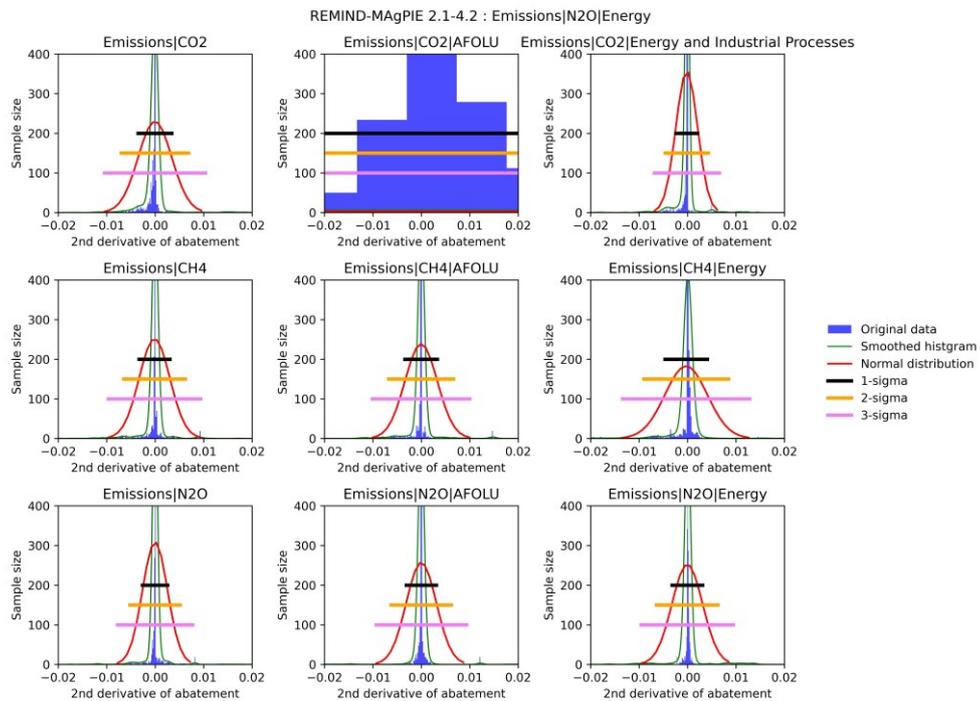

**Figure S55.    Global REMIND - Distribution of second derivative of abatement levels**

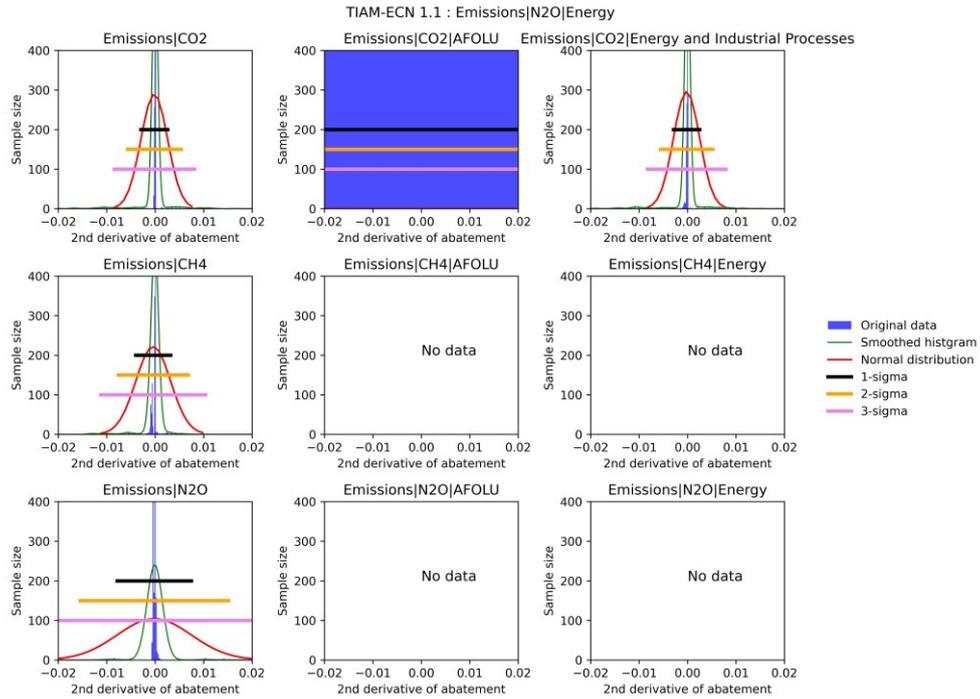

**Figure S56.    Global TIAM - Distribution of second derivative of abatement levels**

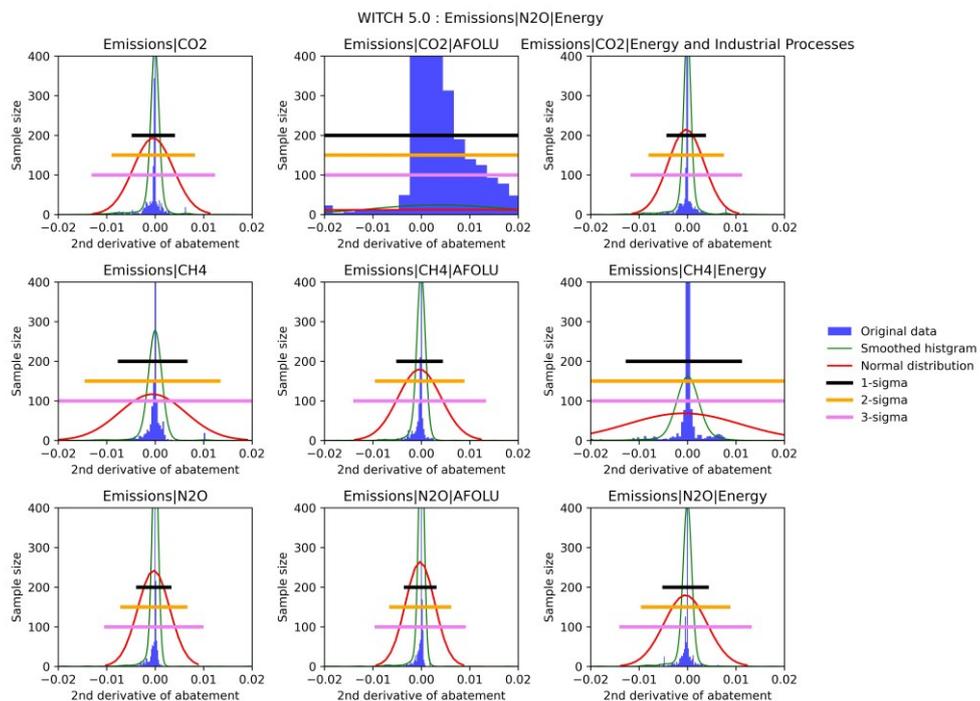

**Figure S57.    Global WITCH - Distribution of second derivative of abatement levels**

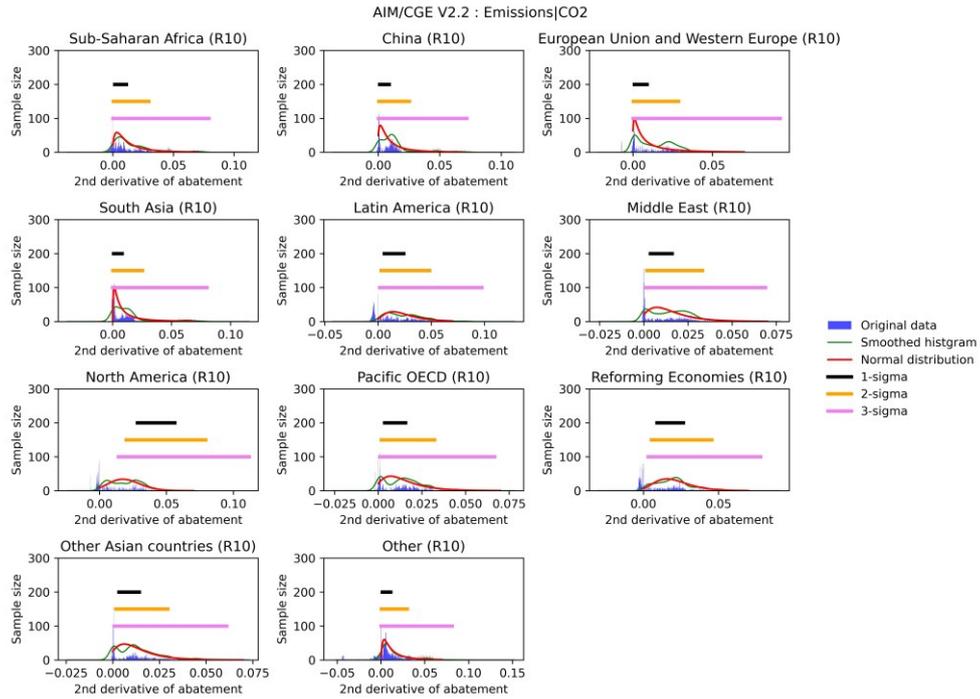

**Figure S58. Regional AIM CO₂ - Distribution of first derivative of abatement levels**

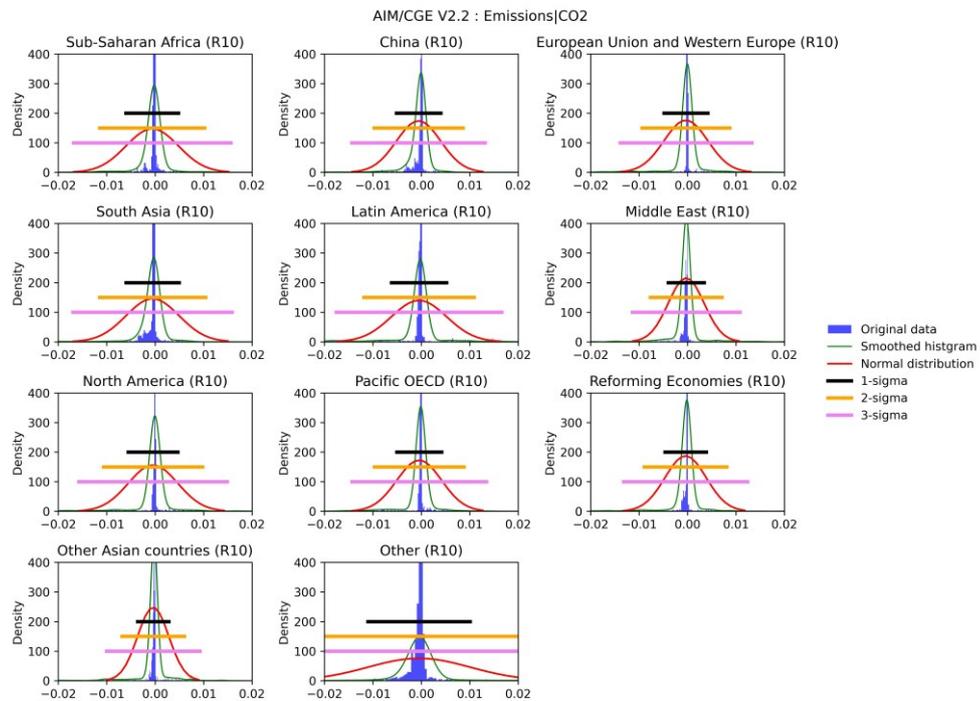

**Figure S59. Regional AIM CO₂ - Distribution of first derivative of abatement levels**

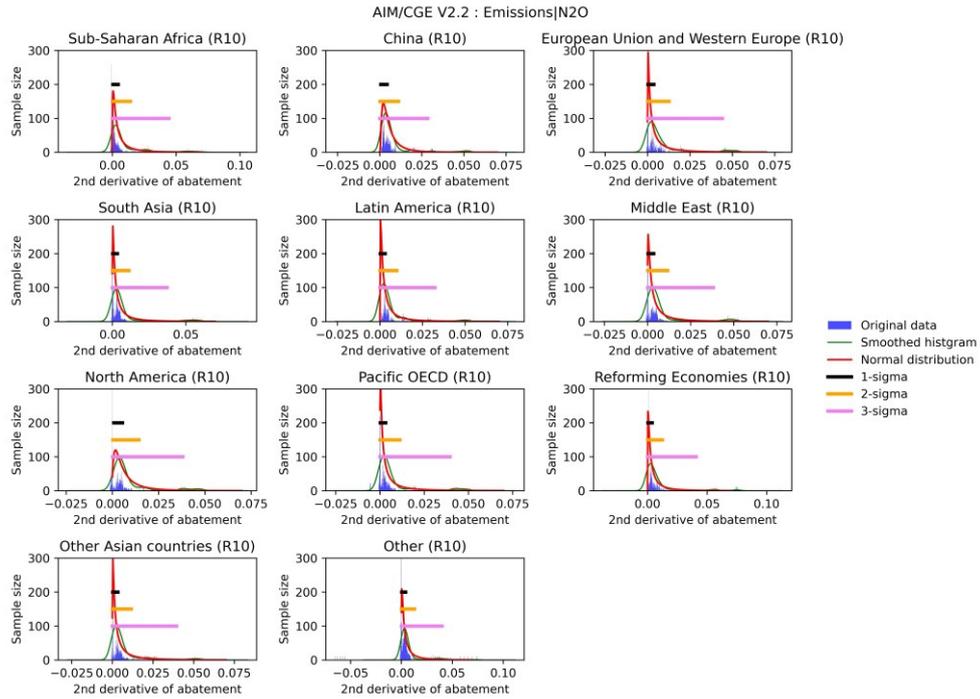

**Figure S60. Regional AIM CH₄- Distribution of first derivative of abatement levels**

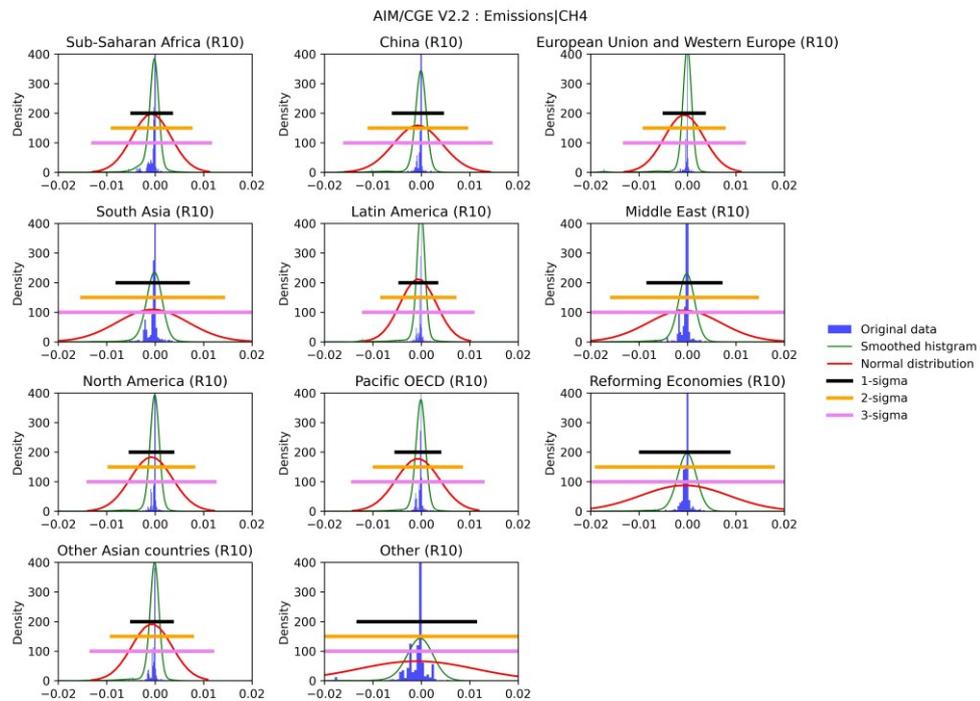

**Figure S61. Regional AIM CH₄ - Distribution of second derivative of abatement levels**

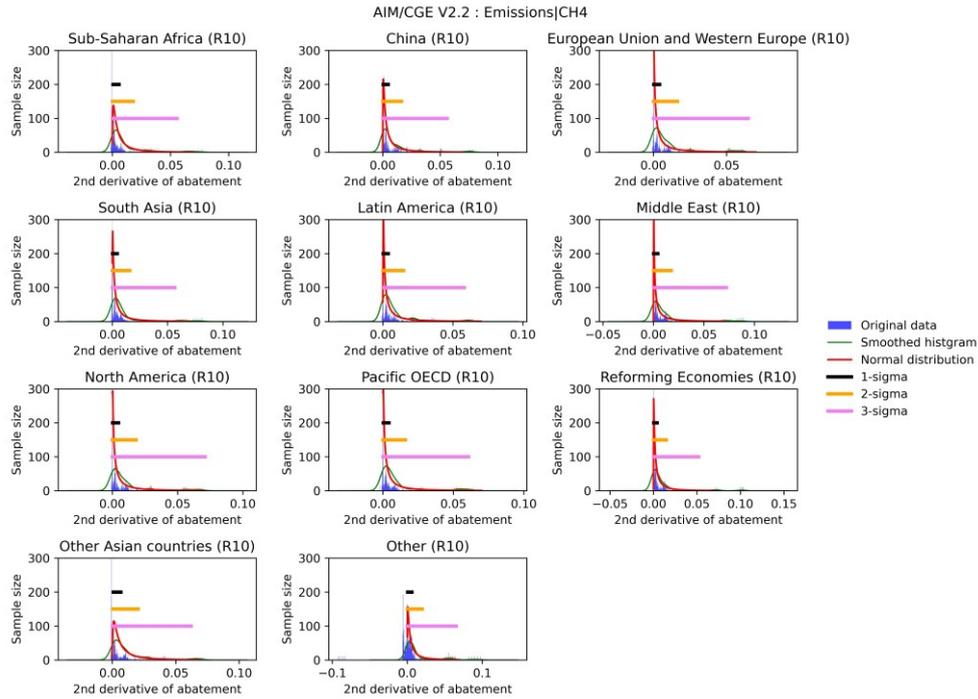

**Figure S62. Regional AIM N₂O - Distribution of first derivative of abatement levels**

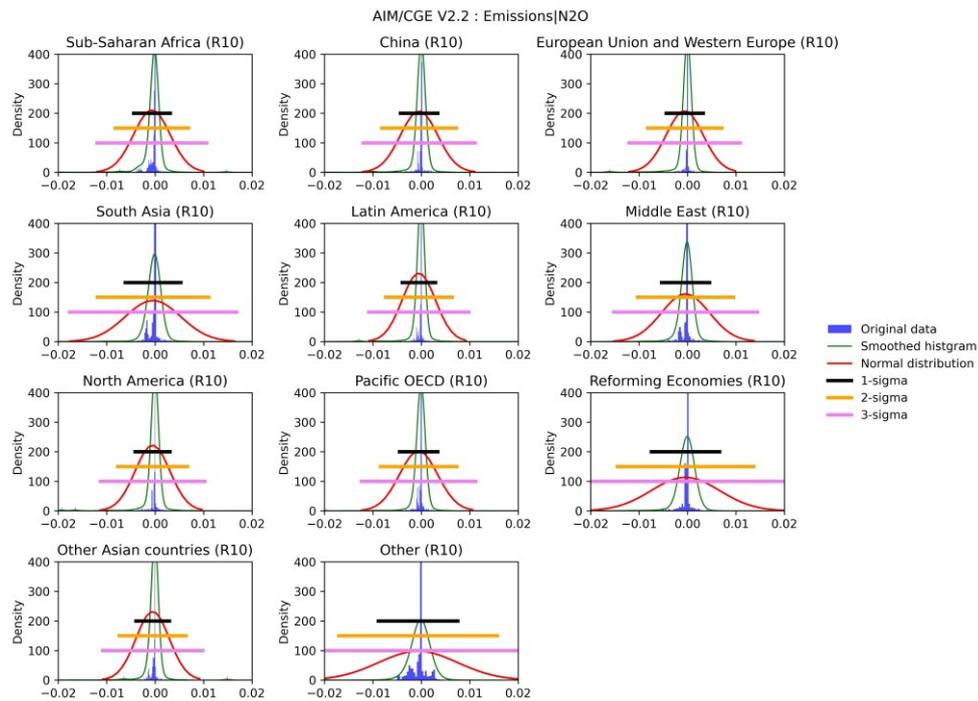

**Figure S63. Regional AIM N₂O - Distribution of second derivative of abatement levels**

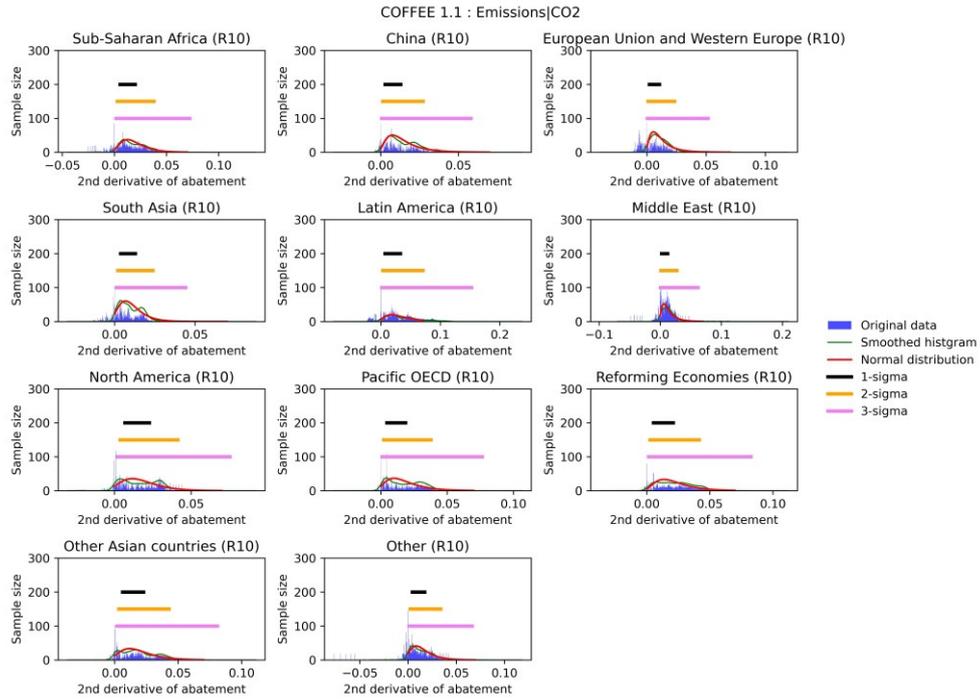

**Figure S64. Regional COFFEE CO$_2$ - Distribution of first derivative of abatement levels**

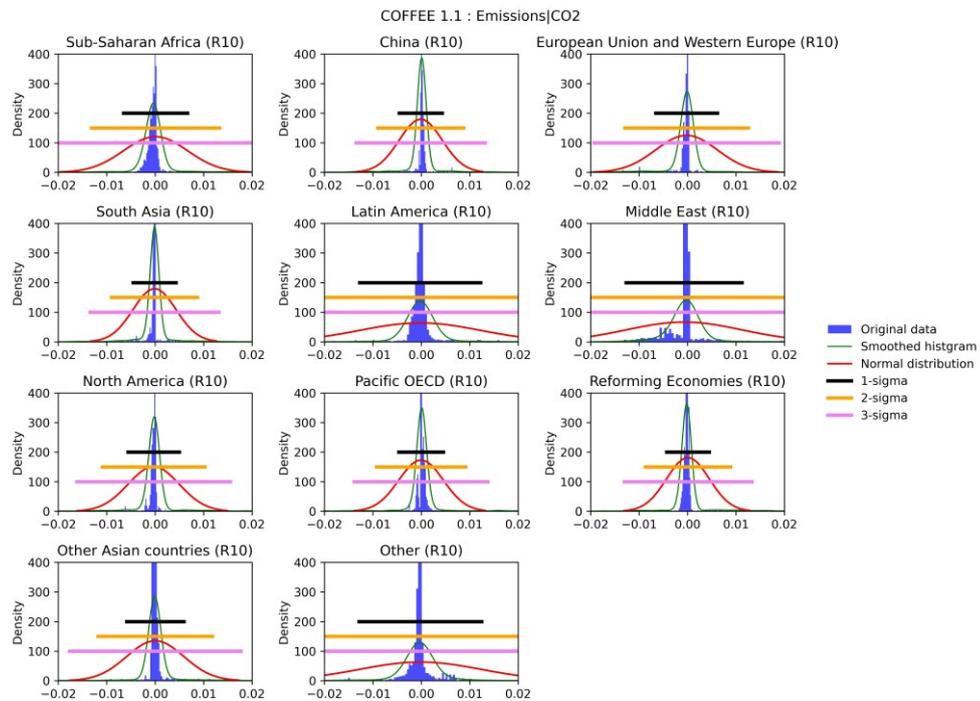

**Figure S65. Regional COFFEE CO$_2$ - Distribution of second derivative of abatement levels**

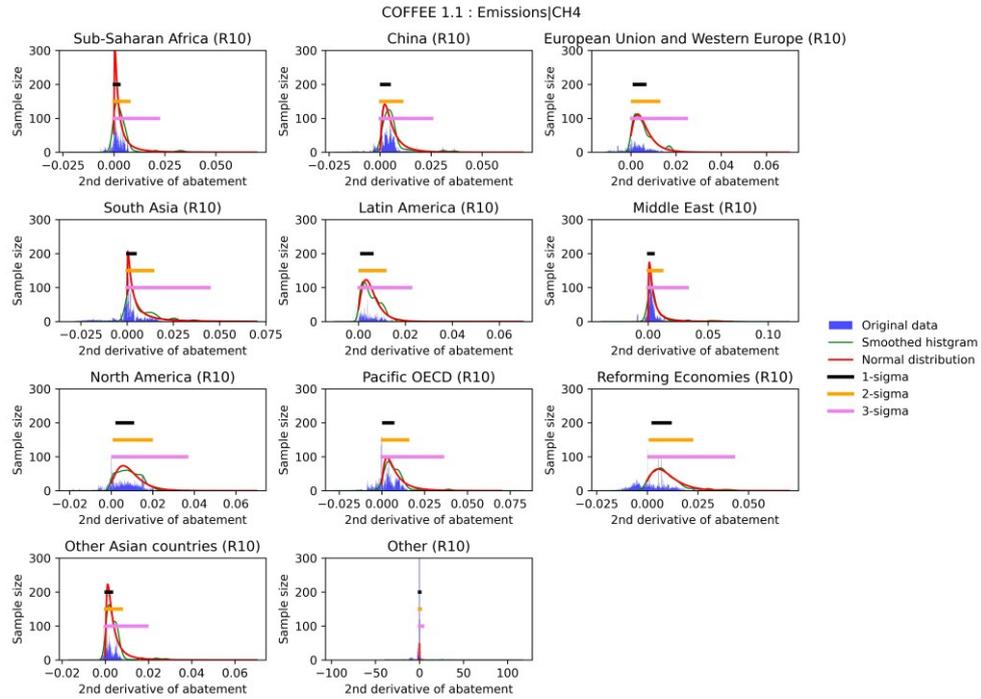

**Figure S66. Regional COFFEE CH₄- Distribution of first derivative of abatement levels**

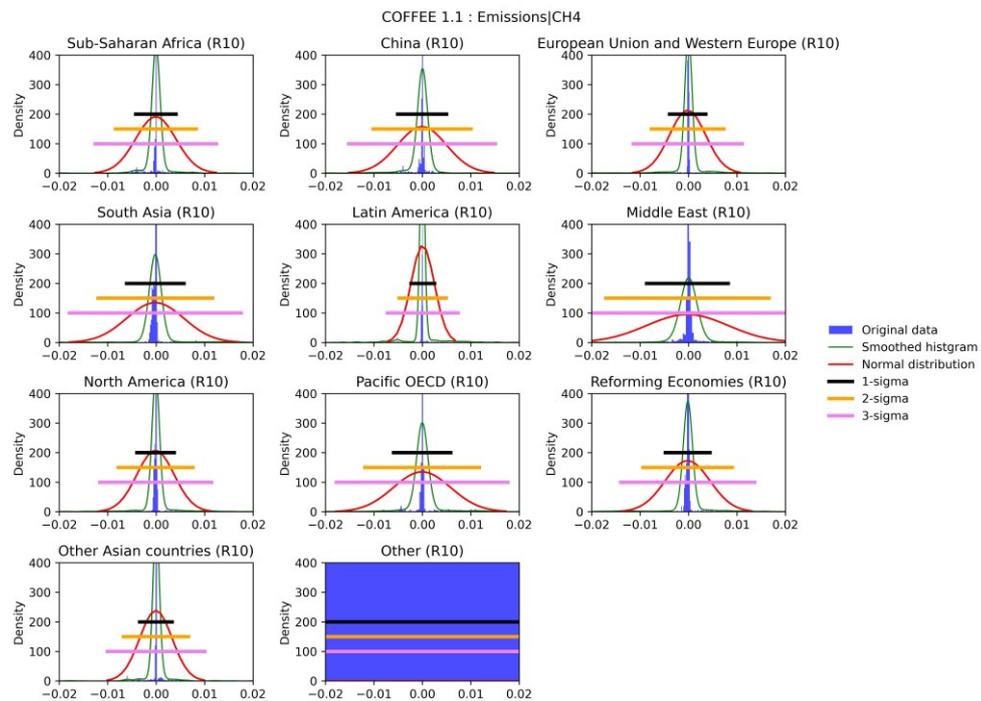

**Figure S67. Regional COFFEE CH₄ - Distribution of second derivative of abatement levels**

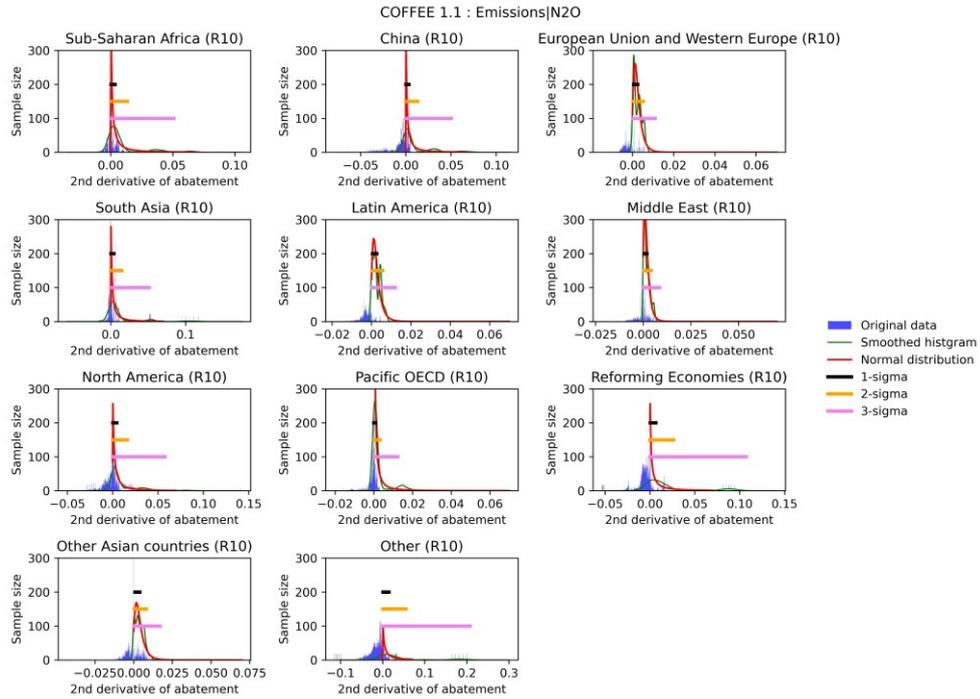

**Figure S68. Regional COFFEE N$_2$O - Distribution of first derivative of abatement levels**

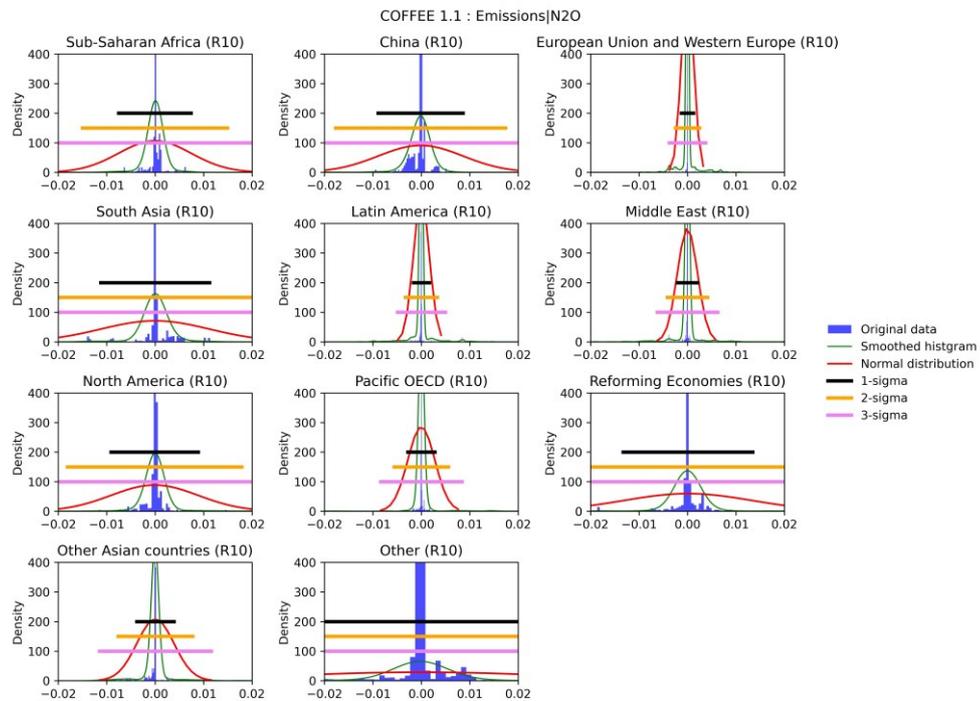

**Figure S69. Regional COFFEE N$_2$O - Distribution of second derivative of abatement levels**

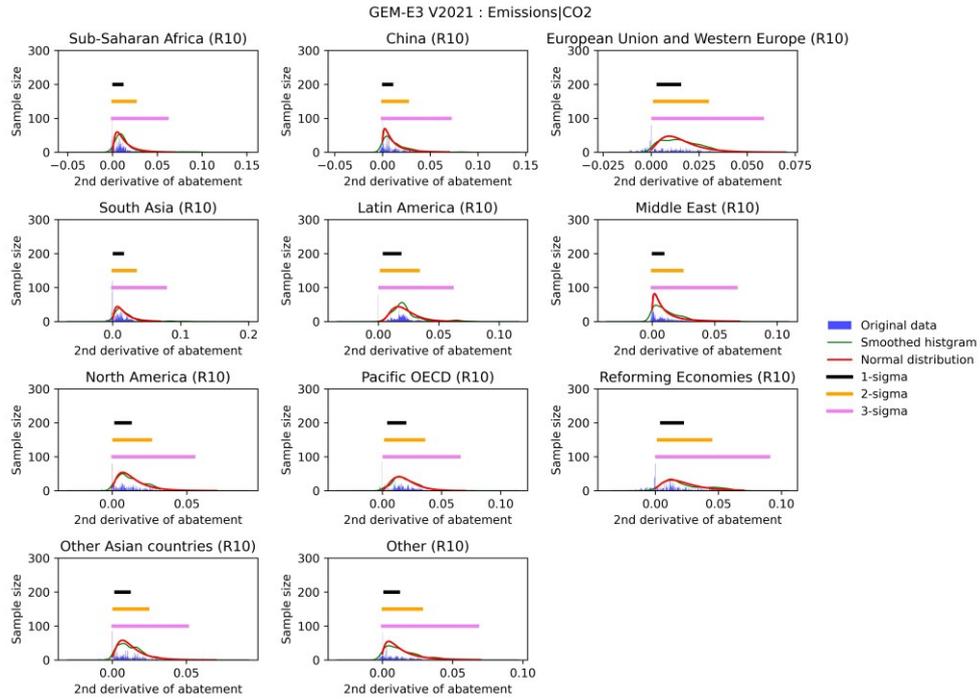

**Figure S70. Regional GEM CO₂ - Distribution of first derivative of abatement levels**

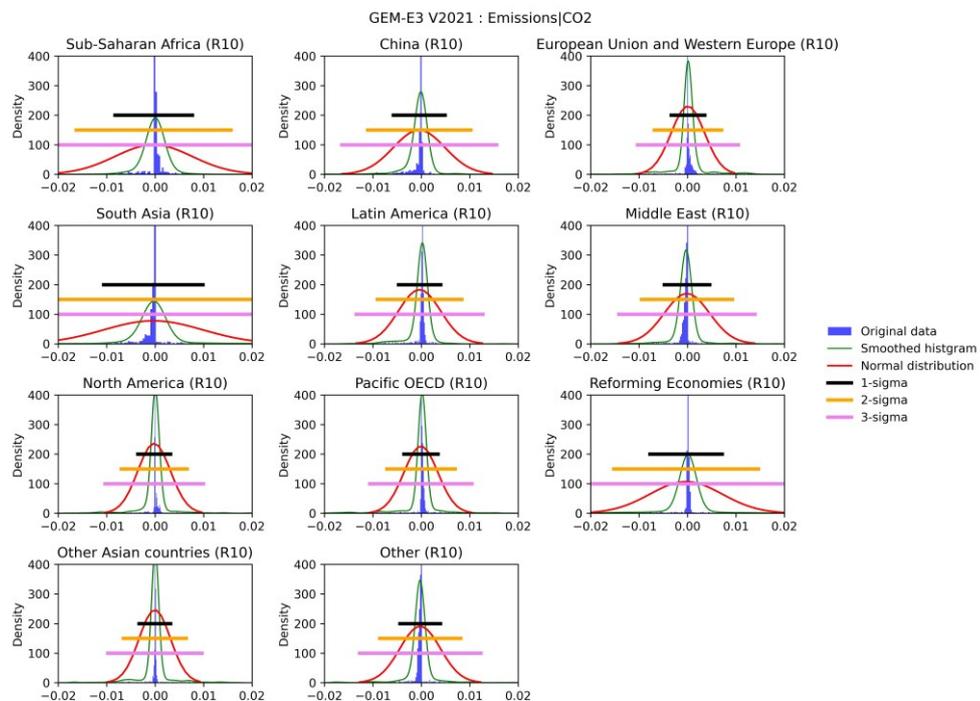

**Figure S71. Regional GEM CO₂ - Distribution of second derivative of abatement levels**

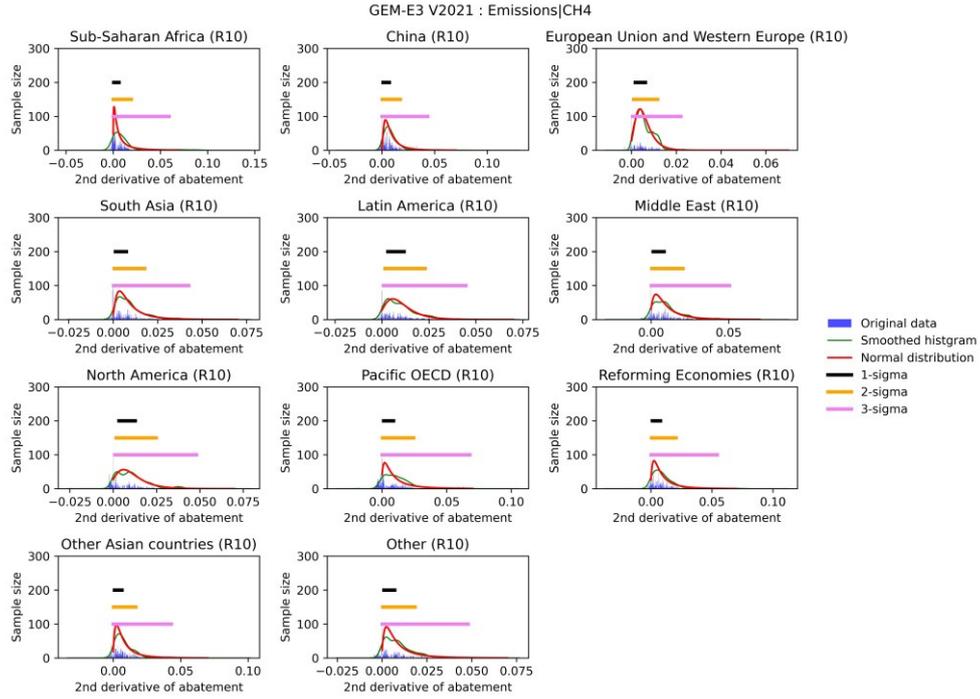

**Figure S72. Regional GEM CH₄- Distribution of first derivative of abatement levels**

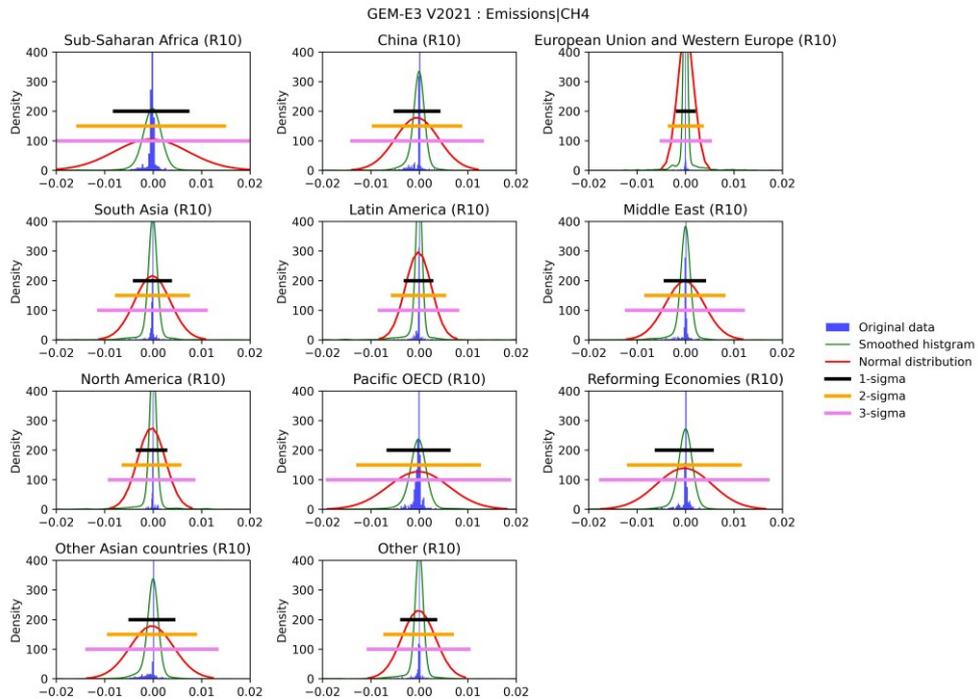

**Figure S73. Regional GEM CH₄ - Distribution of second derivative of abatement levels**

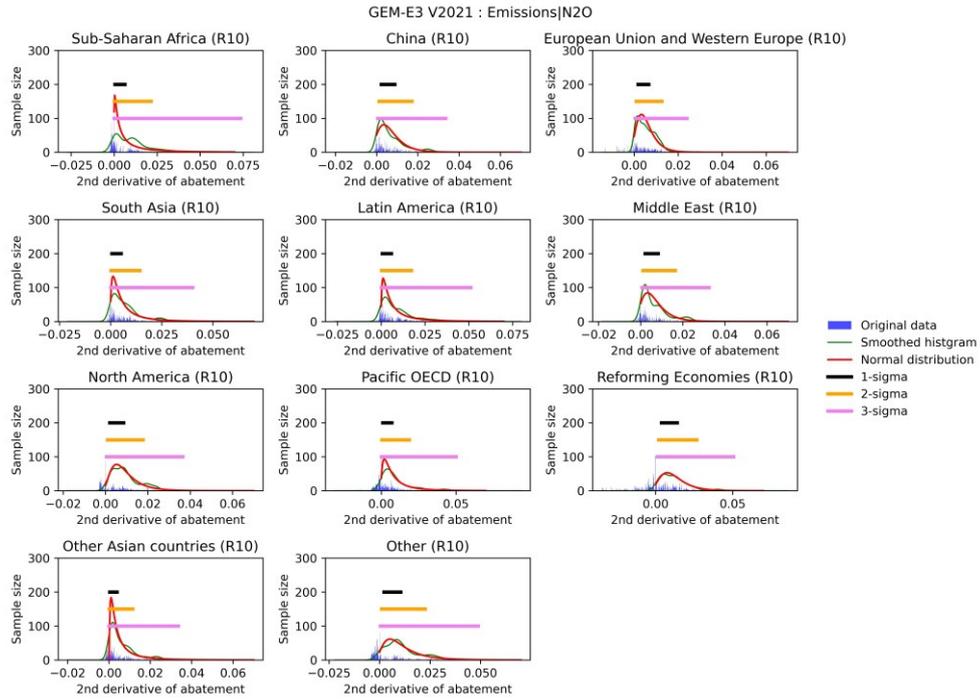

**Figure S74. Regional GEM N₂O - Distribution of first derivative of abatement levels**

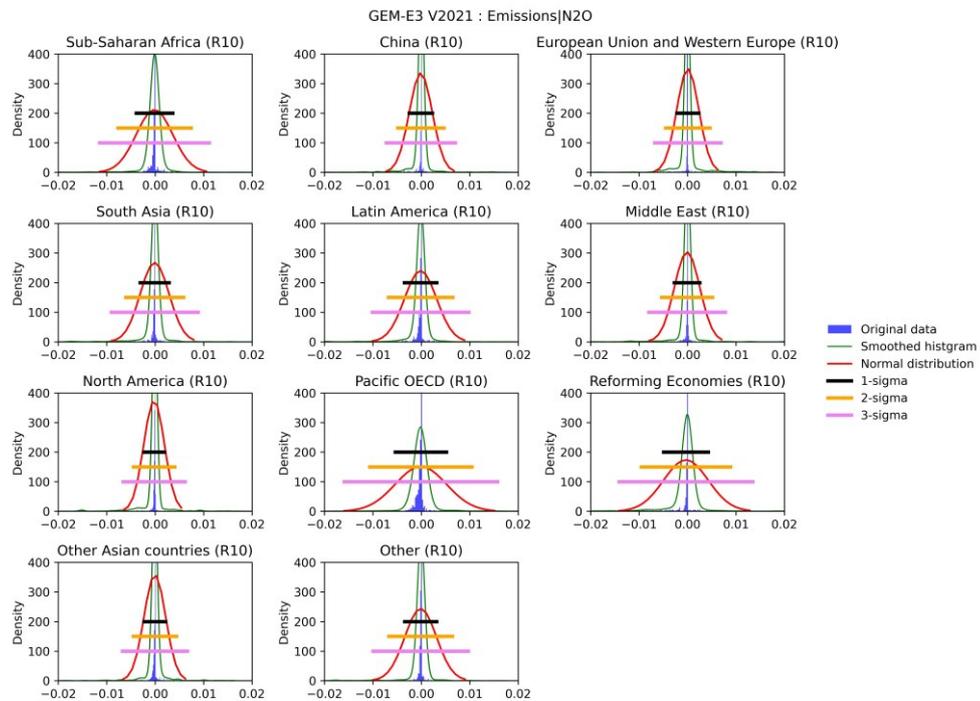

**Figure S75. Regional GEM N₂O - Distribution of second derivative of abatement levels**

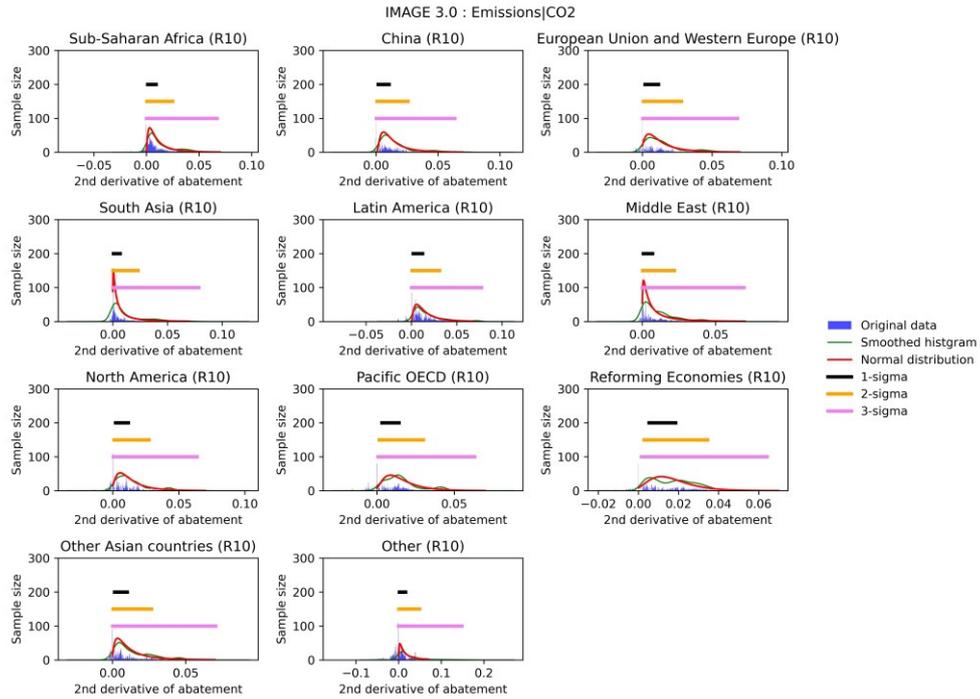

**Figure S76. Regional IMAGE CO₂ - Distribution of first derivative of abatement levels**

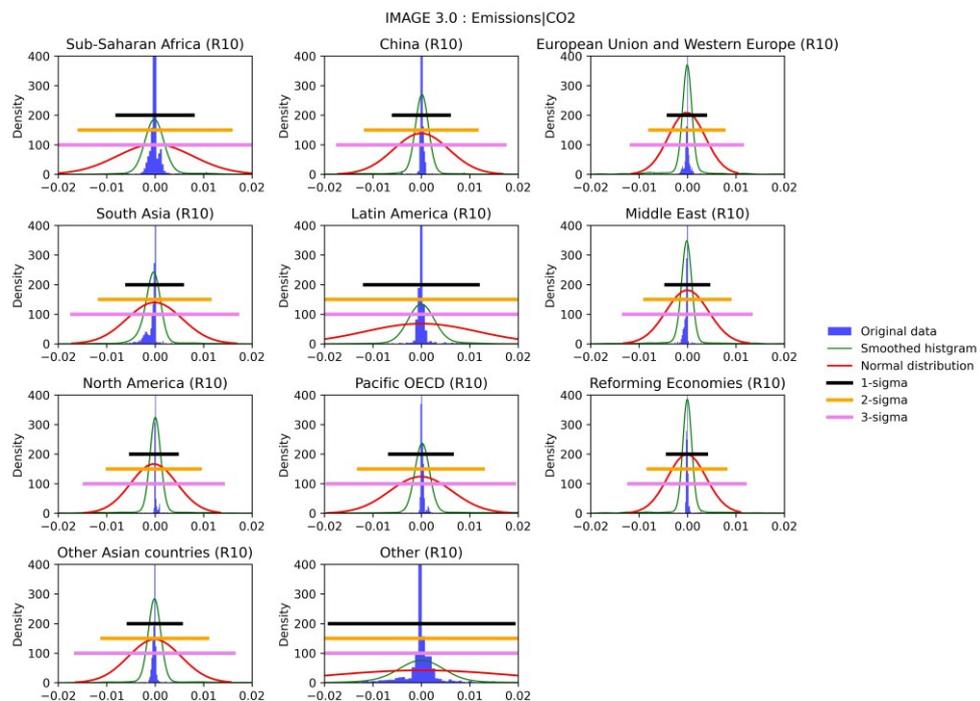

**Figure S77. Regional IMAGE CO₂ - Distribution of second derivative of abatement levels**

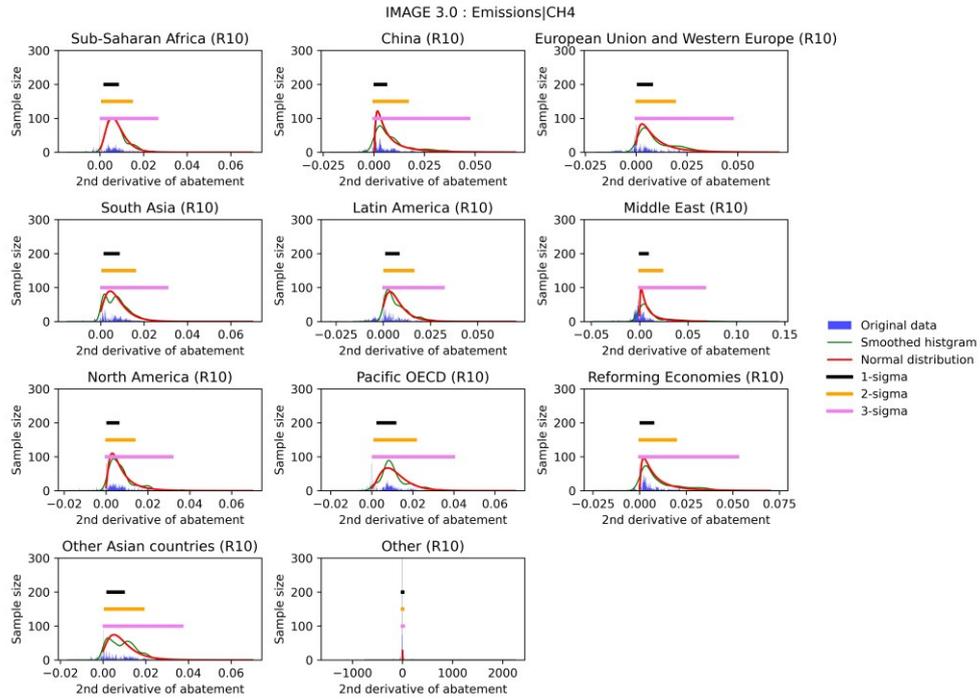

**Figure S78. Regional IMAGE CH₄- Distribution of first derivative of abatement levels**

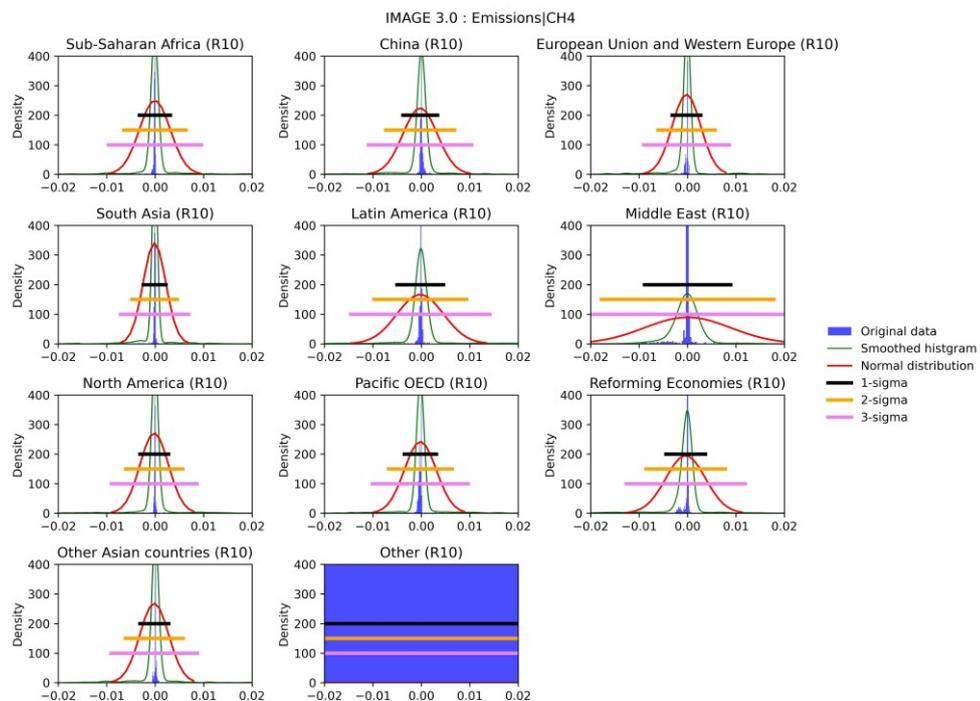

**Figure S79. Regional IMAGE CH₄ - Distribution of second derivative of abatement levels**

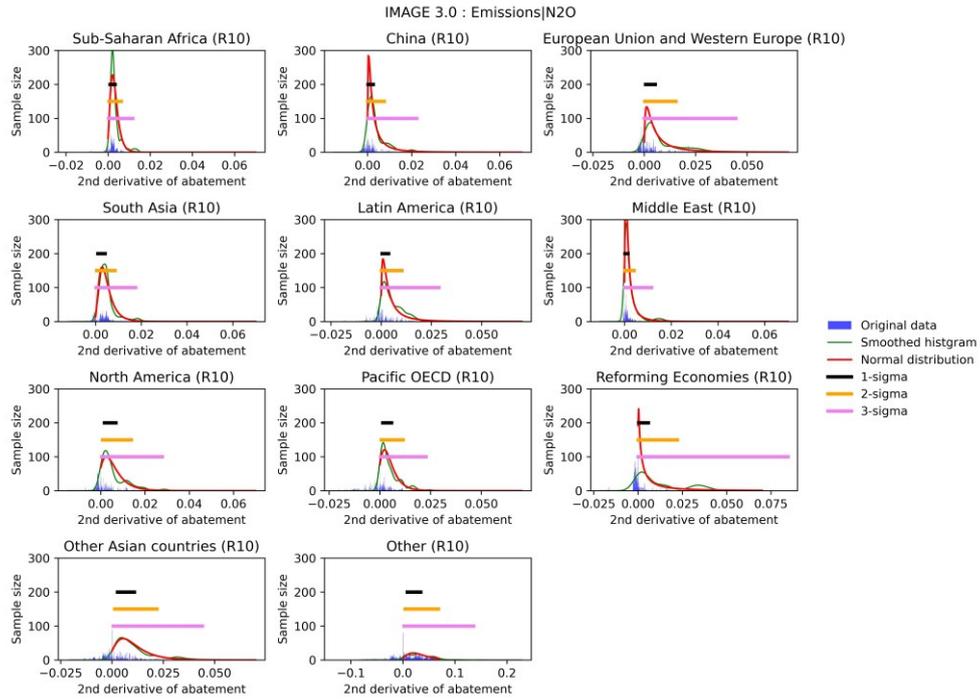

**Figure S80. Regional IMAGE N₂O - Distribution of first derivative of abatement levels**

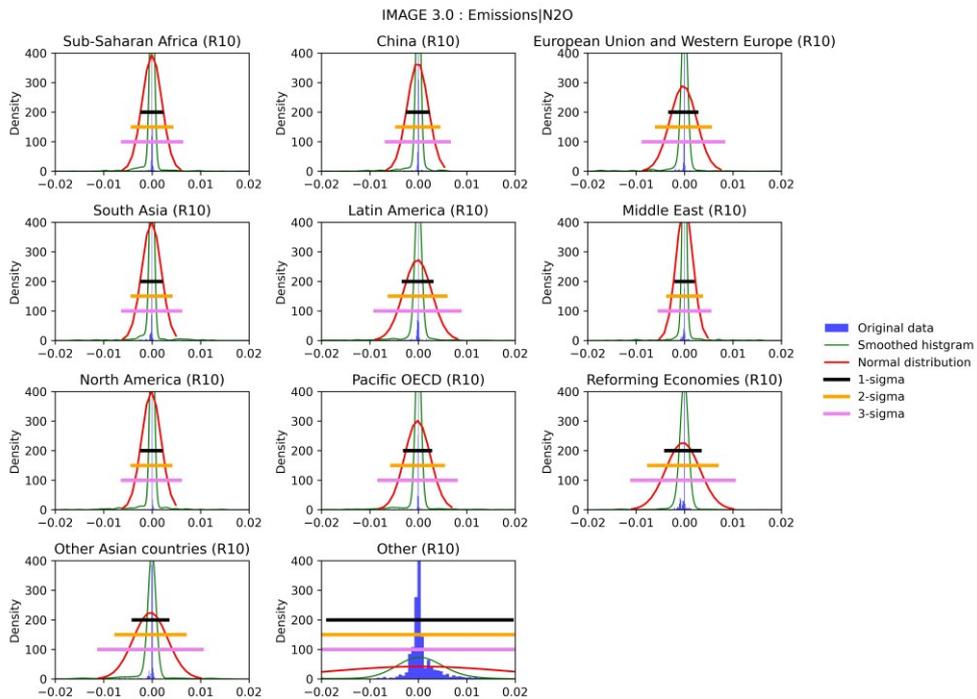

**Figure S81. Regional IMAGE N₂O - Distribution of second derivative of abatement levels**

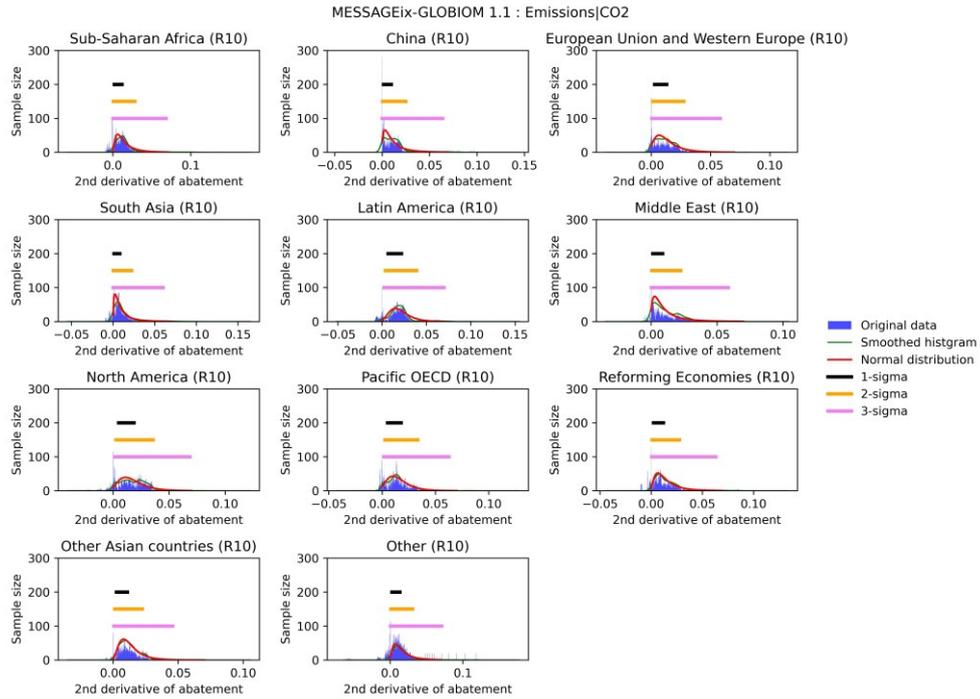

**Figure S82. Regional MESSAGE CO₂ - Distribution of first derivative of abatement levels**

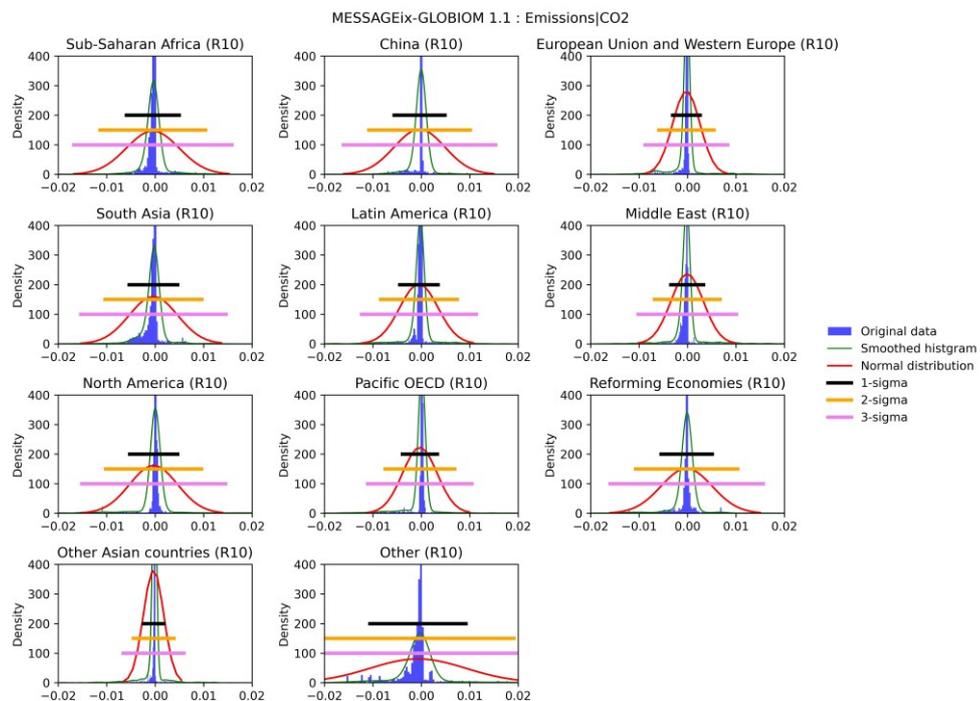

**Figure S83.  Regional MESSAGE CO₂ - Distribution of second derivative of abatement levels**

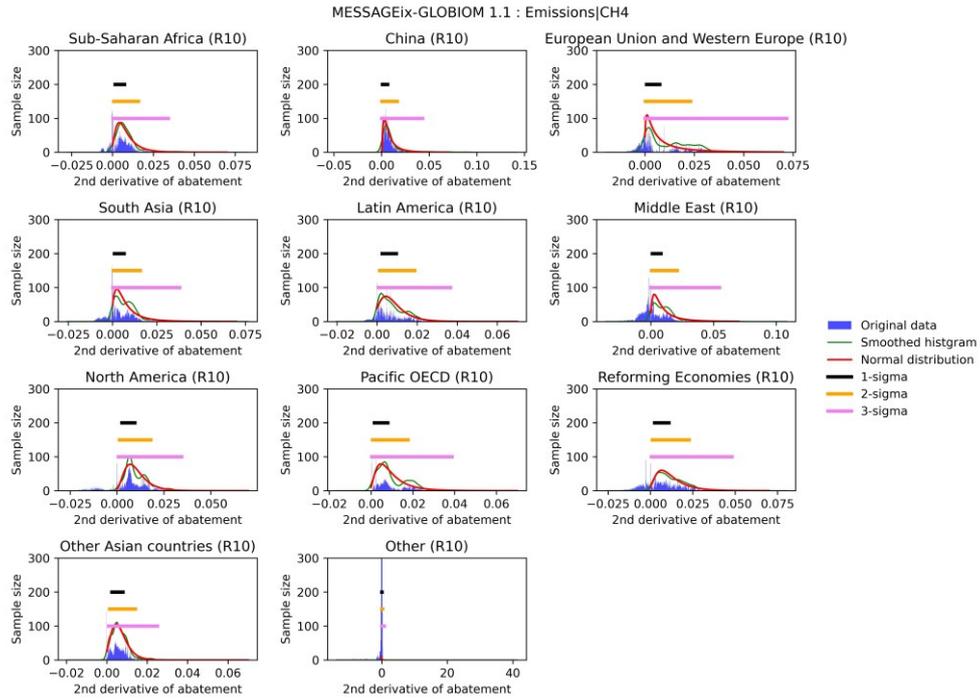

**Figure S84. Regional MESSAGE CH₄ - Distribution of first derivative of abatement levels**

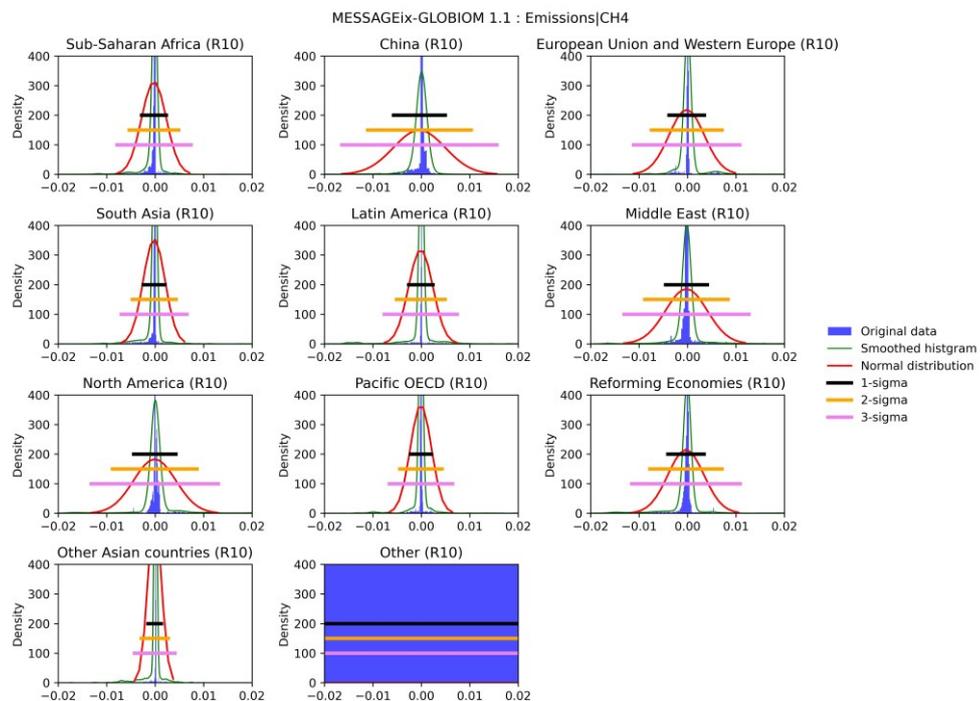

**Figure S85.    Regional MESSAGE CH₄ - Distribution of second derivative of abatement levels**

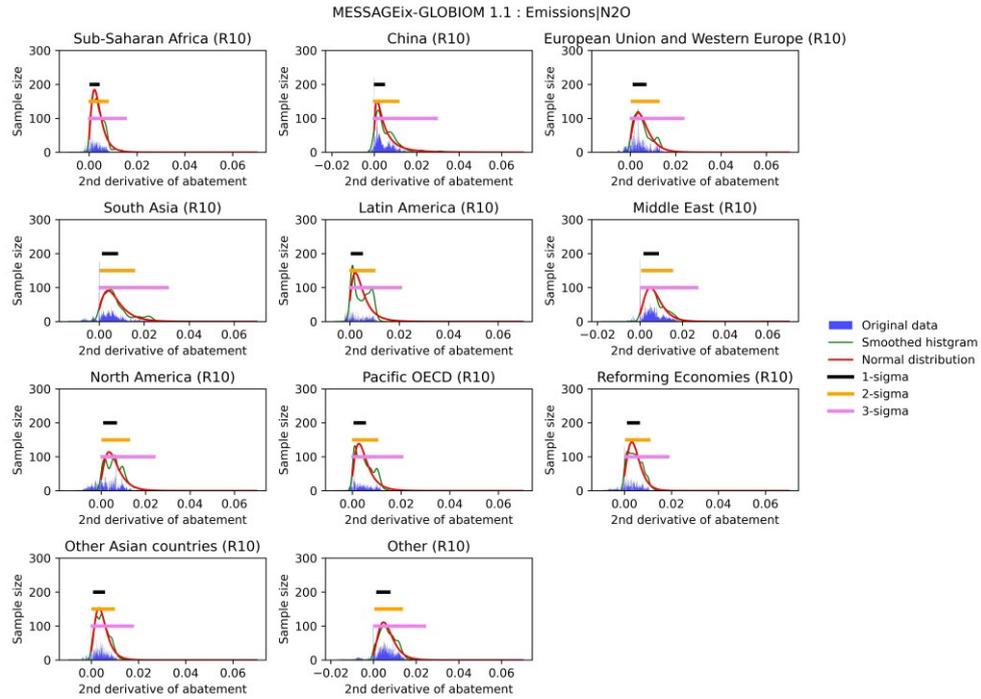

**Figure S86. Regional MESSAGE N₂O - Distribution of first derivative of abatement levels**

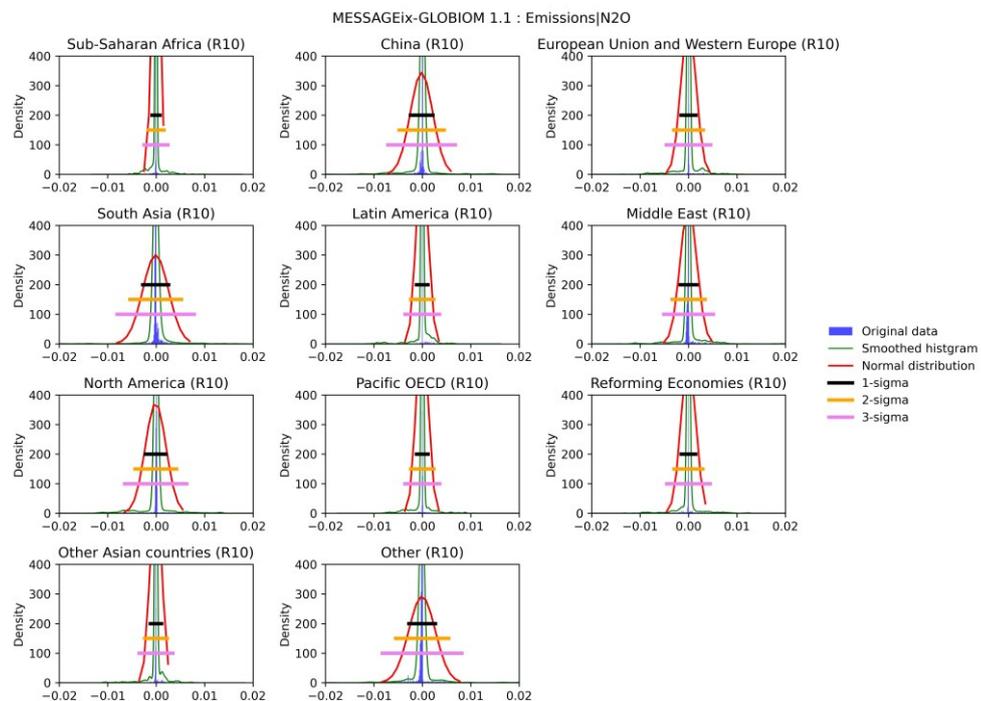

**Figure S87. Regional MESSAGE N₂O - Distribution of second derivative of abatement levels**

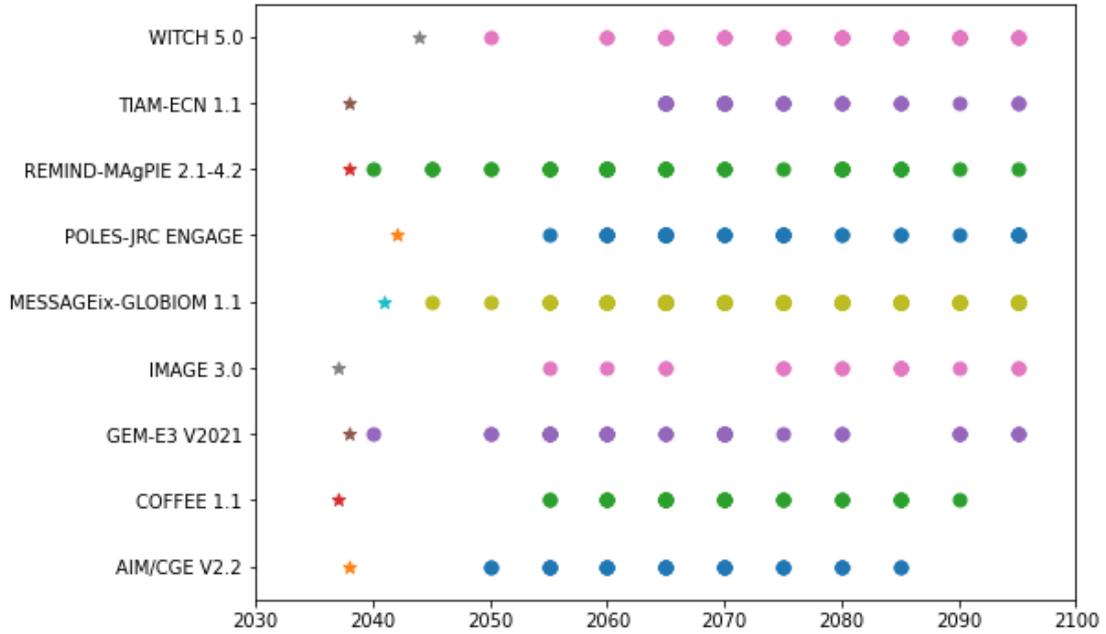

**Figure S88. The earliest year to reach net zero for ENGAGE IAMs and MAC curves we derived.** The circle points are the results from ENGAGE IAMs, and the star points are the results from MAC curves. We calculated the earliest time for different MACs to reach net zero using the largest 1st and 2nd derivatives.

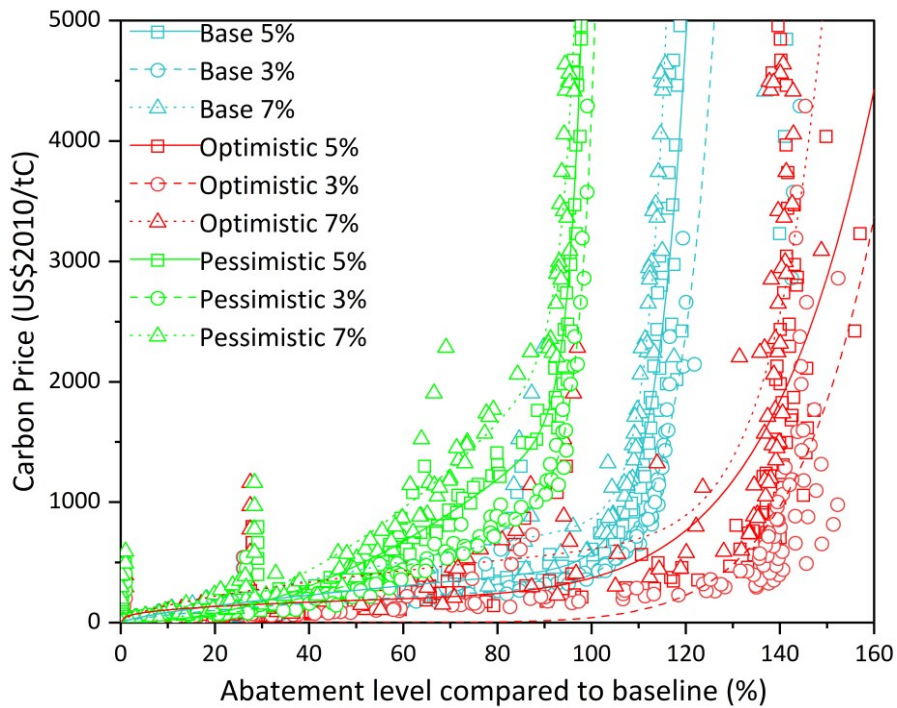

**Figure S89. The relationship between abatement level and carbon price with different discount rates for policy portfolios in GET model.** For individual policy portfolios, only the discount rate in the GET model is changed as 3%, 5% (default), and 7%, respectively.

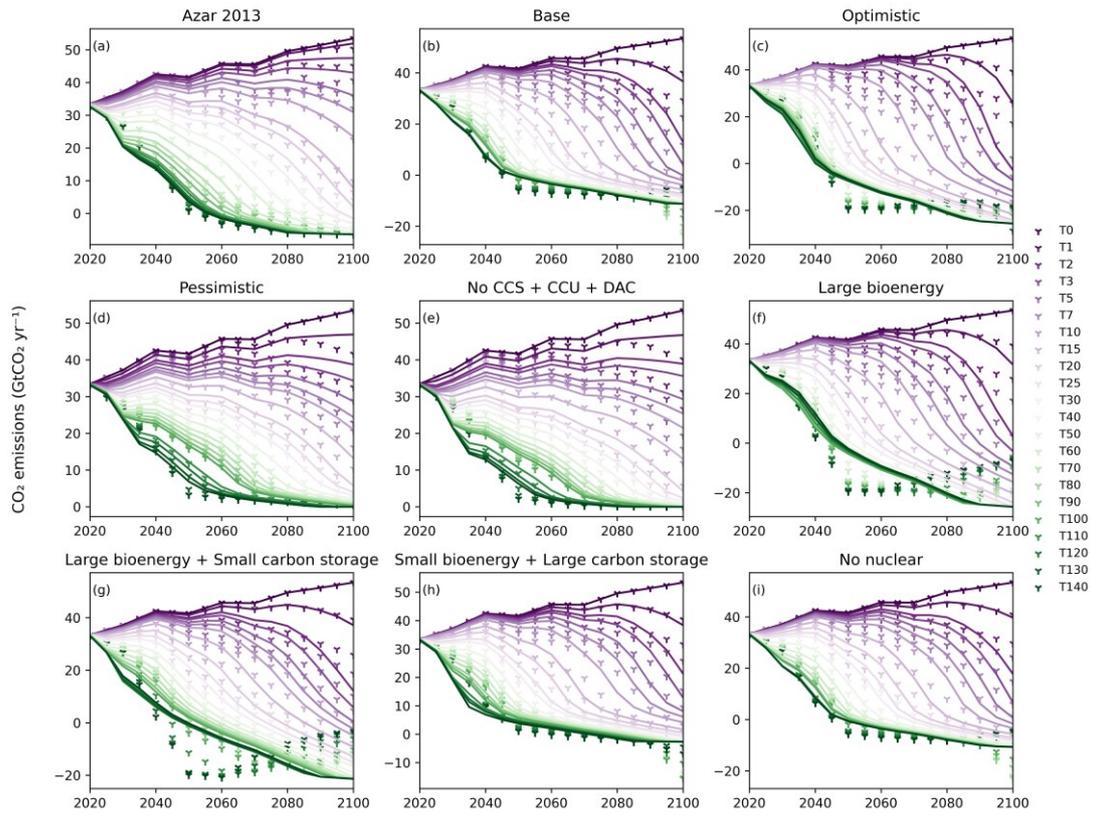

**Figure S90. Test 1 – GET 9 portfolios energy-related CO₂ validation result**

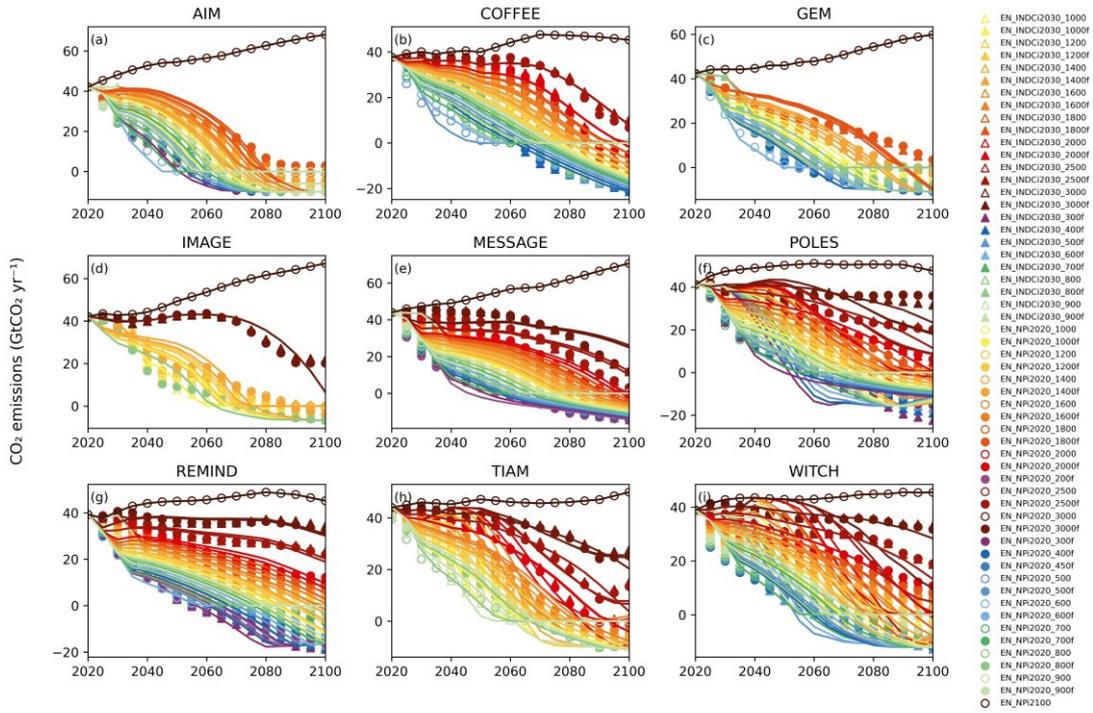

**Figure S91. Test 1 – Global 9 models total anthropogenic CO₂ validation result**

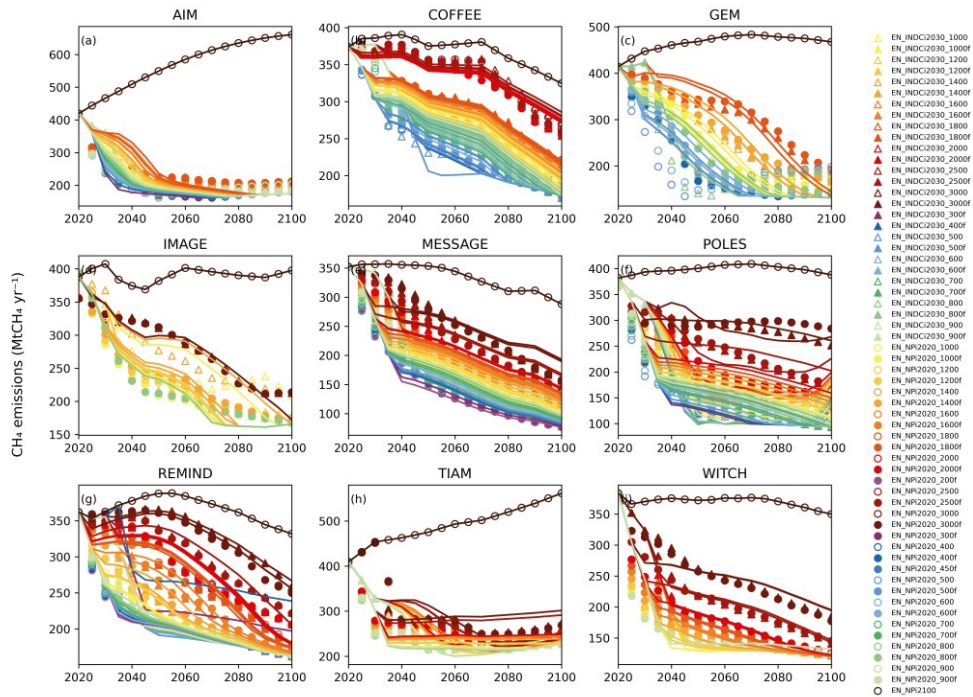

**Figure S92. Test 1 – Global 9 models total anthropogenic CH₄ validation result**

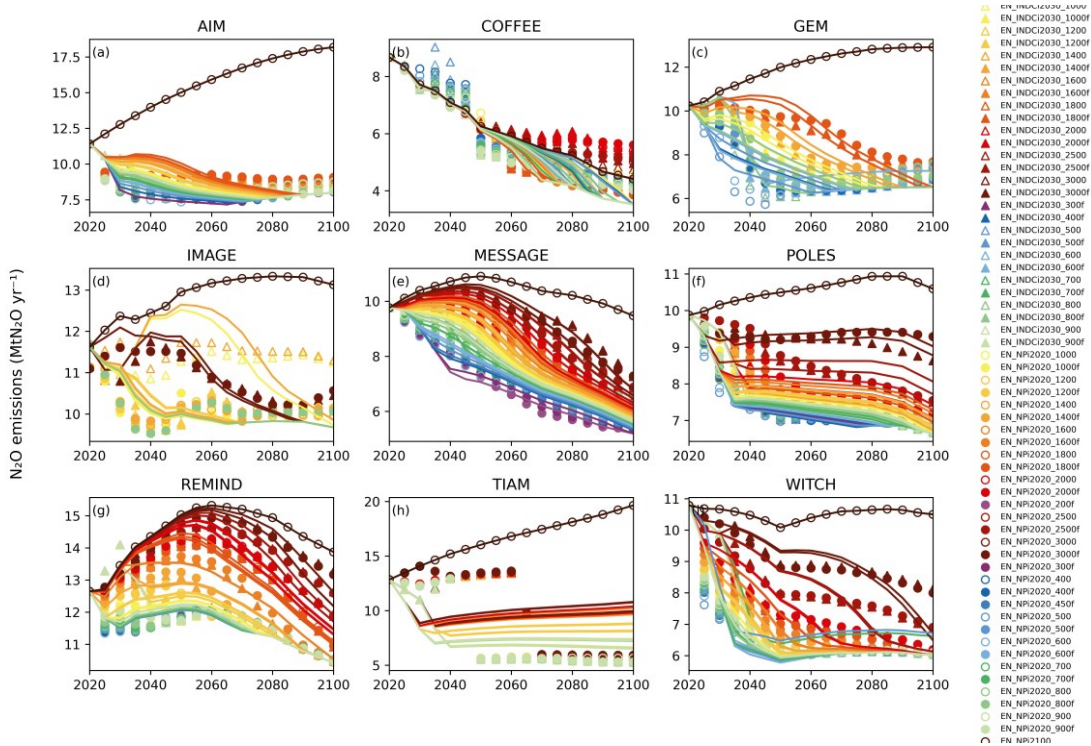

**Figure S93. Test 1 – Global 9 models total anthropogenic N₂O validation result**

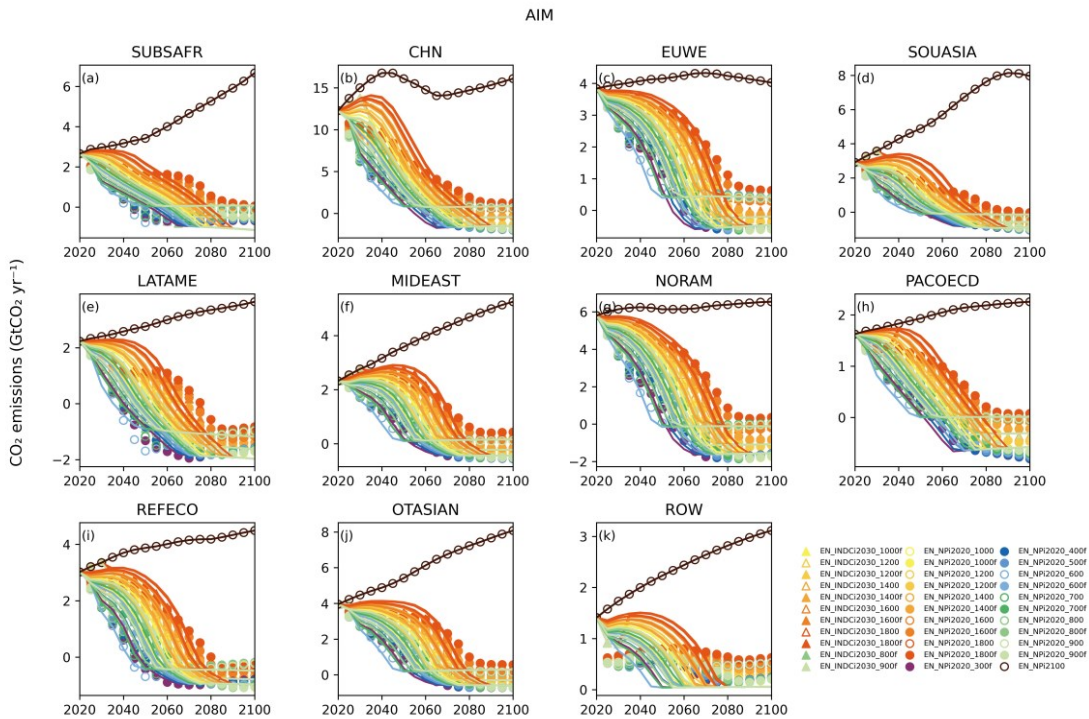

**Figure S94. Test 1 - Regional AIM total anthropogenic CO₂ validation result**

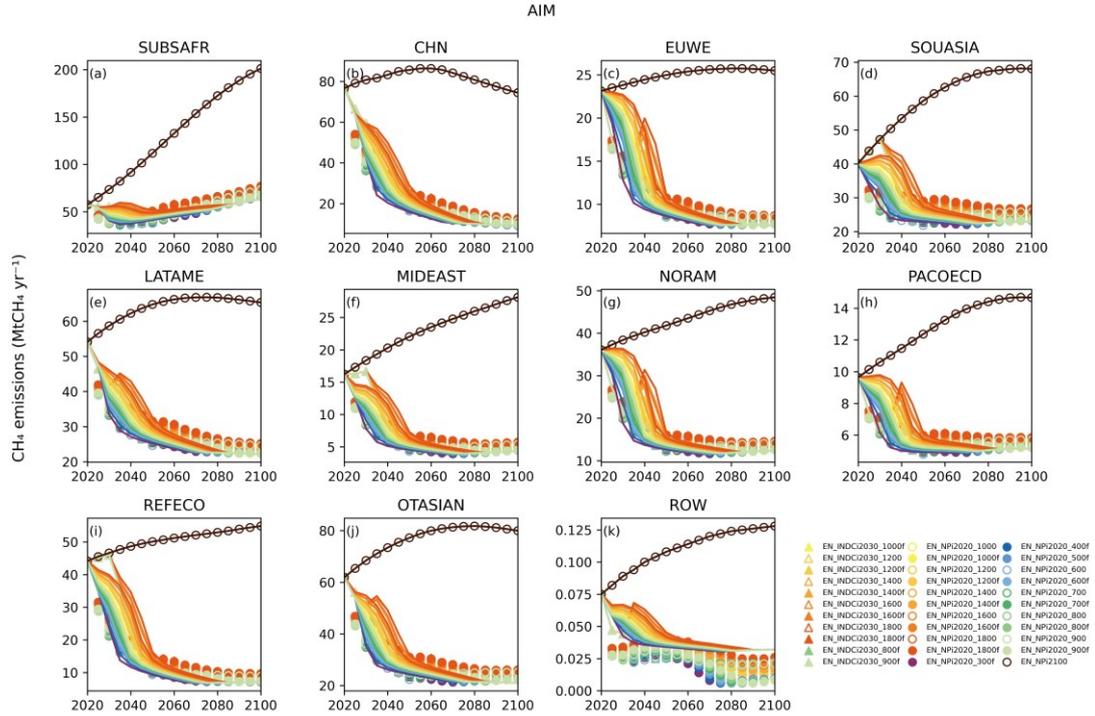

**Figure S95. Test 1 - Regional AIM total anthropogenic CH₄ validation result**

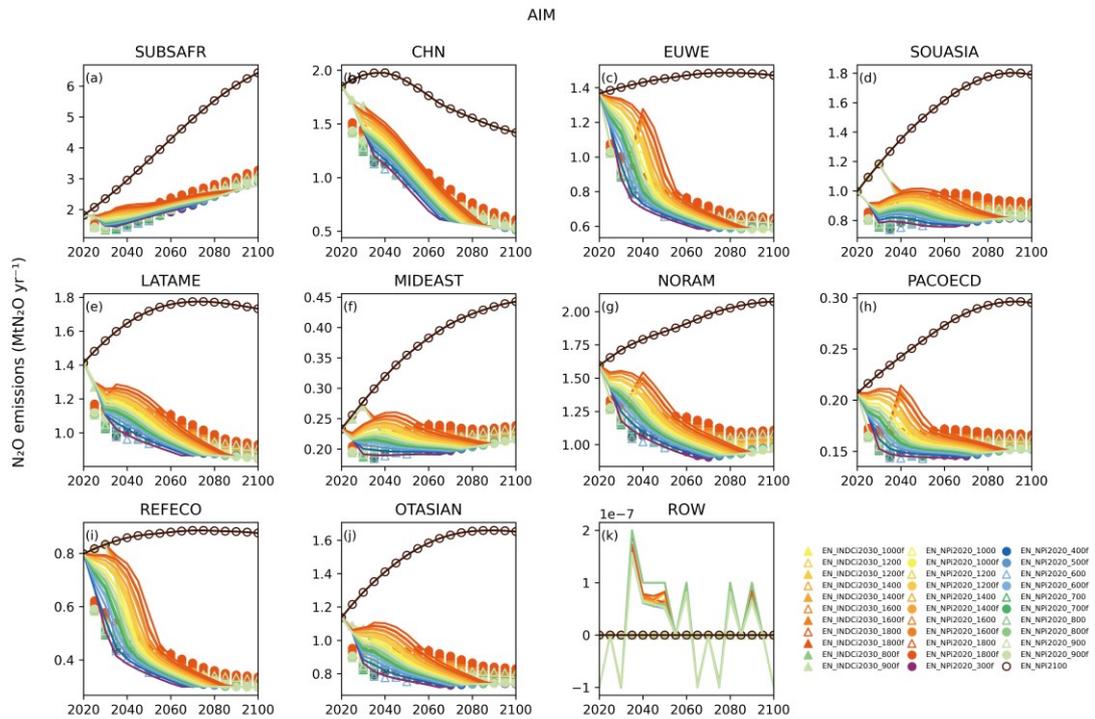

**Figure S96. Test 1 - Regional AIM total anthropogenic N₂O validation result**

**Figure S97. Test 1 - Regional COFFEE total anthropogenic CO₂ validation result**

**Figure S98. Test 1 - Regional COFFEE total anthropogenic CH₄ validation result**

**Figure S99. Test 1 - Regional COFFEE total anthropogenic N₂O validation result**

**Figure S100. Test 1 - Regional GEM total anthropogenic CO₂ validation result**

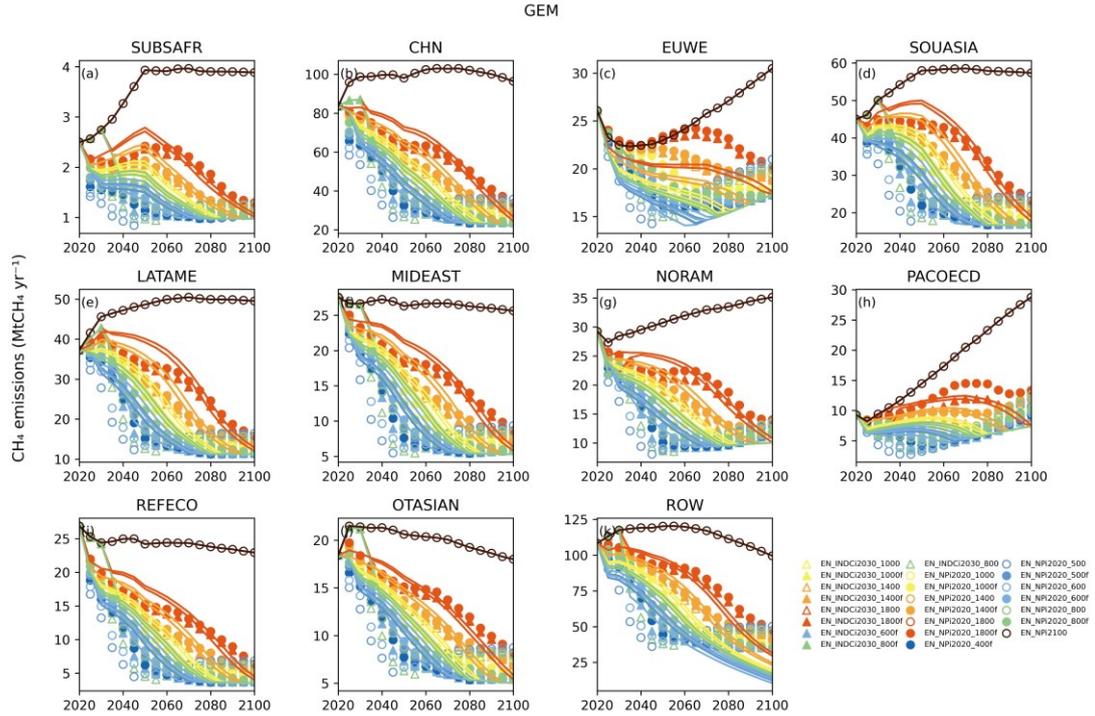

**Figure S101. Test 1 - Regional GEM total anthropogenic CH₄ validation result**

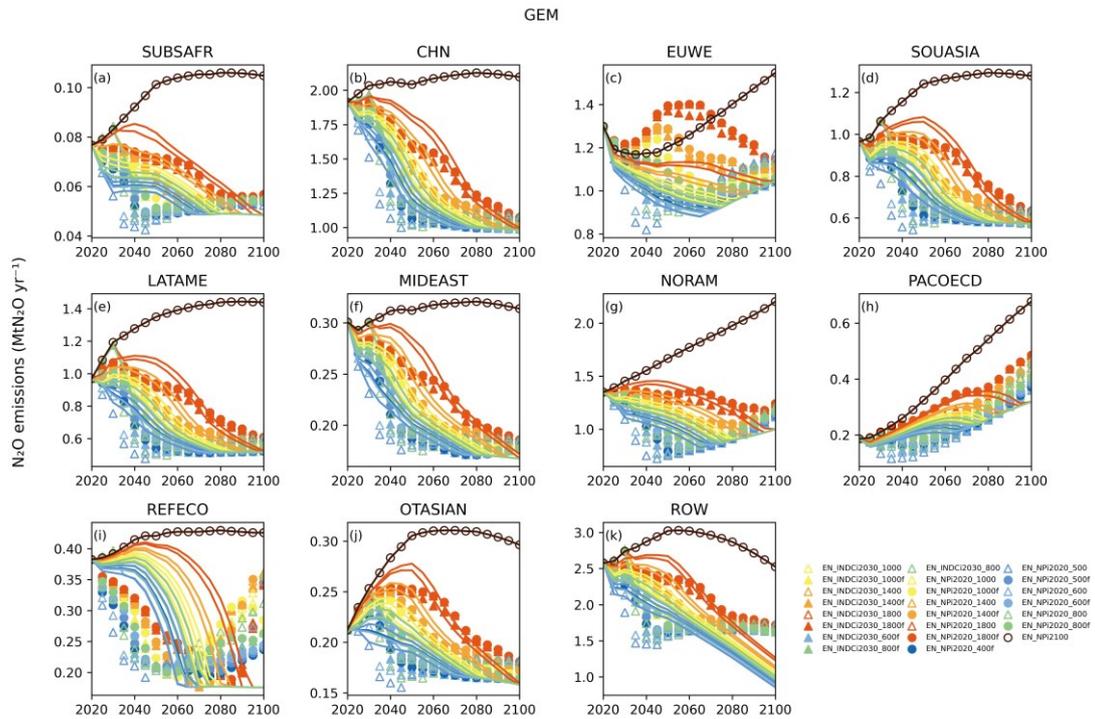

**Figure S102. Test 1 - Regional GEM total anthropogenic N₂O validation result**

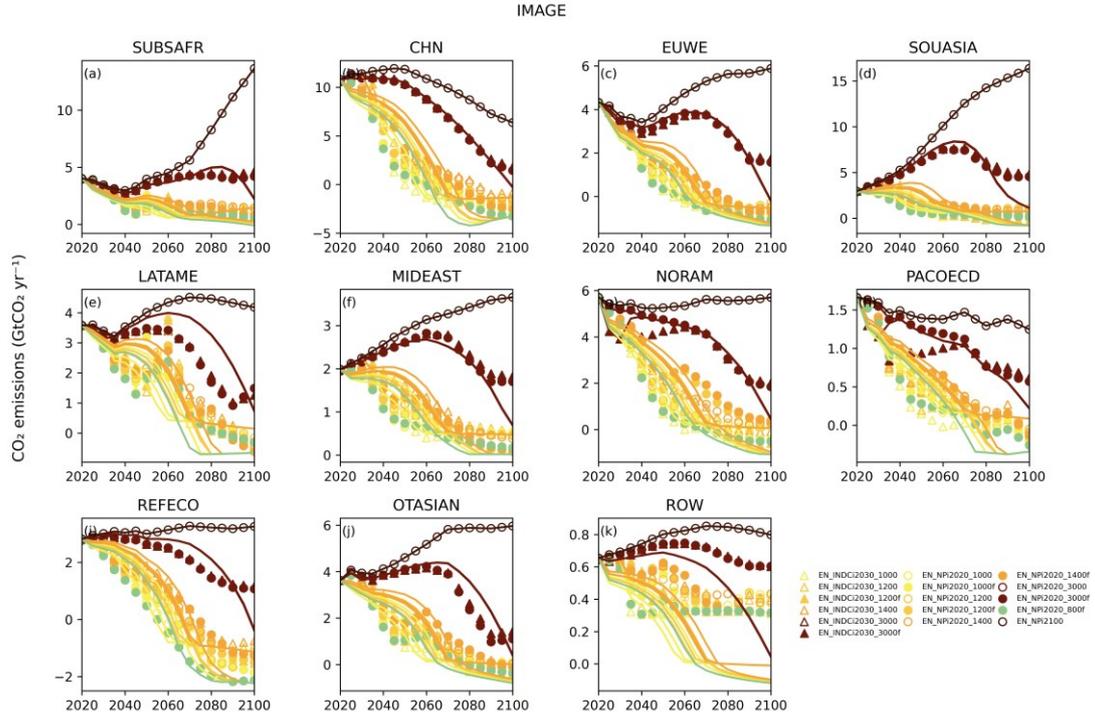

**Figure S103. Test 1 - Regional IMAGE total anthropogenic CO$_2$ validation result**

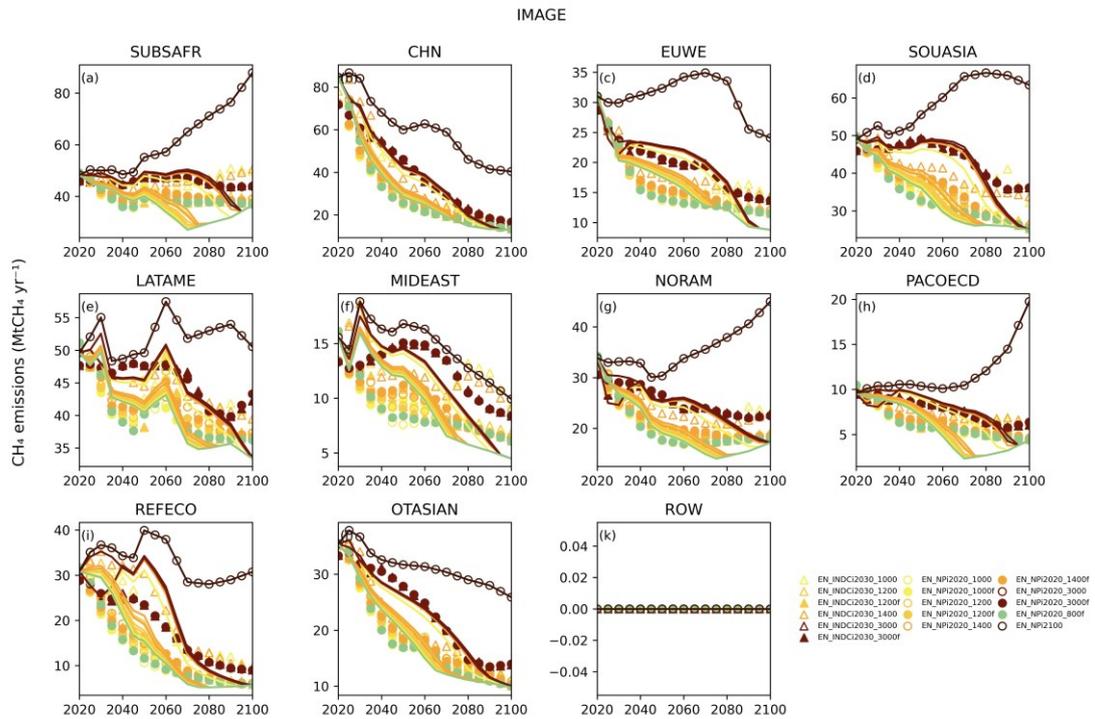

**Figure S104. Test 1 - Regional IMAGE total anthropogenic CH$_4$ validation result**

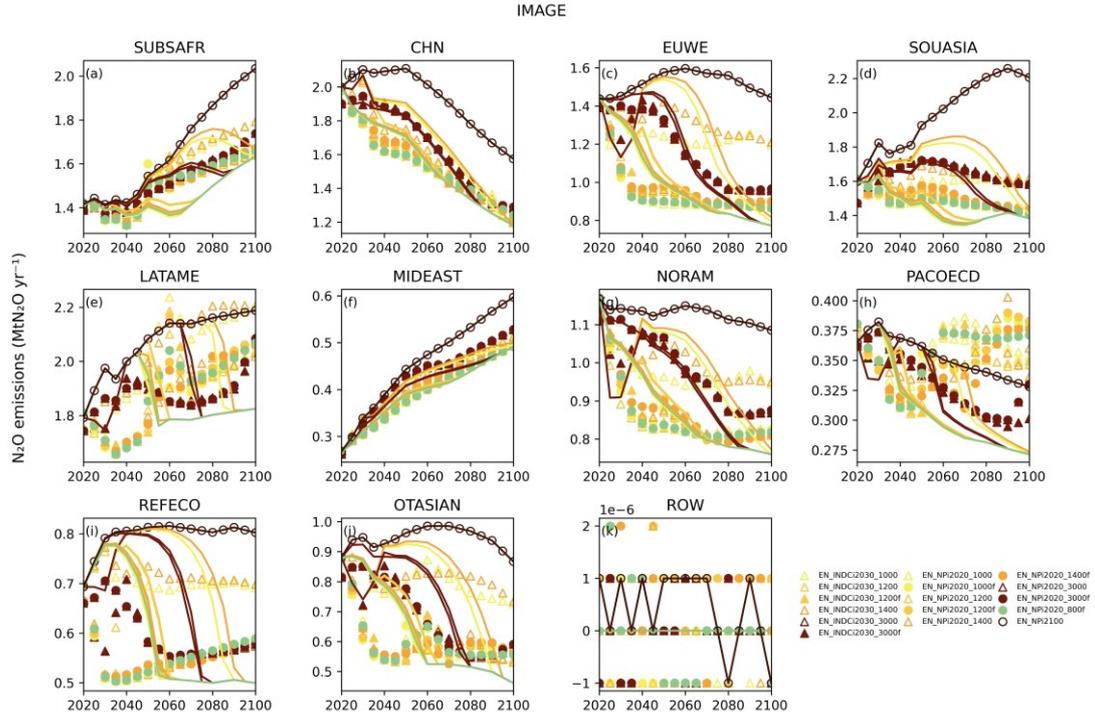

**Figure S105. Test 1 - Regional IMAGE total anthropogenic N₂O validation result**

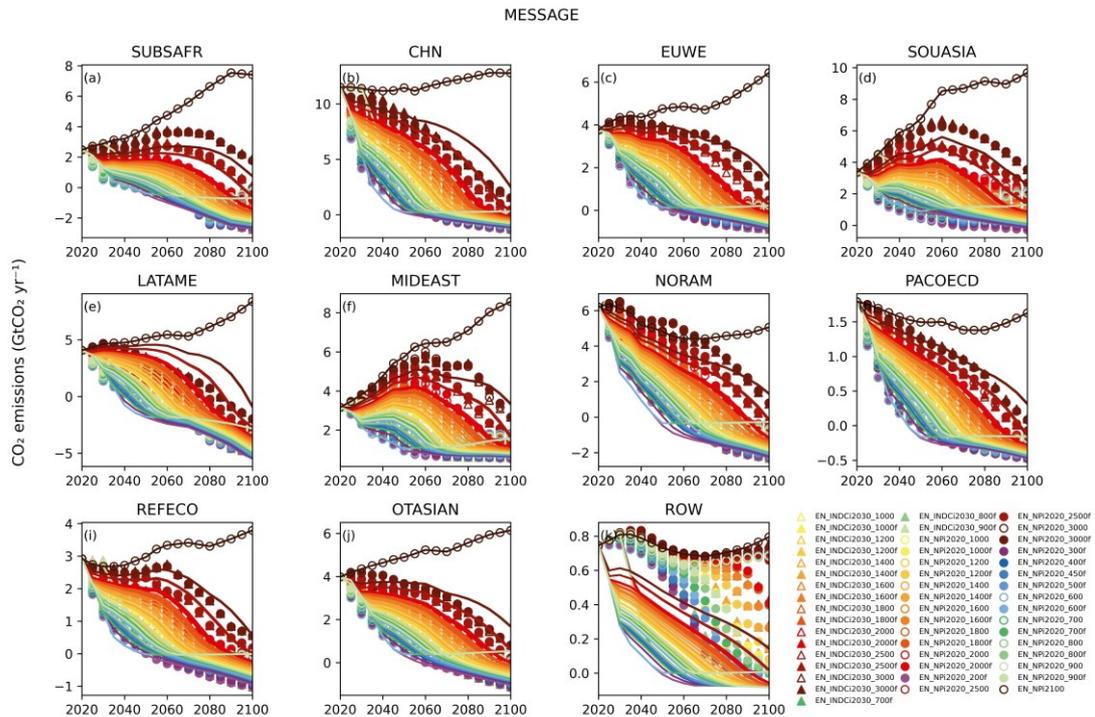

**Figure S106. Test 1 - Regional MESSAGE total anthropogenic CO₂ validation result**

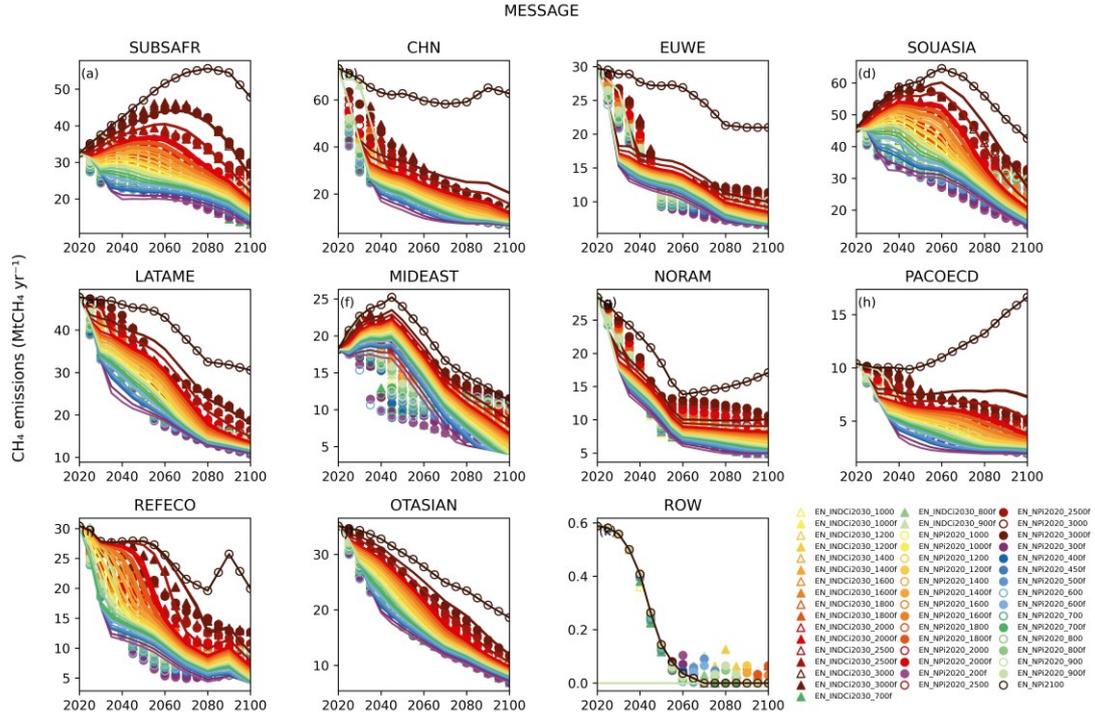

**Figure S107. Test 1 - Regional MESSAGE total anthropogenic CH₄ validation result**

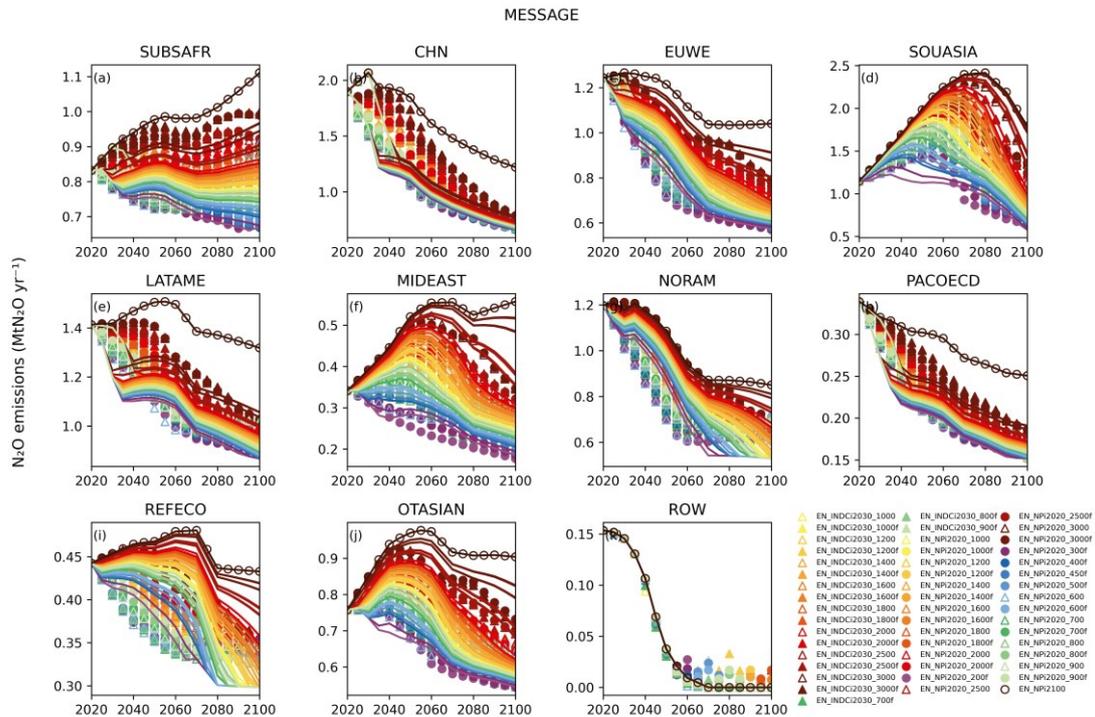

**Figure S108. Test 1 - Regional MESSAGE total anthropogenic N₂O validation result**

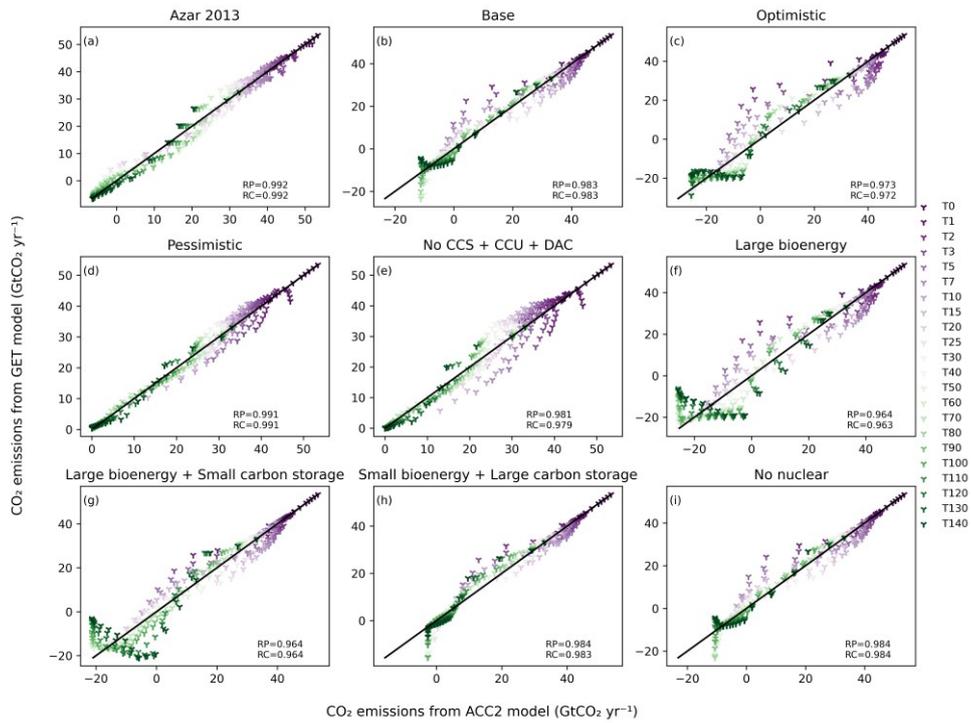

**Figure S109. Test 1 - GET Reproducibility of total anthropogenic CO₂**

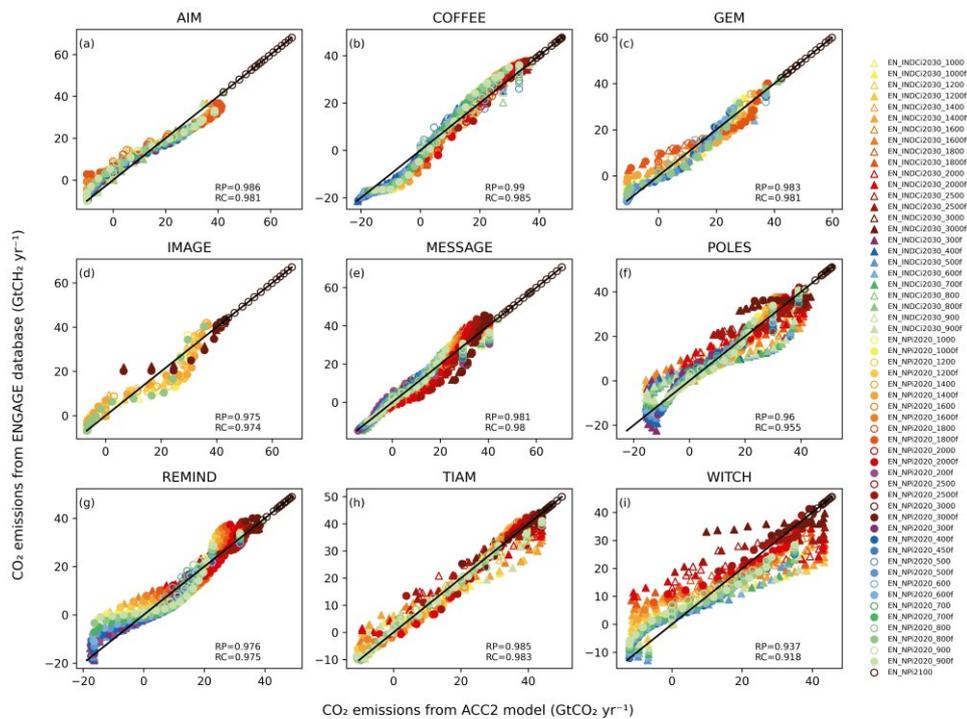

**Figure S110. Test 1 - Global 9 models - Reproducibility of total anthropogenic CO₂**

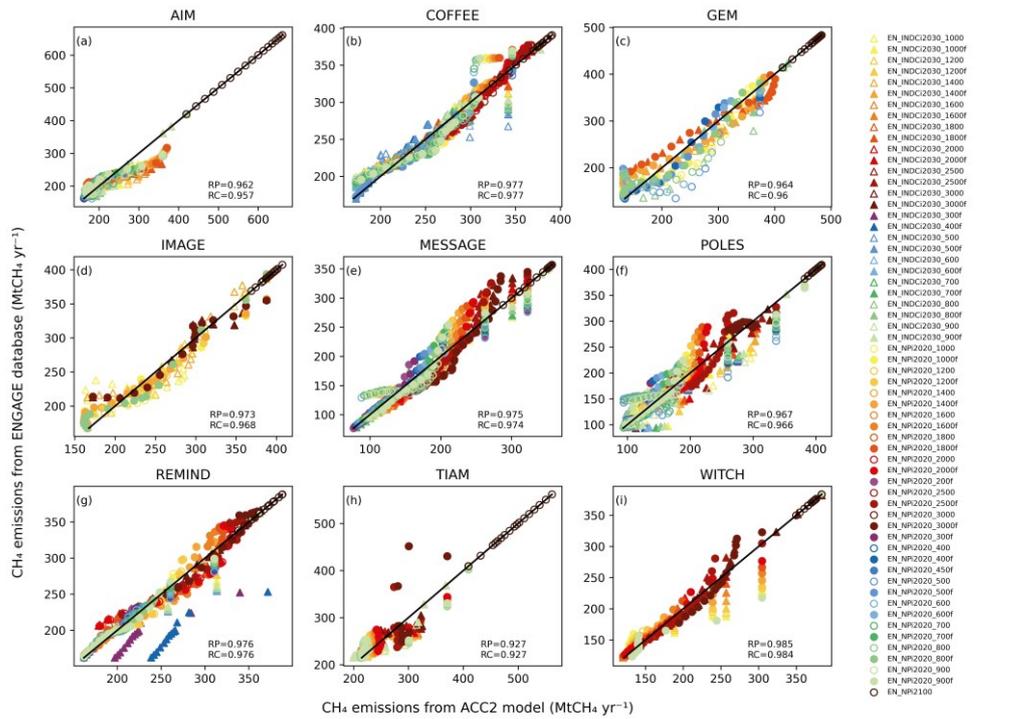

**Figure S111. Test 1 - Global 9 models - Reproducibility of total anthropogenic CH₄**

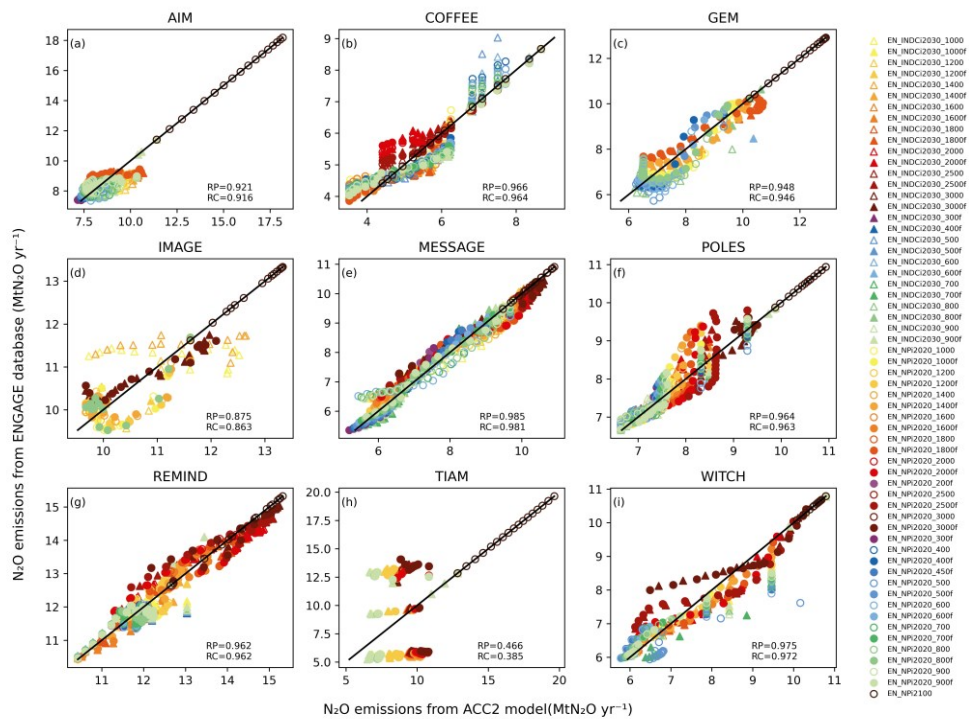

**Figure S112. Test 1 - Global 9 models - Reproducibility of total anthropogenic N₂O**

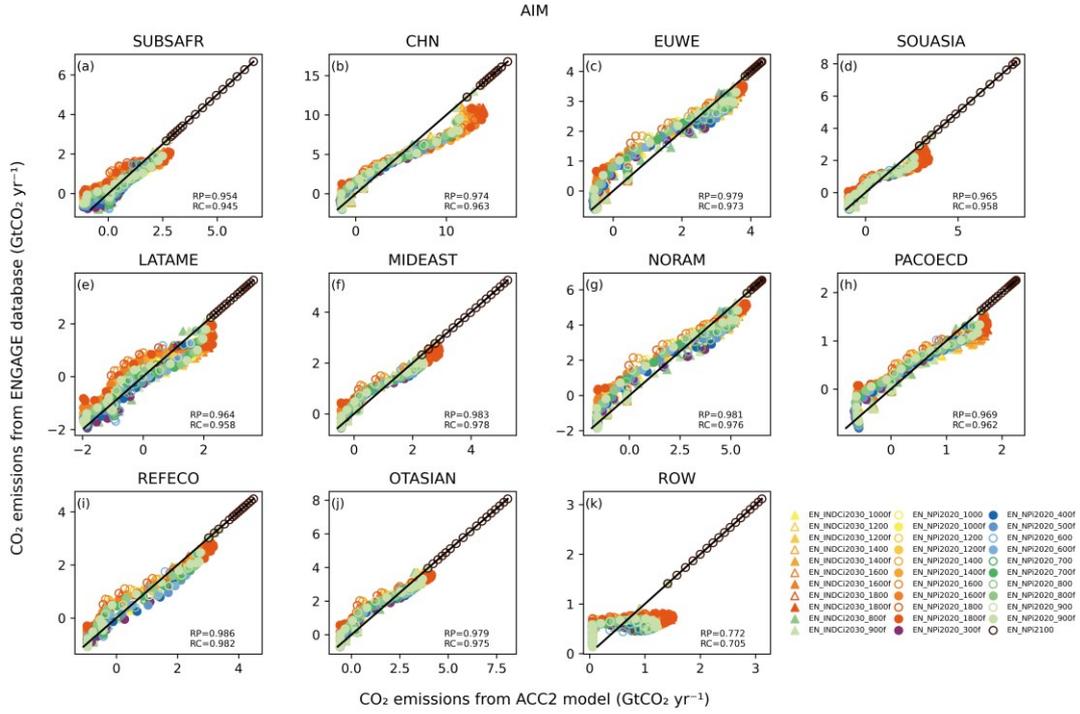

**Figure S113. Test 1 - Regional AIM - Reproducibility of total anthropogenic $CO_2$**

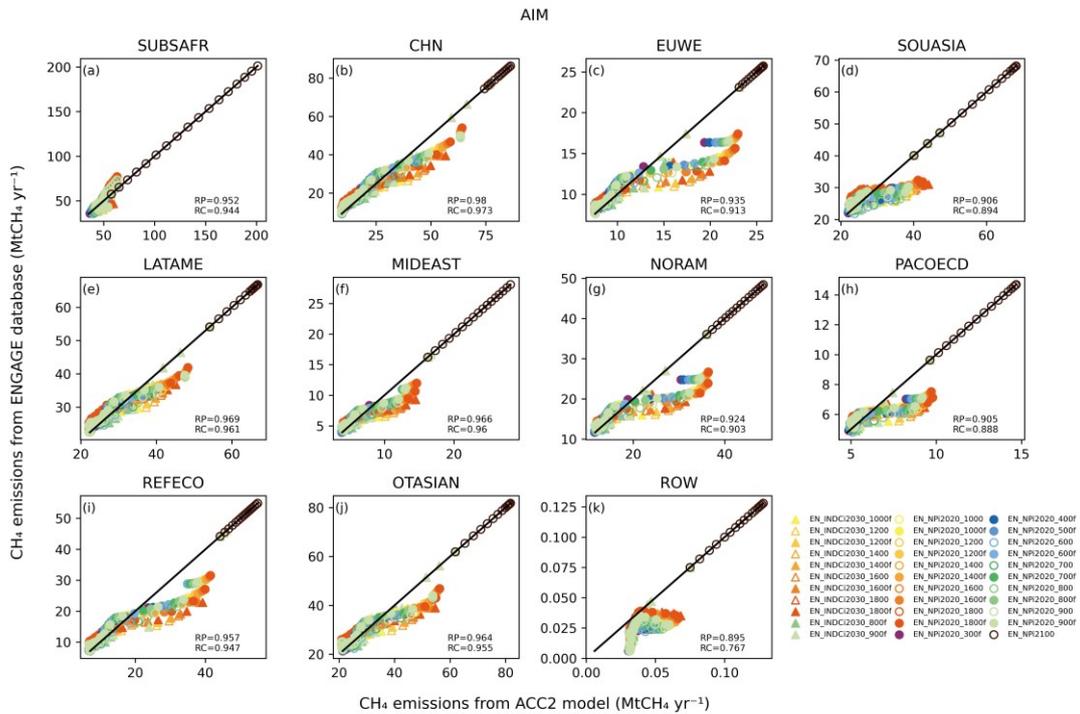

**Figure S114. Test 1 - Regional AIM - Reproducibility of total anthropogenic $CH_4$**

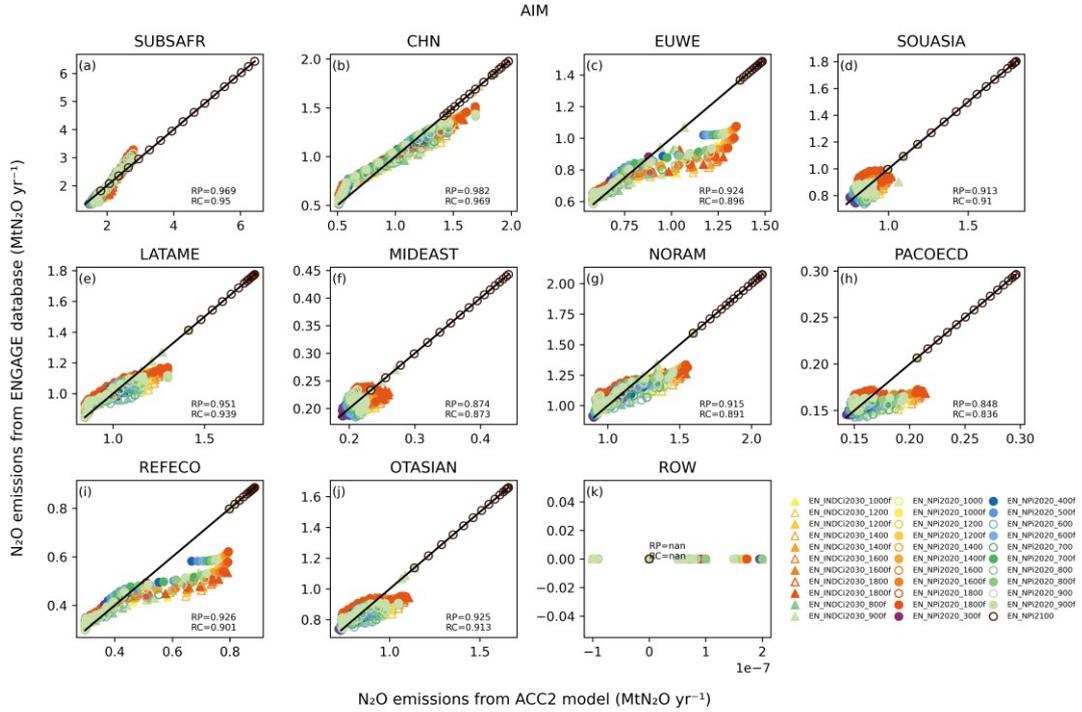

**Figure S115. Test 1 - Regional AIM - Reproducibility of total anthropogenic N₂O**

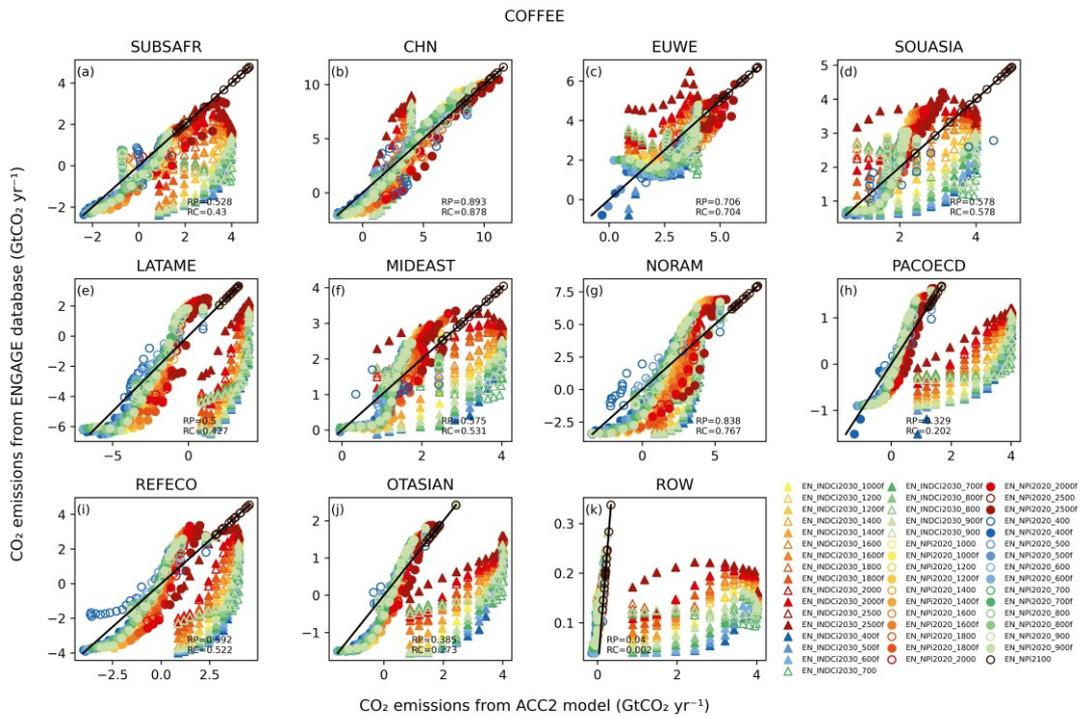

**Figure S116. Test 1 - Regional COFFEE - Reproducibility of total anthropogenic CO₂**

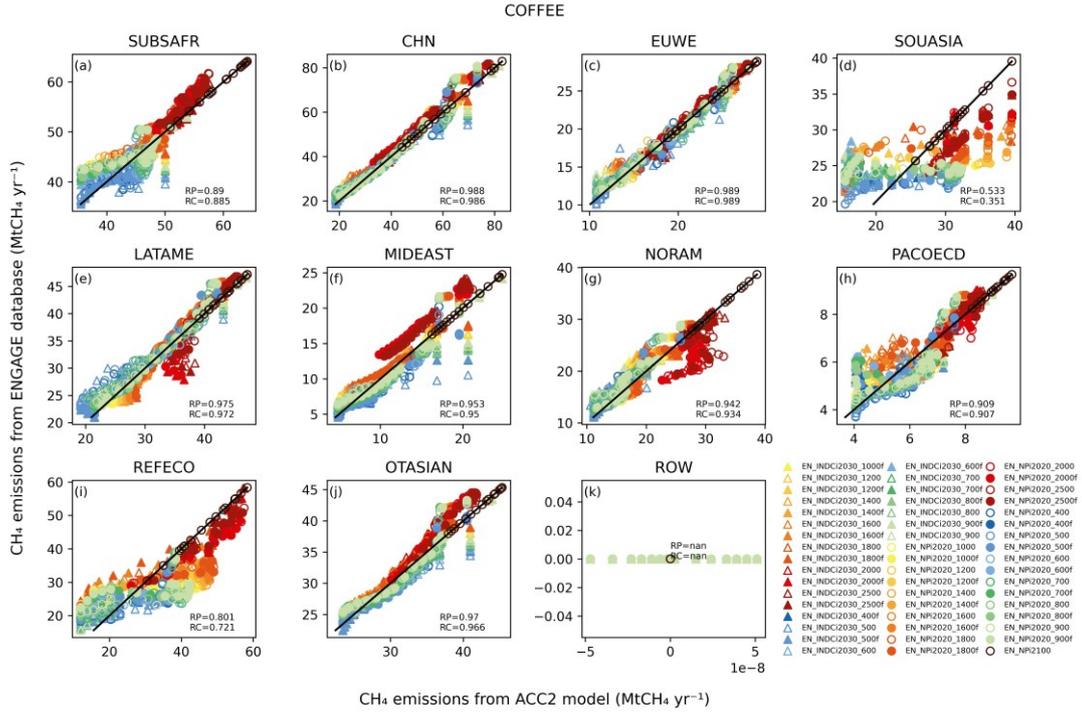

**Figure S117. Test 1 - Regional COFFEE - Reproducibility of total anthropogenic CH$_4$**

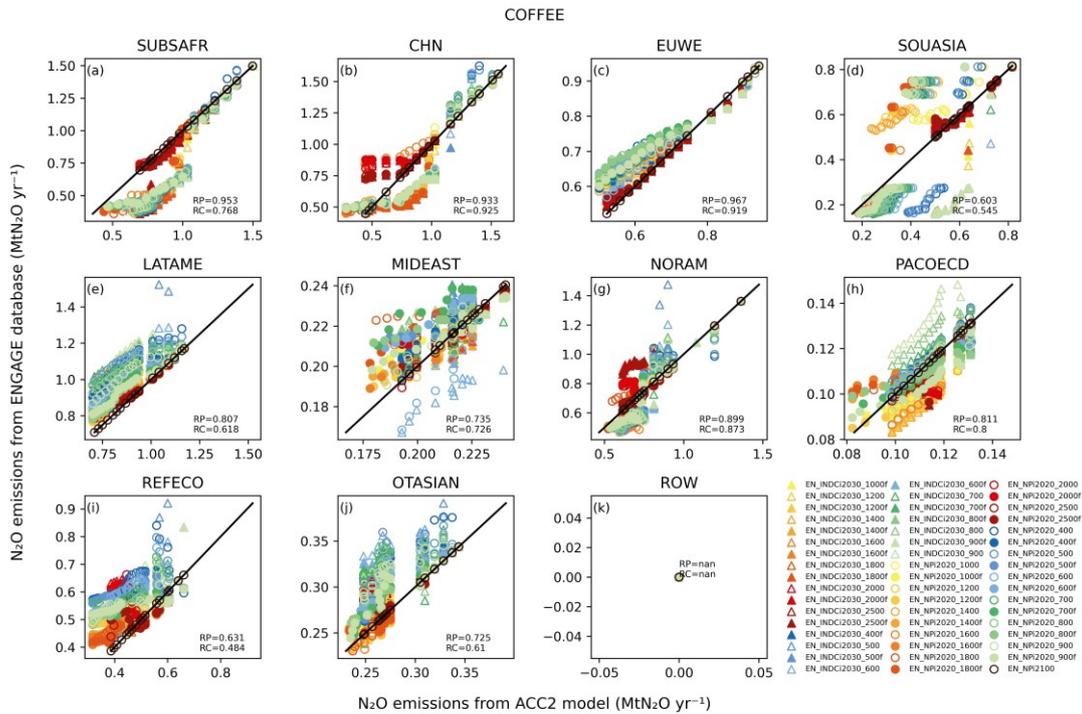

**Figure S118. Test 1 - Regional COFFEE - Reproducibility of total anthropogenic N$_2$O**

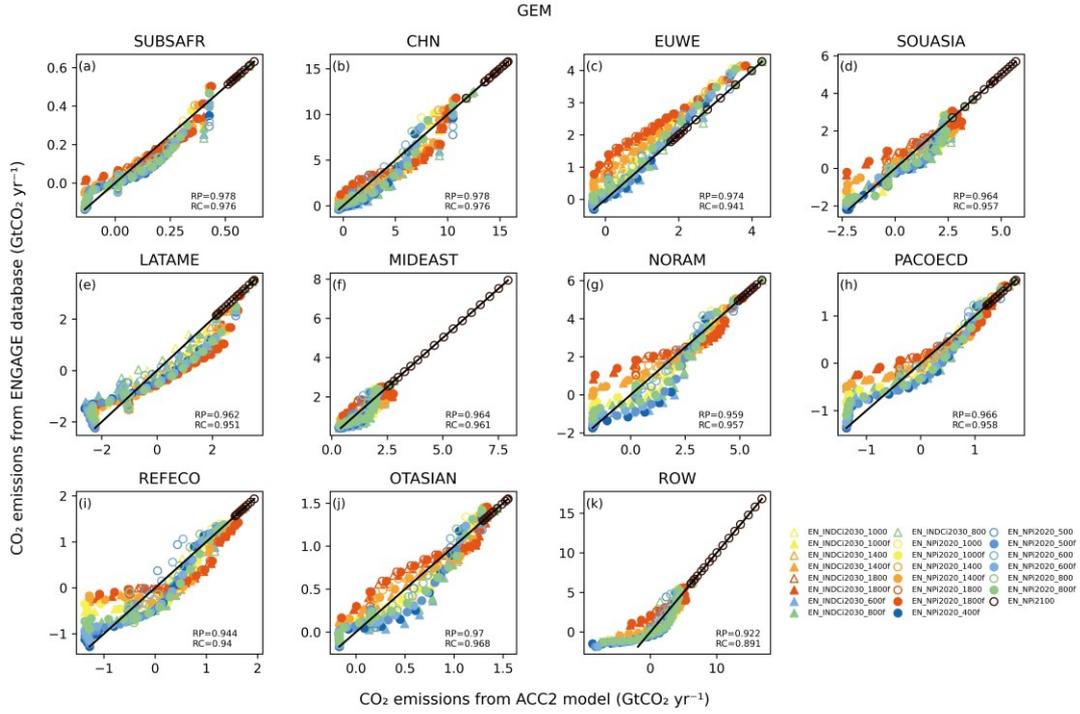

**Figure S119. Test 1 - Regional GEM - Reproducibility of total anthropogenic CO₂**

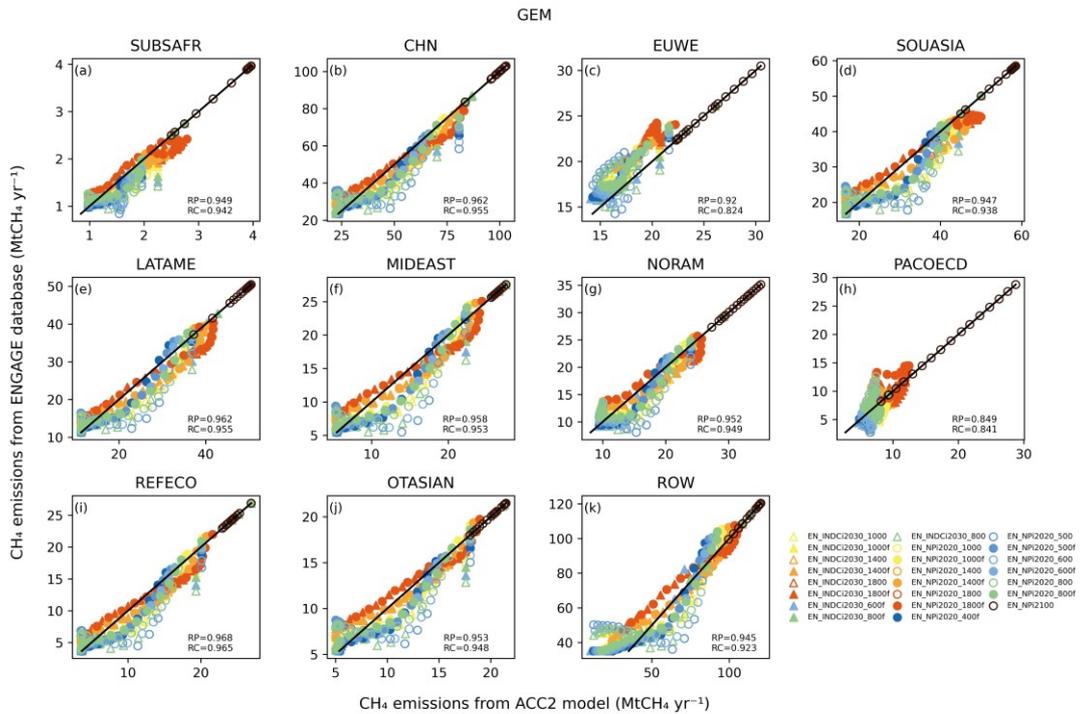

**Figure S120. Test 1 - Regional GEM - Reproducibility of total anthropogenic CH₄**

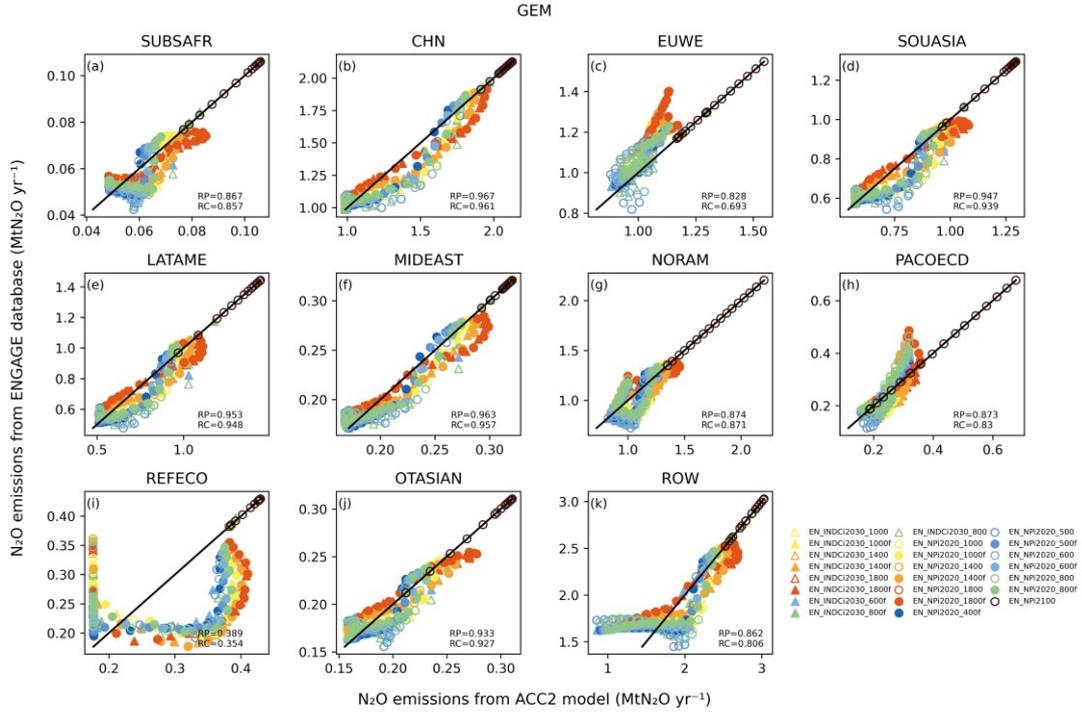

**Figure S121. Test 1 - Regional GEM - Reproducibility of total anthropogenic N₂O**

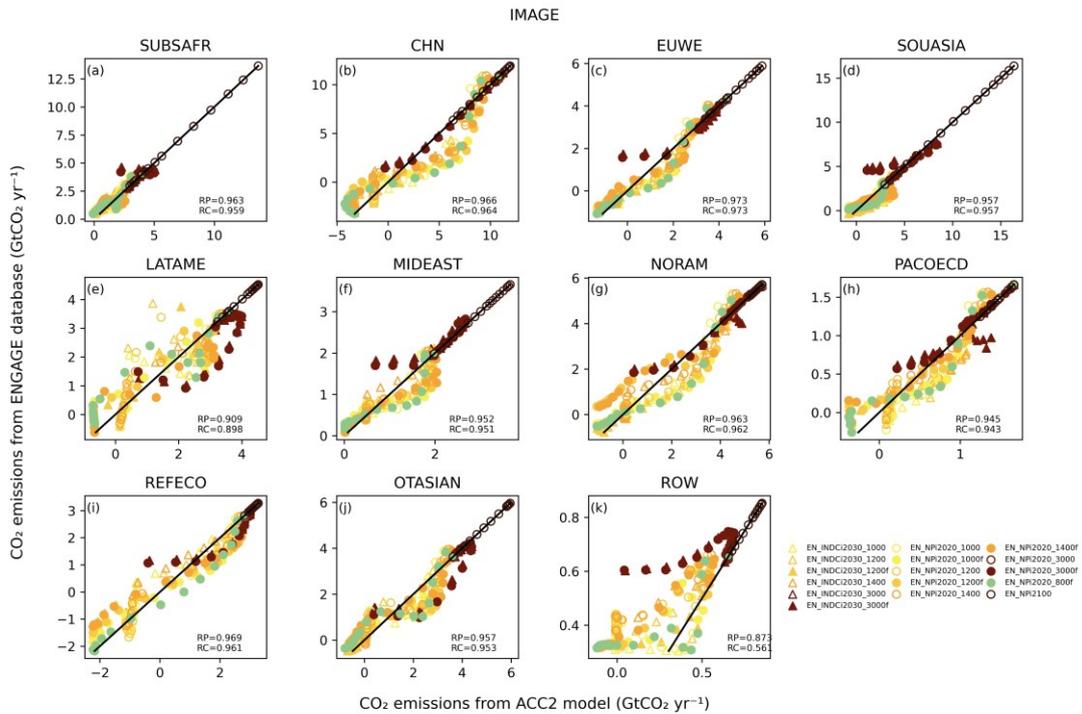

**Figure S122. Test 1 - Regional IMAGE - Reproducibility of total anthropogenic CO₂**

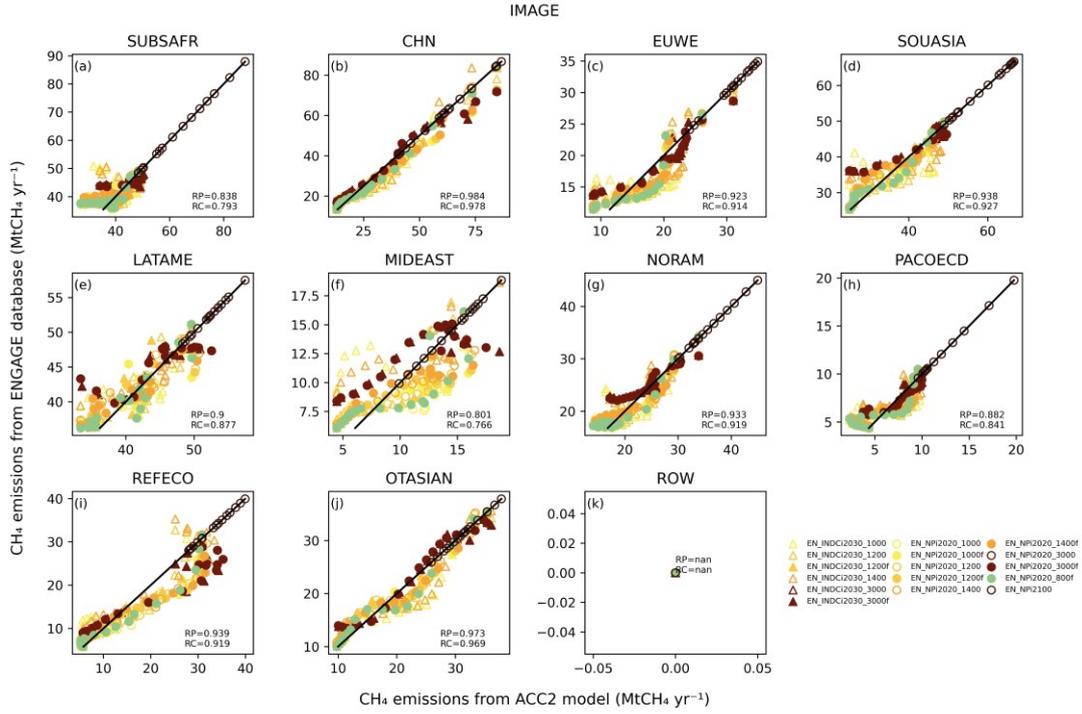

**Figure S123. Test 1 - Regional IMAGE - Reproducibility of total anthropogenic CH₄**

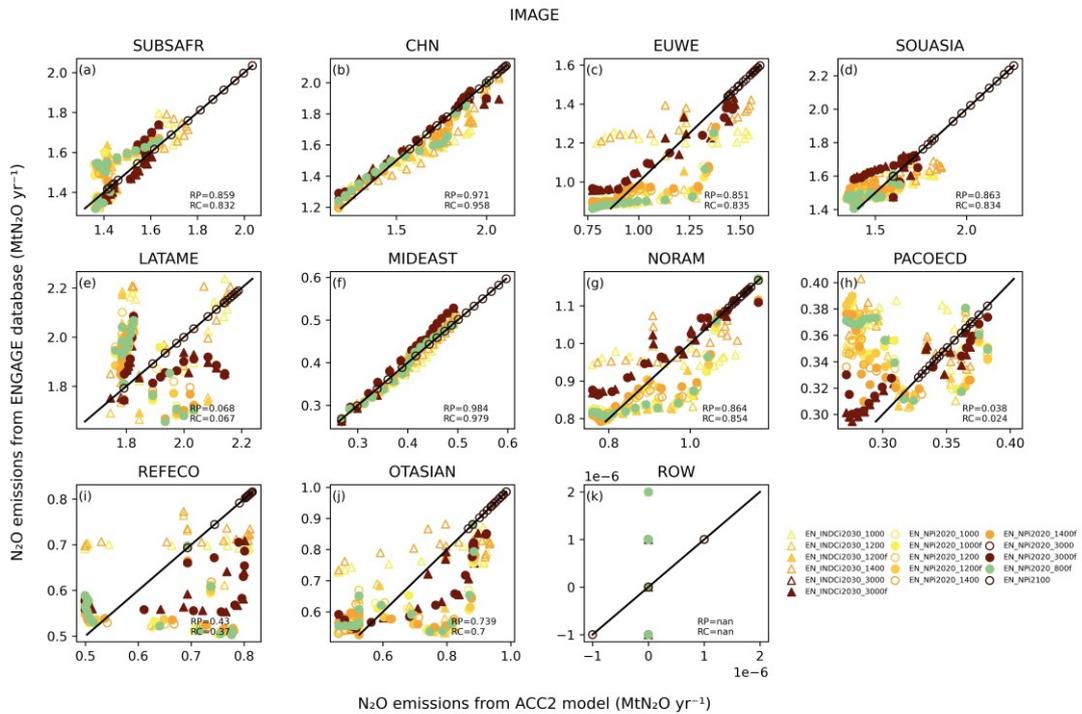

**Figure S124. Test 1 - Regional IMAGE - Reproducibility of total anthropogenic N₂O**

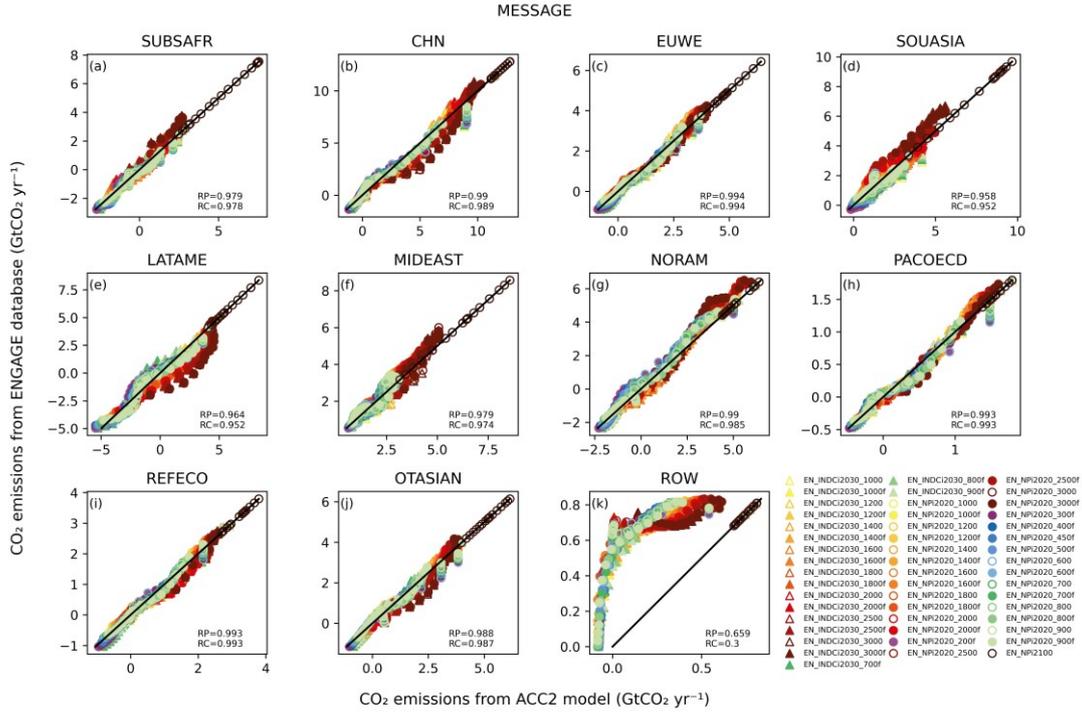

**Figure S125. Test 1 - Regional MESSAGE - Reproducibility of total anthropogenic CO₂**

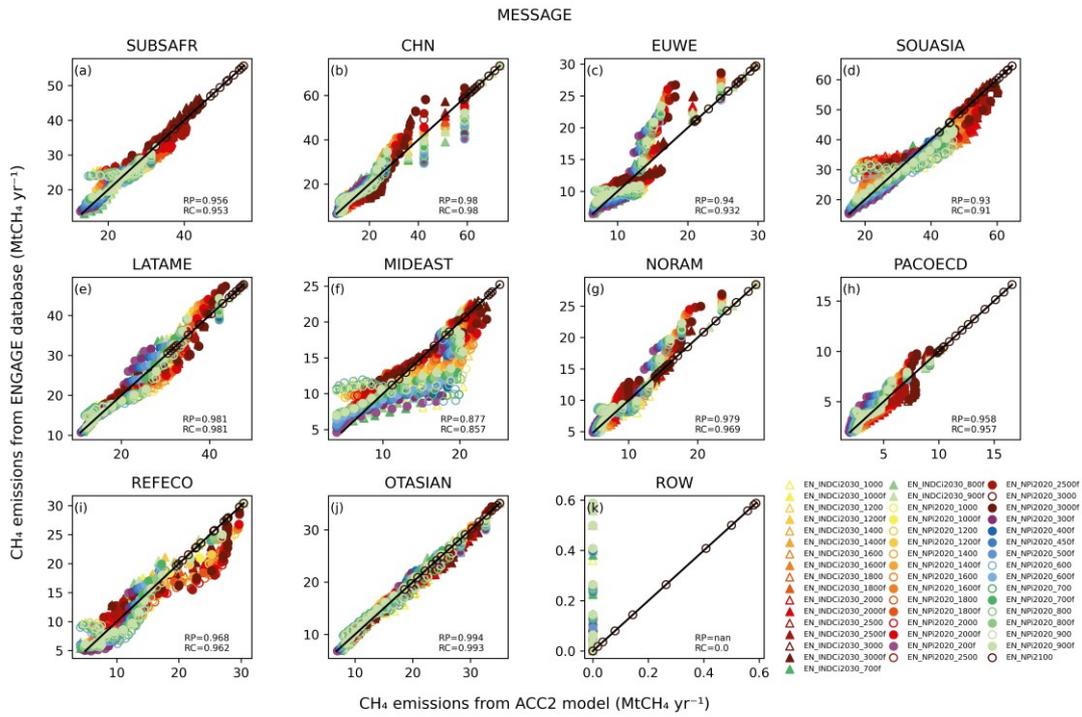

**Figure S126. Test 1 - Regional MESSAGE - Reproducibility of total anthropogenic CH₄**

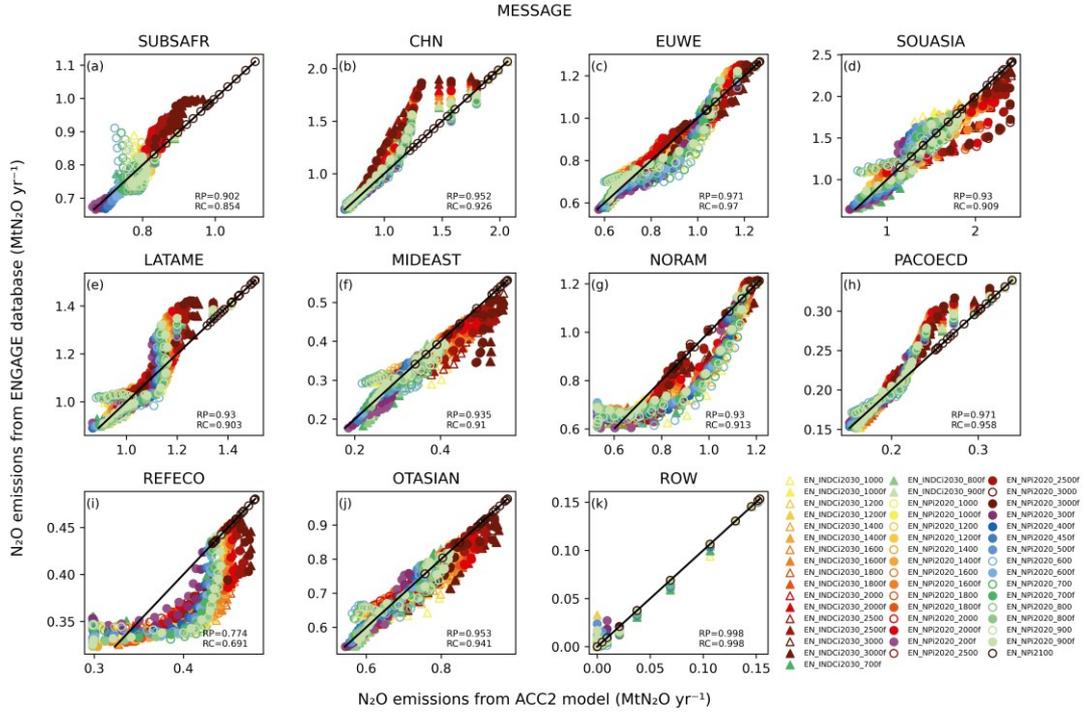

**Figure S127. Test 1 - Regional MESSAGE - Reproducibility of total anthropogenic N₂O**

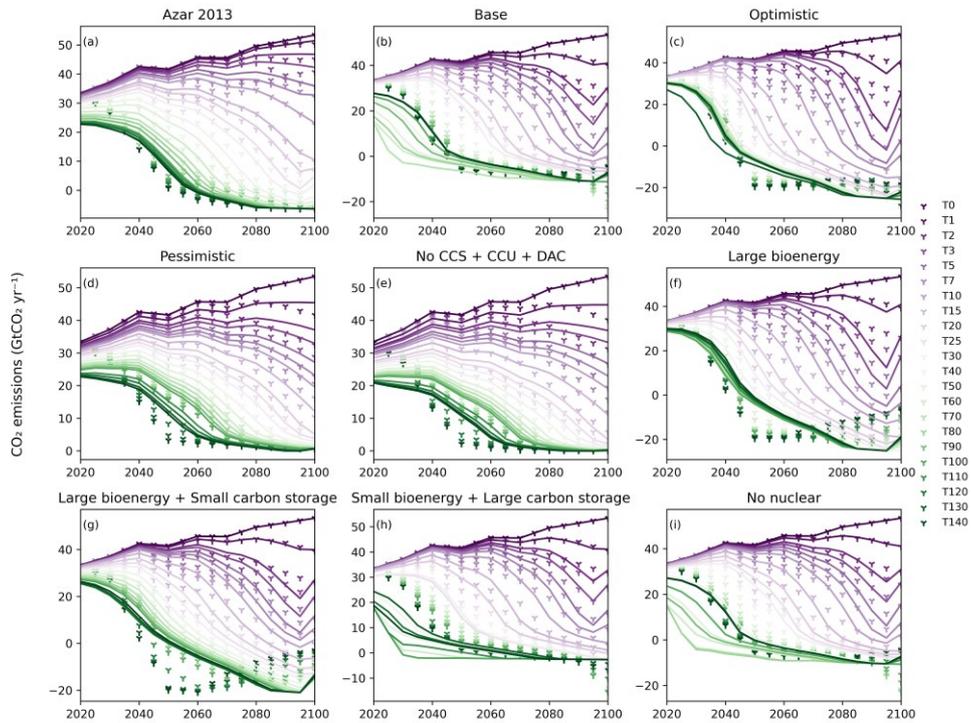

**Figure S128. Test 2 – GET 9 portfolios energy-related CO₂ validation result**

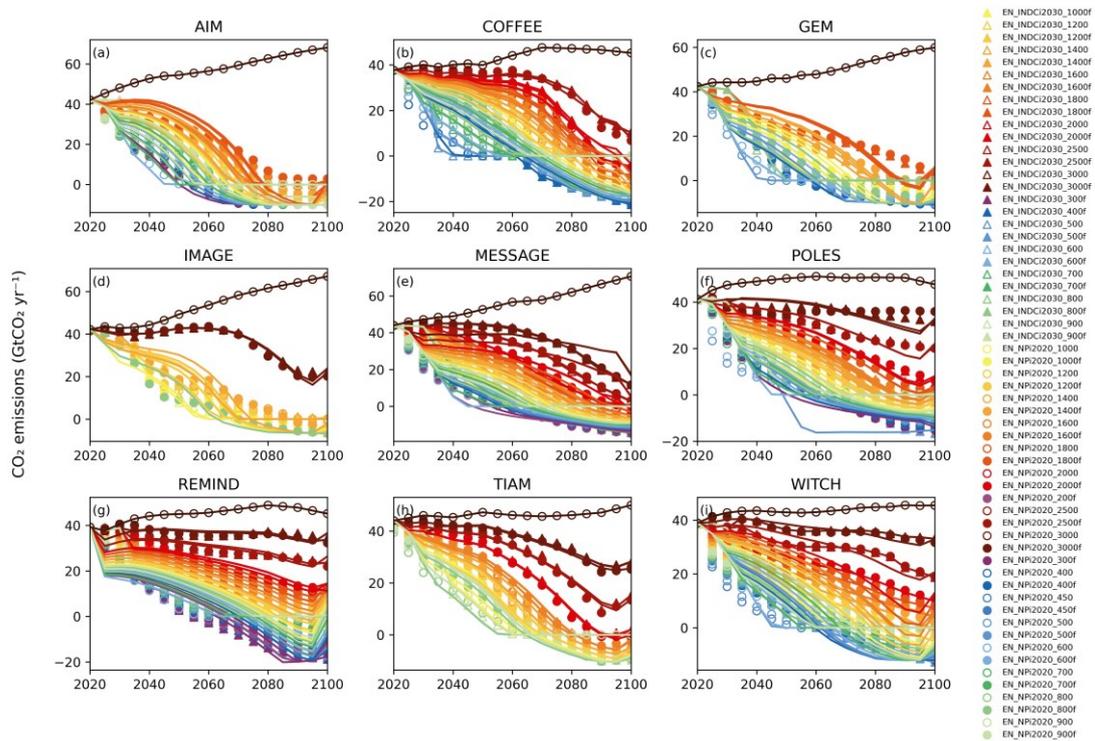

**Figure S129. Test 2 – Global 9 models total anthropogenic CO₂ validation result**

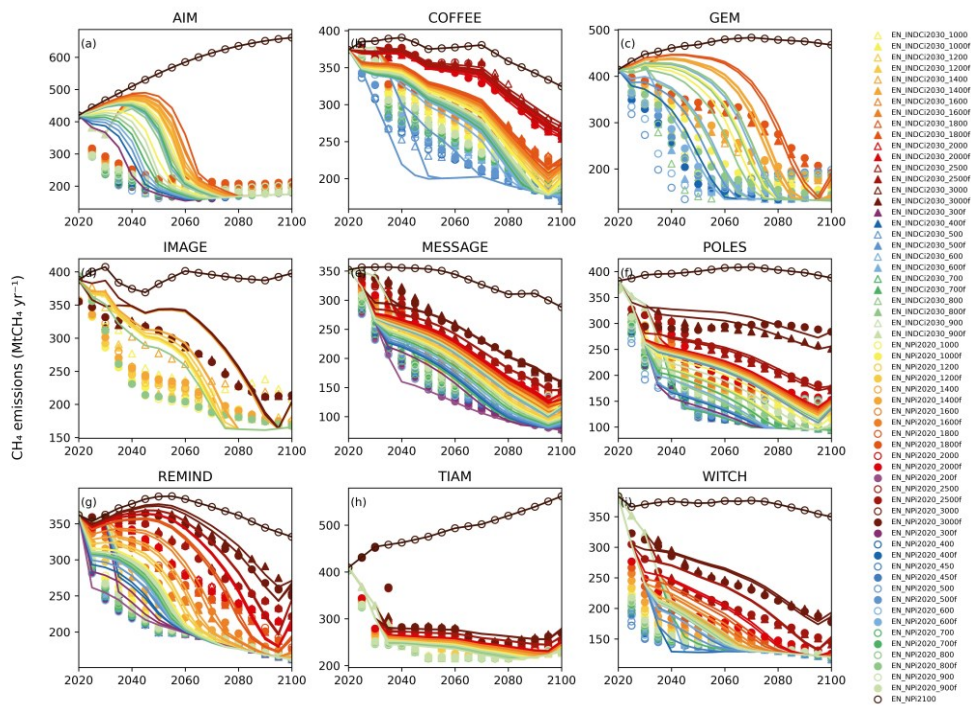

**Figure S130. Test 2 – Global 9 models total anthropogenic CH₄ validation result**

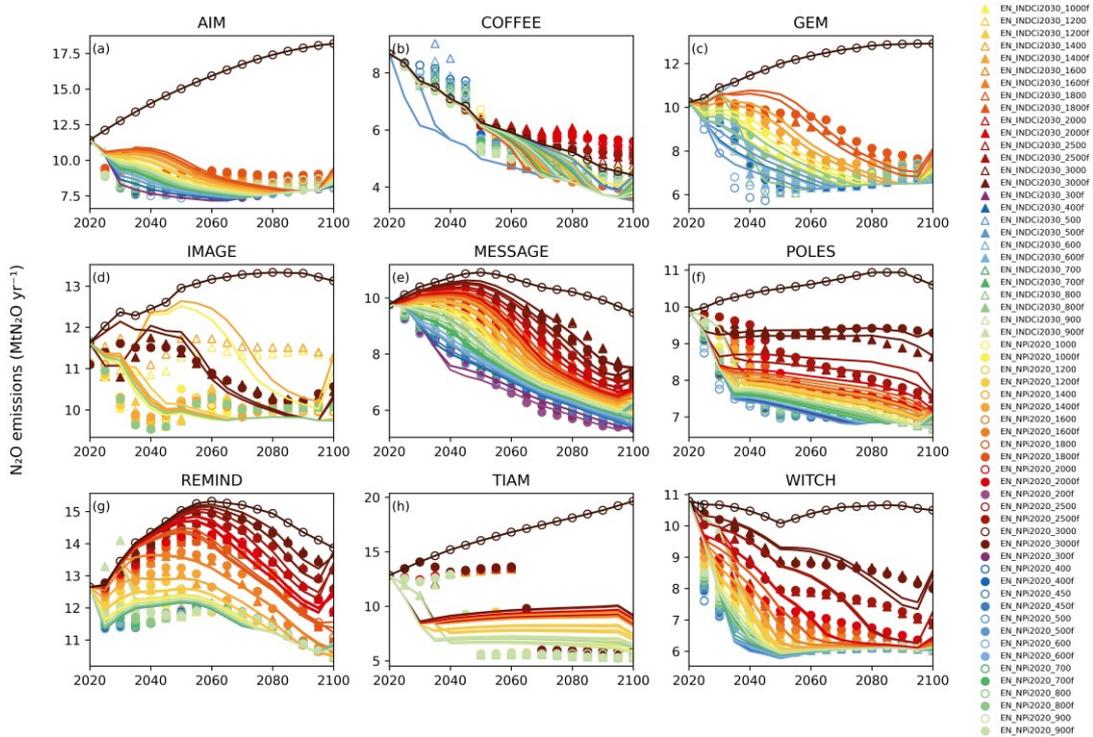

**Figure S131. Test 2 – Global 9 models total anthropogenic N₂O validation result**

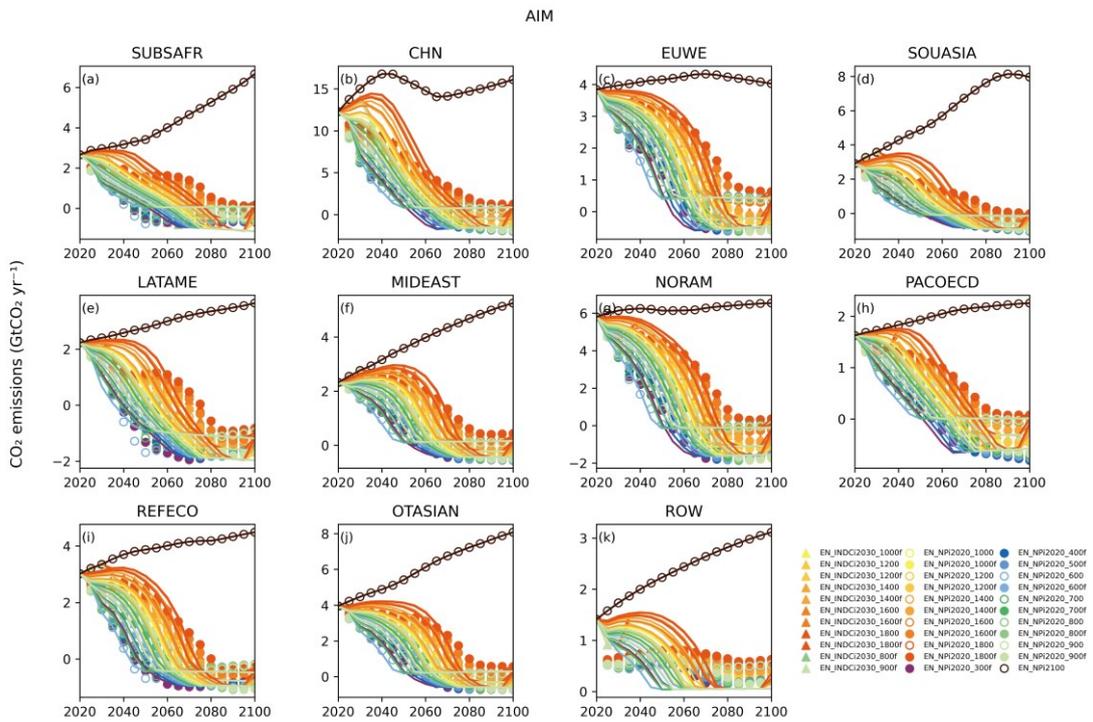

**Figure S132. Test 2 - Regional AIM total anthropogenic CO₂ validation result**

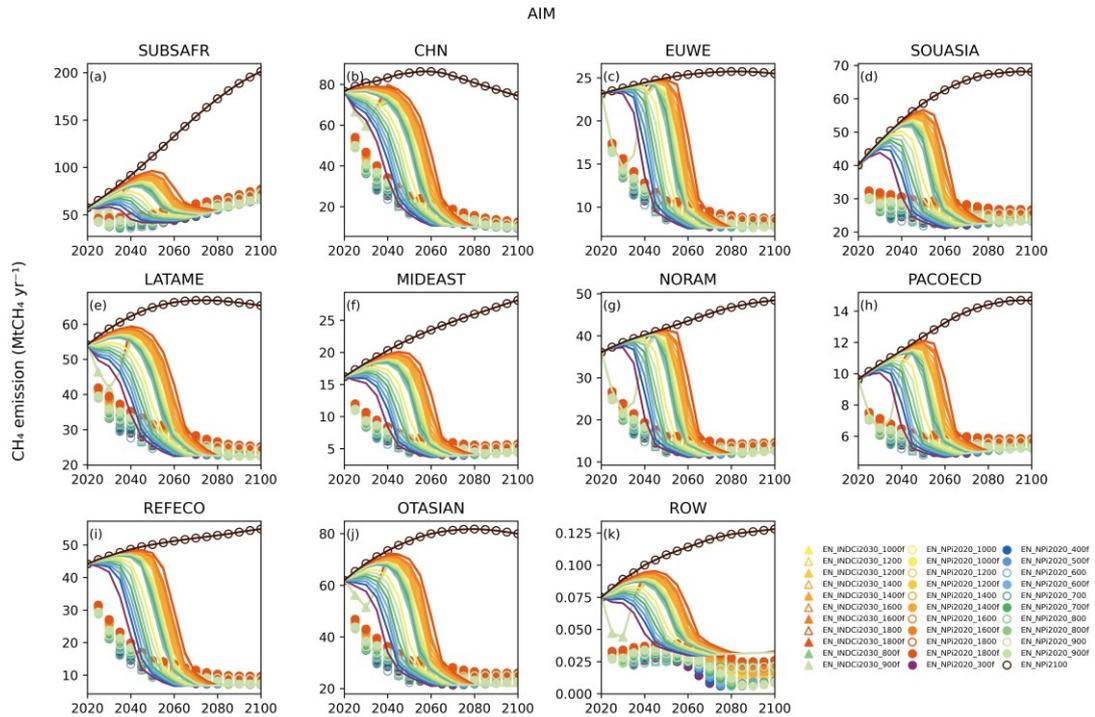

**Figure S133. Test 2 - Regional AIM total anthropogenic CH₄ validation result**

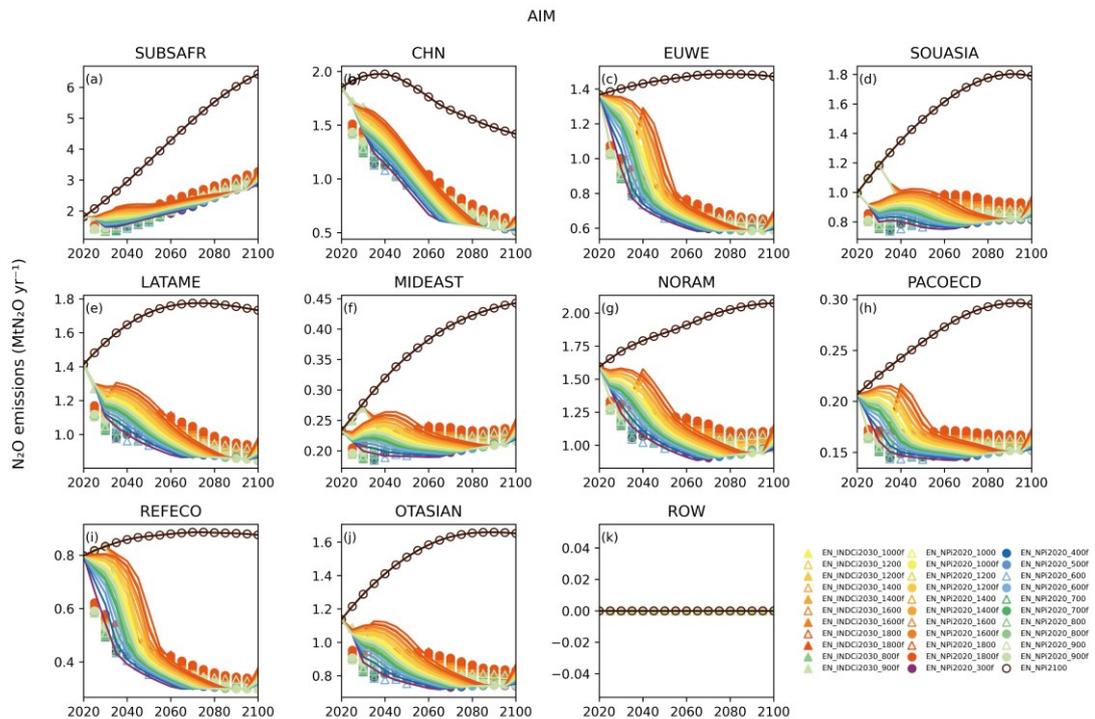

**Figure S134. Test 2 - Regional AIM total anthropogenic N₂O validation result**

**Figure S135. Test 2 - Regional COFFEE total anthropogenic CO₂ validation result**

**Figure S136. Test 2 - Regional COFFEE total anthropogenic CH₄ validation result**

**Figure S137. Test 2 - Regional COFFEE total anthropogenic N₂O validation result**

**Figure S138. Test 2 - Regional GEM total anthropogenic CO₂ validation result**

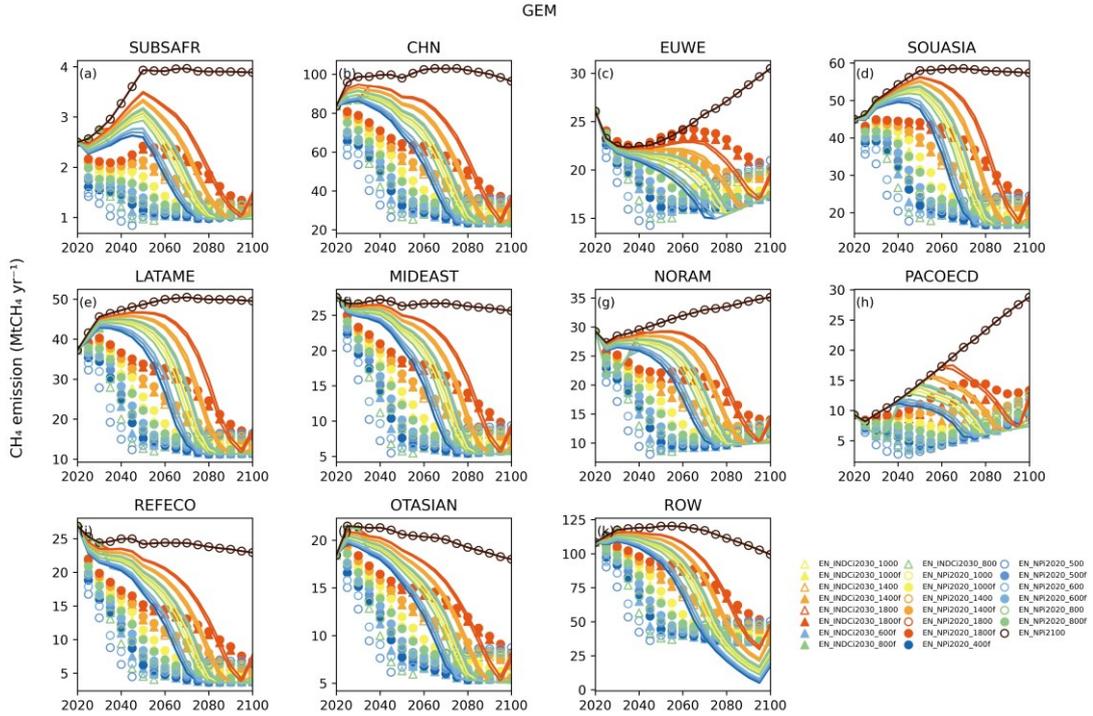

**Figure S139. Test 2 - Regional GEM total anthropogenic CH₄ validation result**

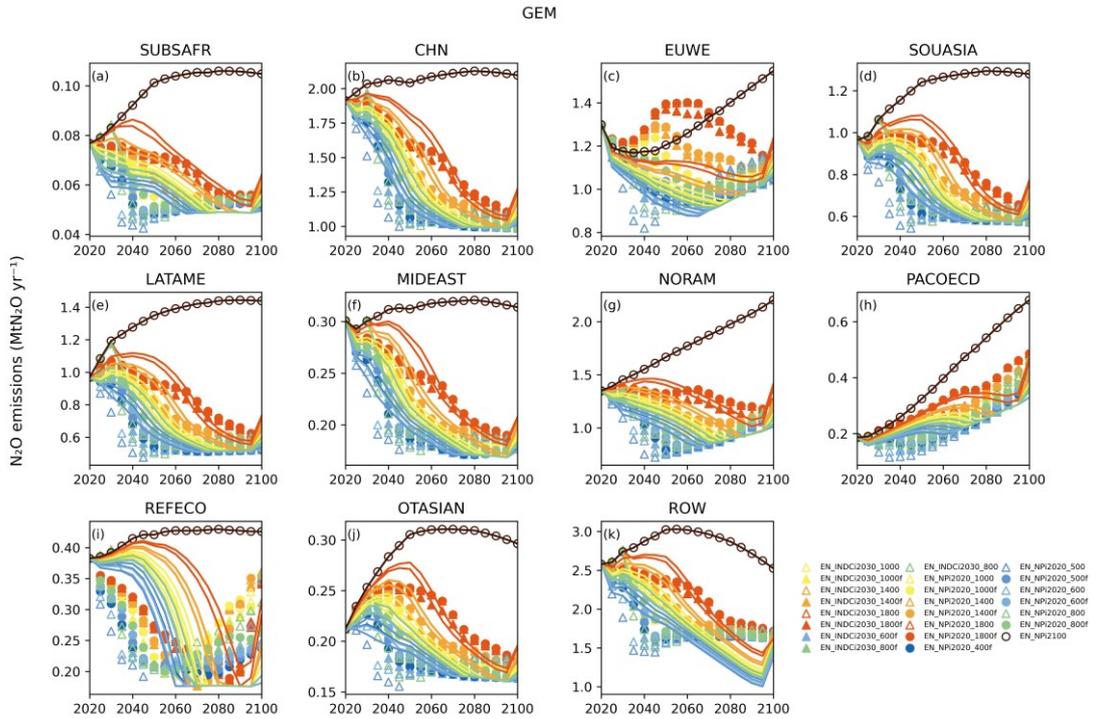

**Figure S140. Test 2 - Regional GEM total anthropogenic N₂O validation result**

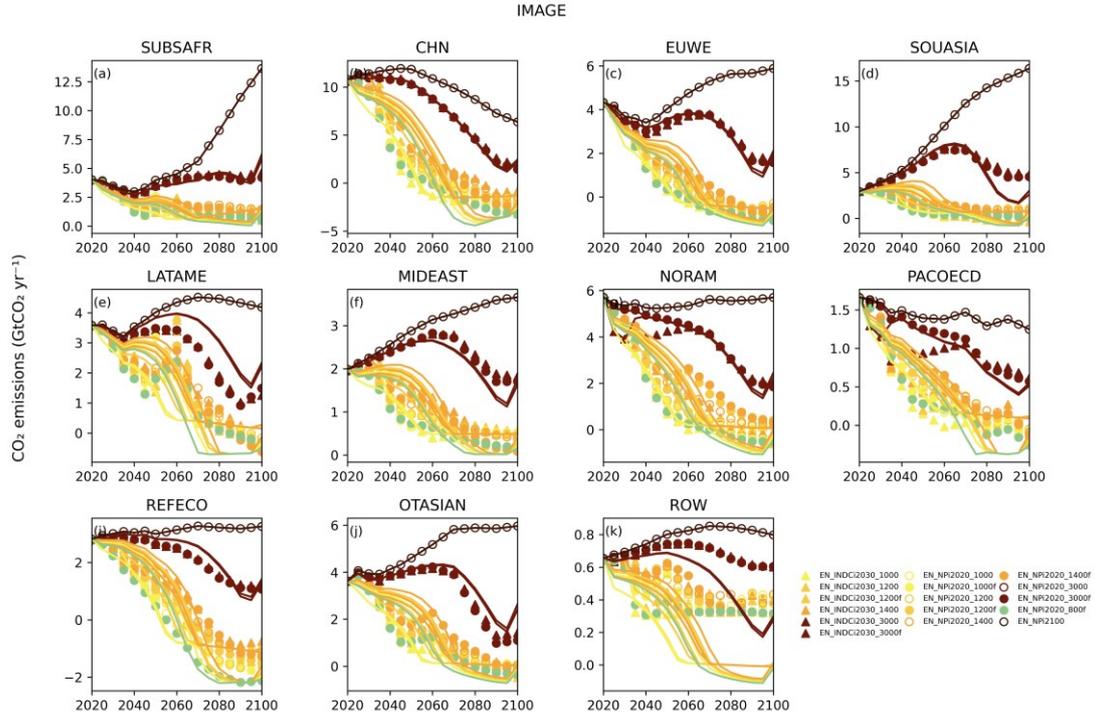

**Figure S141. Test 2 - Regional IMAGE total anthropogenic CO₂ validation result**

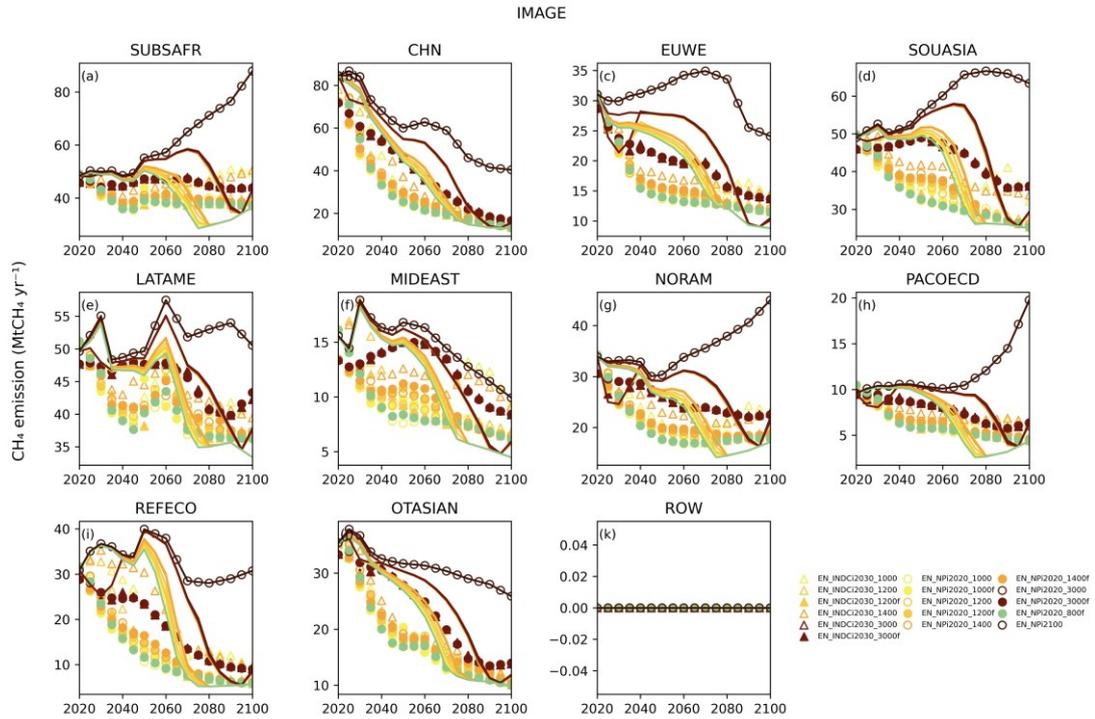

**Figure S142. Test 2 - Regional IMAGE total anthropogenic CH₄ validation result**

**Figure S143. Test 2 - Regional IMAGE total anthropogenic N₂O validation result**

**Figure S144. Test 2 - Regional MESSAGE total anthropogenic CO₂ validation result**

**Figure S145.** Test 2 - Regional MESSAGE total anthropogenic CH₄ validation result

**Figure S146.** Test 2 - Regional MESSAGE total anthropogenic N₂O validation result

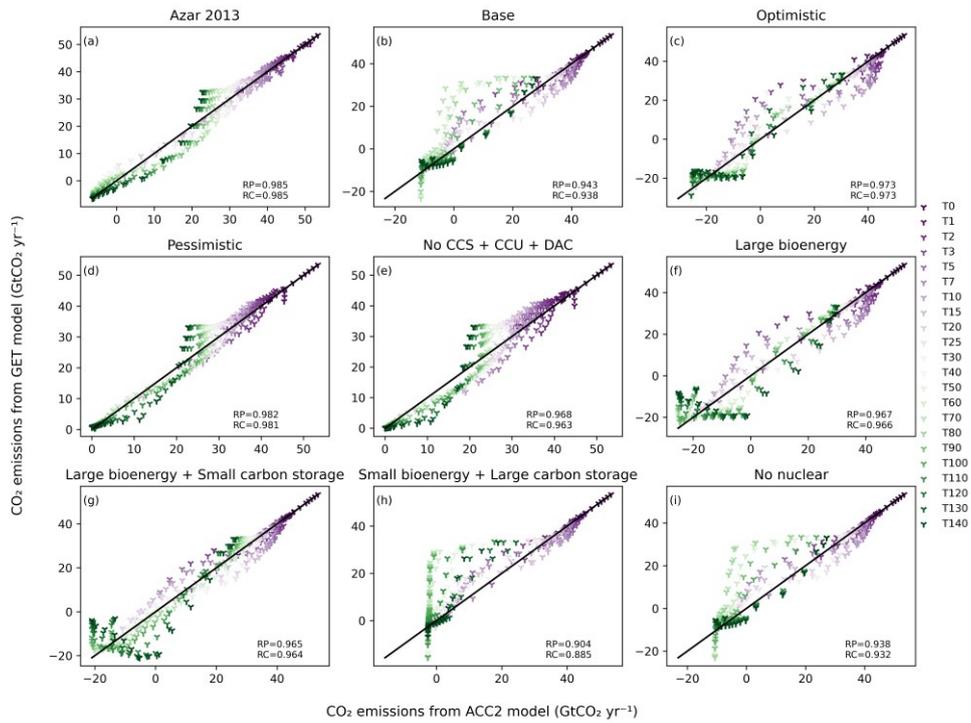

**Figure S147. Test 2 - GET Reproducibility of energy-related CO₂**

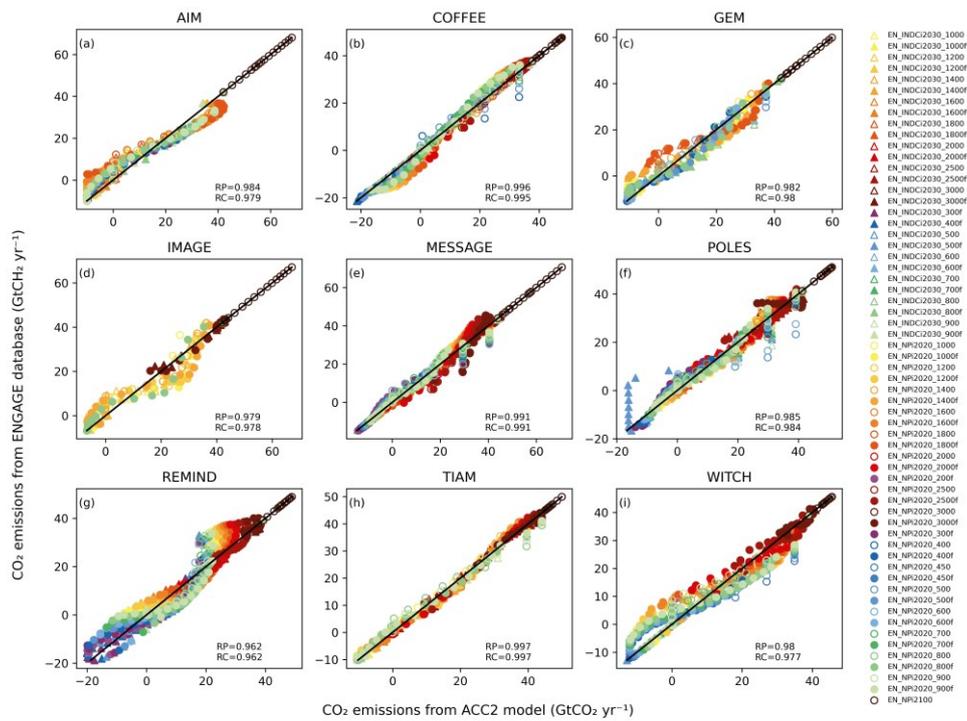

**Figure S148. Test 2 - Global 9 models - Reproducibility of total anthropogenic CO₂**

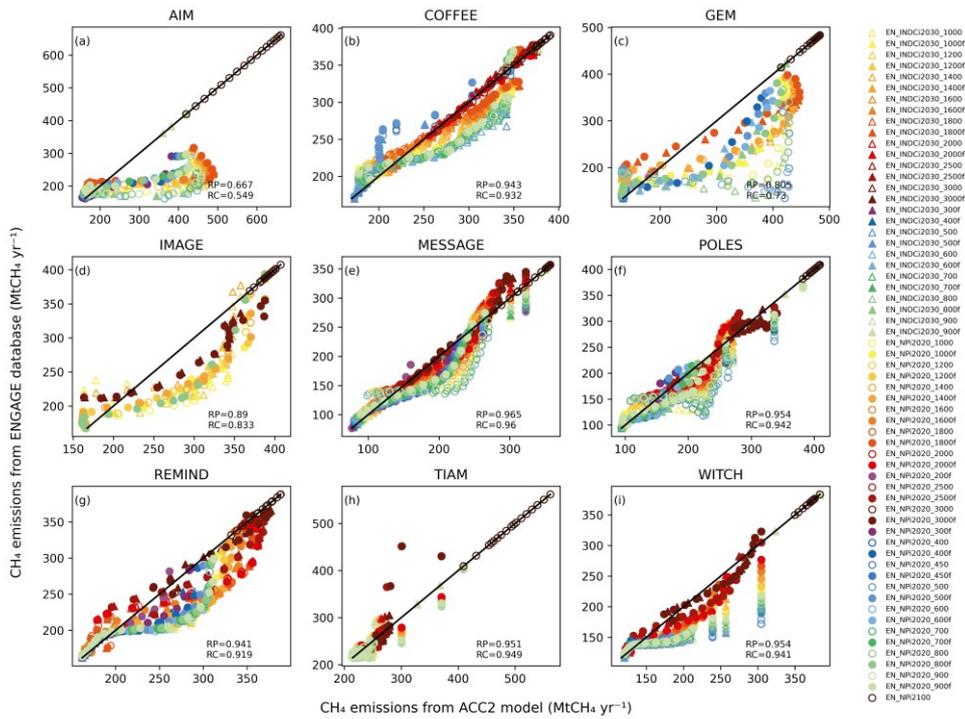

**Figure S149. Test 2 - Global 9 models - Reproducibility of total anthropogenic CH₄**

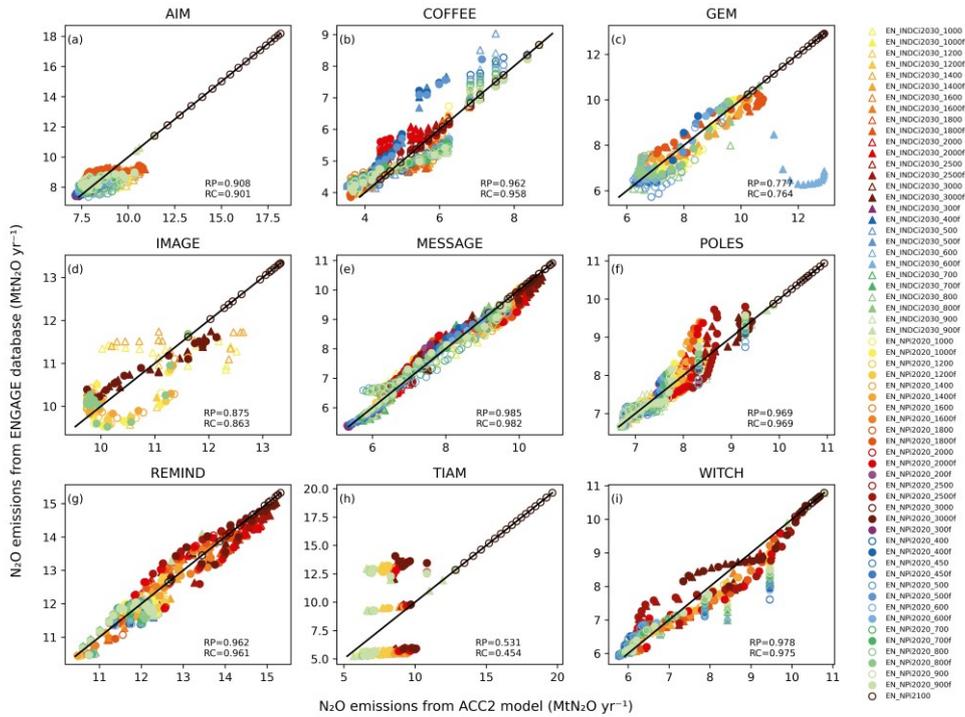

**Figure S150. Test 2 - Global 9 models - Reproducibility of total anthropogenic N₂O**

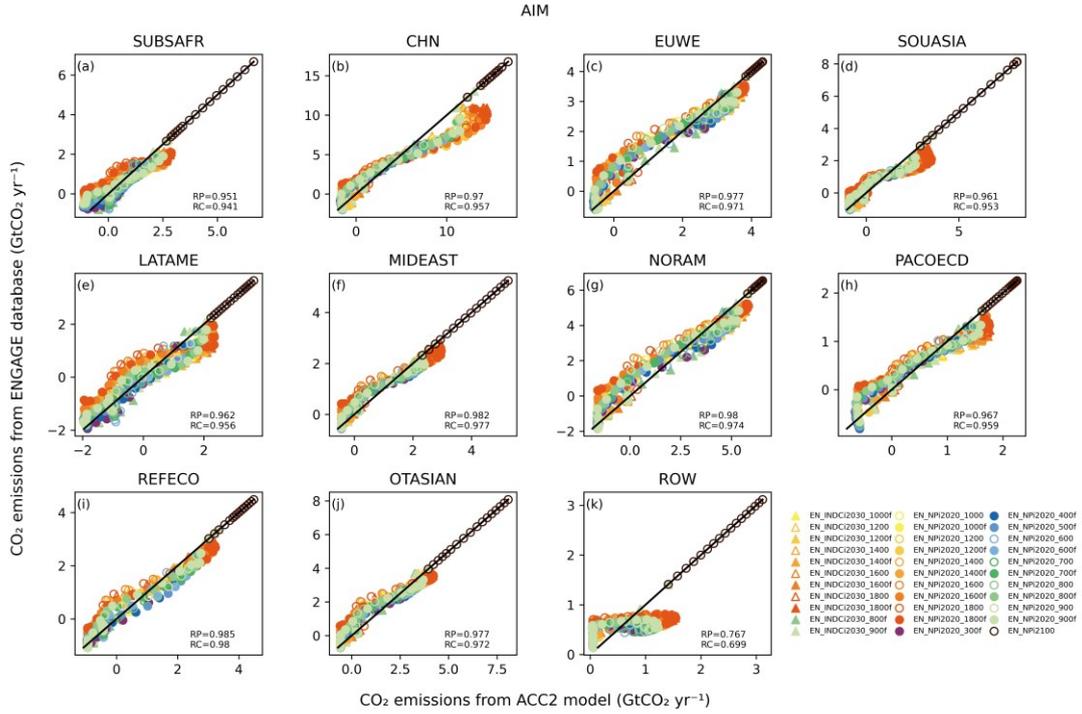

**Figure S151. Test 2 - Regional AIM - Reproducibility of total anthropogenic CO₂**

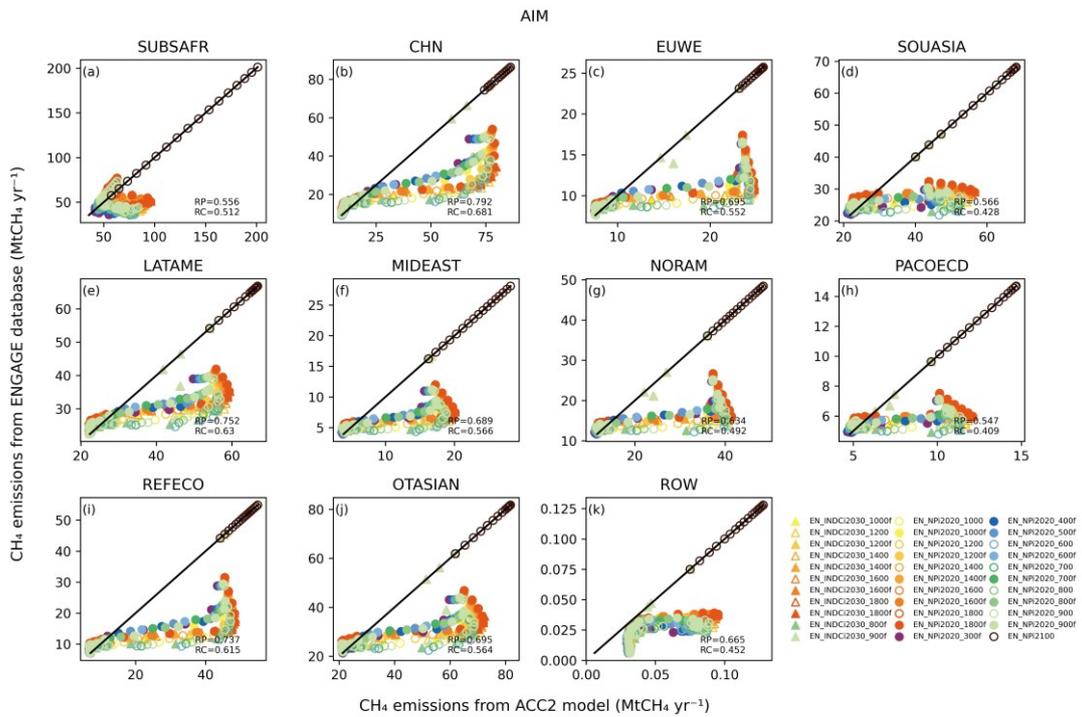

**Figure S152. Test 2 - Regional AIM - Reproducibility of total anthropogenic CH₄**

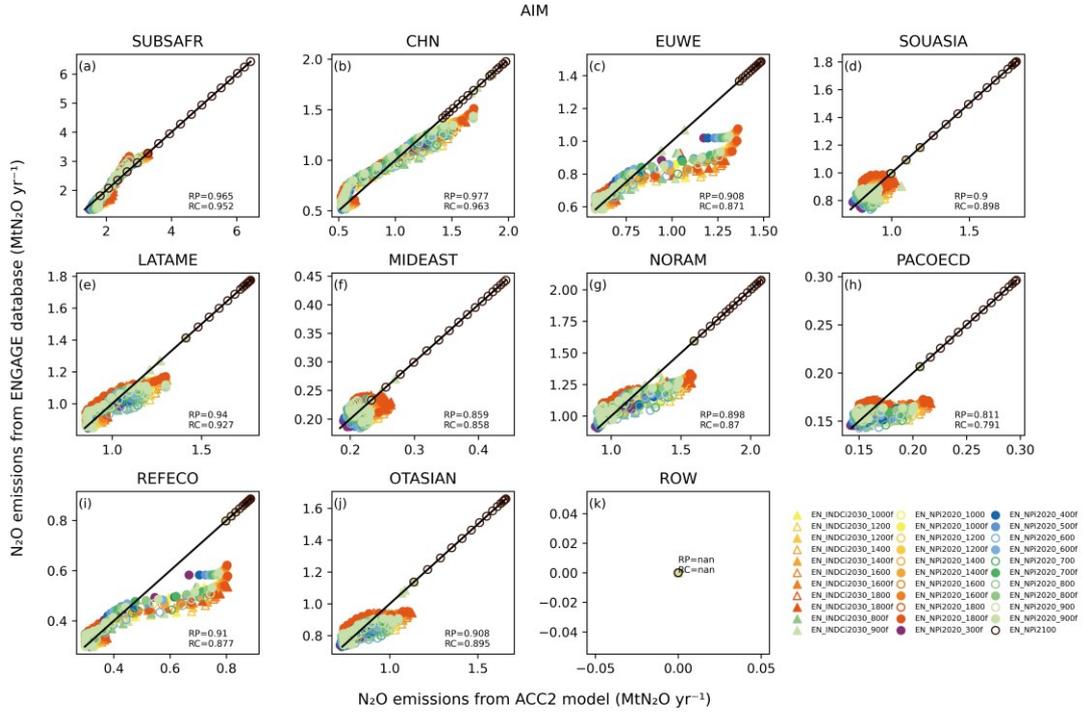

**Figure S153. Test 2 - Regional AIM - Reproducibility of total anthropogenic N₂O**

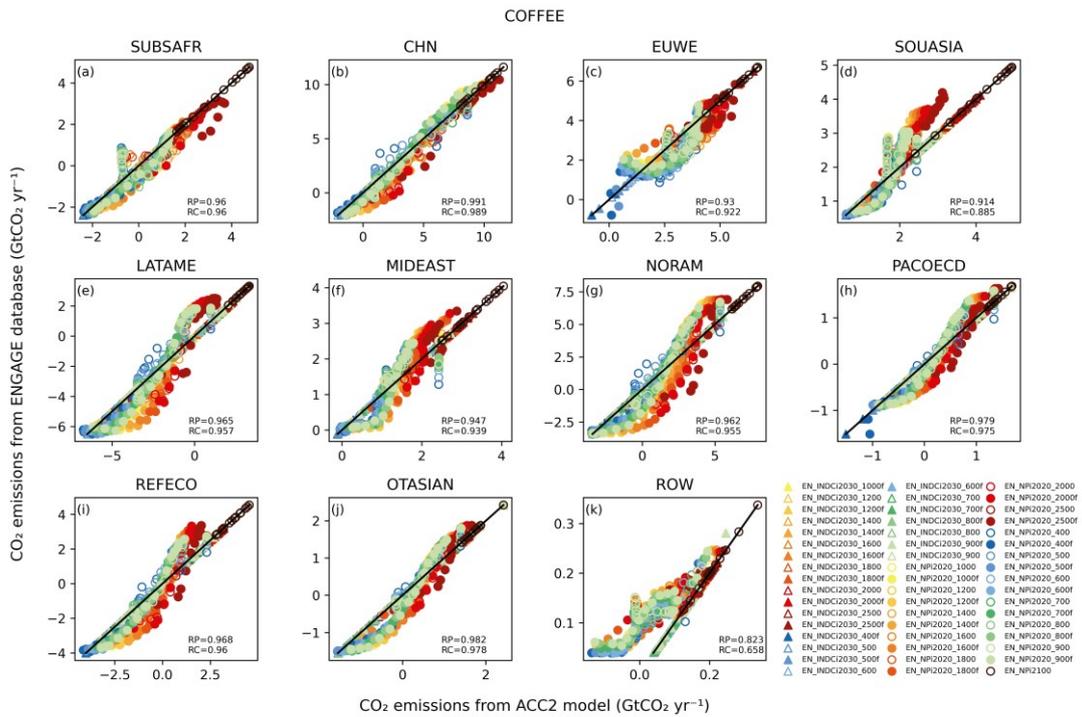

**Figure S154. Test 2 - Regional COFFEE - Reproducibility of total anthropogenic CO₂**

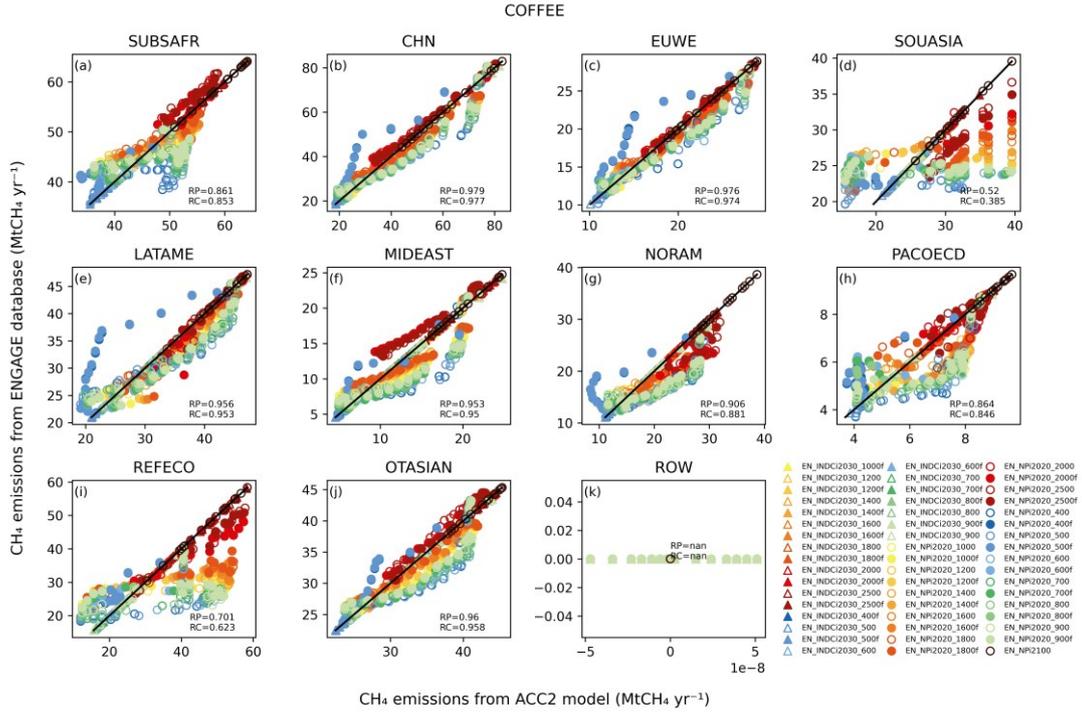

**Figure S155. Test 2 - Regional COFFEE - Reproducibility of total anthropogenic CH$_4$**

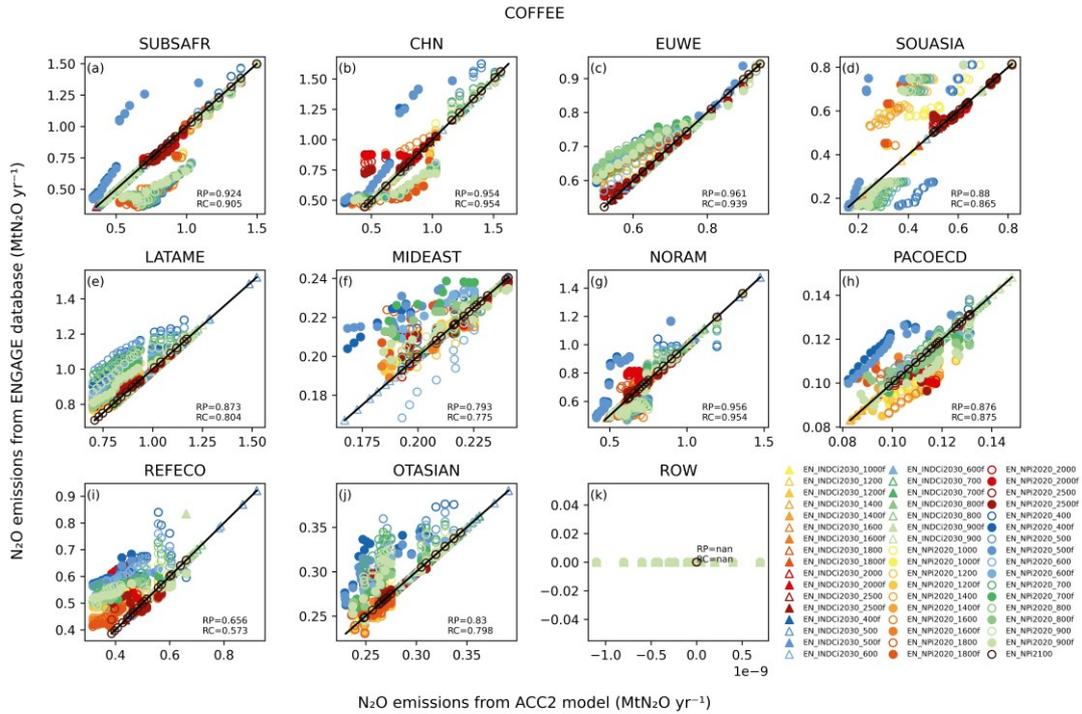

**Figure S156. Test 2 - Regional COFFEE - Reproducibility of total anthropogenic N$_2$O**

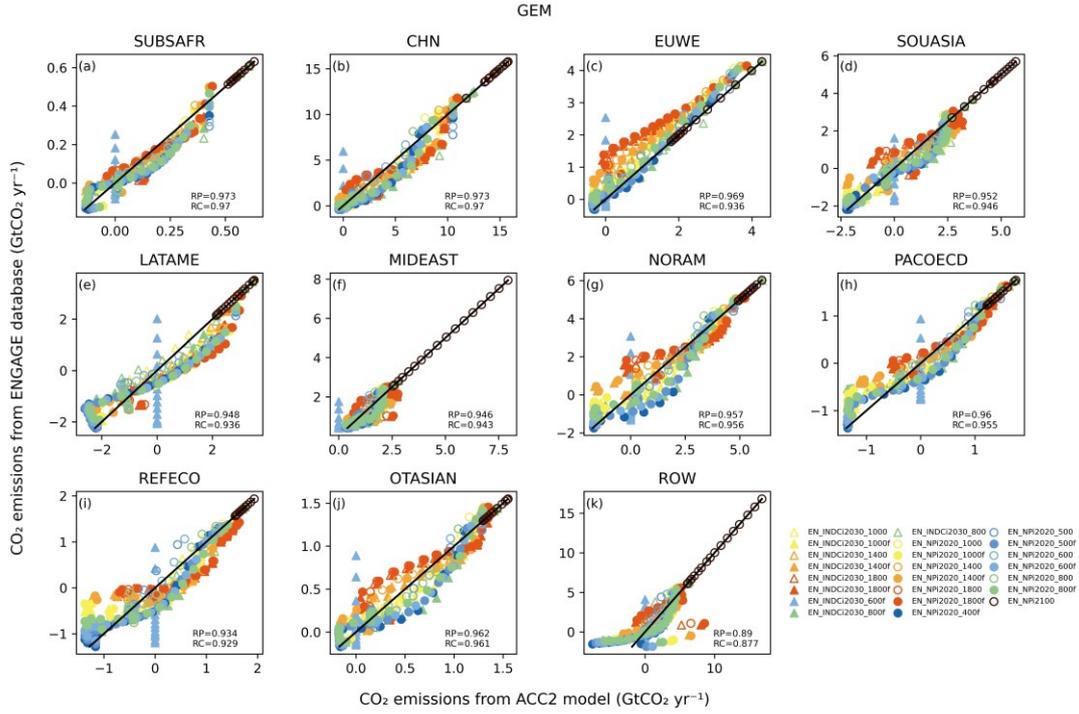

**Figure S157. Test 2 - Regional GEM - Reproducibility of total anthropogenic CO₂**

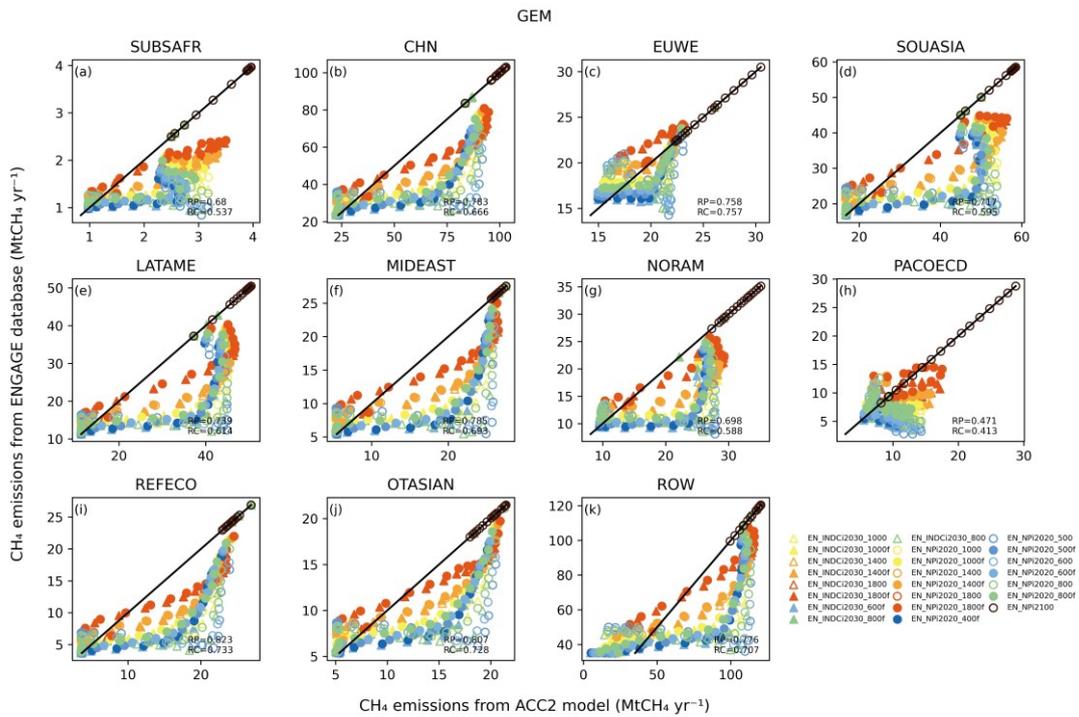

**Figure S158. Test 2 - Regional GEM - Reproducibility of total anthropogenic CH₄**

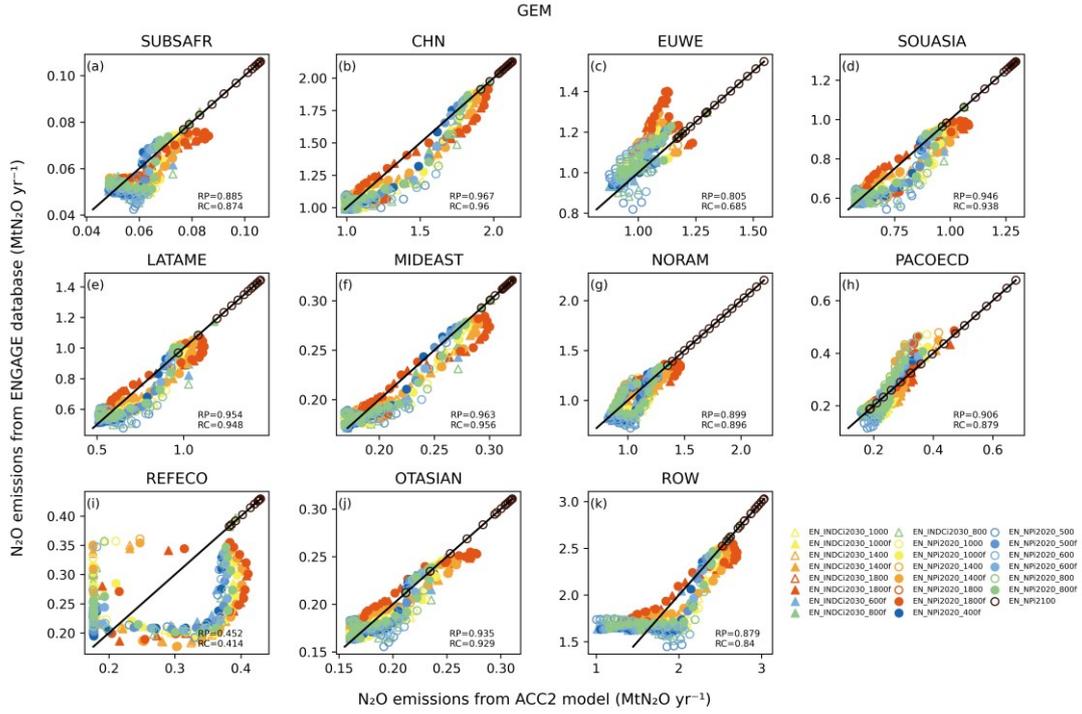

**Figure S159. Test 2 - Regional GEM - Reproducibility of total anthropogenic N₂O**

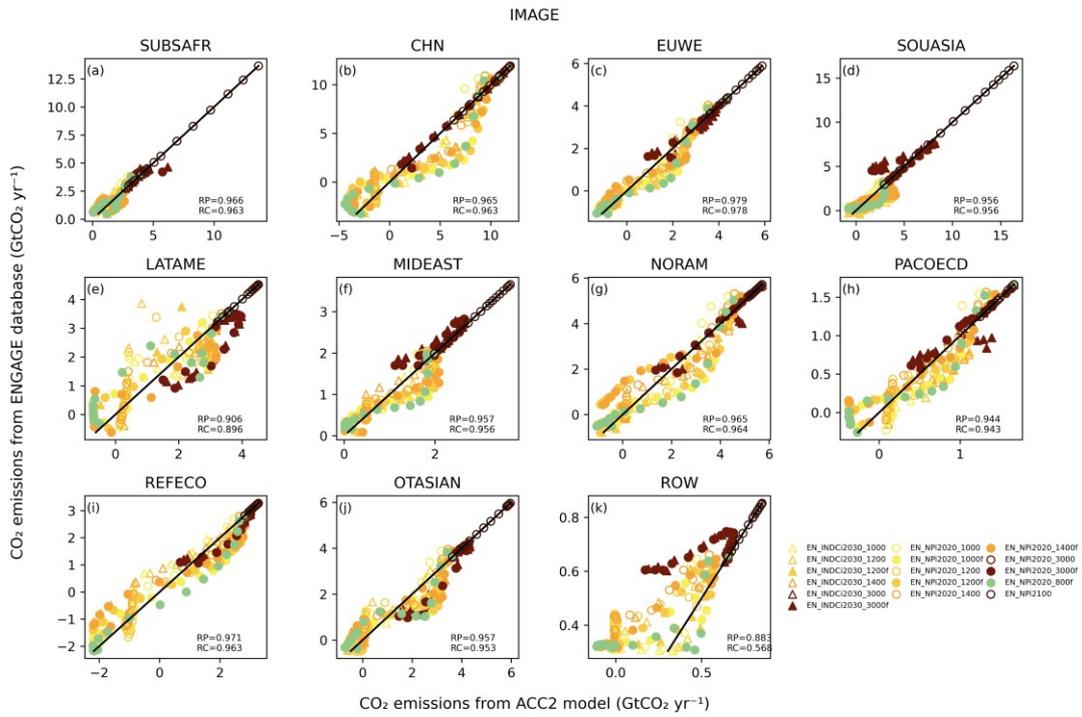

**Figure S160. Test 2 - Regional IMAGE - Reproducibility of total anthropogenic CO₂**

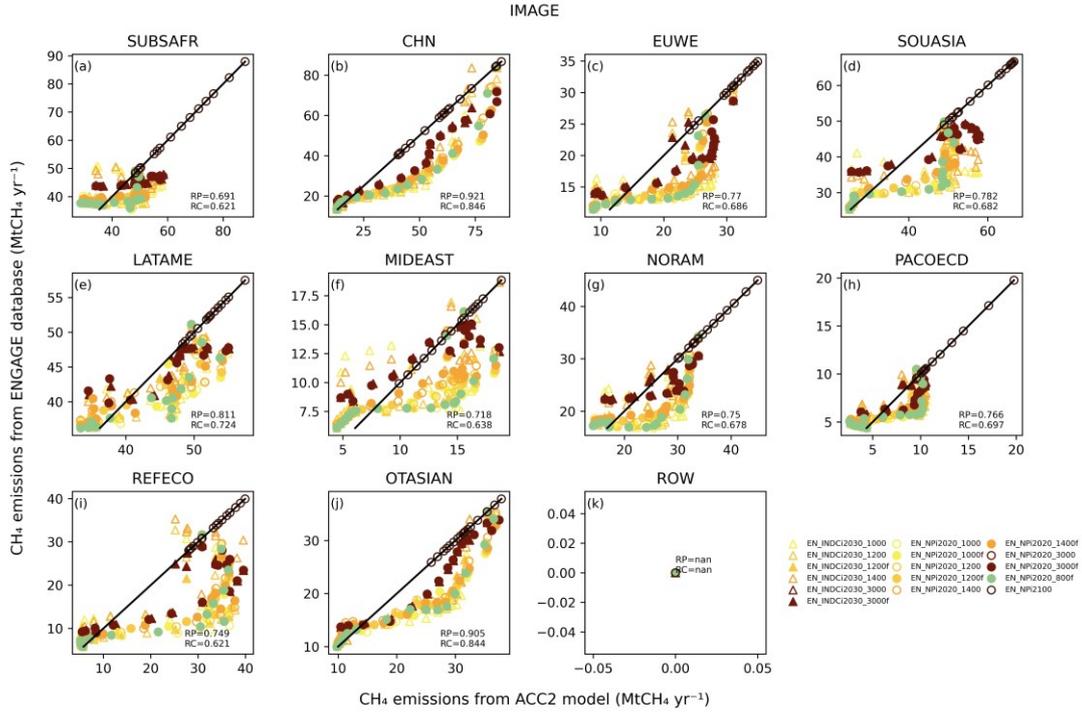

**Figure S161. Test 2 - Regional IMAGE - Reproducibility of total anthropogenic CH₄**

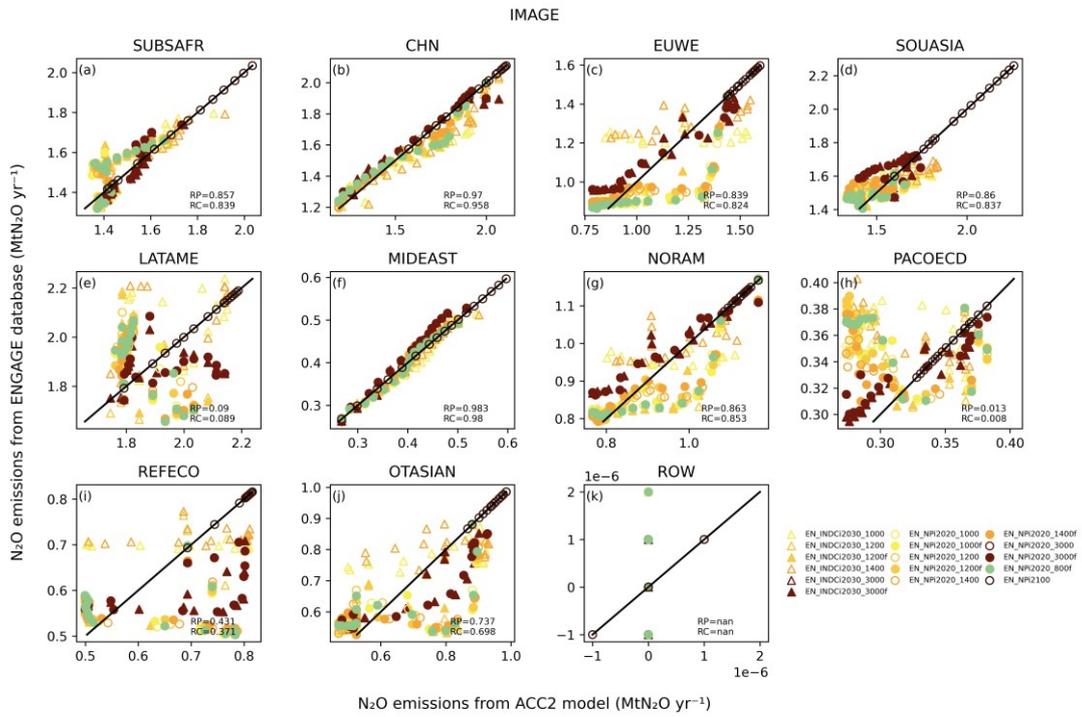

**Figure S162. Test 2 - Regional IMAGE - Reproducibility of total anthropogenic N₂O**

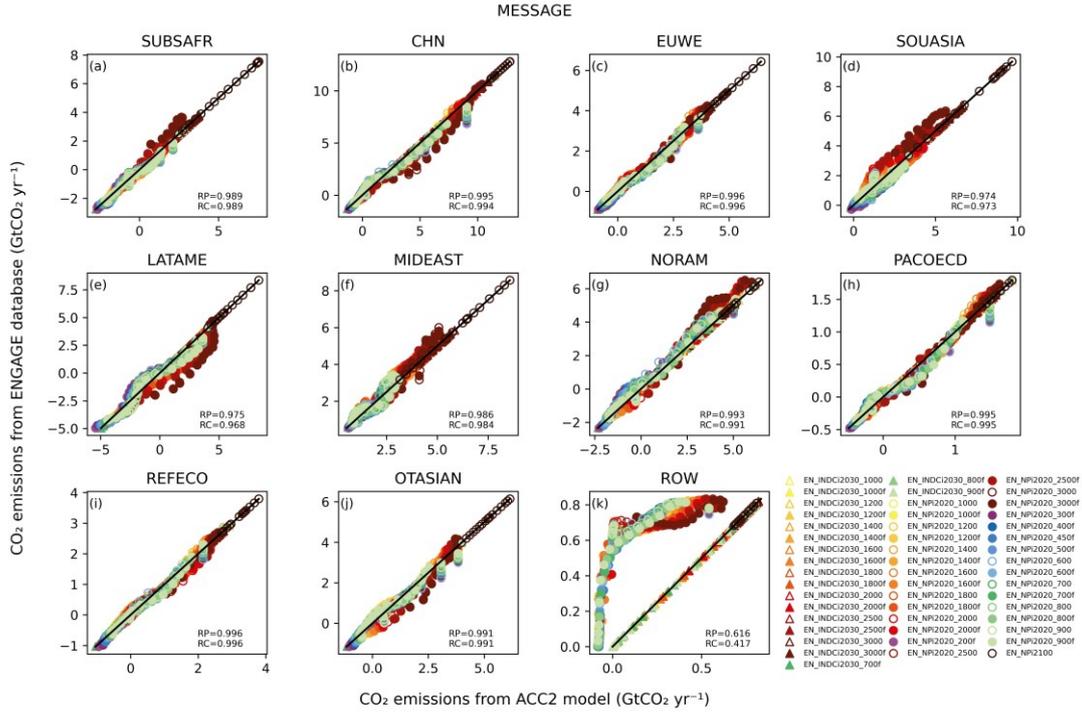

**Figure S163. Test 2 - Regional MESSAGE - Reproducibility of CO₂**

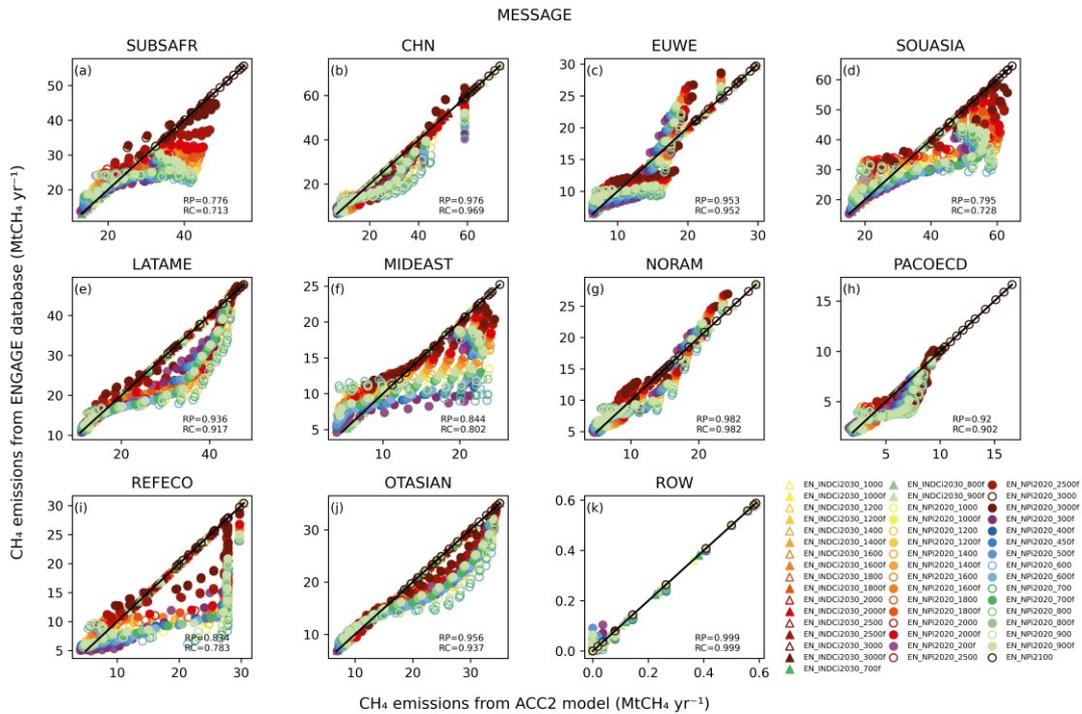

**Figure S164. Test 2 - Regional MESSAGE - Reproducibility of total anthropogenic CH₄**

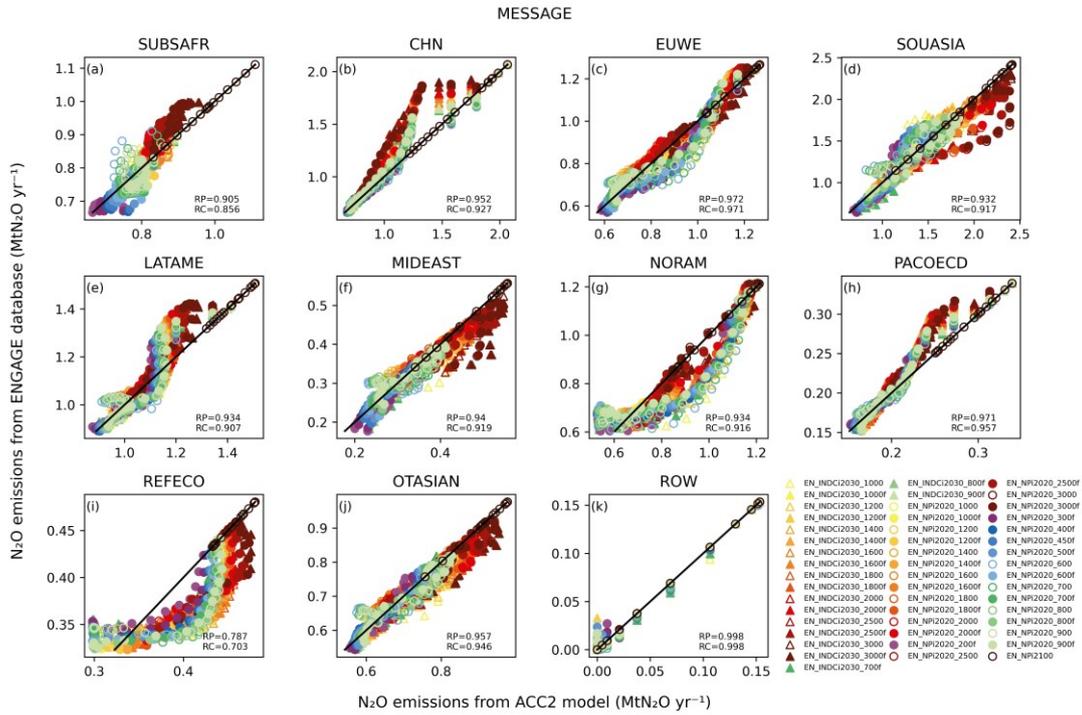

**Figure S165. Test 2 - Regional MESSAGE - Reproducibility of total anthropogenic N₂O**

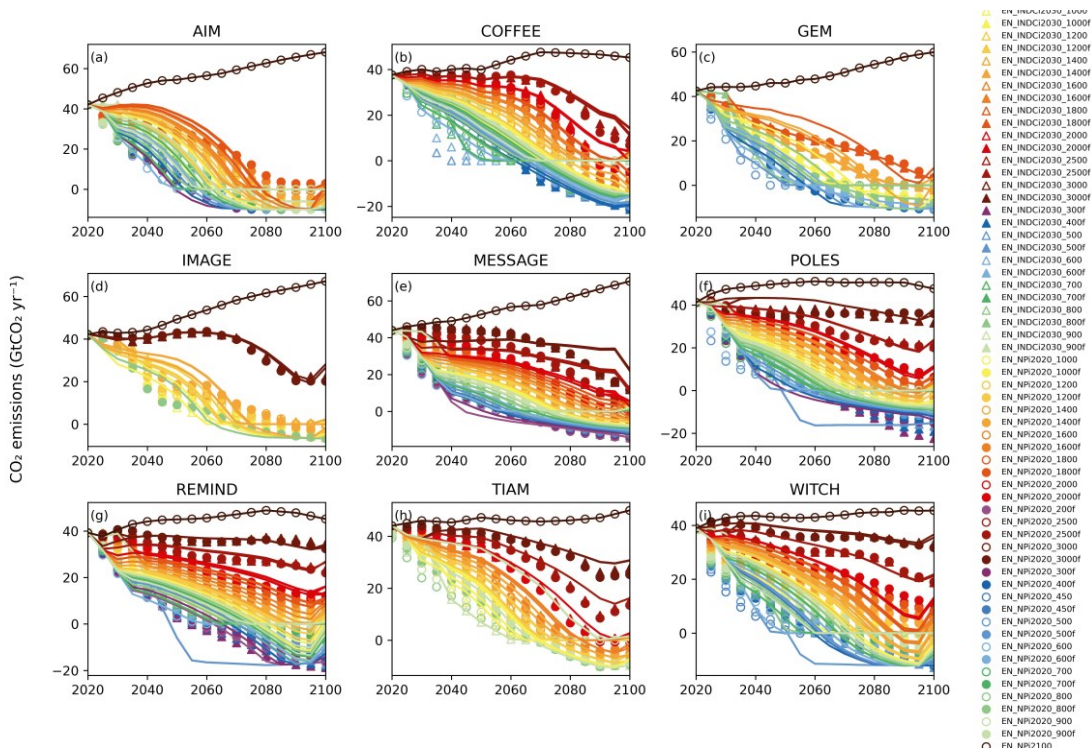

**Figure S166. Test 3 – Global 9 models total anthropogenic CO₂ validation result**

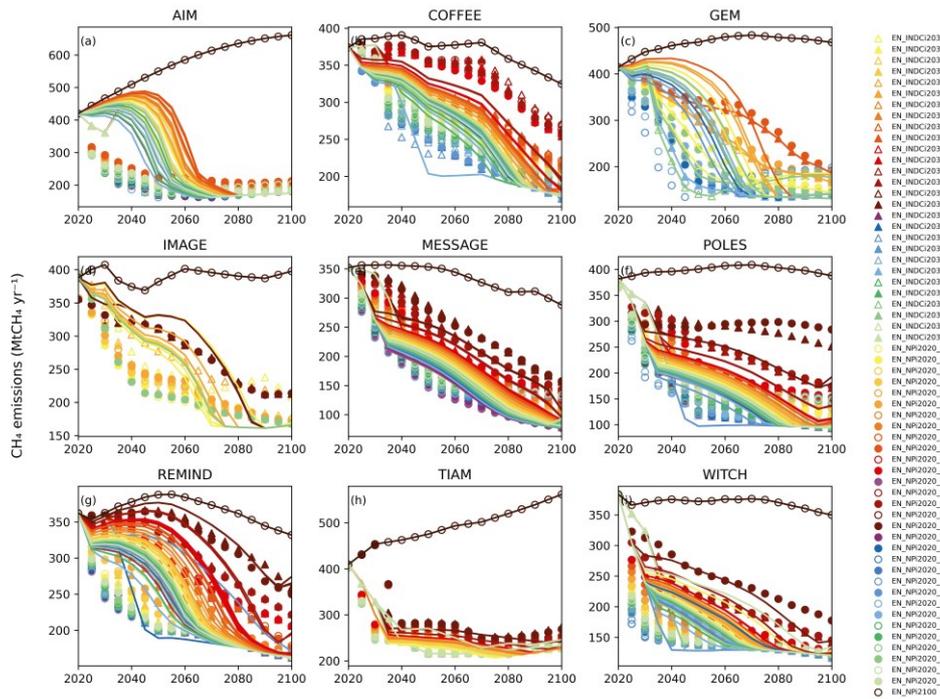

**Figure S167. Test 3 – Global 9 models total anthropogenic CH₄ validation result**

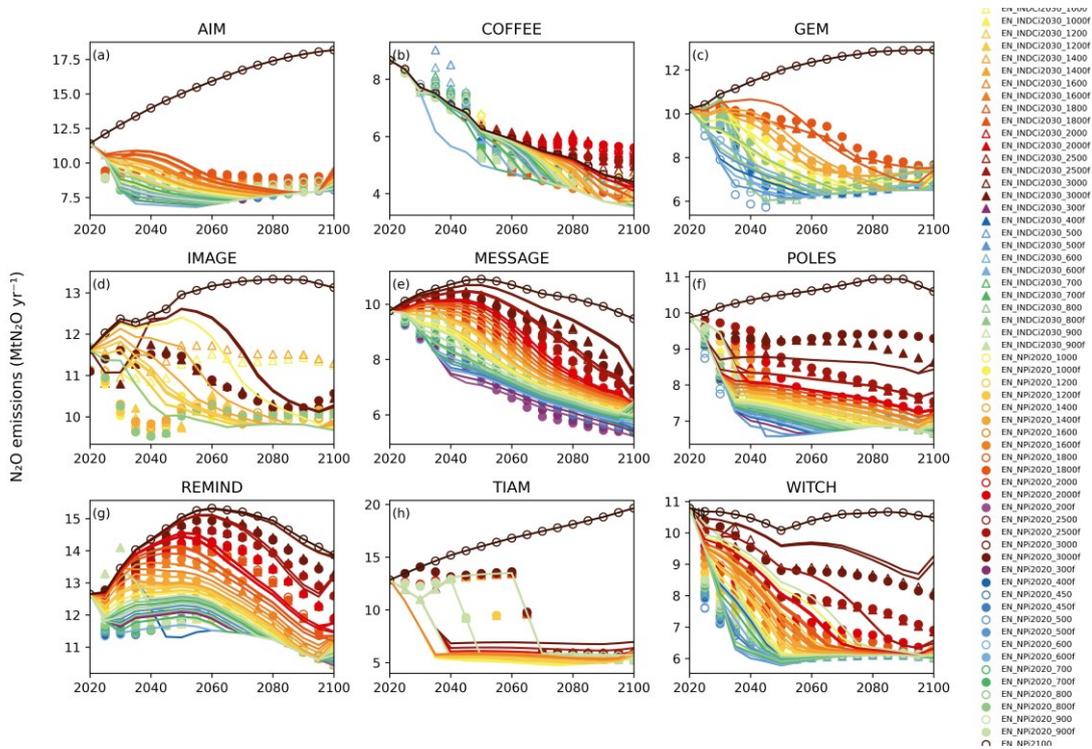

**Figure S168. Test 3 – Global 9 models total anthropogenic N₂O validation result**

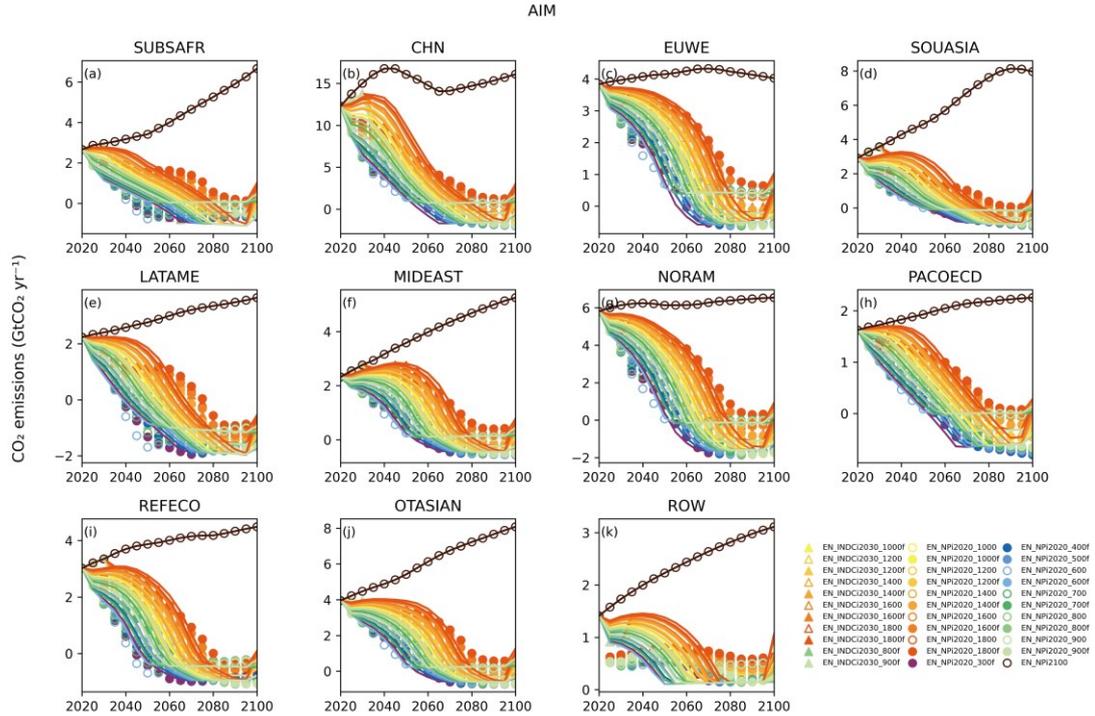

**Figure S169. Test 3 - Regional AIM total anthropogenic CO₂ validation result**

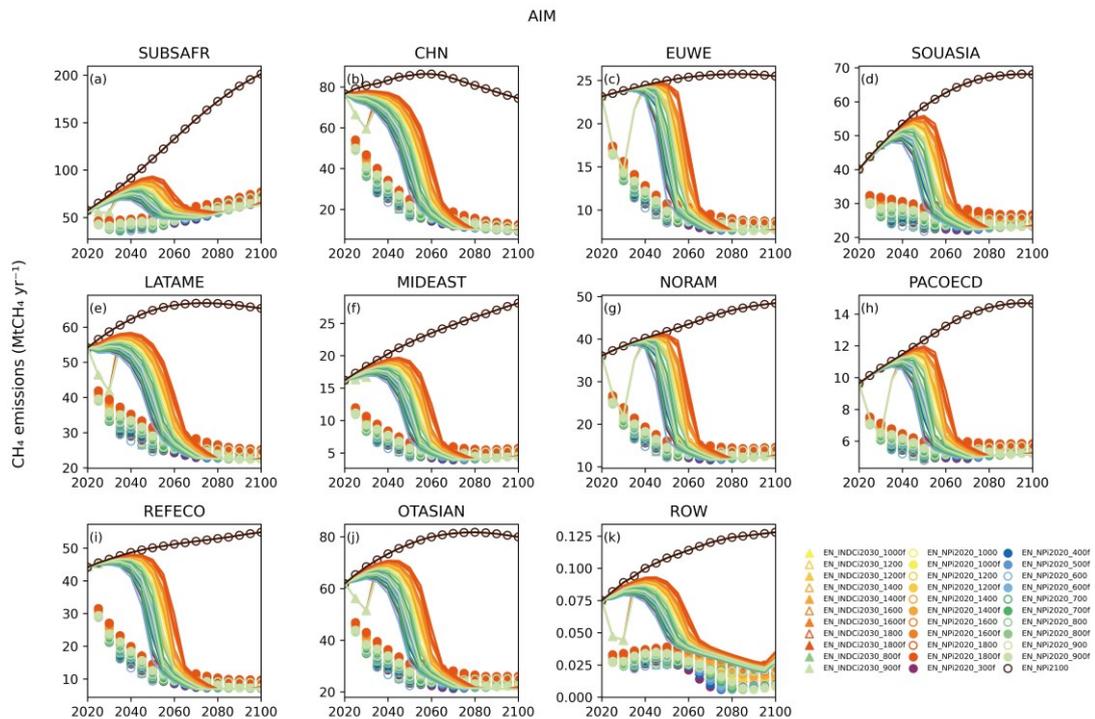

**Figure S170. Test 3 - Regional AIM total anthropogenic CH₄ validation result**

**Figure S171. Test 3 - Regional AIM total anthropogenic N₂O validation result**

**Figure S172. Test 3 - Regional COFFEE total anthropogenic CO₂ validation result**

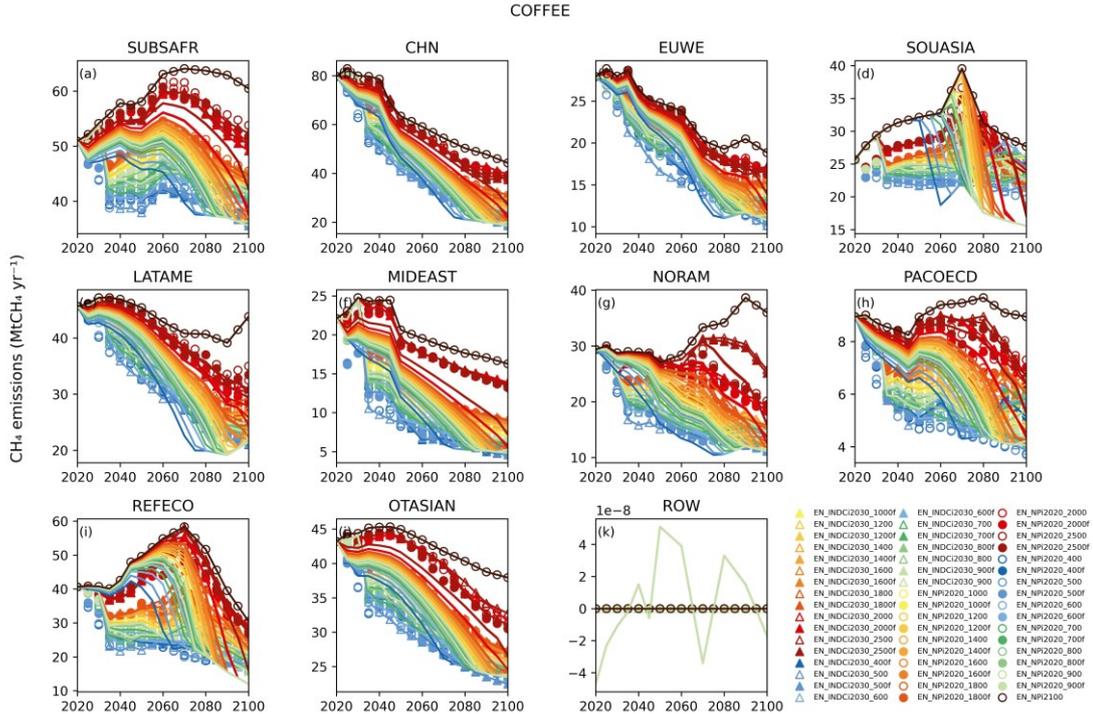

**Figure S173. Test 3 - Regional COFFEE total anthropogenic CH₄ validation result**

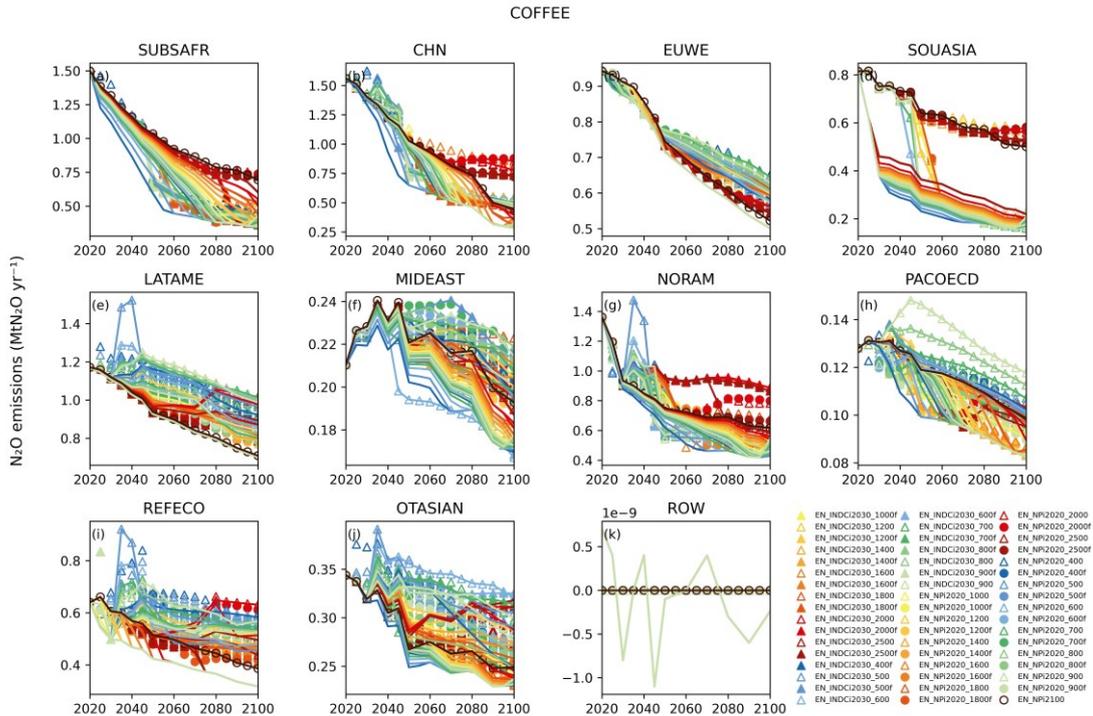

**Figure S174. Test 3 - Regional COFFEE total anthropogenic N₂O validation result**

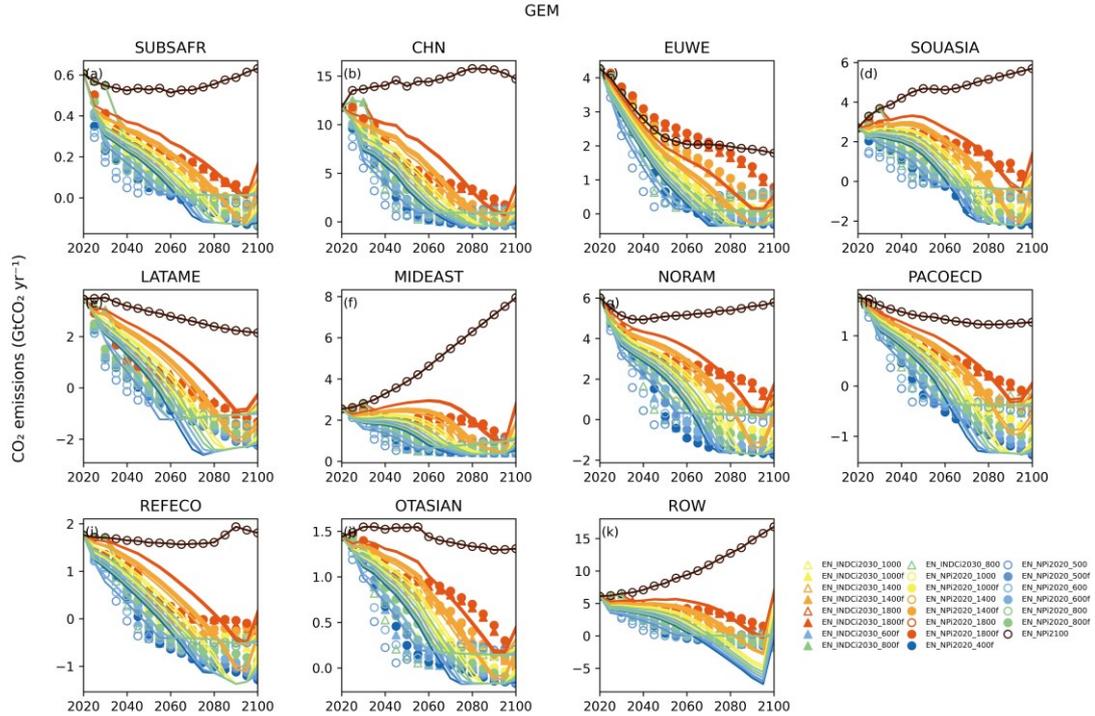

**Figure S175. Test 3 - Regional GEM total anthropogenic CO₂ validation result**

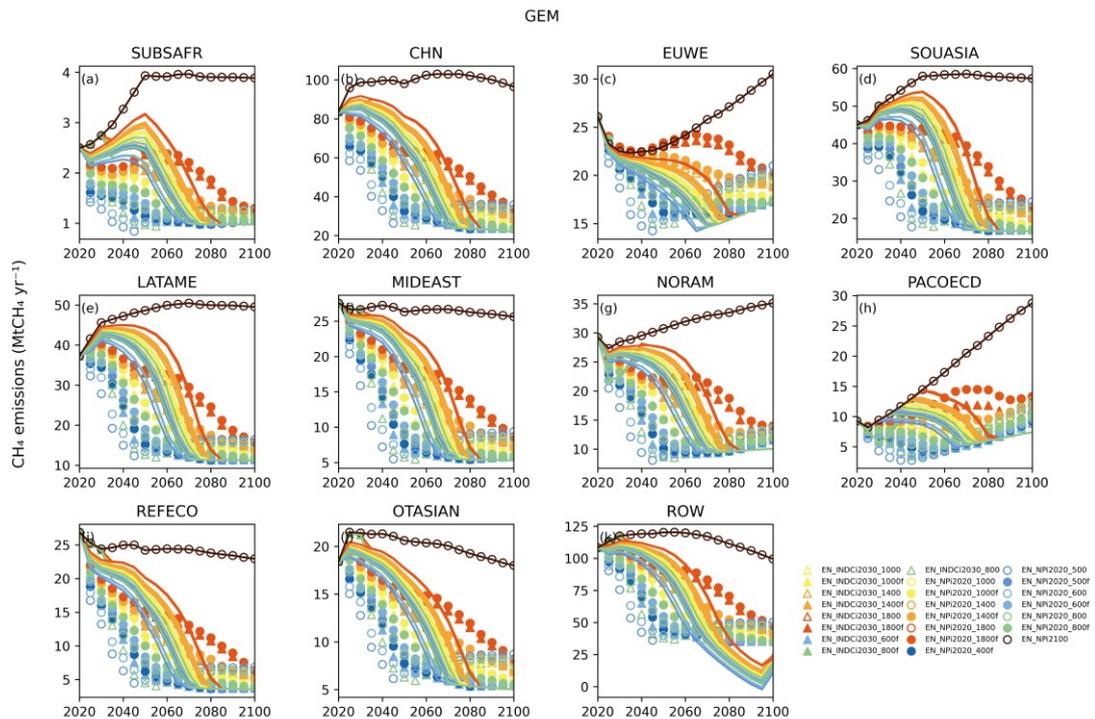

**Figure S176. Test 3 - Regional GEM total anthropogenic CH₄ validation result**

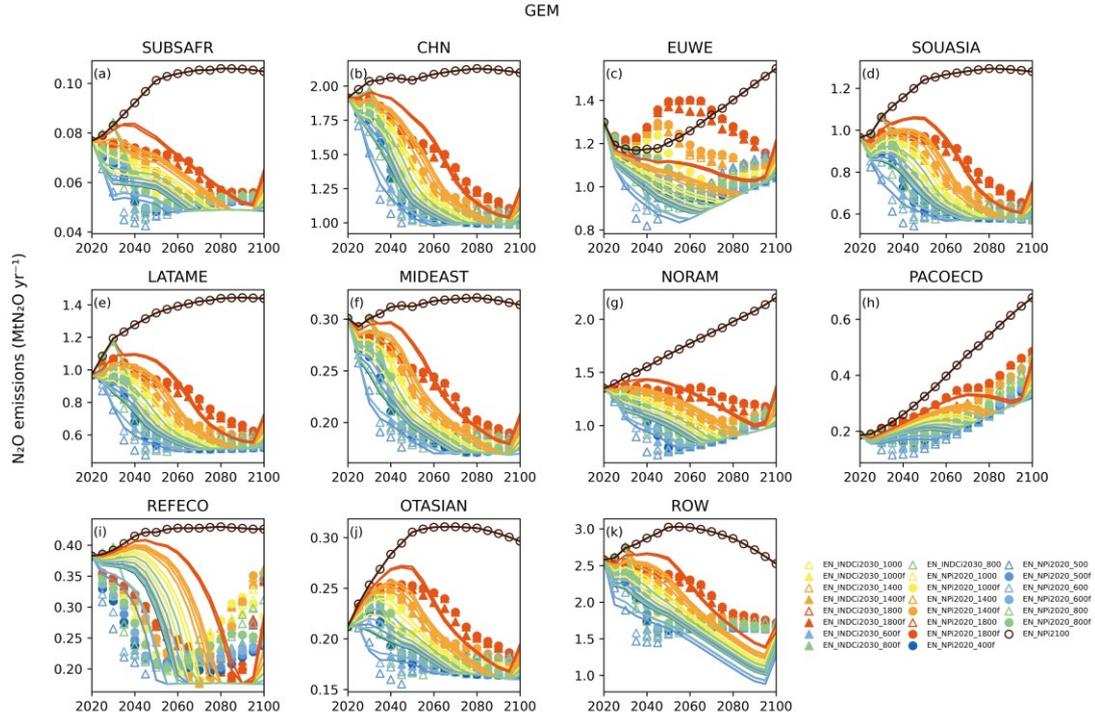

**Figure S177. Test 3 - Regional GEM total anthropogenic N₂O validation result**

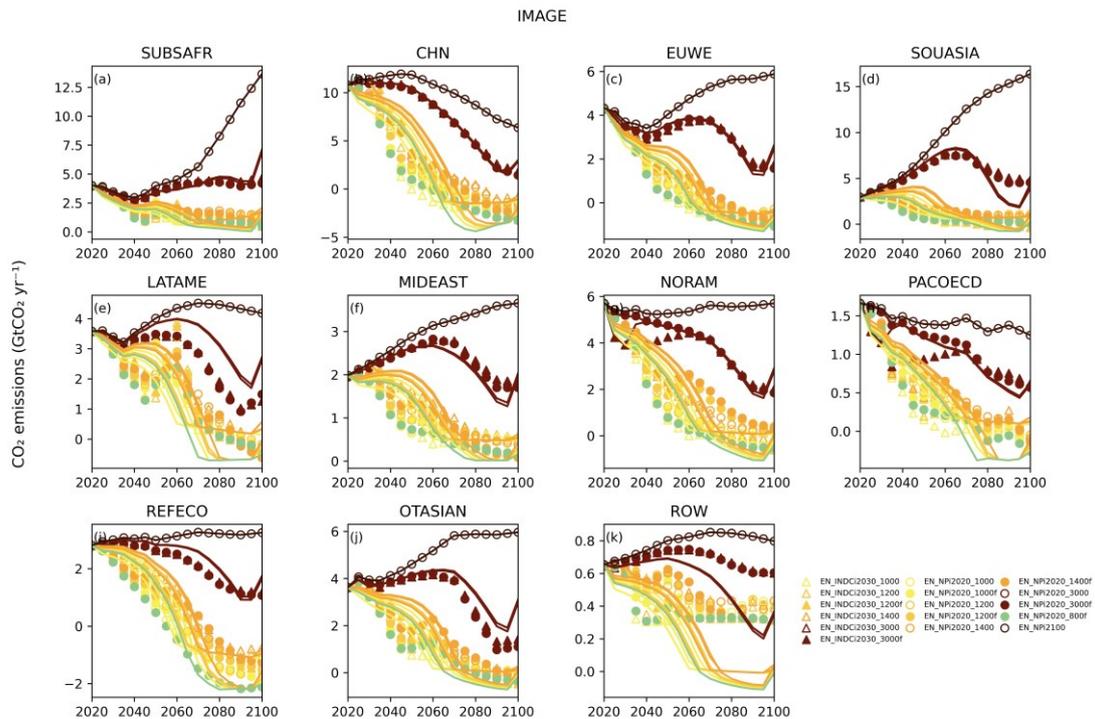

**Figure S178. Test 3 - Regional IMAGE total anthropogenic CO₂ validation result**

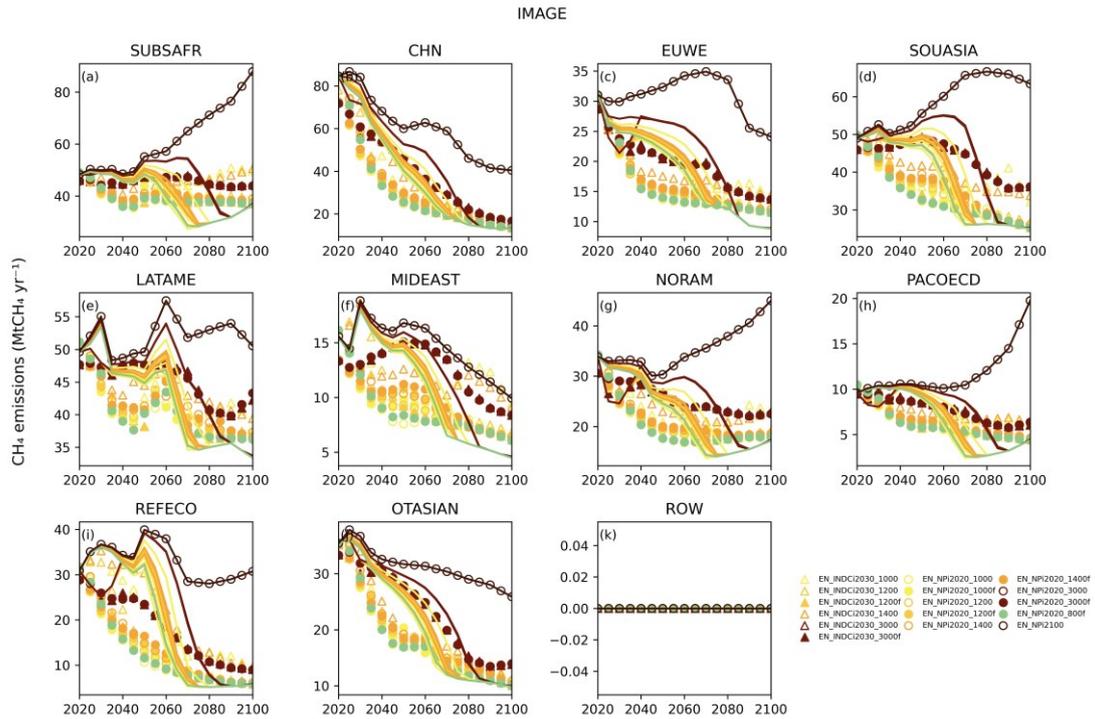

**Figure S179. Test 3 - Regional IMAGE total anthropogenic CH₄ validation result**

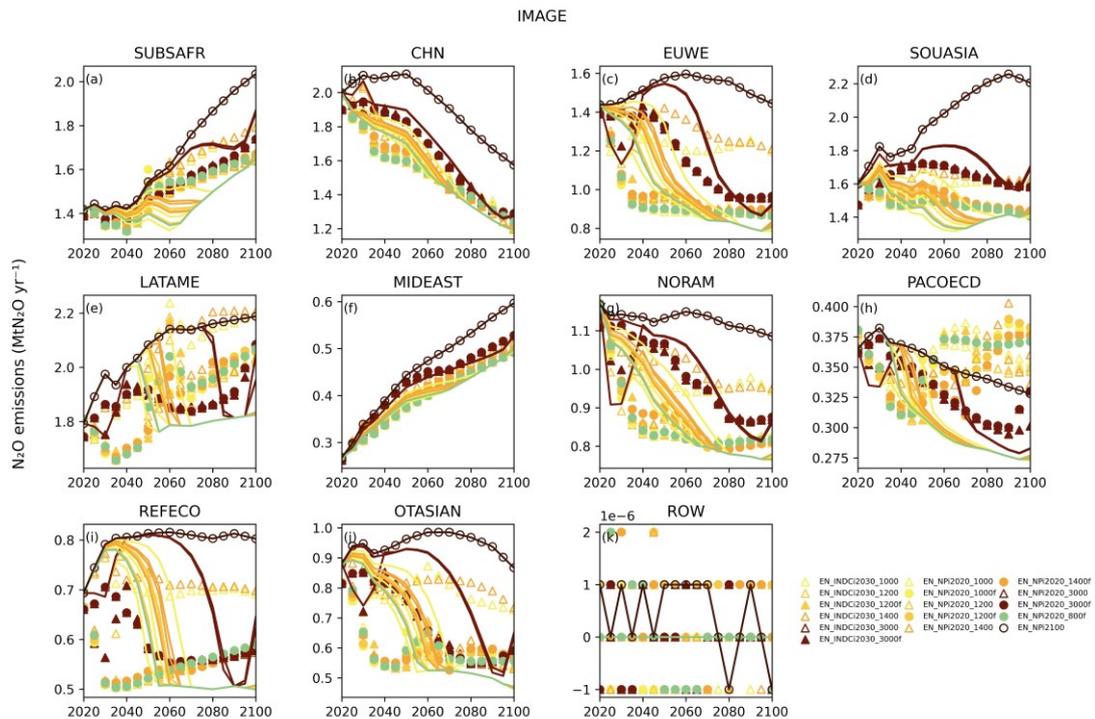

**Figure S180. Test 3 - Regional IMAGE total anthropogenic N₂O validation result**

**Figure S181.** Test 3 - Regional MESSAGE total anthropogenic CO₂ validation result

**Figure S182.** Test 3 - Regional MESSAGE total anthropogenic CH₄ validation result

**Figure S183. Test 3 - Regional MESSAGE total anthropogenic N$_2$O validation result**

**Figure S184. Test 3 - Global 9 models - Reproducibility of total anthropogenic CO$_2$**

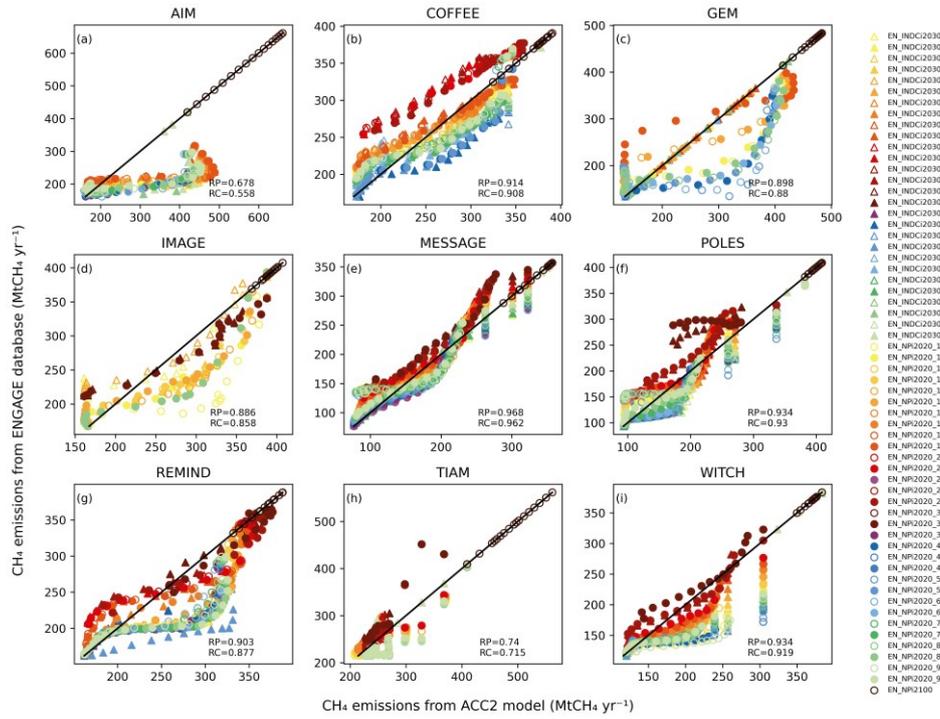

**Figure S185. Test 3 - Global 9 models - Reproducibility of total anthropogenic CH₄**

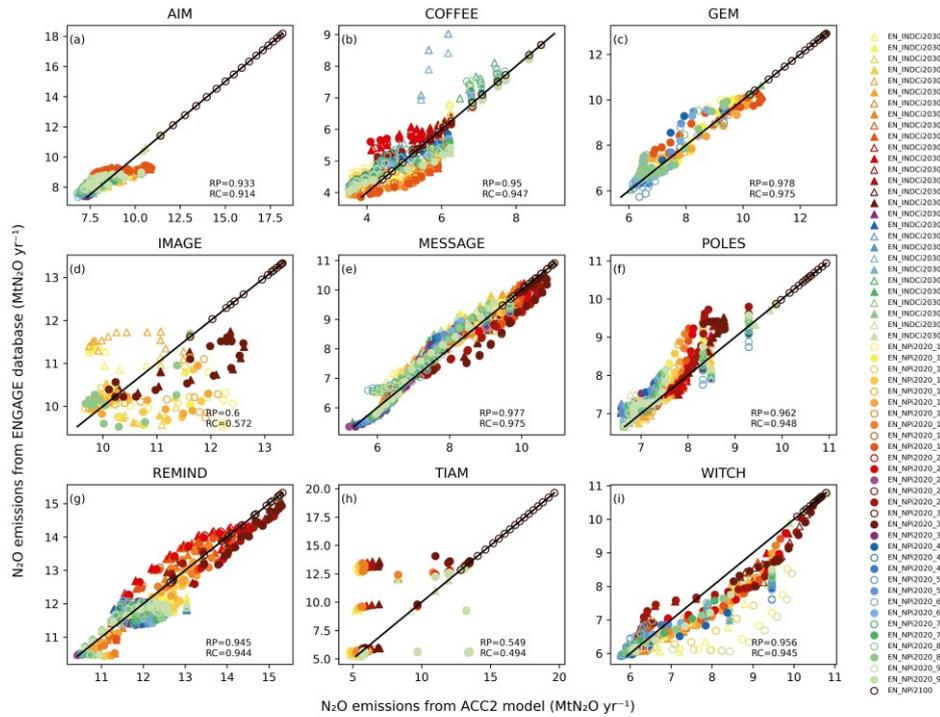

**Figure S186. Test 3 - Global 9 models - Reproducibility of total anthropogenic N₂O**

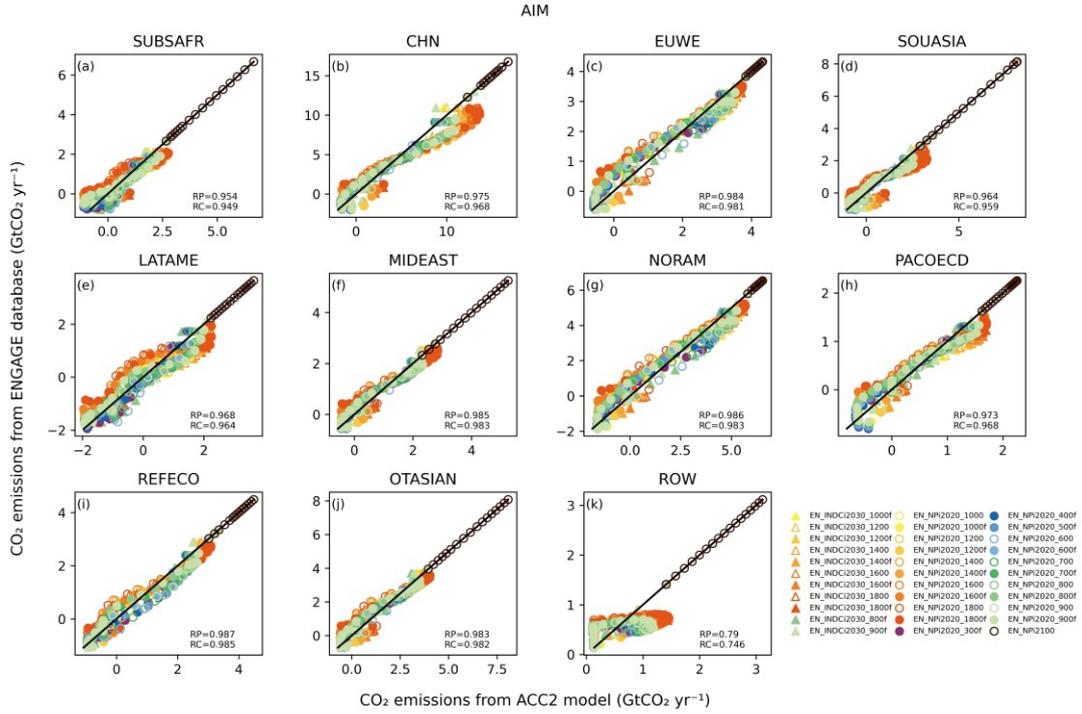

**Figure S187. Test 3 - Regional AIM - Reproducibility of total anthropogenic CO₂**

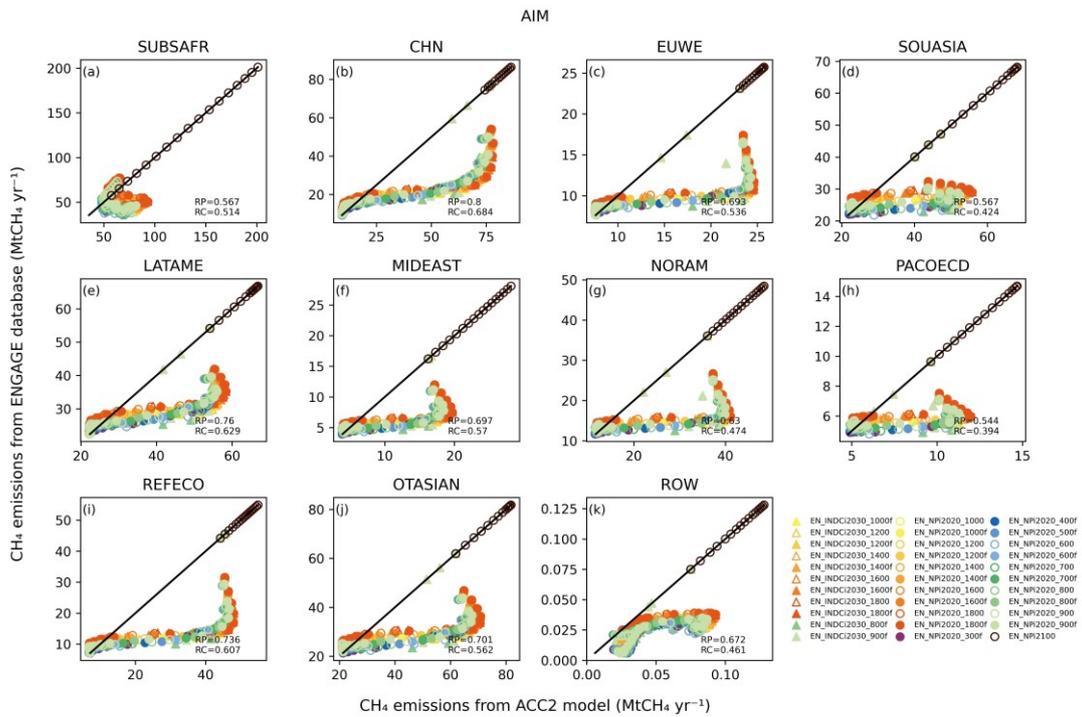

**Figure S188. Test 3 - Regional AIM - Reproducibility of total anthropogenic CH₄**

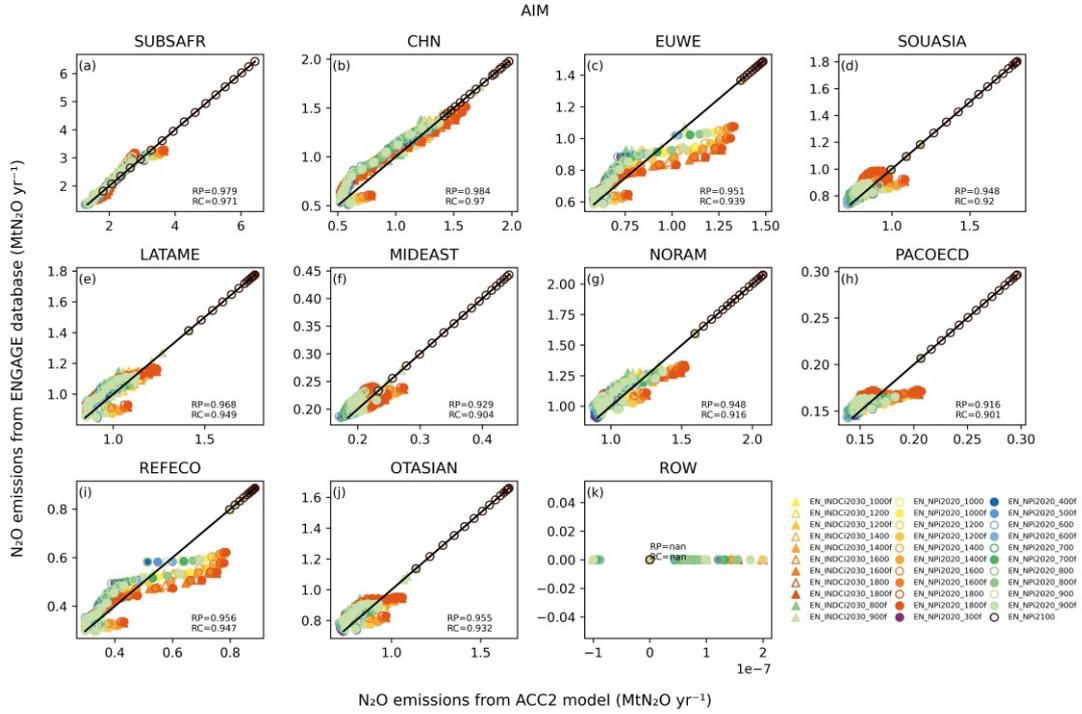

**Figure S189. Test 3 - Regional AIM - Reproducibility of total anthropogenic N₂O**

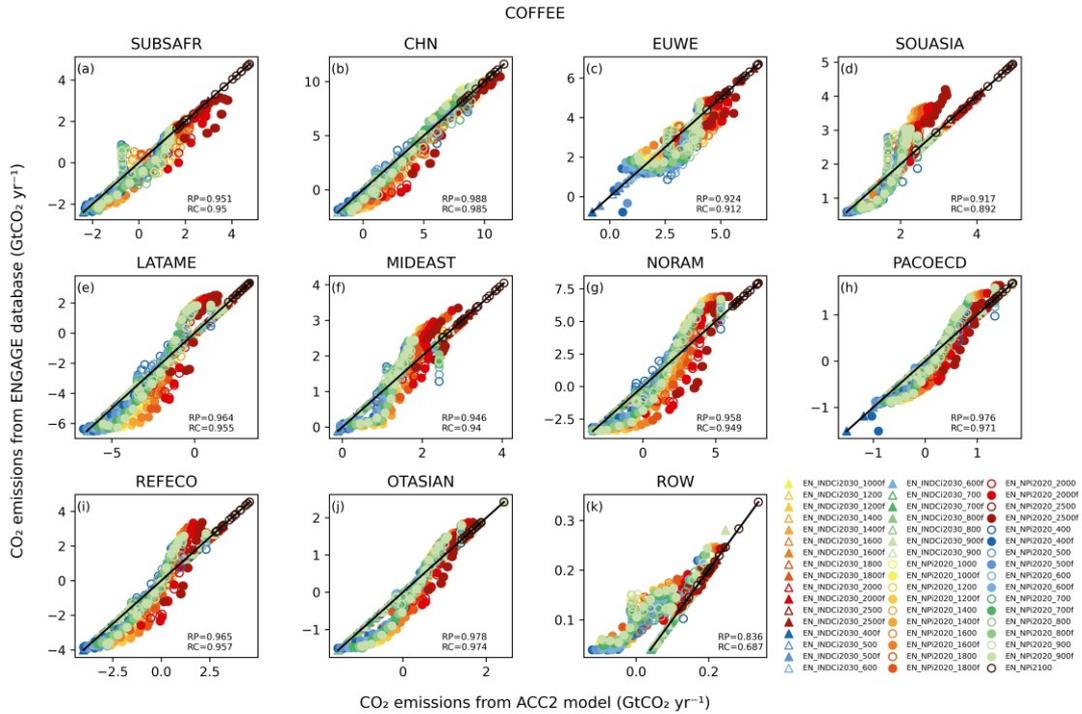

**Figure S190. Test 3 - Regional COFFEE - Reproducibility of total anthropogenic CO₂**

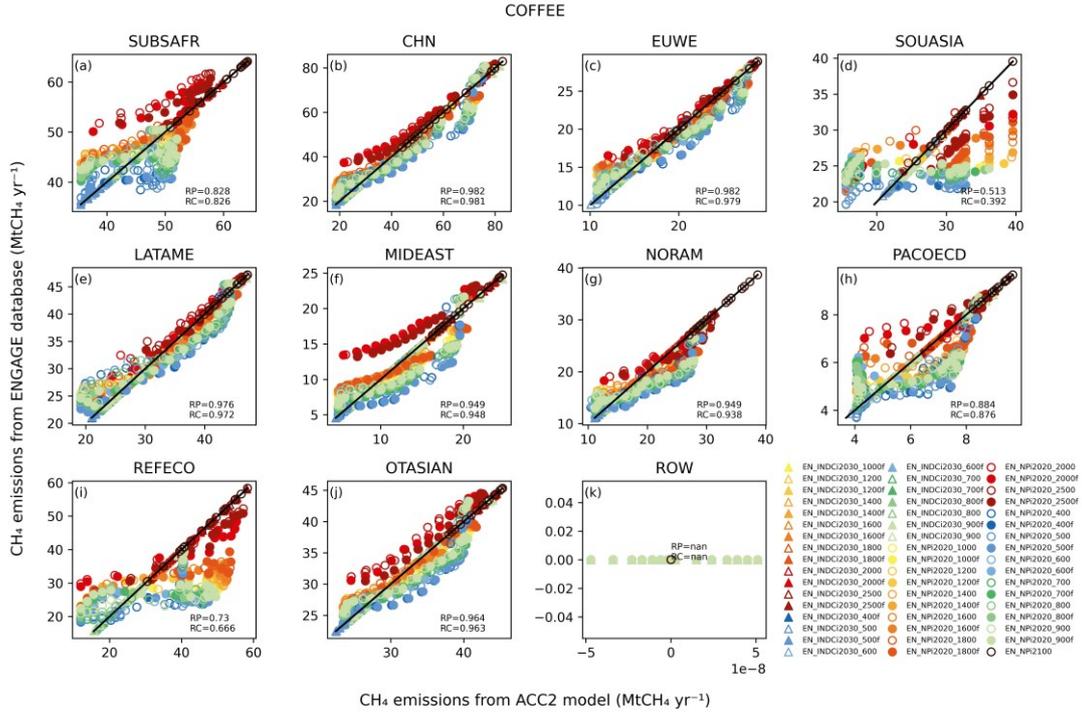

**Figure S191. Test 3 - Regional COFFEE - Reproducibility of total anthropogenic CH₄**

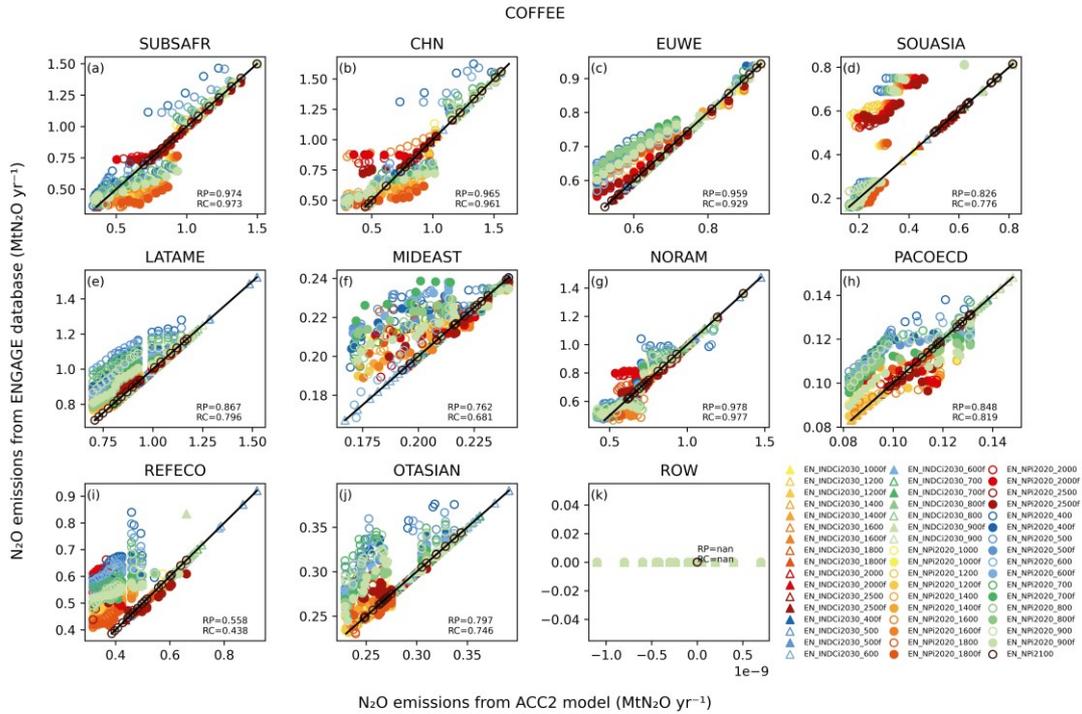

**Figure S192. Test 3 - Regional COFFEE - Reproducibility of total anthropogenic N₂O**

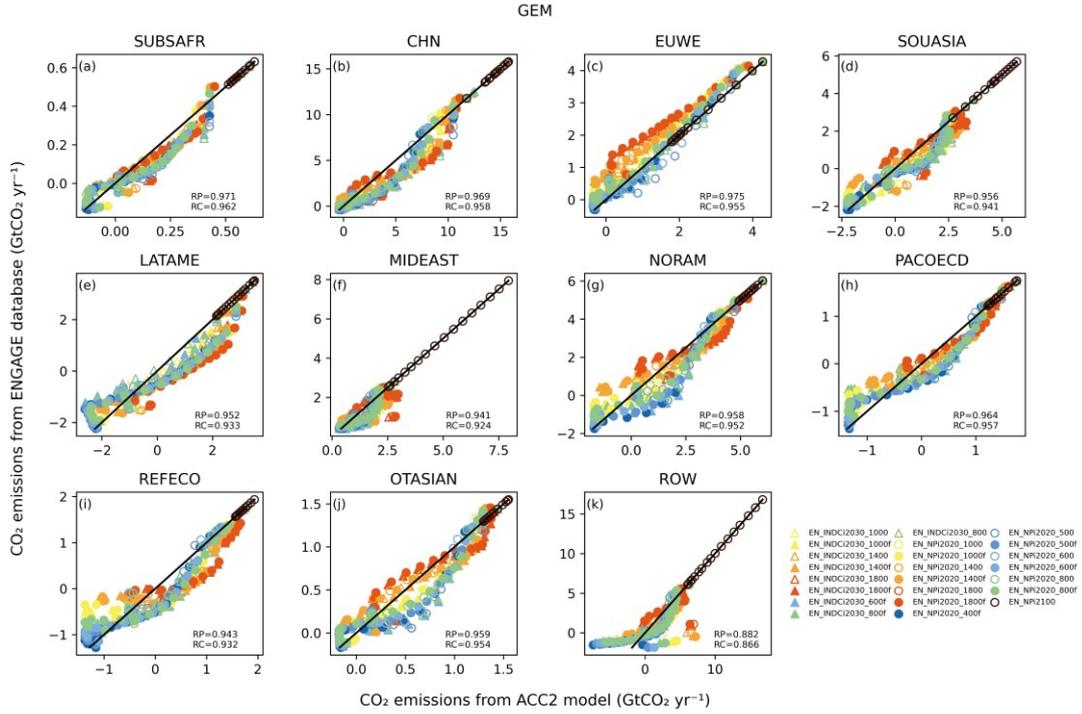

**Figure S193. Test 3 - Regional GEM - Reproducibility of total anthropogenic CO₂**

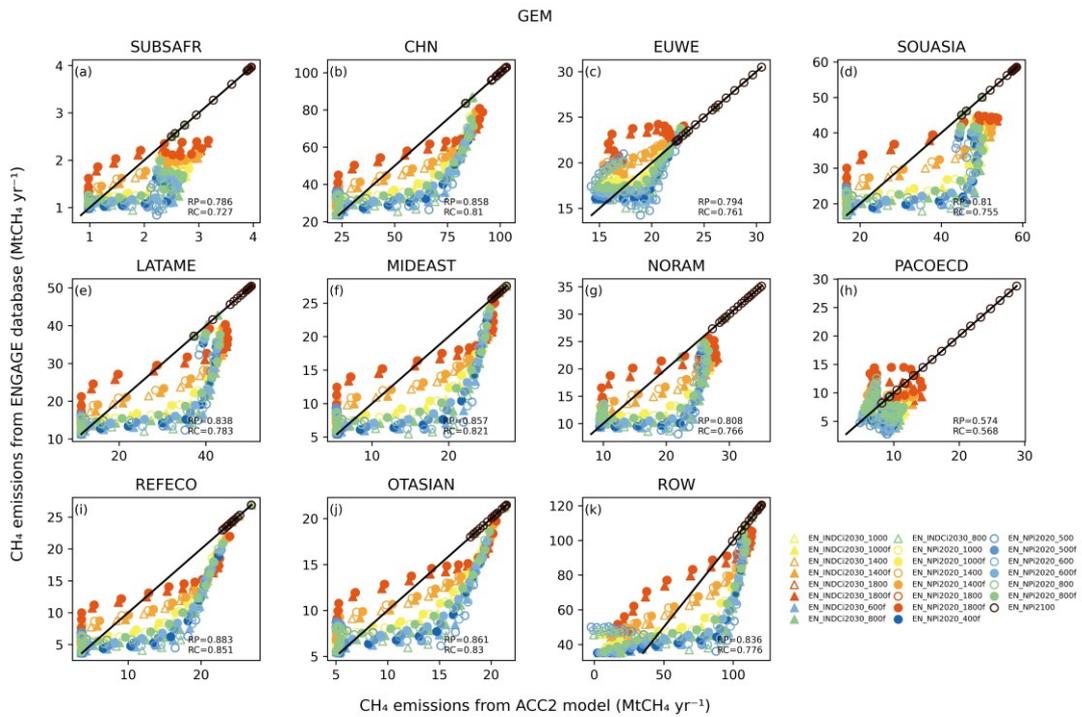

**Figure S194. Test 3 - Regional GEM - Reproducibility of total anthropogenic CH₄**

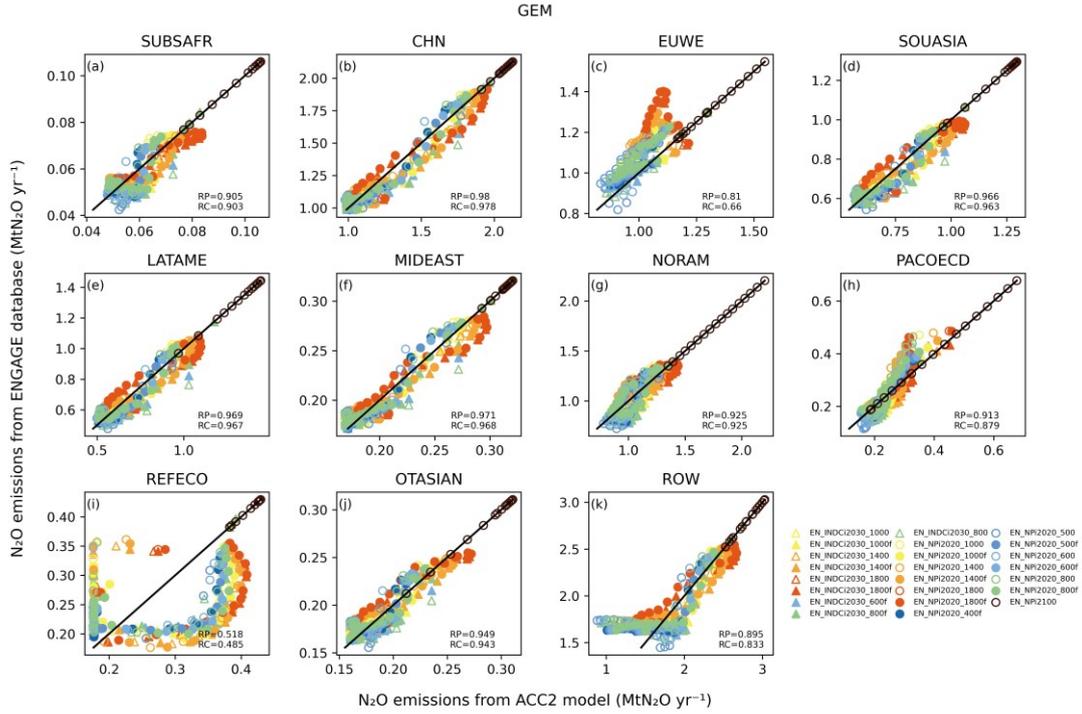

**Figure S195. Test 3 - Regional GEM - Reproducibility of total anthropogenic N₂O**

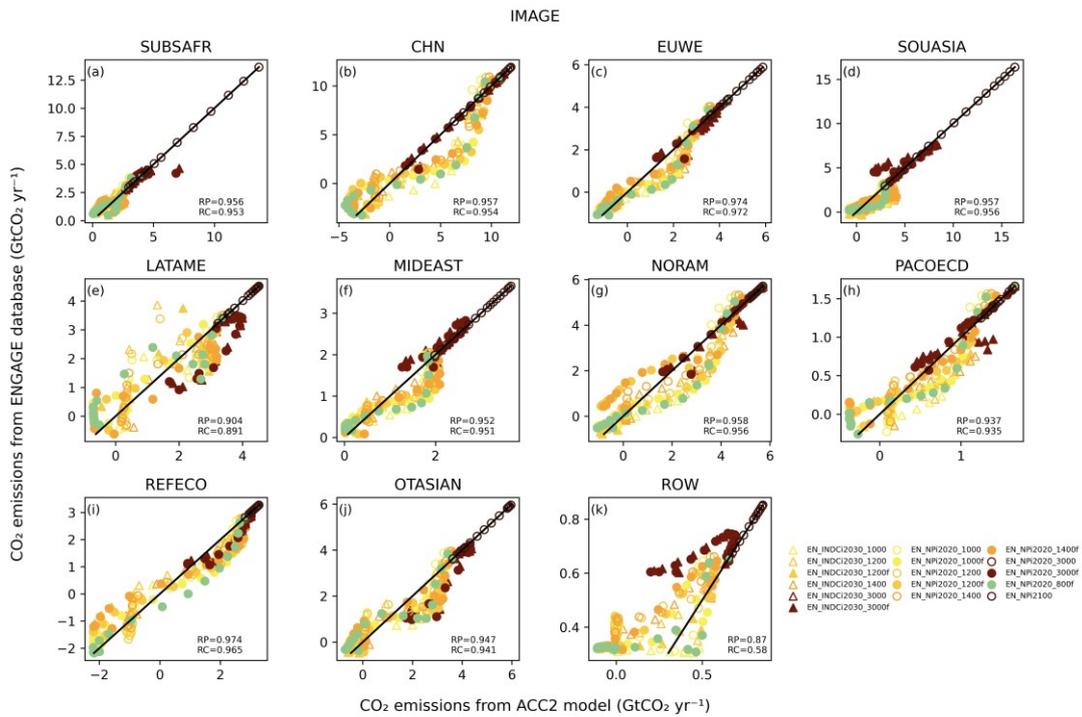

**Figure S196. Test 3 - Regional IMAGE - Reproducibility of total anthropogenic CO₂**

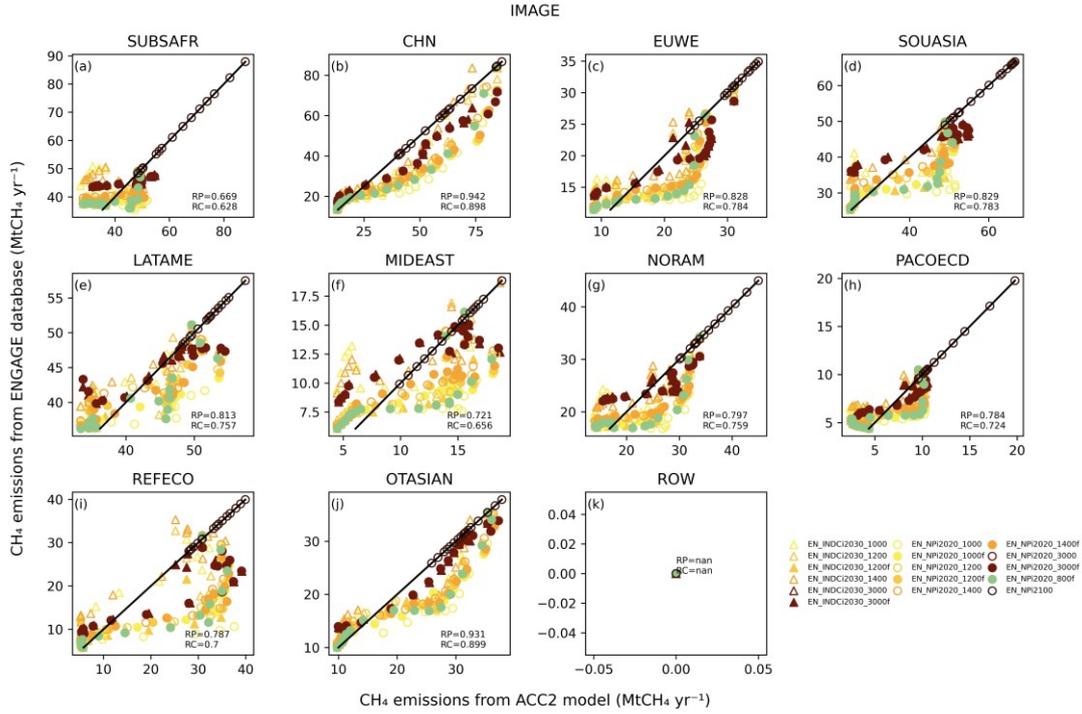

**Figure S197. Test 3 - Regional IMAGE - Reproducibility of total anthropogenic CH₄**

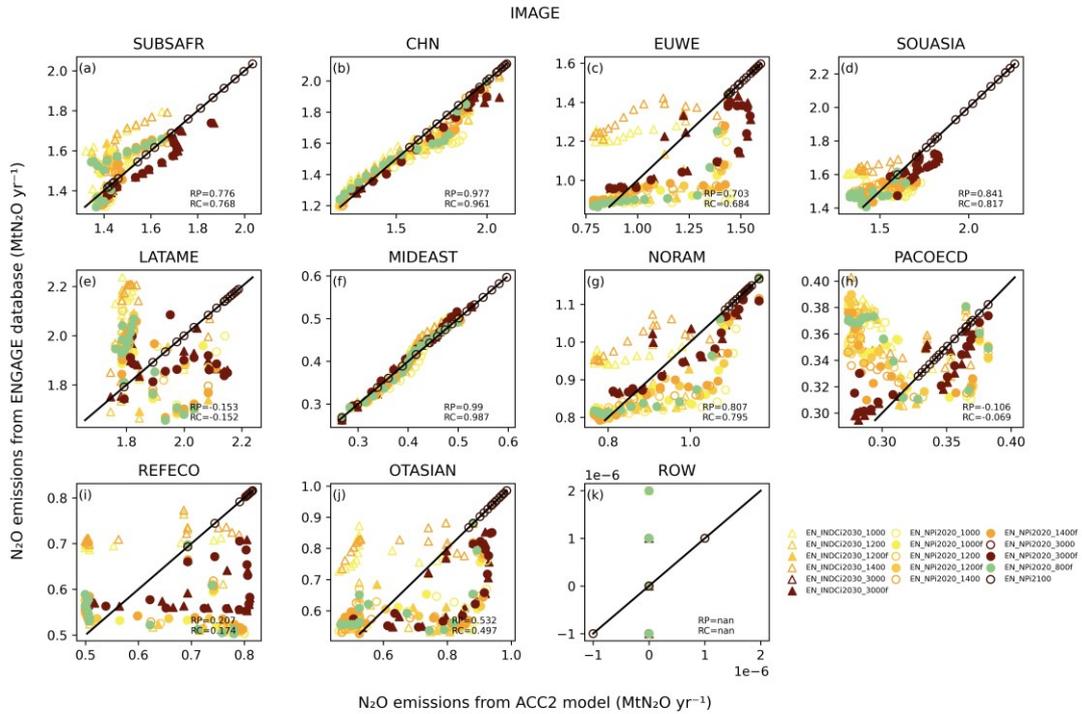

**Figure S198. Test 3 - Regional IMAGE - Reproducibility of total anthropogenic N₂O**

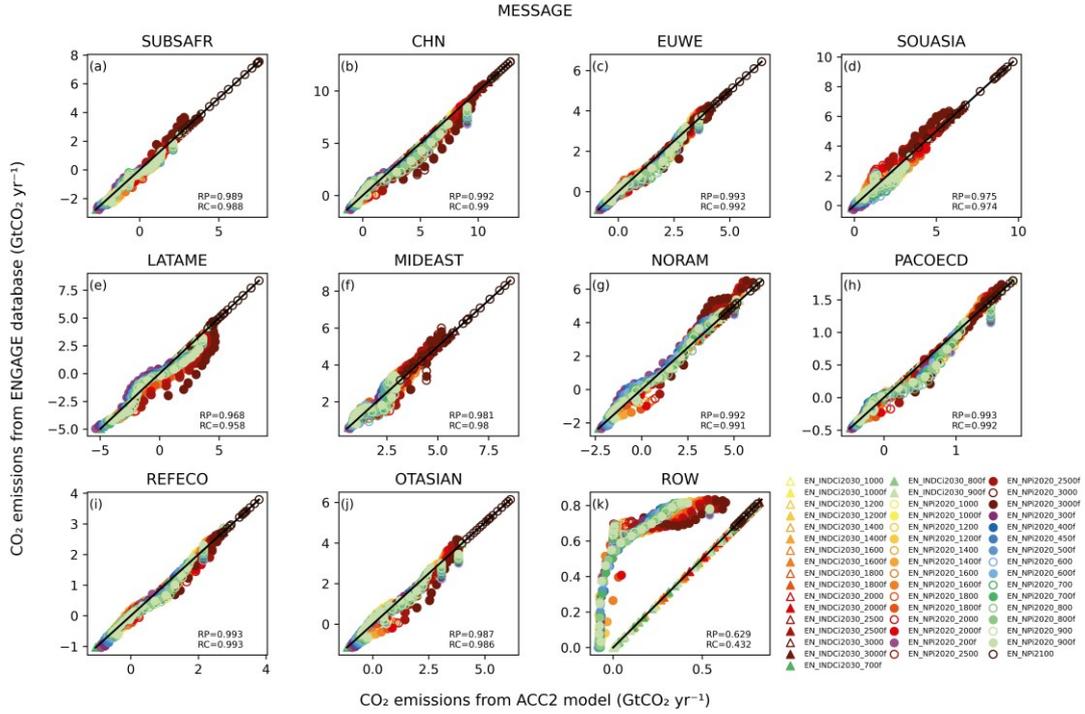

**Figure S199. Test 3 - Regional MESSAGE - Reproducibility of total anthropogenic CO₂**

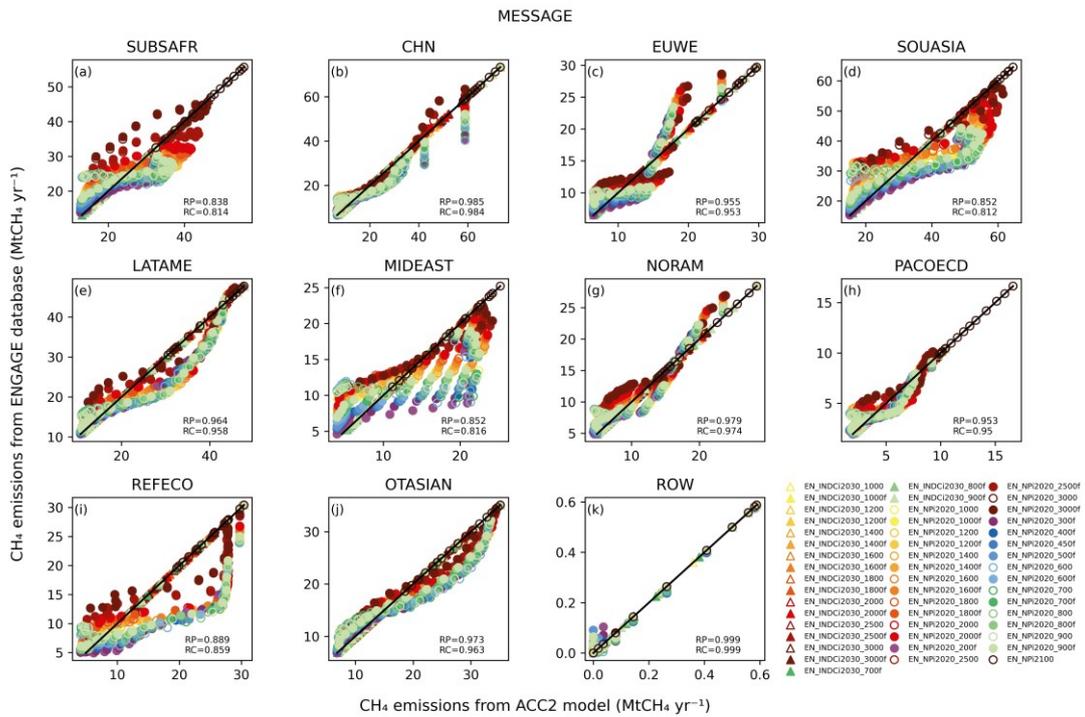

**Figure S200. Test 3 - Regional MESSAGE - Reproducibility of total anthropogenic CH₄**

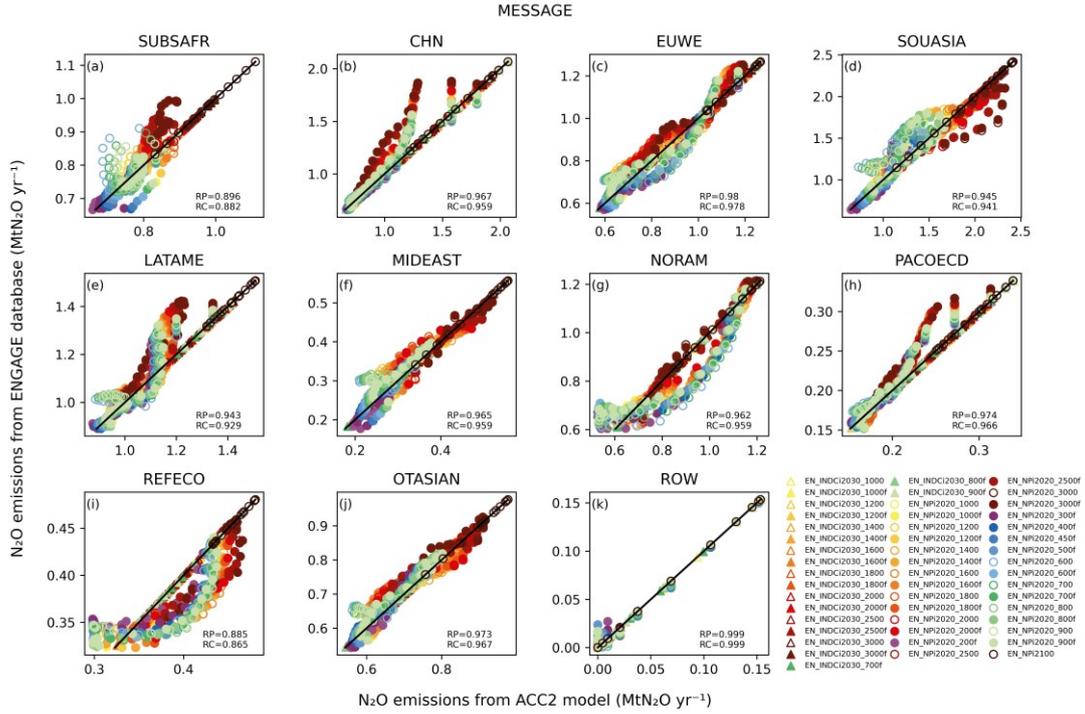

**Figure S201. Test 3 - Regional MESSAGE total anthropogenic N₂O validation result**

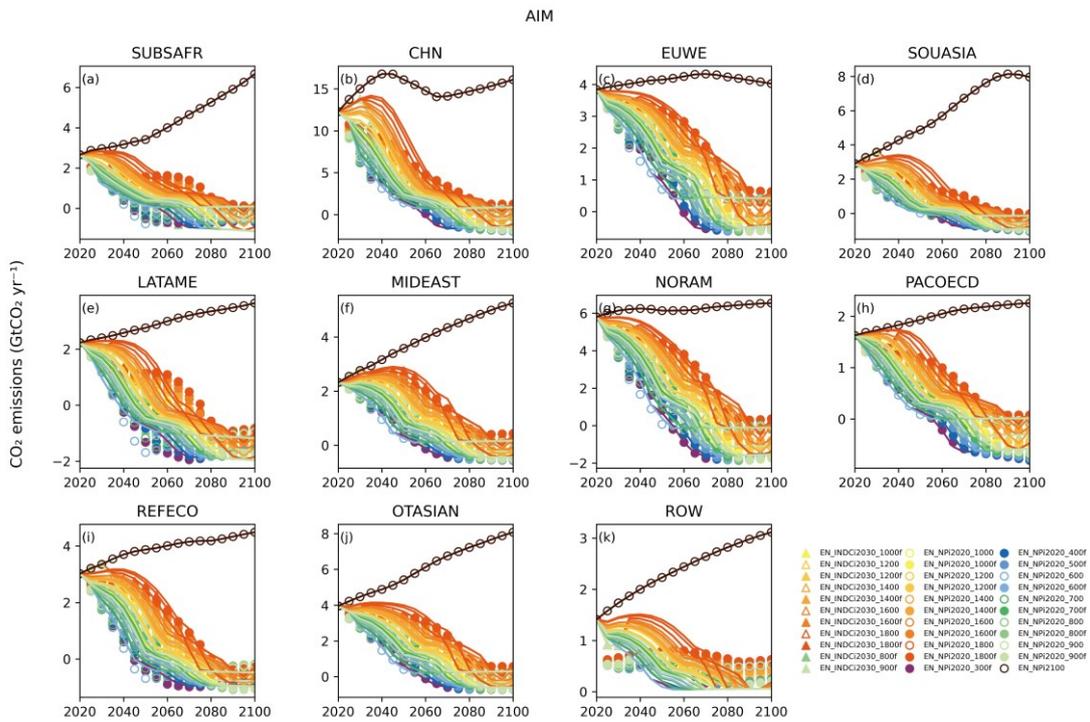

**Figure S202. Test 4 - Regional AIM total anthropogenic CO₂ validation result**

**Figure S203. Test 4 - Regional AIM total anthropogenic CH₄ validation result**

**Figure S204. Test 4 - Regional AIM total anthropogenic N₂O validation result**

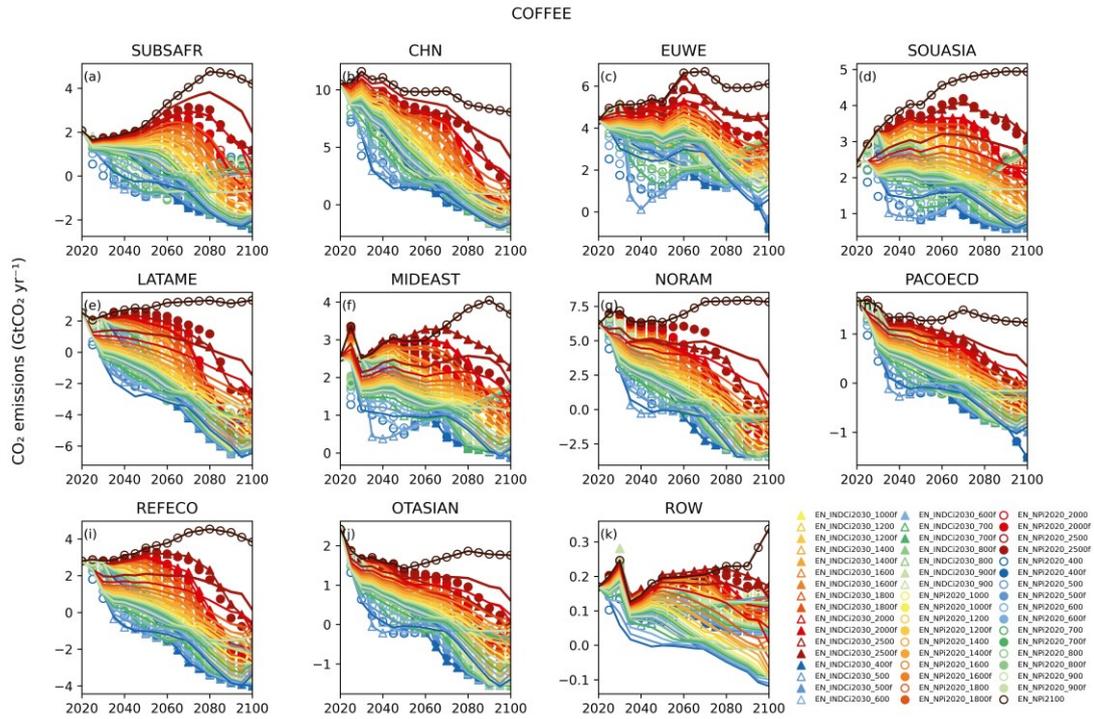

**Figure S205. Test 4 - Regional COFFEE total anthropogenic CO₂ validation result**

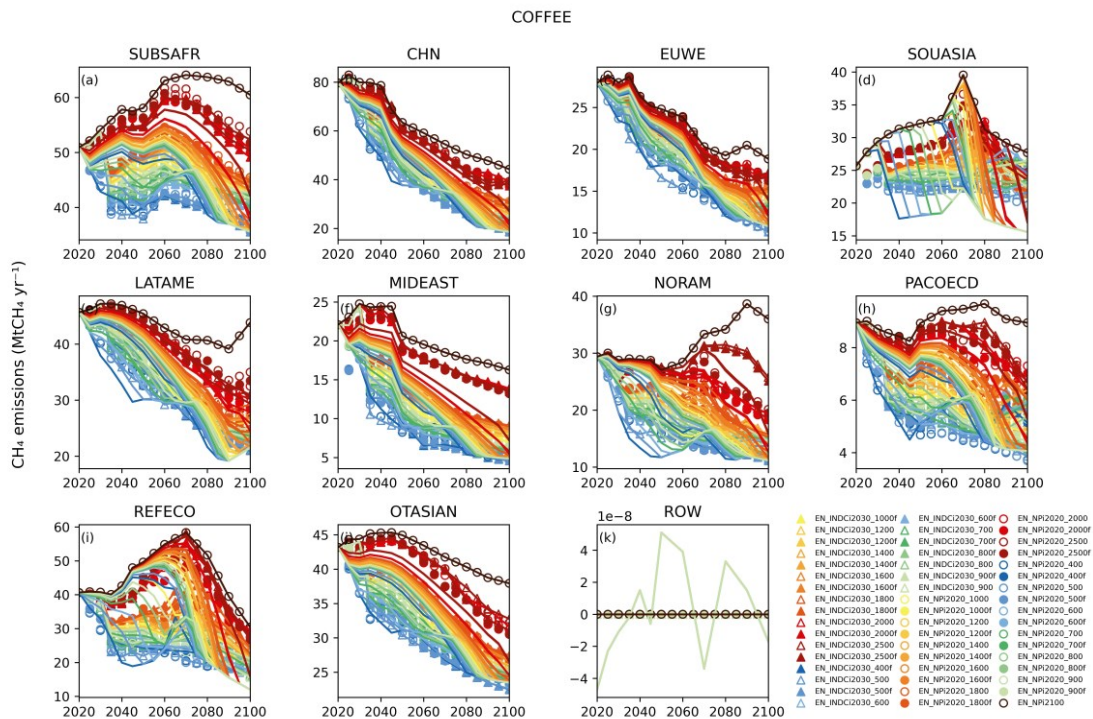

**Figure S206. Test 4 - Regional COFFEE total anthropogenic CH₄ validation result**

**Figure S207. Test 4 - Regional COFFEE total anthropogenic N₂O validation result**

**Figure S208. Test 4 - Regional GEM total anthropogenic CO₂ validation result**

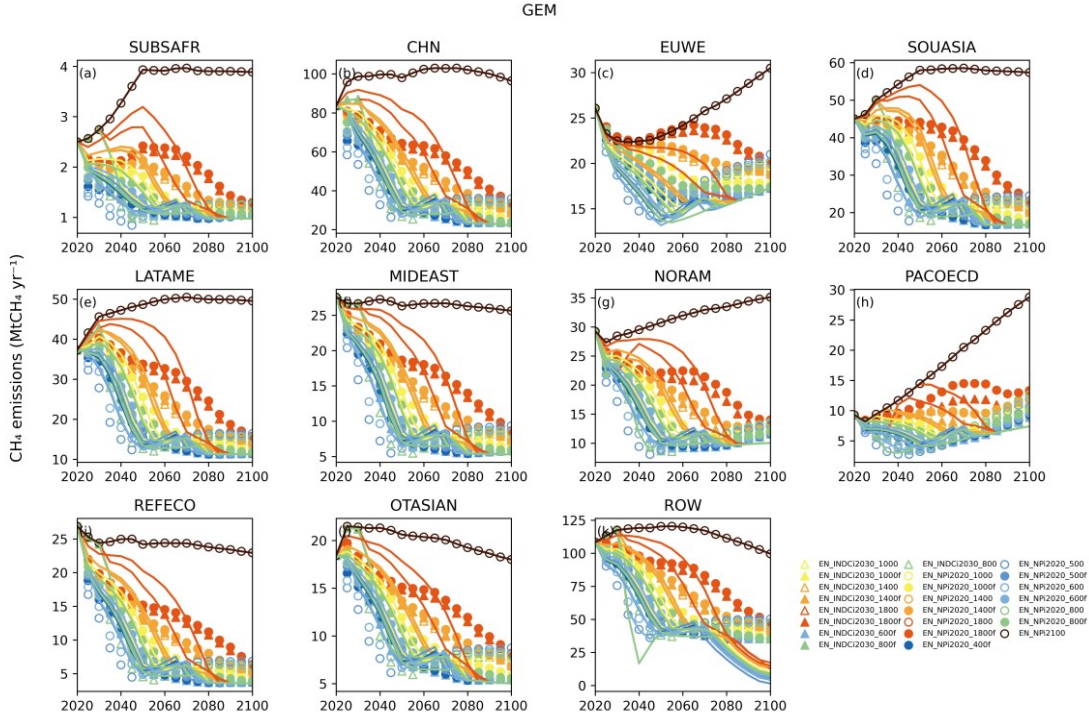

**Figure S209. Test 4 - Regional GEM total anthropogenic CH₄ validation result**

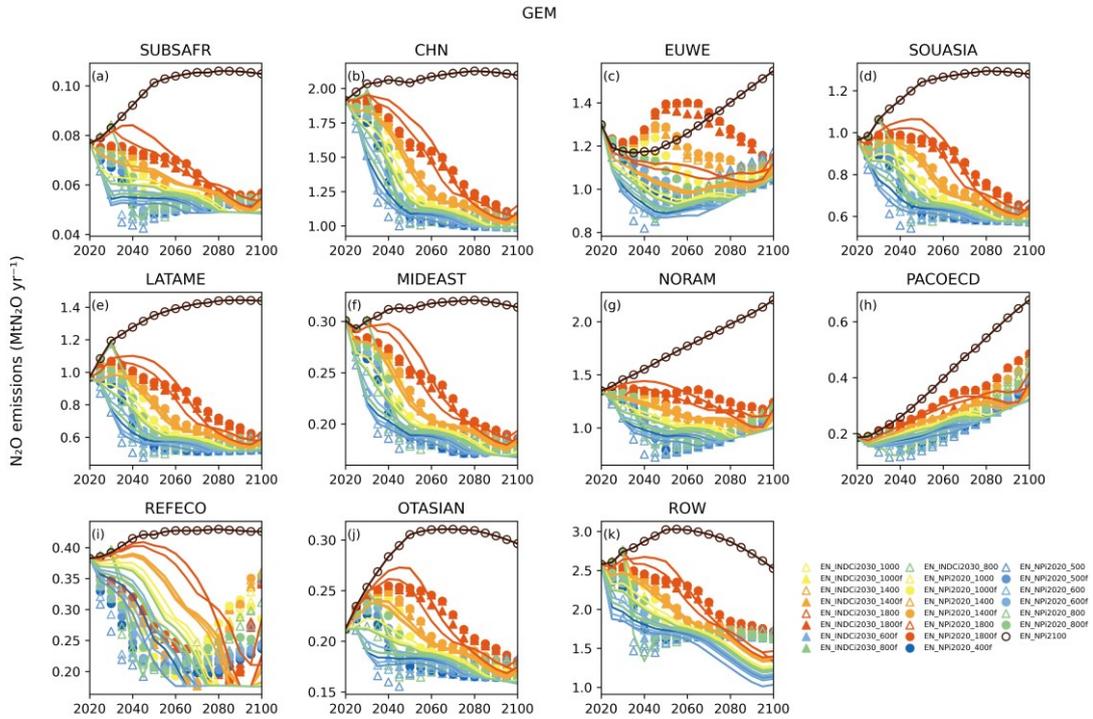

**Figure S210. Test 4 - Regional GEM total anthropogenic N₂O validation result**

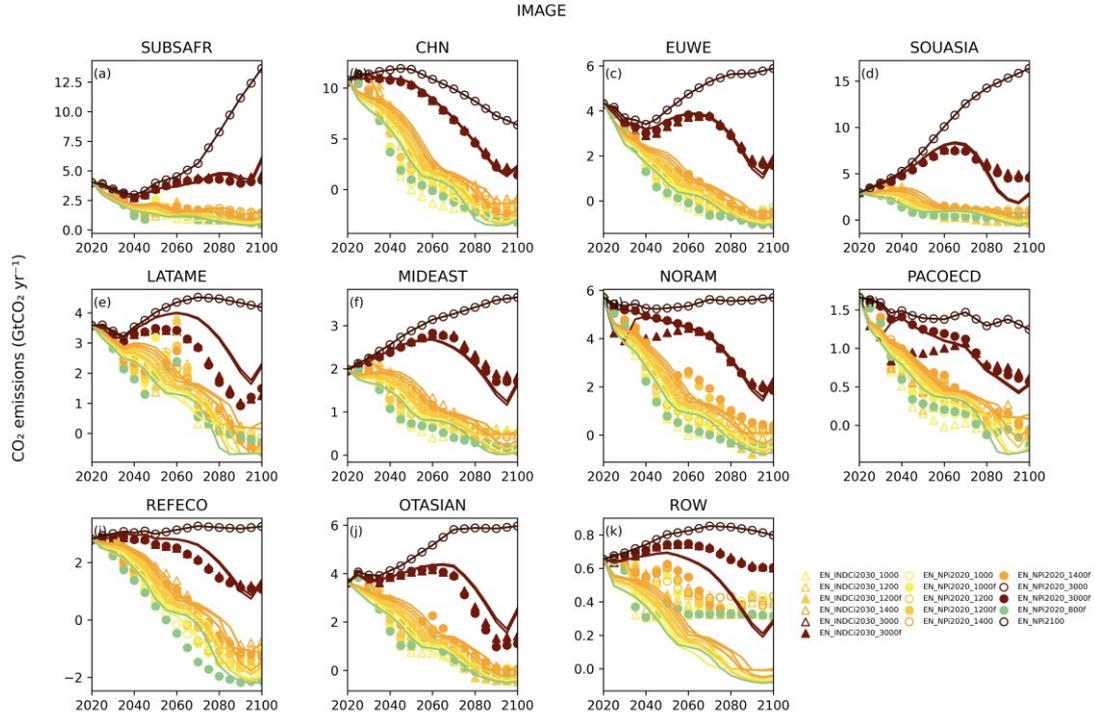

**Figure S211. Test 4 - Regional IMAGE total anthropogenic CO₂ validation result**

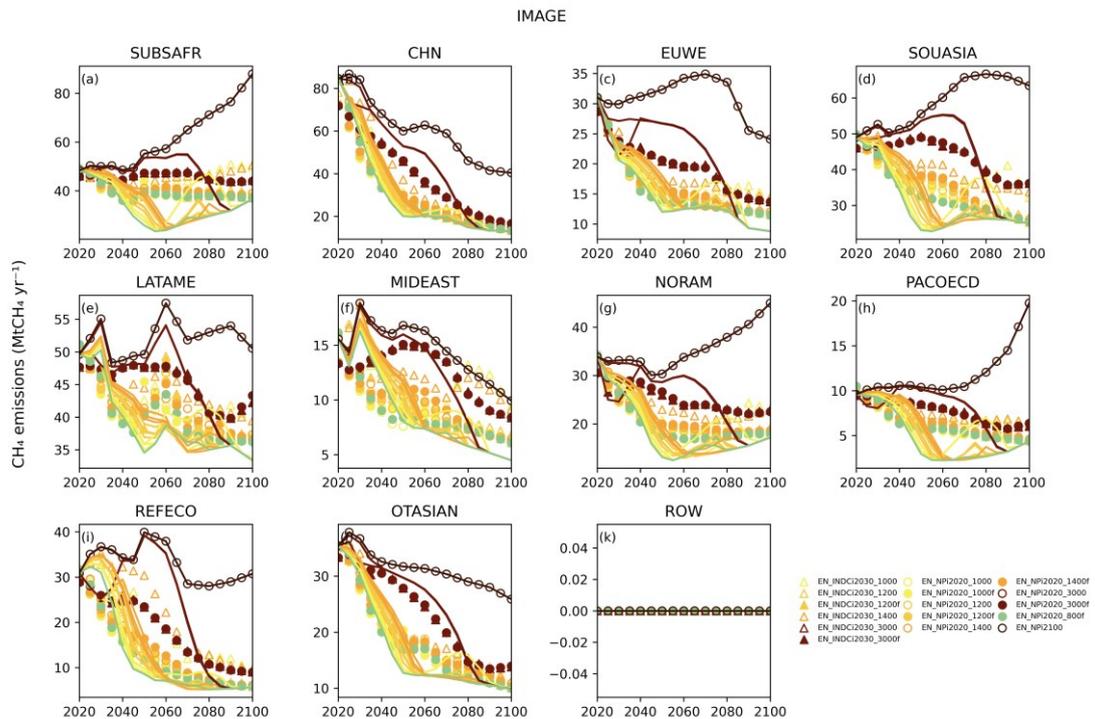

**Figure S212. Test 4 - Regional IMAGE total anthropogenic CH₄ validation result**

**Figure S213. Test 4 - Regional IMAGE total anthropogenic N₂O validation result**

**Figure S214. Test 4 - Regional MESSAGE total anthropogenic CO₂ validation result**

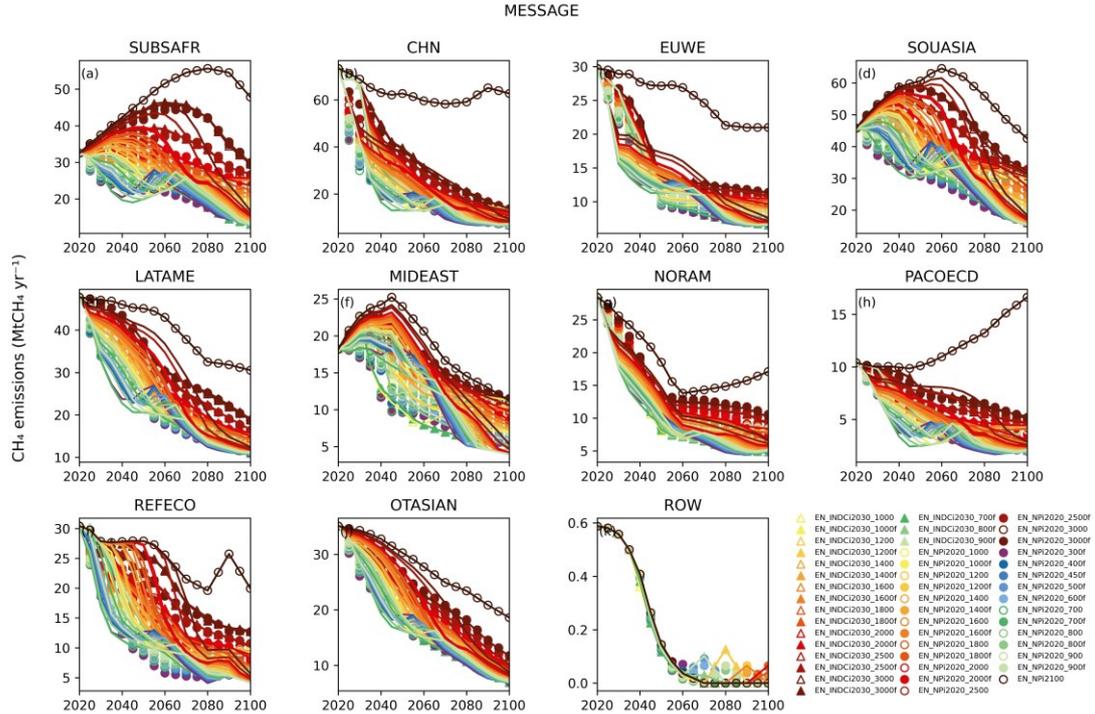

**Figure S215. Test 4 - Regional MESSAGE total anthropogenic CH₄ validation result**

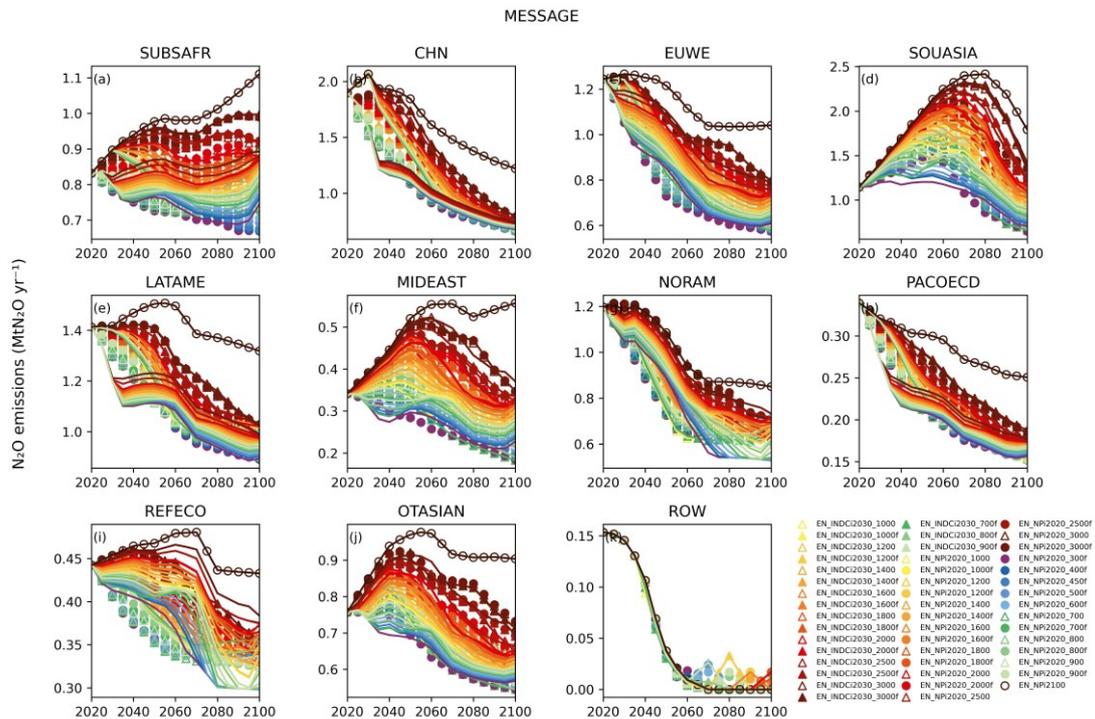

**Figure S216. Test 4 - Regional MESSAGE total anthropogenic N₂O validation result**

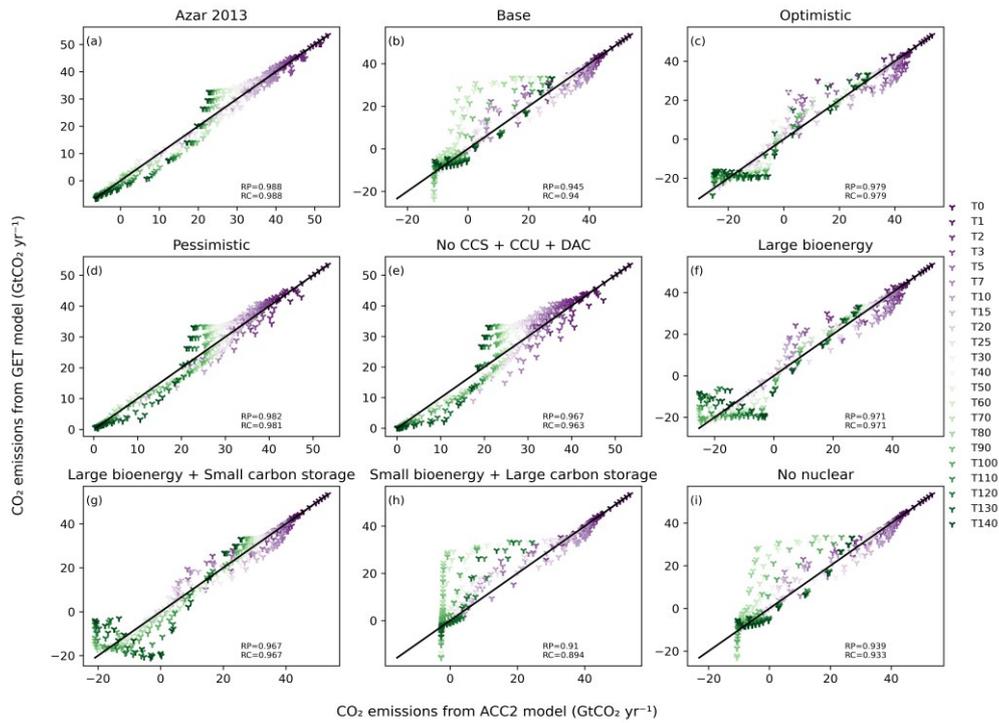

**Figure S217. Test 4 - GET Reproducibility of energy-related CO₂**

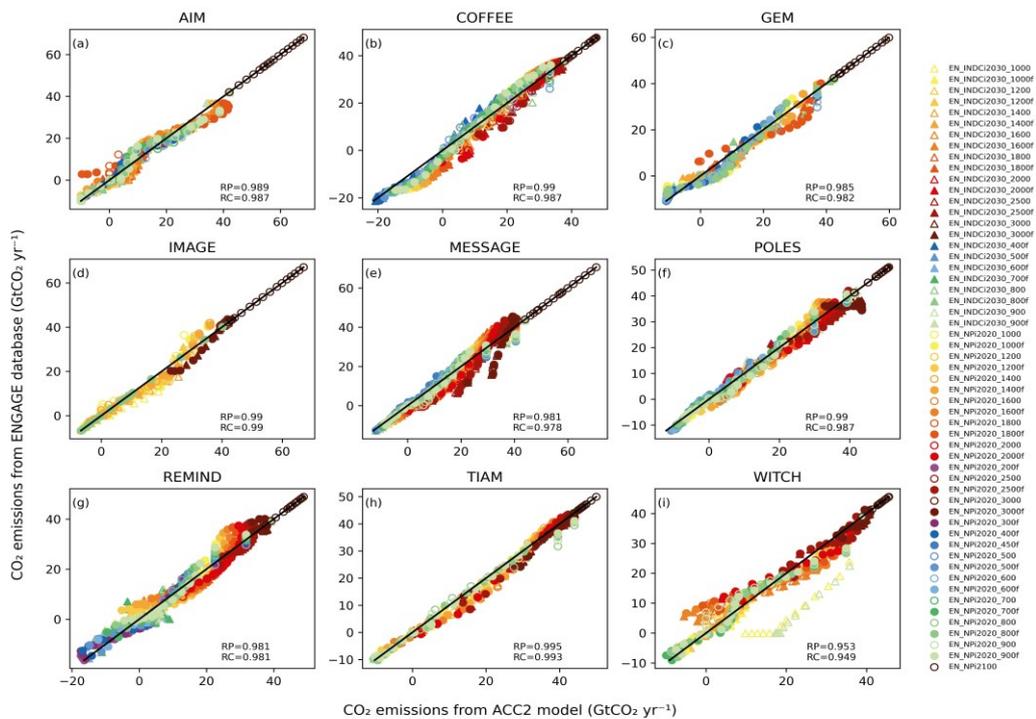

**Figure S218. Test 4 - Global 9 models - Reproducibility of total anthropogenic CO₂**

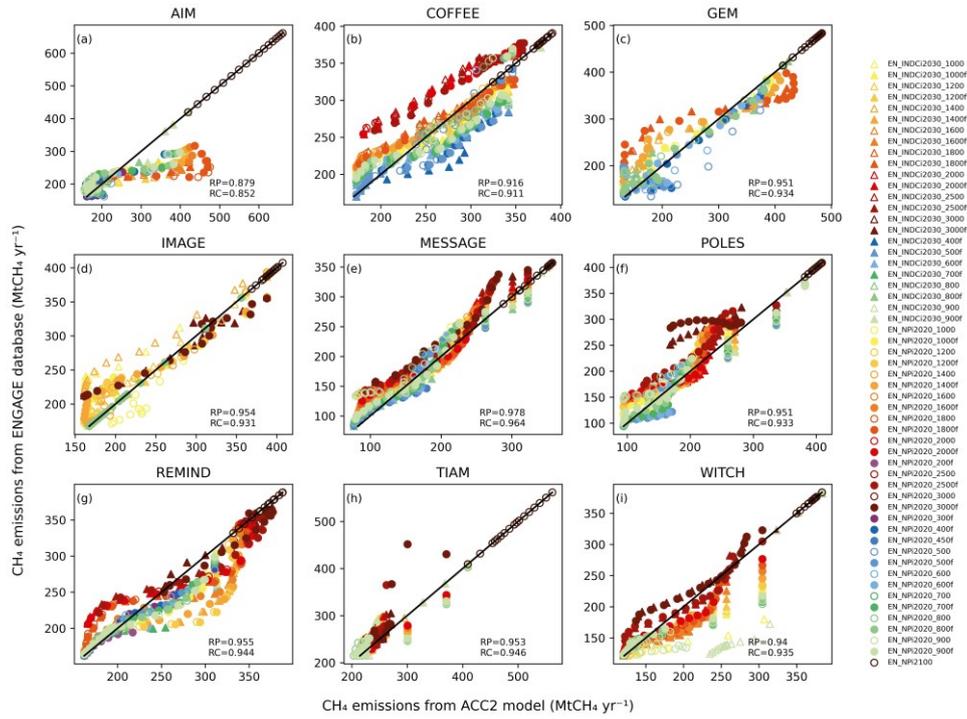

**Figure S219. Test 4 - Global 9 models - Reproducibility of total anthropogenic CH₄**

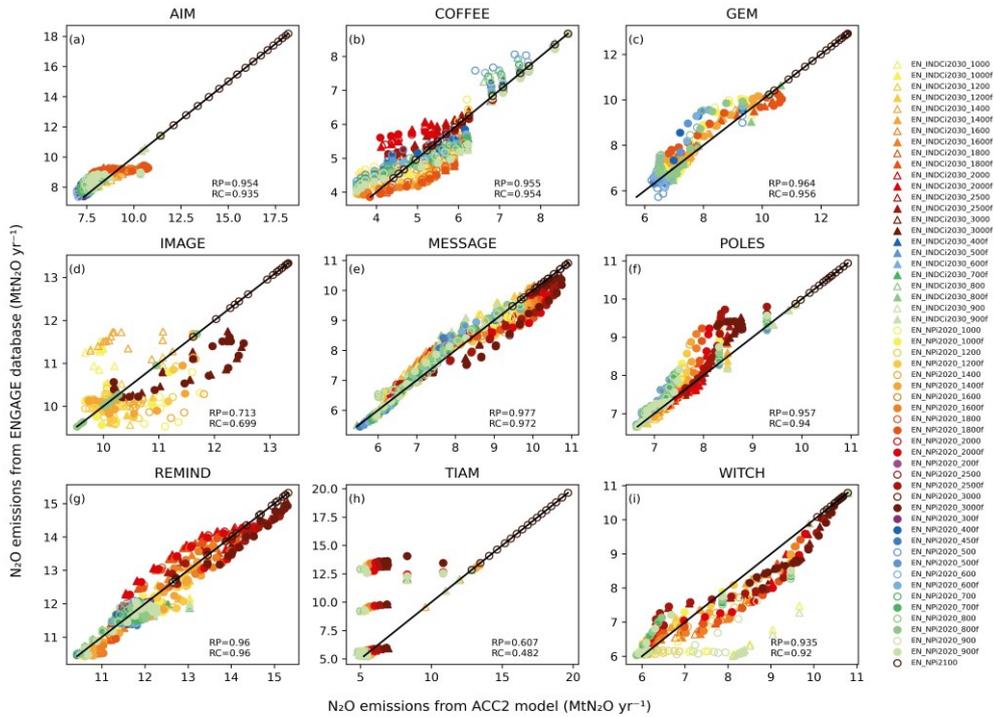

**Figure S220. Test 4 - Global 9 models - Reproducibility of total anthropogenic N₂O**

**Figure S221. Test 4 - Regional AIM - Reproducibility of total anthropogenic CO$_2$**

**Figure S222. Test 4 - Regional AIM - Reproducibility of total anthropogenic CH$_4$**

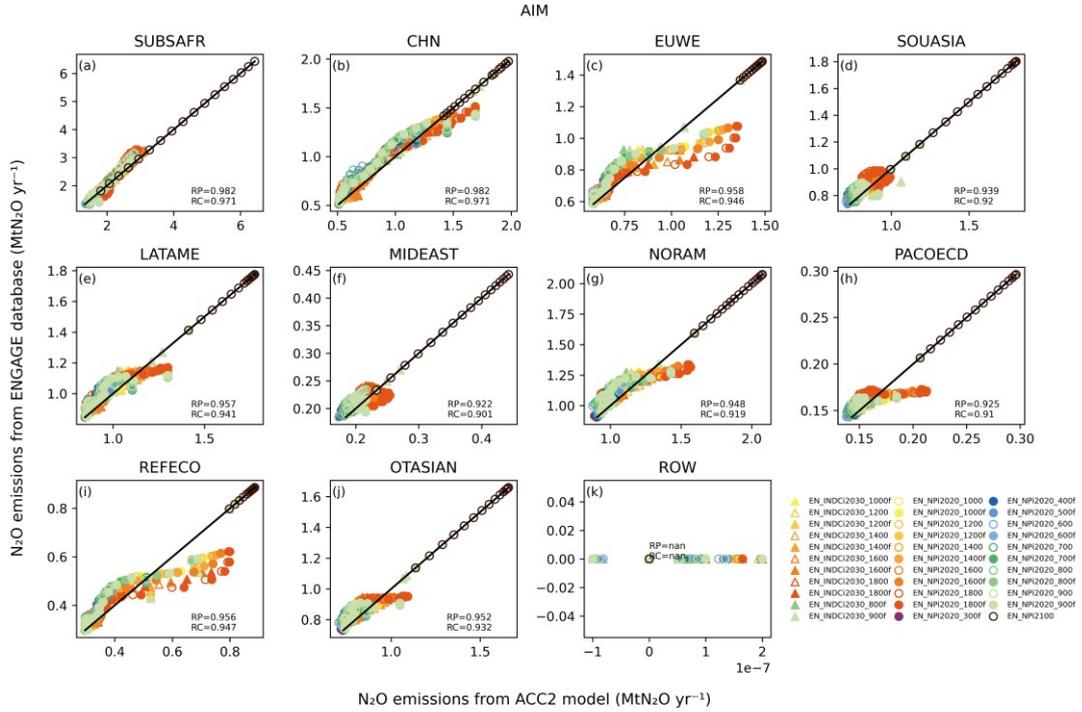

**Figure S223. Test 4 - Regional AIM - Reproducibility of total anthropogenic N₂O**

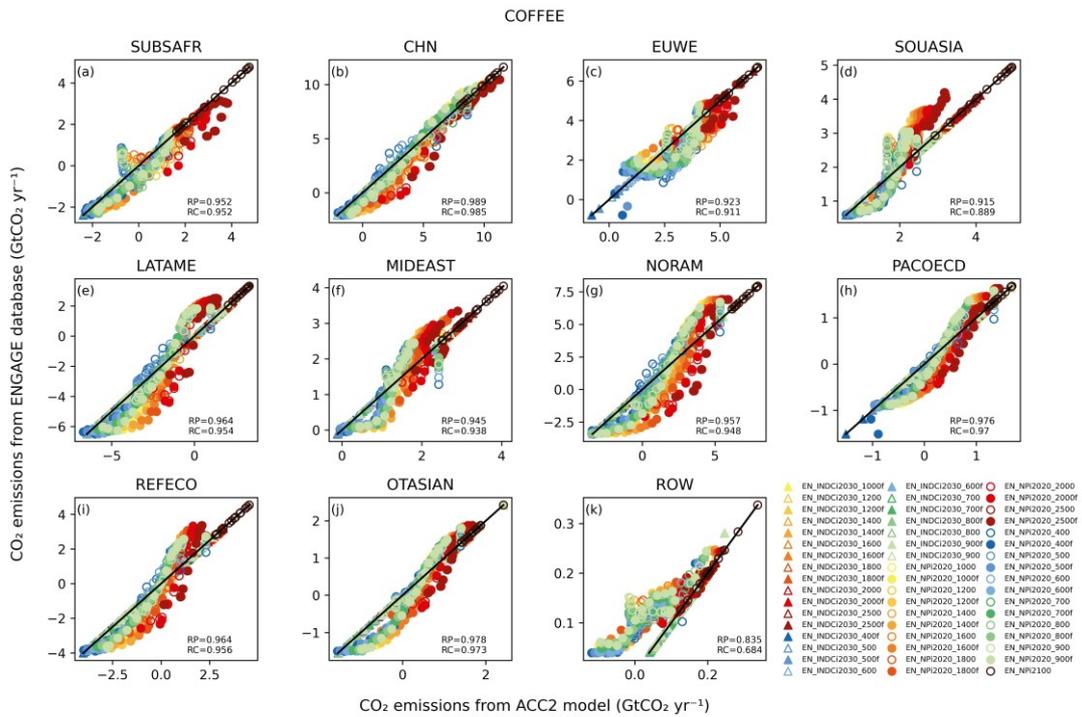

**Figure S224. Test 4 - Regional COFFEE - Reproducibility of total anthropogenic CO₂**

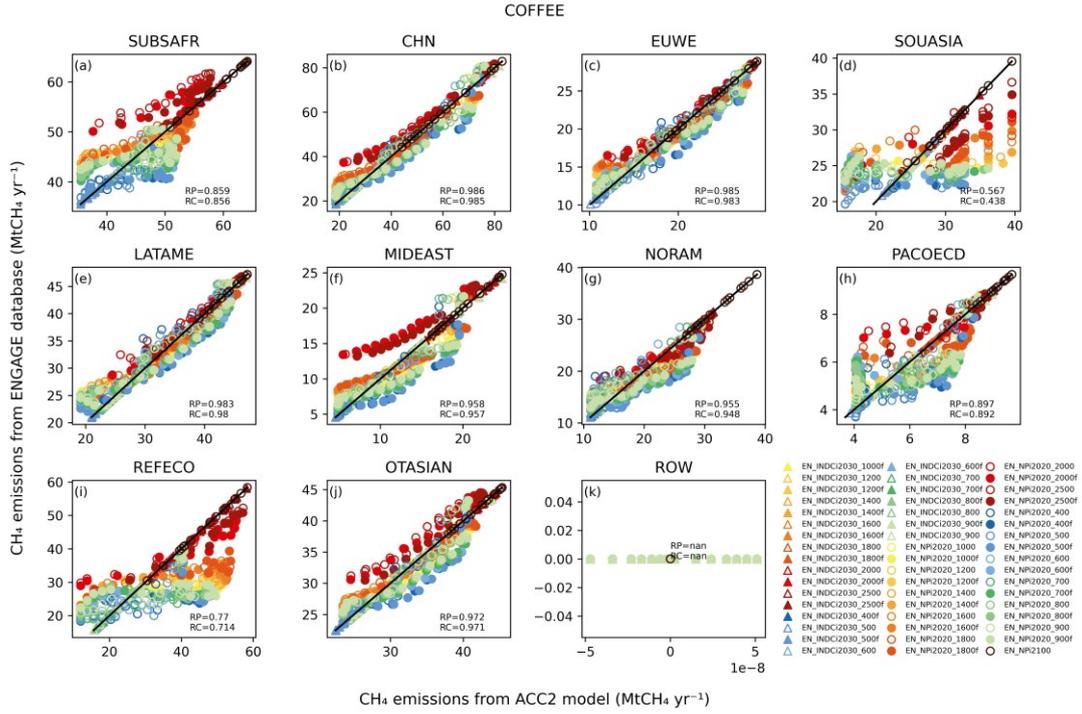

**Figure S225. Test 4 - Regional COFFEE - Reproducibility of total anthropogenic CH₄**

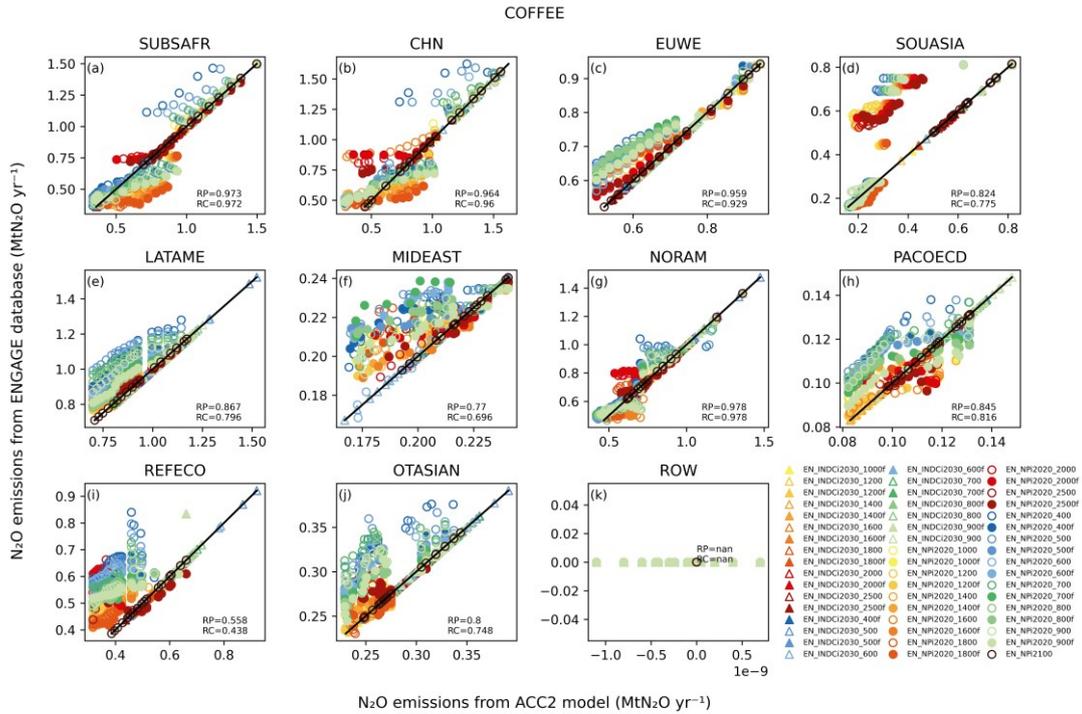

**Figure S226. Test 4 - Regional COFFEE - Reproducibility of total anthropogenic N₂O**

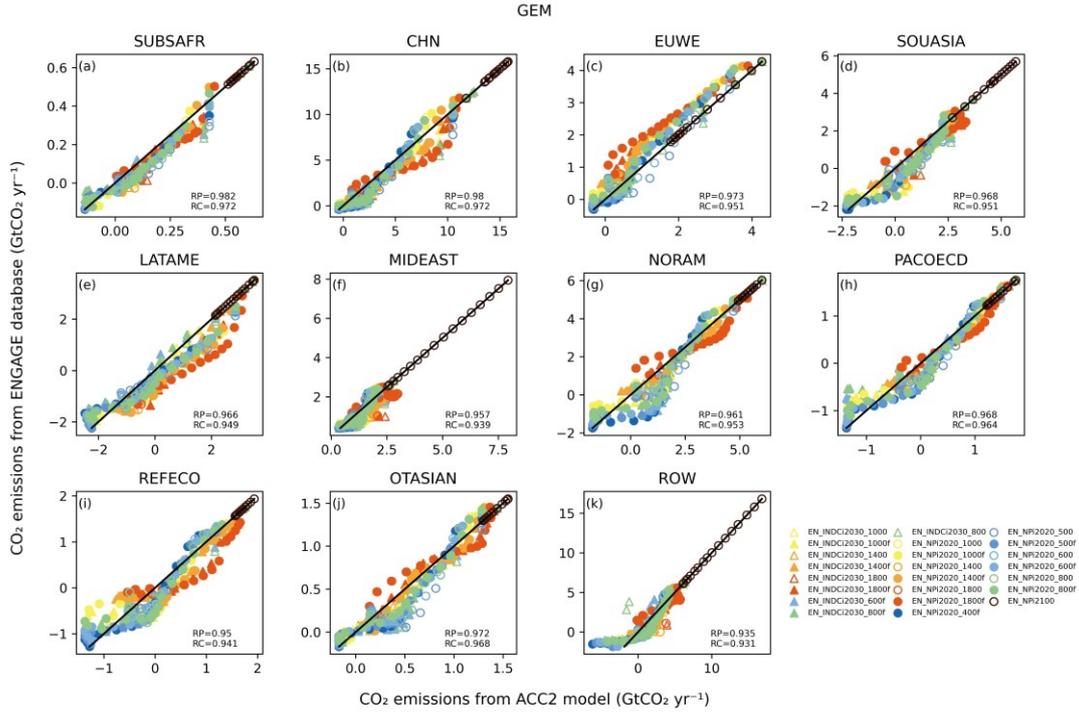

**Figure S227. Test 4 - Regional GEM - Reproducibility of total anthropogenic CO₂**

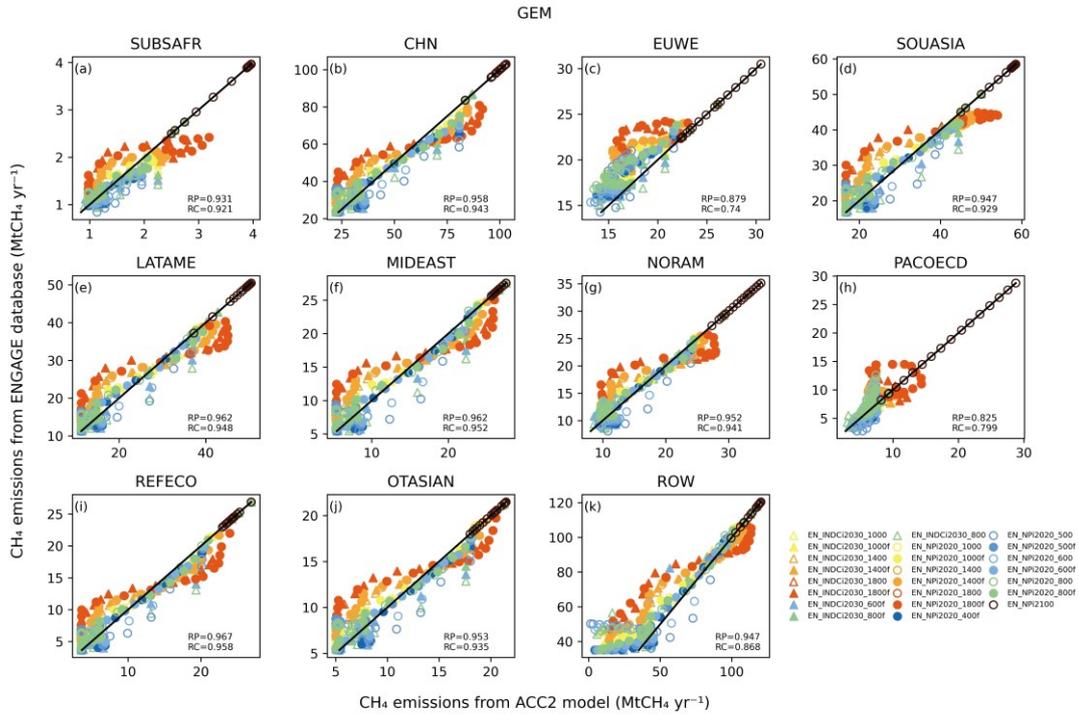

**Figure S228. Test 4 - Regional GEM - Reproducibility of total anthropogenic CH₄**

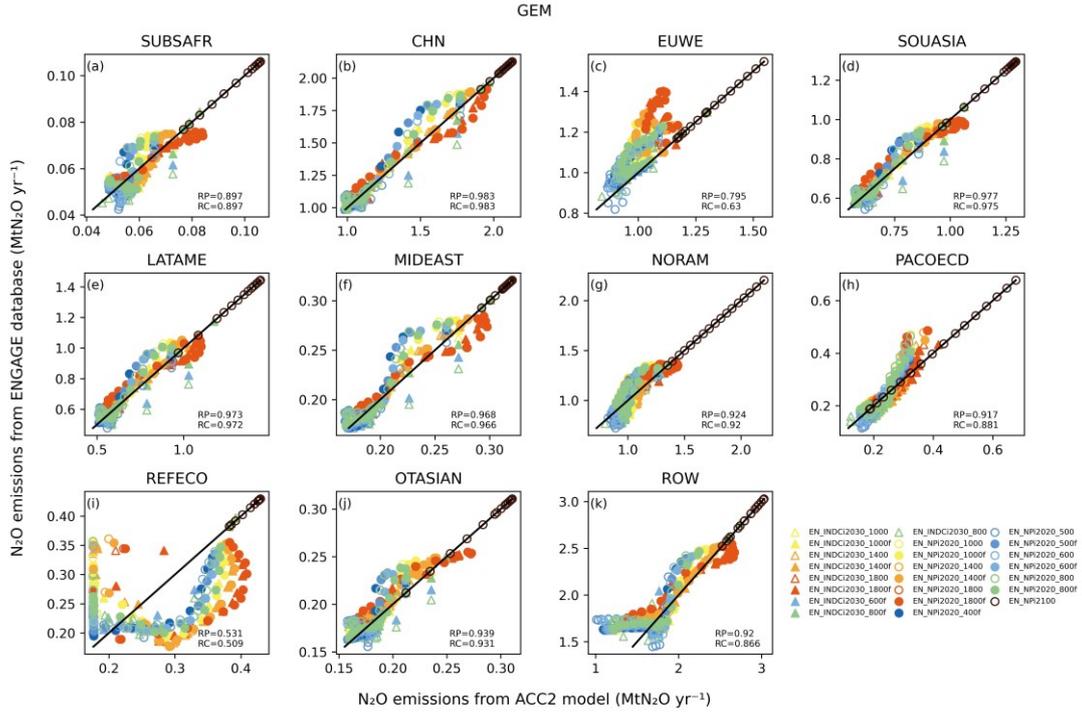

**Figure S229. Test 4 - Regional GEM - Reproducibility of total anthropogenic N₂O**

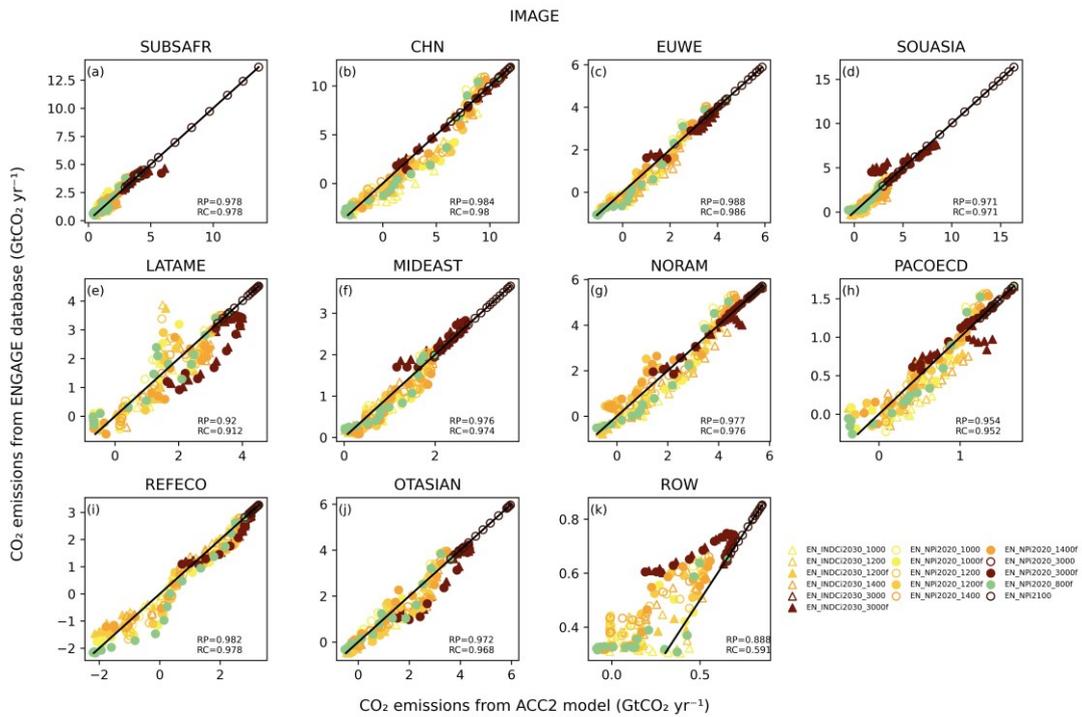

**Figure S230. Test 4 - Regional IMAGE - Reproducibility of total anthropogenic CO₂**

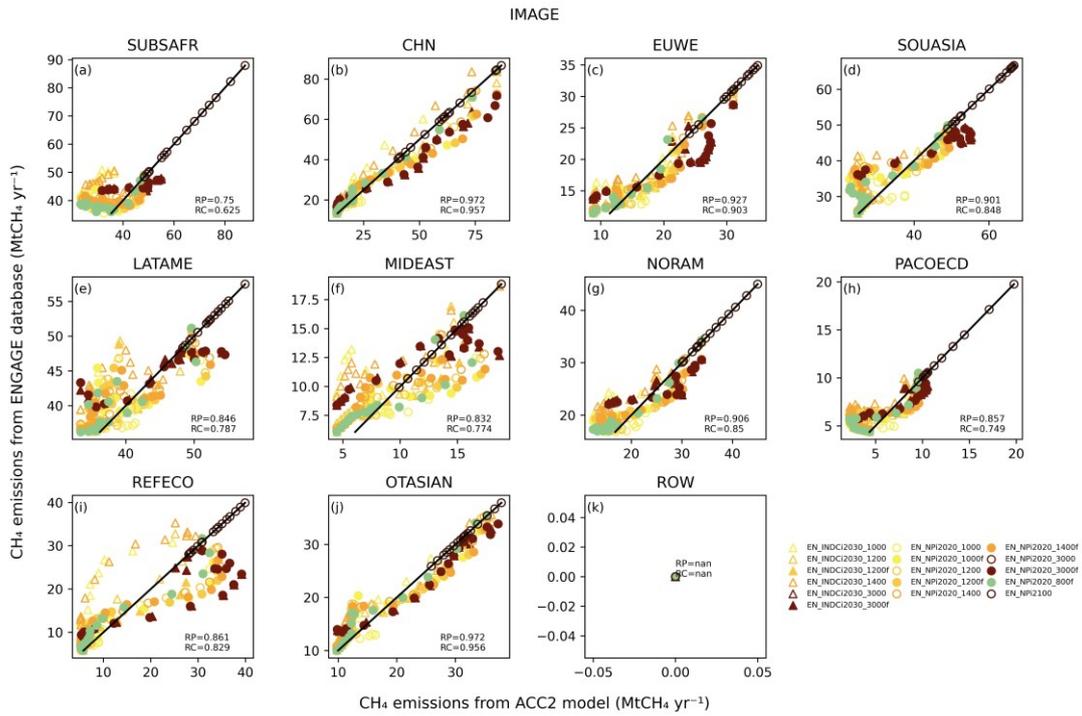

**Figure S231. Test 4 - Regional IMAGE - Reproducibility of total anthropogenic CH₄**

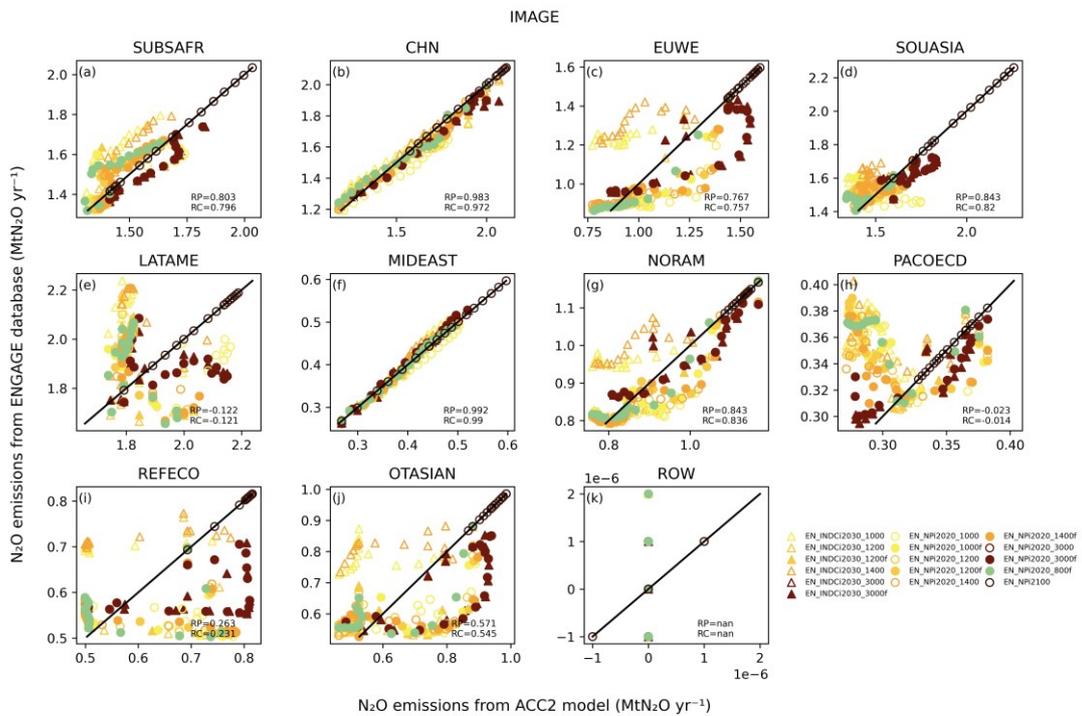

**Figure S232. Test 4 - Regional IMAGE - Reproducibility of total anthropogenic N₂O**

**Figure S233. Test 4 - Regional MESSAGE - Reproducibility of total anthropogenic CO$_2$**

**Figure S234. Test 4 - Regional MESSAGE - Reproducibility of total anthropogenic CH$_4$**

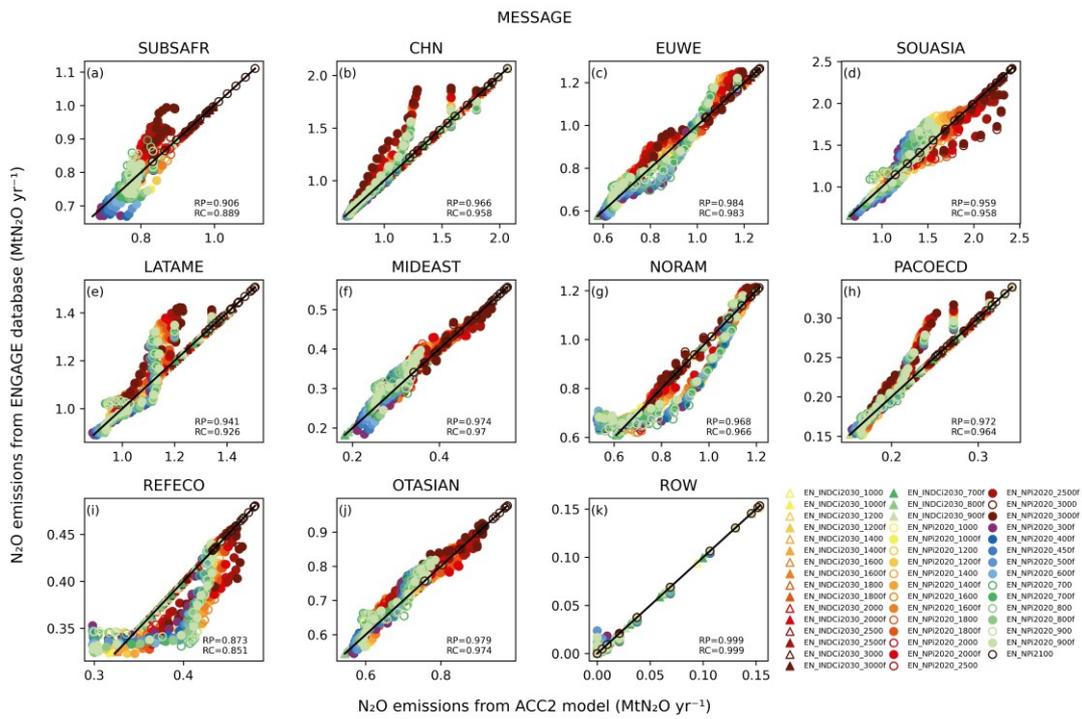

**Figure S235. Test 4 - Regional MESSAGE - Reproducibility of total anthropogenic N₂O**